%% file: thesis.tex
\documentclass[11pt, a4paper, twoside]{Thesis} 

\usepackage{graphicx}
\usepackage{amssymb,amsfonts,amsmath}
\usepackage{epsf}
\usepackage{subfigure}
\usepackage{epstopdf}
\usepackage{mathrsfs}

\usepackage[square, numbers, comma, sort&compress]{natbib} 
\hypersetup{urlcolor=black, colorlinks=true} 
\title{\ttitle} 

\newcommand{\be}{\begin{equation}}
\newcommand{\ee}{\end{equation}}
\newcommand{\bea}{\begin{eqnarray}}
\newcommand{\eea}{\end{eqnarray}}

\makeatletter
\renewenvironment{subarray}[2][c]{%
  \if#1c\vcenter\else\vbox\fi\bgroup
  \Let@ \restore@math@cr \default@tag
  \baselineskip\fontdimen10 \scriptfont\tw@
  \advance\baselineskip\fontdimen12 \scriptfont\tw@
  \lineskip\thr@@\fontdimen8 \scriptfont\thr@@
  \lineskiplimit\lineskip
  \ialign\bgroup\ifx c#2\hfil\fi
    $\m@th\scriptstyle##$\hfil\crcr
}{%
  \crcr\egroup\egroup
}
\makeatother

\begin{document}

\frontmatter 

\setstretch{1.2} 

\fancyhead{} 
\rhead{\thepage} 
\lhead{} 

\pagestyle{fancy} 

\newcommand{\HRule}{\rule{\linewidth}{0.5mm}} 

\hypersetup{pdftitle={\ttitle}}
\hypersetup{pdfsubject=\subjectname}
\hypersetup{pdfauthor=\authornames}
\hypersetup{pdfkeywords=\keywordnames}


\begin{titlepage}
\begin{center}

\textsc{\LARGE \univname}\\[1.5cm] 
\textsc{\Large Doctoral Thesis}\\[0.5cm] 

\HRule \\[0.4cm] 
{\huge \bfseries \ttitle}\\[0.4cm] 
\HRule \\[1.5cm] 
 
\begin{minipage}{0.4\textwidth}
\begin{flushleft} \large
\emph{Author:}\\
\href{http://www.johnsmith.com}{\authornames} 
\end{flushleft}
\end{minipage}
\begin{minipage}{0.4\textwidth}
\begin{flushright} \large
\emph{Supervisor:} \\
\href{http://www.jamessmith.com}{\supname} 
\end{flushright}
\end{minipage}\\[2.5cm]

\large \textit{A thesis submitted in partial fulfilment of the requirements\\ for the degree of \degreename}\\[0.3cm] 
\textit{in the}\\[0.4cm]
\groupname\\\deptname\\[1cm] 
 
{\large \today}\\[1cm] 
\includegraphics[width=4cm]{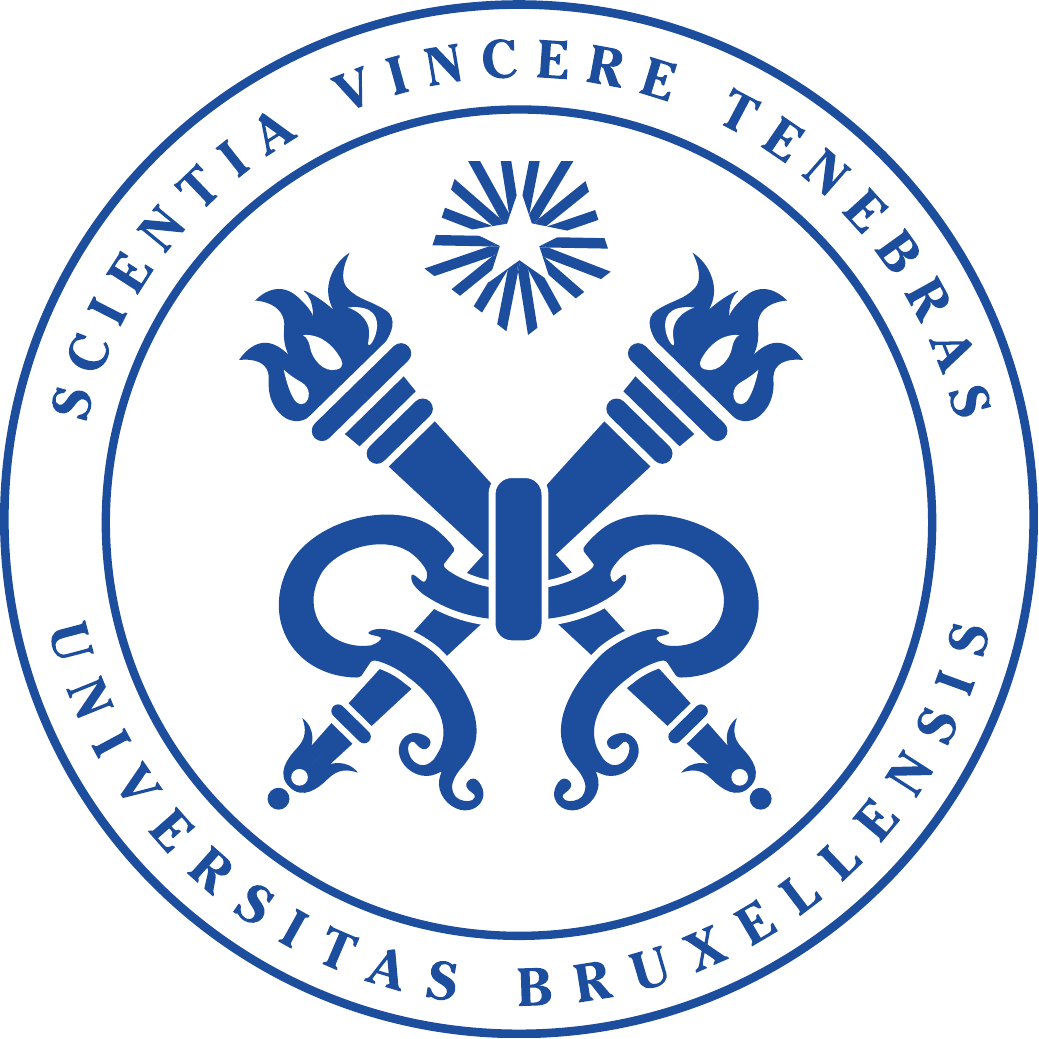} 
 
\vfill
\end{center}

\end{titlepage}

\pagestyle{empty} 

\null\vfill 

\textit{``Whenever sleep overcame me or I became conscious of weakening, I would turn aside to drink a cup of wine, so that my strength would return to me. Then I would return to reading."}

\begin{flushright}
Ibn Sina (Avicenna)
\end{flushright}

\vfill\vfill\vfill\vfill\vfill\vfill\null 

\clearpage 


\addtotoc{Abstract} 

\abstract{\addtocontents{toc}{\vspace{1em}} 

In this thesis, we study the statistical properties of currents in mesoscopic systems. We use the formalism of counting statistics in order to characterize the substantial current fluctuations at this scale. The full counting statistics of the transport processes is obtained starting from a microscopic Hamiltonian describing the electron dynamics in the studied circuits and in the quantum regime.
We consider two particular systems of capacitively coupled parallel transport channels. In the first system, each transport channel contains a single quantum dot in contact with two electron reservoirs. The second system we study is constituted of a double quantum dot coupled to two electrodes and probed by a quantum point contact detector sensitive to the electronic occupation of the double quantum dot via Coulomb interaction. In both systems, chemical potential differences, or thermodynamic forces, are applied to each transport channel that generate fluctuating stationary currents.
The current statistics for these two models is obtained by using a master equation for the probability of occupation in the quantum dots and the number of electron transfers in the electrodes. We verify that the joined probability distribution of the currents in each channel satisfies a fluctuation theorem in the long-time limit, involving the thermodynamic forces of both channels.
The issue of single-current fluctuation theorems for the marginal distribution of the currents in one of the two channels is also investigated. We show that in the limit of large current ratio between both channels, a single-current fluctuation theorem is satisfied individually for the slower circuit in agreement with experimental observations. This theorem involves an effective affinity which depends on the thermodynamic forces applied to both channels and the specific features of the system considered. A detailed study of the effective affinity is made for the two aforementioned systems.
Besides, we introduce a criteria on the initial condition of the transport channels for the observation of a fluctuation theorem at any time. This criteria is also extended to the case of single-current fluctuation theorems.
Finally, we perform the nonequilibrium thermodynamic analysis of the system composed of a double quantum dot probed by a quantum point contact in the presence of temperature and chemical potential differences between the electrodes. A thermal machine is studied and shown to reach highest efficiencies at maximum power by fine tuning the double quantum dot spectrum.

\clearpage 


\setstretch{1.2} 

\acknowledgements{\addtocontents{toc}{\vspace{1em}} 

First of all, I am grateful to Pierre Gaspard for being my supervisor during this thesis.

I am particularly indebted to Massimiliano Esposito who has been of great influence on me. I thank him for his technical support but also and most of all for his advice and guidance all along this thesis.

I would like to thank my colleagues for the very pleasant interaction we had during these four last years, with particular emphasis on the \emph{workshops} organized at l'Atelier. I also thank my colleague Nathan Goldman for his encouragement as well as the enjoying and stimulating discussions we had.

Furthermore, I wish to thank the groups that received me during my PhD with particular mention to the group of Massimiliano Esposito at the University of Luxembourg, the group of Tobias Brandes at the Technische Universit$\ddot{\mbox{a}}$t Berlin and the group of Peter Zoller at the Universit$\ddot{\mbox{a}}$t Innsbruck. I particularly thank Sebastian Diehl for his kind thought and encouraging words.

I am very grateful to the Universit\'e Libre de Bruxelles and the professors for the quality of their teaching, with particular mention to Petr Tiniakov. 

I acknowledge the FRIA for the financial support of my thesis.

Finally, special thanks go to Florence Susant for her continuous support, as well as to my family and friends, who gave me the motivation and confidence that have been essential in the fulfillment of my studies and research.

}
\clearpage 


\pagestyle{fancy} 
\lhead{\emph{Contents}} 
\tableofcontents 

\setstretch{1.2} 

\pagestyle{empty} 

\dedicatory{A Florence} 

\addtocontents{toc}{\vspace{2em}} 


\mainmatter 

\pagestyle{fancy} 


\input{Chapter1}

\input{Chapter2}

\input{Chapter3}

\input{Chapter4} 
\input{Chapter5} 
\input{Chapter6}

\input{Chapter7}


\addtocontents{toc}{\vspace{2em}} 

\appendix 


\input{AppendixA}
\input{AppendixB}
\input{AppendixC}

\input{AppendixD}

\addtocontents{toc}{\vspace{2em}} 

\backmatter


\label{Bibliography}

\lhead{\emph{Bibliography}} 

\bibliographystyle{apsrev4-1} 

\bibliography{thesis} 

\end{document}

%% file: Chapter1.tex

\chapter{Introduction} 

\label{Chapter1} 

\lhead{Chapter 1. \emph{Introduction}} 


\section{Motivations}


Most recent advances in the field of nanotechnology enable us to engineer functional devices down to the molecular scale. In the same way as the first thermal machines motivated fundamental developments in classical thermodynamics, these new devices open new insights and constitute at present day a challenge for researchers working on the extension of thermodynamics to small systems.

Lying between the nanoscale and the macroscopic world, the so-called mesoscopic devices are of particular interest. Indeed, these systems exhibit intermediate properties due to the presence of strong fluctuations while still involving a large number of constituents such as molecules or atoms, thus presenting a worthy challenge for statistical mechanics theory. On top of that, the critical size of these systems together with low temperatures often imposed on them lead to non trivial behavior due to quantum effects \cite{Datta1997, Imry2008}. A typical example is given by semiconductor nanostructures \cite{2004cond.mat.12664B} which can be structured to contain a thin layer of mobile electrons subjected to virtually arbitrary potential energy landscapes. The study of their quantum transport properties reveals new physical behavior such as Coulomb blockade \cite{2005cond.mat..8454V} or quantized conductance \cite{PhysRevLett.60.848, PhysRevB.43.12431}.

Along the same line, quantum dot circuits have recently attracted a lot of interest, for both experimentalists and theoreticians. Indeed, due to quantum confinement along the three dimensions, semiconducting quantum dots exhibit highly tunable and sharp discrete spectra \cite{RevModPhys.75.1} with properties fairly similar to those of atoms and molecules, which is the reason they are sometimes referred to as artificial atoms.

Since their discovery in 1980 by Russian physicist Alexey I. Ekimov \cite{1981JETPL..34..345E}, researchers have suggested a variety of applications in photovoltaic, light emitting and photodetector devices or even as quantum bits for quantum computing \cite{Trauzettel2007}.

Most remarkably, these mesoscopic circuits have recently been used as charge counting devices \cite{Schoelkopf1998, Wei2003,Fujisawa2004, bylander2005current, PhysRevLett.96.076605, fujisawa2006, PhysRevLett.99.206804, gustavsson2009, PhysRevB.81.125331, PhysRevX.2.011001} in order to measure current fluctuations down to the order of the $aA \, (10^{-18} A)$. By performing the counting statistics of electron transfers through mesoscopic devices, these experiments give us access to the fluctuations of the current around its mean, challenging most recent results of nonequilibrium statistical physics as we will here discuss.


\section{Historical developments}


The production of work by heating is still at present one of the primary sources of energy through the use of thermal machines. This concept dates back to 1690 and the works of French physicist and mathematician Denis Papin, who used the vacuum created by heating a piston to lift a weight \cite{papinrecueil, pluviose1997machines}. Starting form these ideas, steam engines were developed, first by Thomas Newcomer and Thomas Savery and later by James Watt.

These technical developments attracted the interest of physicists such as Sadi Carnot, whose aim at improving the efficiency of such machines led to a definition of useful work and the introduction of its famous \emph{Carnot cycle} of maximal efficiency, thereby seeding the premises of modern thermodynamics \cite{carnot1978reflexions}. However, it is James Joule in 1843, who made the connection between heat and work as forms of energy transfer \cite{joule1843xxxii}, ultimately leading to the principle of energy conservation by Hermann von Helmholtz, a principle known today as the first law of thermodynamics.

This insight eventually led William Thomson (Lord Kelvin) in 1848 to rewrite Carnot's principle as
\be
\sum_{i=1}^{n} \frac{Q_i}{T_i} \leq 0
\ee
for a thermal engine connected to $n$ reservoirs at temperature $T_i$ and delivering an amount of heat $Q_i$ to the engine \cite{grandy2008entropy}. However, it was finally left to German physicist Rudolf Clausius in 1865, to introduce the entropy state function $S$ \cite{clausius1867mechanical} satisfying the second law of thermodynamics
\be
\frac{dQ}{T} \leq dS,
\ee
with equality holding only in the limit of a reversible step, i.e. for an infinitesimal transition between two thermodynamic equilibrium states. These advances laid down the core elements for the subsequent developments of today modern thermodynamics, with leading contributions by physicists such as Willard Gibbs, James Clerk Maxwell or Max Planck to mention just a few.

Though the idea that heat could be related to a form of motion was raised by Francis Bacon as early as 1620 in his \emph{Noven Organum}, it is not until the middle of the nineteenth century that this was formalized by physicists such as Rudolf Clausius, James Clerk Maxwell, Willard Gibbs and, maybe most notably, Ludwig Boltzmann. It is the Austrian physicist Ludwig Boltzmann who laid down the cornerstone of statistical mechanics setting up the bridge between the microscopic world and the thermodynamic description, through  his entropy formula
\be
S= k_{B} \ln W
\ee
linking the entropy thermodynamic function $S$ of a given macrostate to the number $W$ of its corresponding microstates in terms of the constant bearing his name $k_{B} = 1.3806488 \times 10^{-23} \, {\rm m}^{2} \cdot {\rm kg} \cdot  {\rm s}^{-2} \cdot {\rm K}^{-1}$.

These theoretical advances, together with the observation by Robert Brown in 1827 of the random movements of pollen grains through water \cite{brown1828xxvii}, motivated the investigation of fluctuations at the microscopic scales. Thereafter, Albert Einstein showed in 1905 an early form of the fluctuation-dissipation theorem linking the linear response of the Brownian particle to an external force, to the size of its position fluctuations at thermodynamic equilibrium \cite{einstein1905movement}. Along the same line, Johnson and Nyquist derived in 1928 a similar relation between the resistance of a circuit and its thermal current fluctuations at equilibrium \cite{nyquist1928thermal}. These works would later be systematized by Ry$\overline{{\rm o}}$go Kubo starting from a Hamiltonian description of the microscopic dynamics, ultimately leading to the fluctuation-dissipation theorem \cite{callen1951irreversibility, kubo1957statistical, de2013non}. This latter was first obtained by Callen and Welton in 1951, relating the energy dissipated in a system due to the action of an external force, to its equilibrium fluctuations.

Subsequently, generalization of these results beyond linear response would follow \cite{bochkov1977general, bernard1959irreversible, efremov1969fluctuation, hanggi1982nonlinear}, culminating in the derivation of the celebrated fluctuation theorems constraining the microscopic fluctuations of physical quantities for systems arbitrarily far from equilibrium. The first versions of these fluctuation theorems dealt with the entropy or irreversible work fluctuations in closed systems described by thermostatted equations of motion \cite{evans1993probability, evans1996causality} or in stochastic systems \cite{crooks1998nonequilibrium, kurchan1998fluctuation, crooks1999entropy, lebowitz1999gallavotti, searles1999fluctuation, seifert2005entropy, andrieux2007fluctuation, esposito2007entropy}. Similar fluctuation relations have been obtained for the irreversible work in driven isolated Hamiltonian systems \cite{cleuren2006fluctuation, horowitz2007comparison}. In particular, Gavin E. Crooks derived in 1999 \cite{crooks1999entropy} a fluctuation relation, today known as the \emph{Crooks fluctuation theorem}, for the work probability distributions $p(W)$ and $\tilde{p} (W)$ during a time-dependent driving process and the reversed process
\be \label{CrooksFlucTheorem}
p(W) = \tilde{p} (-W) \, \mbox{e}^{\beta (W - \Delta F)},
\ee
where $\Delta F$ is the free energy difference of the initial and final equilibrium states with inverse temperature given by $\beta$. These work fluctuation theorems generalize the well-known Jarzynski equality \cite{PhysRevE.56.5018, jarzynski1997nonequilibrium, jarzynski2004nonequilibrium} which is obtained by averaging both sides of Eq. (\ref{CrooksFlucTheorem}) leading to
\be \label{jarzeq}
\langle \mbox{e}^{-\beta W} \rangle =\mbox{e}^{-\beta \Delta F} .
\ee
Remarkably enough, the work fluctuation theorems relying on the hypothesis of microscopic time reversibility of the underlying Hamiltonian dynamics also imply inequality
\be
\langle W \rangle \geq \Delta F,
\ee
which is nothing but a form of the second law of thermodynamics \cite{clausius1867mechanical} laying at the heart of the irreversibility observed at the macroscale.

Regarding experimental investigations, both work and heat fluctuations have been measured in order to validate some of the fluctuation theorems described above in the classical regime \cite{trepagnier2004experimental, wang2005experimental, schuler2005experimental, tietz2006measurement}. Moreover, the results of the fluctuation theorems have been magnificently used to obtain equilibrium thermodynamic parameters, such as free energy differences, from measurements taken arbitrarily far from equilibrium \cite{liphardt2002equilibrium, collin2005verification}. On the other hand, experimental investigation of the fluctuation theorems in the quantum regime is an active field of research \cite{RevModPhys.83.771}. An experimental setup of uttermost importance consists of open quantum systems exchanging energy or matter with reservoirs \cite{fujisawa2006, PhysRevB.81.125331, PhysRevX.2.011001}. As mentioned earlier, it is nowadays possible to resolve experimentally the fluctuations in small mesoscale devices at the single electron level \cite{fujisawa2006, bylander2005current, PhysRevLett.99.206804, gustavsson2009, PhysRevB.81.125331, PhysRevX.2.011001}. These experimental possibilities and the need to describe non-equilibrium fluctuations in these systems have motivated the rapid developments of these issues in the field of electron counting statistics \cite{levitov1993charge, levitov1996electron, gurvitz1997measurements, nazarov2002circuit, bagrets2003full, nazarov2003quantum, nazarov2003full, pilgram2003stochastic, shelankov2003charge, levitov2004counting, 2004PhRvB70k5327R, 2005EL69475F, nazarov2007full, RevModPhys.81.1665}. Most notably, these works led to the \emph{Levitov-Lesovik formula} \cite{levitov1993charge, levitov1996electron, S07, klich2003elementary, GK06} for the cumulant generating function of the number of charges transferred between two electrodes
\be \label{levles1}
\mathcal{G}(i \lambda) = \int \frac{d\omega}{2 \pi} \ln \left\{ 1+ \gamma (\omega) \left( f_L (\omega) (1- f_R (\omega ) )( \mbox{e}^{i \lambda} -1 ) +  (1- f_L (\omega ) )f_R (\omega)( \mbox{e}^{-i \lambda} -1 )  \right) \right\}
\ee
in terms of the Fermi-Dirac distribution $f_{i} (\omega)$ in the electrode $i = L$ or $R$ and the transmission coefficient $\gamma (\omega)$ of the channel.

In analogy with the work fluctuation relations, constraints on the probability distribution of the currents of heat and matter through open quantum systems have also been derived \cite{wojcik2004classical, PhysRevB.78.115429}. The most general form of this fluctuation theorem \cite{andrieux2009monnai} can be written in terms of the probability distribution of the energy and matter fluctuations as
\be \label{fluctintr}
p(\Delta E_i , \Delta N_i ) = p(-\Delta E_i , -\Delta N_i ) \, \mbox{e}^{\sum_{i'=1}^{n} \beta_{i'} (\Delta E_{i'} - \mu_{i'} \Delta N_{i'})} ,
\ee
where $\Delta E_i$ and $\Delta N_i$ are respectively the energy and matter changes in the $i$th reservoir with inverse temperature $\beta_i$ and chemical potential $\mu_i$, connected to the open quantum system\footnote{The thermodynamic forces applied to the system do not appear explicitly in relation \ref{fluctintr}. However, a fluctuation theorem can be be established from (\ref{fluctintr}) in the long-time limit which depends explicitly on the thermodynamic affinities. This point will be discussed in further detail in Chapters \ref{Chapter2} and \ref{Chapter4} of this thesis.}. Such relations are sometimes referred to as \emph{exchange fluctuation relations} in order to distinguish them from the work fluctuation theorems mentioned above. Remarkable relations can be derived for the non-linear transport coefficients by using these relations \cite{AG07JSM, andrieux2009monnai, RevModPhys.81.1665}, thus generalizing and synthesizing previous work. The theoretical investigation and experimental validation of the quantum version of the exchange fluctuation relation is a subject at the forefront of current research in the field of non-equilibrium statistical physics.


\section{Single-electron counting}


\begin{figure}[htbp]
	\centering
		\includegraphics[width=14cm]{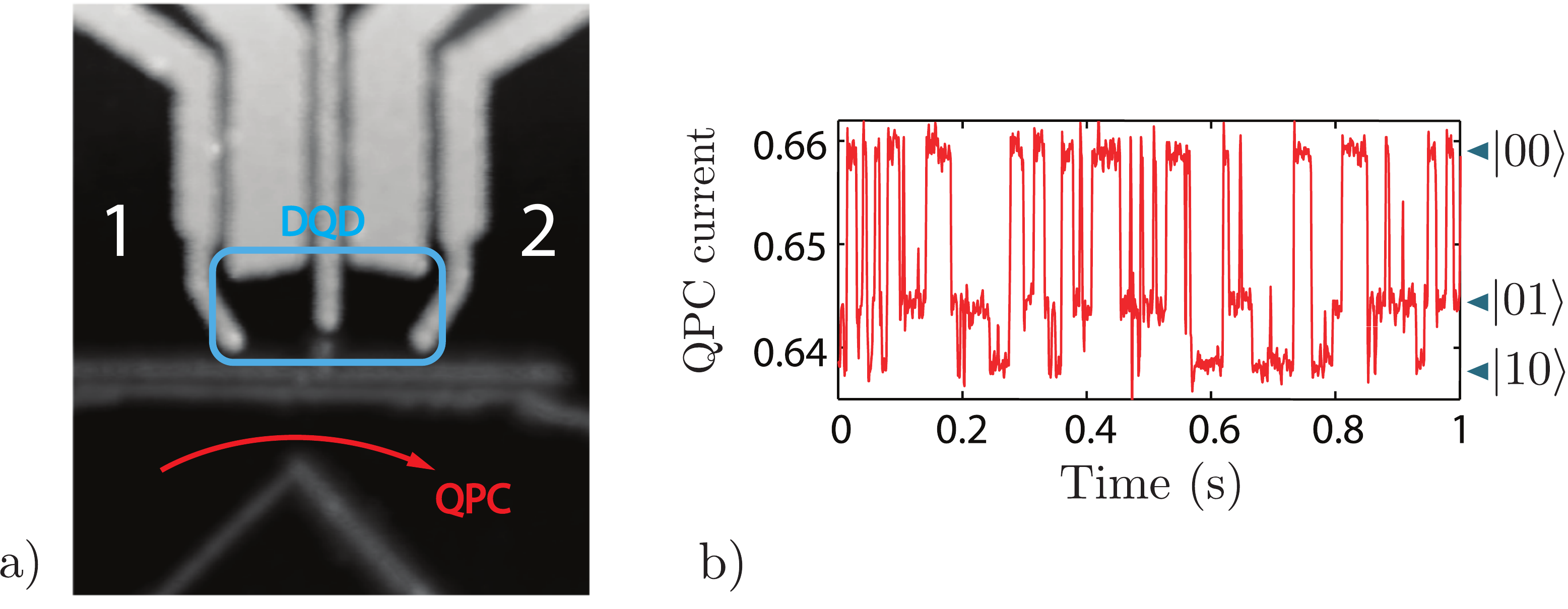}
	\caption{a) Atomic-force micrograph of the setup described with the text, composed of a double quantum dot (DQD) channel and a quantum point contact (QPC) detector. Electrons can travel between reservoirs $1$ and $2$ via the DQD marked by the blue rectangle. b) Current trajectory in the QPC for a particular experimental realisation, the current takes three different values depending on the state occupation in the DQD. Both figures where adapted from Ref. \cite{PhysRevX.2.011001}.}
	\label{figureDQDQPC}
\end{figure}

Modern technology enables the real-time detection of electrons tunneling in quantum dot circuits \cite{Schoelkopf1998, Wei2003,Fujisawa2004, bylander2005current, PhysRevLett.96.076605, fujisawa2006, PhysRevLett.99.206804, gustavsson2009, PhysRevB.81.125331, PhysRevX.2.011001}, allowing the measurement of currents down to the order of the ${\rm aA}$ as well as their fluctuations. In these experiments, the real-time detection of the charge on the quantum dots of the monitored circuit is performed by tracking the time resolved current in a secondary circuit whose macroscopic current of the order of the ${\rm nA}$ can be measured with conventional ampere meters. Indeed, the electrical charge on the quantum dots is strongly correlated to the current in the detector due to Coulomb interaction, leading to experimentally accessible discrete current jumps in the detector as discrete charges leave or enter the quantum dots.

Fujisawa \emph{et al.} recently used this method to perform the bidirectional counting of electrons accross a double quantum dot circuit \cite{fujisawa2006}. Experimental results were then used by Utsumi \emph{et al.} to investigate the validity of particle exchange fluctuation relations in the double quantum dot circuit \cite{PhysRevB.81.125331}. The experimental setup used consists of a double quantum dot connected to two electrodes submitted to an electric bias $\Delta V = {\rm e} ( \mu_{1} - \mu_{2})$ where ${\rm e}$ is the electron charge while $\mu_1$ and $\mu_2$ denote the chemical potentials of electrons in the electrodes $1$ and $2$ respectively. The electronic occupation in the double quantum dot is monitored by use of a quantum point contact circuit whose current fluctuations are used to track the electrons tunneling in the double quantum dot and its electrodes. An atomic-force micrograph of one of the circuits used in Ref. \cite{PhysRevX.2.011001} is shown in \textsc{Figure}~\ref{figureDQDQPC}~a).

During the counting experiment, the double quantum dot may occupy three of the four possible charge states , namely $|00 \rangle$, $| 10 \rangle$, $|01\rangle$ and $|11\rangle$, depending on the adjustment of experimental parameters. As a consequence, the current in the quantum point contact can take three different values depending on the double quantum dot state. The current in the detector is then recorded along several trajectories during a given time interval $\tau$ while the monitored double quantum dot is subject to the electric bias $\Delta V$.

As can be seen in \textsc{Figure} \ref{figureDQDQPC} b), the recorded current trajectories exhibit sudden jumps corresponding to tunneling events in the double quantum dot channel which are in one-to-one correspondence with the real-time state of the double quantum dot. The number of electrons exchanged between the reservoirs and the double quantum dot is inferred from these trajectories. As an example, let us consider a recorded trajectory where the three allowed states for the double quantum dot are given by $|00\rangle$, $|10 \rangle$ and $|01 \rangle$. A sequence of the form
\be
|00\rangle \rightarrow |10\rangle \rightarrow |01 \rangle \rightarrow |00\rangle
\ee
along the trajectory is interpreted as a net transfer of one electron from the electrode $1$ to the electrode $2$ and contributes to the number $n_{e}^{12}$. Conversely, a sequence of the form
\be
|00\rangle \rightarrow |01\rangle \rightarrow |10 \rangle \rightarrow |00\rangle
\ee
is interpreted as a net transfer of one electron from the electrode $2$ to the electrode $1$ which contributes to the number $n_{e}^{21}$. Finally, the net transfer of electrons $n_e$ transfered from the electrode $1$ to the electrode $2$ during time $\tau$ is given by the difference
\be
n_e = n_{e}^{12} - n_{e}^{21}.
\ee
By performing this procedure over several trajectories, the statistics of net electron transfers through the double quantum dot channel is obtained.

By using experimental results, Utsumi \emph{et al.} showed that the resulting probability distribution $p (n_e)$ for the number of electrons $n_e$ transferred across the double quantum dot channel obeys the exchange fluctuation relation
\be \label{effintro}
p (n_e) = p(-n_e) \mbox{e}^{\tilde{A} n_e}
\ee
with an effective affinity $\tilde{A}$ that proved to be one order of magnitude lower than to the expected theoretical affinity of the isolated double quantum dot channel $A_{DQD} = \beta \frac{ \Delta V}{e} $ within experimental precision \cite{PhysRevB.81.125331}. This result is interpreted as the back-action of the quantum point contact detector on the monitored circuit \cite{PhysRevB.81.125331, cuetara2011fluctuation, cuetara2013effective}. One of the main results of this thesis is the theoretical investigation of effective fluctuation theorems of the form (\ref{effintro}) on the basis of the microscopic Hamiltonian dynamics of the whole setup, including the detection circuit.


\section{Goal of the thesis}


Our aim in this thesis is to give a theoretical study of the statistical properties of currents in mesoscopic circuits. The importance of current fluctuations at this scale requires a full characterization of their statistics in order to apprehend the transport processes in these systems. Moreover, due to the small size together with the low temperature typically imposed on these systems, a careful treatment of quantum effects is mandatory. One of the key issues regarding the counting experiments described in the previous Section is the characterization of the back-action of the detection setup and its consequences on the exchange fluctuation relations in the measured circuit.

In this thesis, we develop the theory of counting statistics starting from a microscopic Hamiltonian description of the circuits in the quantum regime. Calculations are made within the density matrix formalism which has already been successfully applied to the problem of electron counting statistics \cite{harbola2006quantum, timm2008tunneling, esposito2009thermoelectric, esposito2007entropy, pedersen2007coherent, 2004PhRvB70k5327R, 2008PhRvL100o0601F, 2005EL69475F, 2005PhRvB72p5347W, 2006PhRvL96b6805B, 2006PhRvE73d6129E, 2006PhRvB73c3312K, 2006RvMaP18619D, 2007PhRvB76p1404E, 2007PhRvB75o5316E, PhysRevB.76.085408, PhysRevB.77.195315, RevModPhys.81.1665}. We show that the total system composed of the detector coupled to the measured circuits obeys a bivariate fluctuation theorem in consistency with Eq.~(\ref{fluctintr}). Though single channel exchange fluctuation relations of the form (\ref{effintro}) for the sole detected circuit do not hold in general, we show that such relations are recovered with respect to a model-dependent modified affinity $\tilde{A}$ in the limit of a large current ratio between the detector and the probed circuit in consistency with the experimental observations, which constitutes the main result of the thesis.

This thesis is organized as follows. In Chapter \ref{Chapter2}, we apply the results of Ref. \cite{andrieux2009monnai} to a  double channel circuit and show the existence of a bivariate fluctuation theorem for the energy and particle currents in the long-time limit. The derivation is established on the basis of microreversibility for the underlying dynamics as well as on the assumption of initial and final equilibrium states in the reservoirs. We end this chapter by introducing the cumulant generating function and making the connection to the large deviation function characterizing the current fluctuations in the long time limit.

Chapter \ref{Chapter3} is devoted to the introduction of the density matrix formalism as applied to the counting statistics of a given physical observable in an open quantum system connected to macroscopic reservoirs \cite{2006PhRvE73d6129E, nazarov2003full, RevModPhys.81.1665}. This is done through the introduction of a modified density matrix, which depends on a counting parameter keeping track of the fluctuations of the corresponding observable. The macroscopic character of the circuit electrodes together with their weak coupling to the quantum dots enable us to perform the Born and Markov approximations \cite{JChemphys1992, PhysRevLett.73.1060, JChemphys1997, JChemphys1999, JPhysChemB2005, PhysRevA.77.032104, breuer2007theory} as discussed in section \ref{pertexpandbmapprox}. Furthermore, the fast oscillations due to the large Bohr frequencies in the quantum dots as compared to the relaxation rate induced by the electrodes enable us to perform the rotating wave approximation \cite{gardinerzoller, tannoudjiintproc, PhysRevA.78.022106, vankampen} and get a Lindblad master equation \cite{Linblad, RevModPhys.52.569} for the modified reduced density matrix of the open quantum system. As a consequence, the counting statistics is fully accounted for by the evolution of the diagonal elements of the modified density matrix. A general and formal solution of the modified master equation is then written down and used to show that the cumulant generating function is given by the leading eigenvalue of the modified rate matrix depending on the counting parameter \cite{RevModPhys.81.1665}.

We apply these results to a specific model composed of two capacitively coupled parallel channels in Chapter \ref{Chapter4} \cite{cuetara2011fluctuation}. Thanks to the results of Chapter \ref{Chapter3}, we perform the full counting statistics in both channels starting from a Hamiltonian quantum description and we establish the bivariate fluctuation theorem introduced in Chapter \ref{Chapter2}. Furthermore, we study the finite-time effects and discuss the existence of exchange fluctuation relations for any measurement times. More specifically, we show that a finite-time fluctuation theorem holds provided the system is prepared in a stationary state with the electrodes over which we perform the counting disconnected from the circuit prior to the counting experiment. Besides, we study the emergence of a single-current fluctuation theorem of the form (\ref{effintro}) in the large current ratio limit between the two channels, i.e. in the experimental detection regime \cite{fujisawa2006, PhysRevX.2.011001}. Under this condition, we study the properties of the corresponding affinity which is shown to be strongly dependent on the parameters of the model and may significantly differ from the affinity of an isolated circuit as experimentally observed \cite{PhysRevB.81.125331}.

In Chapter \ref{Chapter5}, we perform a similar study \cite{cuetara2013effective} on a model composed of a double quantum dot channel probed by a quantum point contact as in experiments \cite{fujisawa2006, PhysRevX.2.011001}. The counting statistics of the double quantum dot channel is obtained by using the modified master equation approach introduced in Chapter \ref{Chapter3} while the coherent transport of electrons in the quantum point contact at non-equilibrium steady state is treated non perturbatively by use of scattering matrix theory, as required by the high transparency of the quantum point contact used in counting statistics experiments. We investigate the emergence of single-current fluctuation theorems in the limit of weak tunneling within the double quantum dot as well as in the limit of strong asymmetrical coupling to the quantum point contact detector.

We next study thermoelectric effects in the model first presented in Chapter \ref{Chapter5}. However, contrary to Chapter \ref{Chapter5}, the two electrodes of the quantum point contact are each assumed to be individually at equilibrium, which is only valid for a low transparency quantum point contact. In this regime we show that the transition rates obey a local detail balance or Kubo-Martin-Schwinger (KMS) condition \cite{Kubo1998, RevModPhys.81.1665} involving the irreversible entropy production in the double quantum dot due to electron tunneling events in the quantum point contact, which constitutes the main result of Chapter \ref{Chapter6}. Next, we consider the case of a heat engine and investigate the efficiency of our setup at maximum power. We show that highest efficiencies at maximum power close to the Curzon-Ahlborn efficiency \cite{curzon1975efficiency,tu2008efficiency,schmiedl2008efficiency,esposito2009thermoelectric,esposito2010efficiency,sanchez2011optimal} are attained by fine tunning the spectrum of the double quantum dot.

We end up with the concluding Chapter \ref{Chapter7}, where we summarize the main results of the thesis and give some perspectives for future research.

%% file: Chapter2.tex

\chapter{Fluctuation relations for currents in quantum systems} 

\label{Chapter2} 

\lhead{Chapter 2. \emph{Fluctuation relations for currents in quantum systems}} 


\section{Microreversibility of quantum systems}


We recall the basic elements of quantum mechanics needed to understand the physics considered in this work. We begin with the time evolution of quantum state vectors through the Schr\"odinger equation and briefly recall the concept of statistical ensemble description by introducing the density matrix operator. After presenting the time-reversed dynamics, we construct the time reversed trajectories whose link with the original trajectories is used to show the microreversibility of quantum dynamics. This central result is one of the basic elements used to prove the so-called fluctuation theorems.


\subsection{Forward evolution}


In quantum mechanics, the state of a physical system at time $t$ is characterized by a state vector $| \Psi (t) \rangle $ of the Hilbert space $\mathcal{H}$ and which evolves according to the Schr\"odinger equation
\be \label{schrodinger}
i \hbar \frac{d}{dt} | \Psi (t) \rangle = H(t)| \Psi (t) \rangle,
\ee
where $H(t)$ is the Hamiltonian operator of the quantum system.

A formal solution of the Schr\"odinger equation with initial condition $| \Psi (t_{0}) \rangle$ at time $t_{0}$ is given by
\be \label{schrodingersol}
| \Psi (t) \rangle = U(t,t_{0}) | \Psi (t_{0}) \rangle
\ee
in terms of the unitary time-evolution operator $U(t,t_{0})$. The unitarity of $U(t,t_{0})$ is ensured by the Hermiticity of the Hamiltonian $H(t)$, so that normalization of state vectors is preserved through quantum evolution. By using equation (\ref{schrodinger}), the time-evolution operator is shown to be a solution of the equation of motion
\be
i \hbar \frac{d}{dt} U(t,t_{0}) = H(t) U(t,t_{0}),
\ee
with the initital condition
\be
U(t_{0}, t_{0}) =I .
\ee

This equation can be alternatively written in integral form as 
\bea \nonumber
U(t,t_{0}) & = & I + \sum_{n=1}^{\infty} \left( -\frac{i}{\hbar} \right)^{n} \\
&& \times \int_{t_{0}}^{t} dt_{1} \,  \int_{t_{0}}^{t_{1}} dt_{2} \, \dots \int_{t_{0}}^{t_{n-1}} dt_{n} \,
 H(t_{1}) H(t_{2}) \dots H(t_{n}) \nonumber \\  
&\equiv & \mbox{T}_{+} \exp{ \left[ -\frac{i}{\hbar} \int_{t_{0}}^{t} d\tau H(\tau) \right] }, \label{unitaryint2}
\eea
where we introduced the time ordering operator $\mbox{T}_{+}$ which orders time-dependent operators on its right in such a way that they appear with increasing time arguments from right to left\footnote{One should remember that the action of the time ordering operator on a product of fermionic operators is given by the ordered sequence multiplied by the parity of the permutation needed to bring them in the ordered form. This is not, however, the case for a sequence of Hamiltonians which are bosonic operators.}.

In a wide range of physical applications, it is often the case that we are not able to control precisely enough the initial condition of the quantum system under study as to uniquely determine its initial state vector. As a consequence, the initial condition can at most be assumed to be picked up from a statistical distribution over a set of quantum states $ \{ | \psi_{\alpha} \rangle \}$. The system is thus said to be in a statistical mixture and is conveniently described by a density matrix operator $\rho_{0}$ \cite{breuer2007theory}. This operator is positive definite and normalized according to $\mbox{Tr} \, \{ \rho_{0} \} = 1$. The density matrix operator can be expanded on the set of quantum states $ \{ | \psi_{\alpha} \rangle \}$ as
\be \label{26} 
\rho_{0} = \sum_{\alpha} P_{\alpha} | \psi_{\alpha} \rangle \langle \psi_{\alpha} | 
\ee
with the eigenvalues $P_{\alpha}$ satisfying $0 \leq  P_{\alpha} \leq 1$ and normalized according to
\be
\sum_{\alpha} P_{\alpha} =1.
\ee
The quantity $P_{\alpha}$ is thus interpreted as the probability weight given to the state vector $ | \psi_{\alpha} \rangle$ in the initial statistical ensemble.

By using equation (\ref{schrodingersol}), we can express the density matrix operator at time $t$ in terms of its initial value $\rho_{0}$ at time $t_{0}$ and the time-evolution operator as
\be \label{evolution}
\rho (t) = U(t,t_{0}) \, \rho_{0} \, U^{\dagger} (t,t_{0}),
\ee
which as unit normalization $\mbox{Tr}  \{ \rho (t)\} = 1$ by virtue of the unitarity of the evolution operator and the normalization of the initial density matrix.

The density matrix operator is well suited to calculate the mean value of an operator $A$ at time $t$ as 
\be \label{average2}
\langle A \rangle_t = \mbox{Tr} \{ \rho (t) A \},
\ee
which is interpreted as the average over an infinite number of measurement results of the operator $A$ at time $t$, starting with initial conditions drawn from the statistical ensemble $\rho_{0}$, i.e. state vectors picked up in the ensemble $\{ | \psi_{\alpha} \rangle  \}$ according to the probability distribution $P_{\alpha}$. Similarly, the probability $p_{\psi} (t)$ to observe the system in the state $| \psi \rangle$ at time $t$ is given by
\bea
p_{\psi} (t)& = & \langle \psi | \rho (t) | \psi \rangle \\
& = & \mbox{Tr} \{ P_{\psi} \, \rho (t) \}
\eea
in terms of the projector onto state $| \psi \rangle$ defined by $P_{\psi} = | \psi \rangle \langle \psi |$.

We obtain an equation of motion for the density matrix operator by differentiating equation (\ref{evolution}) with respect to time $t$, which yields the Landau-von Neumann equation
\begin{equation}
\dot{\rho}(t) = -\frac{i}{\hbar} \left[ H(t) , \rho (t) \right]
\end{equation}
ruling the time evolution of the density matrix operator, together with the initial condition $\rho (t_{0}) = \rho_{0} $.


\subsection{Time-reversed evolution}


Given a quantum trajectory in Hilbert space $\mathcal{H}$ between initial and final conditions $|\Psi (T_{0}) \rangle$ and $|\Psi (T) \rangle$ such that
\be
| \Psi (T) \rangle = U(T,T_{0}) | \Psi (T_{0}) \rangle,
\ee
we can construct a time-reversed trajectory with initial and final conditions chosen as the time reversed state vectors of, respectively, $|\Psi (T) \rangle$ and $| \Psi (T_{0}) \rangle $.

\begin{figure}[htbp]
	\centering
		\includegraphics[width=14cm]{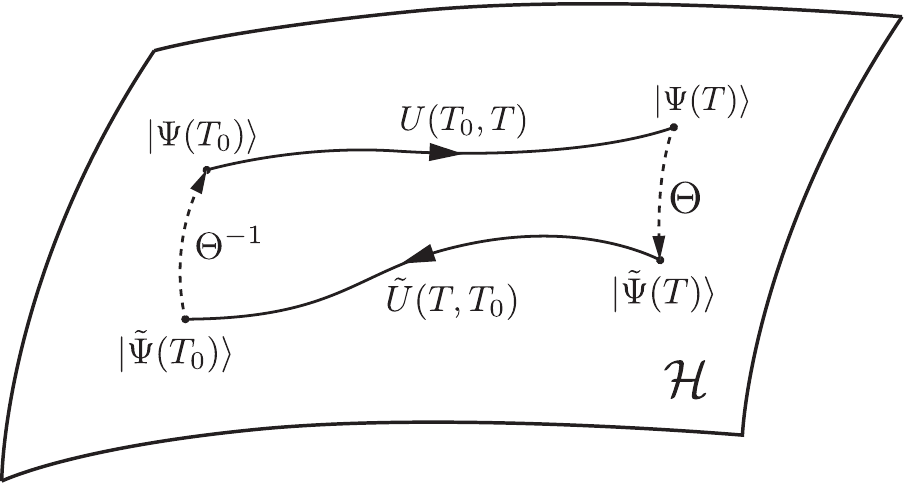}
	\caption{Schematic picture of a trajectory from state vector $| \Psi (T_{0}) \rangle$ to $| \Psi (T) \rangle$ in Hilbert space $\mathcal{H}$ and its associated time-reversed trajectory from time-reversed state $ | \tilde{\Psi} (T) \rangle$ to $ | \tilde{\Psi} (T_{0}) \rangle$.}
	\label{trajectories}
\end{figure}

Such time-reversed states are obtained in quantum mechanics by the application of the time reversal operator $\Theta$ \cite{sakurai} on state vectors
\bea
|\tilde{\Psi} (T) \rangle & = & \Theta | \Psi (T) \rangle, \\
|\tilde{\Psi} (T_{0}) \rangle & = & \Theta | \Psi (T_{0}) \rangle.
\eea

The time reversal operator is anti linear\footnote{This property can be deduced by assuming the existence of a finite minimal energy eigenstate in physical systems \cite{sakurai}.} and changes the sign of all odd parameters such as, for example, the momentum and angular momentum quantum numbers, or an eventual external magnetic field.

In the following, we assume the Hamiltonian operator to be invariant under the time reversal transformation\footnote{We will not consider the case of systems embedded in an external magnetic field in this thesis. However, generalization is straightforward and considered for example in Ref. \cite{andrieux2009monnai}.} which thus obeys the symmetry relation
\be \label{symham}
\Theta H(t) \Theta^{-1} = H(t).
\ee

In order to fully characterize the time-reversed trajectories, we must correctly define the time-reversed dynamics connecting our initial and final conditions $| \tilde{\Psi} (T) \rangle$ and $| \tilde{\Psi} (T_{0}) \rangle$. This can be done by using the time-reversed evolution operator $\tilde{U} (T,T_{0})$ corresponding to the Hamiltonian $H(T-t)$ and satisfying the differential equation
\be \label{timerevU2}
i \hbar \frac{d}{dt} \tilde{U} (t,t_{0}) = H(T-(t-T_0 )) \tilde{U} (t,t_{0}),
\ee
with initial condition $\tilde{U} (t_{0},t_{0}) = I$.

With these definitions, it is easy to show that the initial condition $ | \Psi (T_{0}) \rangle$ is unaffected by the successive application of
\be \label{premicro2}
\Theta^{-1} \tilde{U} (T,T_{0}) \Theta U (T,T_{0}) = I,
\ee
which performs the travel along the considered quantum trajectory followed by its time-reversed, effectively resulting in an identity operation. This is illustrated in \textsc{Figure} \ref{trajectories} by following the closed path starting and ending on the same state vector $| \psi (T_{0}) \rangle$.

This being true for any initial and final states $| \Psi (T) \rangle$ and $| \Psi (T_{0}) \rangle$ and making use of the unitary property of the evolution operator, we deduce the fundamental result \cite{RevModPhys.81.1665, andrieux2009monnai}
\be \label{microreversibility} \boxed{
\tilde{U} (T,t_{0}) = \Theta  \, U^{\dagger} (T,t_{0}) \, \Theta^{-1}
},
\ee
which is usually referred to as the property of \textbf{microreversibility of quantum dynamics}. A formal derivation of this result is given in Appendix \ref{AppendixA}.


\section{Consequences of microreversibility on current fluctuations}


The microreversibility relation (\ref{microreversibility}) is the basic ingredient together with the assumption of equilibrium initial and final states in order to show the fluctuation theorems \cite{PhysRevLett.71.2401, PhysRevLett.74.2694, Kurchan1998, Lebowitz1999, Andrieux2006, Andrieux2007, andrieuxthesis, RevModPhys.83.771}. The work done in this thesis focuses on the current fluctuation theorems relating the probabilities of opposite random values for the currents of energy and matter to the thermodynamic affinities driving their mean values \cite{PhysRevB.72.235328, PhysRevB.75.155316, PhysRevB.76.085408, RevModPhys.81.1665, PhysRevB.78.115429, andrieux2009monnai, PhysRevLett.104.076801}. 

In this Section, we illustrate these known results on a generic four-terminal circuit composed of several reservoirs and finite quantum subsystems. This constitutes the prime example of the models studied in this thesis. The circuit is connected through an interaction Hamiltonian for a fixed amount of time enabling energy and matter exchanges. We make use of the microreversibility of  quantum dynamics to establish a fluctuation theorem for the energy and matter exchanges between the reservoirs. This relation is expressed in terms of the probability distribution of energy and particle changes in the reservoirs during the interaction time and the thermodynamic affinities driving the fluxes. We finally introduce the generating function of the energy and particle fluctuations and express the fluctuation theorem as a symmetry relation for this generating function in terms of the thermodynamic affinities.


\subsection{Hamiltonian}


We consider a system composed of two interacting parallel transport channels. Each transport channel is made of two reservoirs of energy and particles and an intermediate quantum subsystem, which we assume to be small in the sense that it contains a finite amount of energy and matter contrary to the reservoirs.

These elements are initially decoupled from each other and set in contact between times $t=0$ and $t=T$ through a time-dependent interaction $V(t)$, before being decoupled again. The total Hamiltonian of our system is given by
\be
H(t) = \sum_{\alpha=1,2} \sum_{i = L, C ,R} H_{ \alpha i} + V(t),
\ee
where $H_{ \alpha i}$ is the Hamiltonian of the left- and right-hand reservoirs for, respectively, $i = L$ and $R$ in channel $\alpha=1$ or $2$, and $H_{\alpha C}$ is the Hamiltonian of the intermediate quantum system between the reservoirs $L$ and $R$ in channel $\alpha$. The interaction Hamiltonian $V(t)$  is non zero during $0<t<T$ and describes the energy and particle exchanges between the parts of the system\footnote{The more general situation of different initial and final Hamiltonians before and after the interaction has taken place can also be handled but is not considered in the present discussion.}.

The particle number in each reservoir and intermediate system is assumed to be conserved within each part of the system and in the absence of interaction so that 
\bea \label{ennumommutation}
\left[ N_{\alpha i} , H_{\alpha i} \right] = 0,
\eea
where $N_{\alpha i}$ is the particle number operator of the corresponding element in channel $\alpha$.


\subsection{Initial and final conditions}


Our aim lies in describing the nonequilibrium transport properties of these two channels in the long-time limit. These will depend on the thermodynamic affinities given by the temperature and chemical potential differences between the macroscopic reservoirs. In this sense, the initial statistical properties of the small intermediate quantum systems should be irrelevant when considering the current fluctuations in the long-time limit as we will do in the section \ref{longtimelimit}. Taking this point into account, we will consider the intermediate quantum systems as part of the left-hand reservoir of their respective channel \cite{andrieux2009monnai}, which justifies the redefinition
\be \label{systemresabs}
H_{\alpha L} + H_{ \alpha C} \rightarrow H_{\alpha L} .
\ee

We are thus left with four macroscopic reservoirs, and will consider grand-canonical initial and final states in each of these reservoirs. The initial and final density matrix of the system are thus chosen as
\be \label{initialdensity}
\rho(0) = \rho (T) = \prod_{  \alpha , i} \exp{ \left[ -\beta_{\alpha i} \left( H_{\alpha i} - \mu_{\alpha i} N_{\alpha i} - \phi_{\alpha i} \right) \right]},
\ee
where we introduced the inverse temperature $\beta_{\alpha i} = (k_{B} T_{\alpha i})^{-1}$ and chemical potential $\mu_{\alpha i}$ of the reservoir $i$ in channel $\alpha$, as well as its thermodynamic grand-potential 
\be 
\phi_{\alpha i} = -\beta_{\alpha i}^{-1} \ln{ \left[ \mbox{Tr} \left\{ e^{-\beta_{\alpha i} \left( H_{\alpha i} - \mu_{\alpha i} N_{\alpha i}  \right)}  \right\} \right]},
\ee
the trace being taken over the whole Hilbert space of the corresponding reservoir.


\subsection{Two-time quantum measurements}


In order to characterize the fluxes of energy and particles between the reservoirs during the time of interaction $T$, we perform simultaneous quantum measurements of the energy and particle number in each reservoir at the initial and final times chosen as $t=0$ and $t=T$. The possibility of such simultaneous measurements of energy and particle number is ensured by the commutation relations (\ref{ennumommutation}).

We denote by $| \Psi_{k} \rangle$ the eigenstates of the energy and particle number operators of each reservoir
\bea \label{eigenbasis1}
H_{\alpha i} | \Psi_{k} \rangle & = & \epsilon_{ \alpha i}^{k} | \Psi_{k} \rangle, \\ \label{eigenbasis2} 
 N_{\alpha i} | \Psi_{k} \rangle & = & n_{\alpha i}^{k} | \Psi_{k} \rangle .
\eea

According to the laws of quantum mechanics, the probability $p_{kk'}$ to observe the state $| \Psi_{k} \rangle$ at time $t=0$, followed by the observation of the state $| \Psi_{k'} \rangle$ at time $t=T$ is given by\footnote{Note that we do not explicitly write the time dependence of the probability $p_{kk'}$ in order to lighten the notations. This convention will be used in the remaining of this Chapter for the probability distributions and the generating function defined below. The dependence on time should however be clear from the context.}
\be \label{227}
p_{kk'} = | \langle \Psi_{k'} | U(T,0) | \Psi_{k} \rangle |^{2} \langle \Psi_{k} | \rho (0) | \Psi_{k} \rangle
\ee
in terms of the evolution operator $U(T,0)$ from time $t=0$ to time $t=T$ and the initial density matrix introduced in (\ref{initialdensity}).

The probability $p(\Delta \epsilon_{\alpha i}, \Delta n_{\alpha i})$ to observe the changes $\Delta \epsilon_{\alpha i}$ and $\Delta n_{\alpha i}$ of energy and particle number in each reservoir is thus given by
\be \label{enmatfluct}
p(\Delta \epsilon_{\alpha i}, \Delta n_{\alpha i}) = \sum_{kk'} \prod_{\alpha , i} \delta \left[ \Delta \epsilon_{ \alpha i}- (\epsilon_{\alpha i}^{k'}-\epsilon_{\alpha i}^{k})\right] \delta \left[ \Delta n_{\alpha i}^{k}- (n_{  \alpha i}^{k'}-n_{  \alpha i}^{k}) \right] p_{kk'},
\ee
where $\delta (\, \cdot \,)$ denotes the Dirac delta distribution or the Kronecker delta symbol $\delta_{\cdot , 0}$ whether its argument is a continuous or discrete number.

Similarly, we can perform this two-time quantum measurement procedure on time-reversed trajectories, and get the time-reversed probabilities of energy and matter transfers. We thus write the probability $\tilde{p}(-\Delta \epsilon_{ \alpha i}, -\Delta n_{ \alpha i})$ to observe the changes $-\Delta \epsilon_{\alpha i}$ and $-\Delta n_{\alpha i}$ during the time-reversed evolution as
\be \label{revenmatfluct}
\tilde{p}( - \Delta \epsilon_{ \alpha i}, - \Delta n_{  \alpha i}) = \sum_{kk'} \prod_{\alpha i} \delta \left[ -\Delta \epsilon_{ \alpha i}- (\epsilon_{  \alpha i}^{k'}-\epsilon_{ \alpha i}^{k})\right] \delta \left[ -\Delta n_{\alpha i}- (n_{\alpha i}^{k'}-n_{\alpha i}^{k}) \right] \tilde{p}_{kk'}
\ee
in terms of the time reversed probability
\be
\tilde{p}_{kk'} =  | \langle \tilde{\Psi}_{k'} | \tilde{U}(T,0) | \tilde{\Psi}_{k} \rangle |^{2} \langle  \tilde{\Psi}_{k} | \rho (T) | \tilde{\Psi}_{k}\rangle
\ee
to observe initially the time-reversed state $| \tilde{\Psi}_{k}\rangle = \Theta | \Psi_{k}  \rangle$ and finally the time-reversed state $| \tilde{\Psi}_{k'} \rangle= \Theta | \Psi_{k'}\rangle  $ after having let the system evolve with the time-reversed evolution operator $\tilde{U} (T,0)$ during the time interval $T$.


\subsection{Fluctuation relations for the currents}


By making use of microreversibility relation (\ref{microreversibility}) and the anti-linearity of the time reversal operator $\Theta$, we show that
\bea
p_{kk'} & = & | \langle \tilde{\Psi}_{k'} | \Theta U^{\dagger} (T,0) \Theta^{-1}  | \tilde{\Psi}_{k} \rangle |^{2} \langle \Psi_{k} | \rho (0) | \Psi_{k} \rangle \\
& = & | \langle \tilde{\Psi}_{k}| \tilde{U}(T,0) | \tilde{\Psi}_{k'} \rangle |^{2}  \langle \Psi_{k} | \rho (0) | \Psi_{k} \rangle \\
& = & \tilde{p}_{k'k} \, \mbox{e}^{\sum_{\alpha , i} \beta_{\alpha i} \left[( \epsilon_{\alpha i }^{k'} - \epsilon_{ \alpha i }^{k} ) - \mu_{\alpha i} (n_{\alpha i }^{k'} - n_{\alpha i}^{k})\right] },
\eea
where the choice (\ref{initialdensity}) for the initial and final density matrices has been used in order to obtain the last equality. This relation emphasizes the fact that the difference in probability to observe a trajectory and its time reversal is inherent to the different occurrence statistics of the initial and final conditions and not to the dynamics of the quantum system if the latter ones are microreversible. 

By using this last result, we write down a symmetry relation for the probability distribution of the energy and matter fluctuations
\be 
p(\Delta \epsilon_{\alpha i}, \Delta n_{\alpha i}) = \tilde{p}( -\Delta \epsilon_{\alpha i},  -\Delta n_{\alpha i}) \, \mbox{e}^{\sum_{i,\alpha} \beta_{\alpha i} \left[\Delta \epsilon_{\alpha i} - \mu_{\alpha i} \Delta n_{\alpha i} \right] }
\ee
in terms of the temperatures and chemical potentials of the reservoirs. By considering a symmetric interaction under time reversal, $V(t) = V(T-t)$,  the forward and reversed evolution operators between times $t=0$ and $t=T$ are identical
\be
U(T,0) = \tilde{U} (T,0)
\ee 
so that the probability densities of the energy and matter fluctuations along forward and reversed trajectories become equal
\be
p(\Delta \epsilon_{\alpha i}, \Delta n_{\alpha i}) = \tilde{p}(\Delta \epsilon_{\alpha i}, \Delta n_{\alpha i}).
\ee

From now on, we assume that the protocol does not perform work on our closed quantum system so that we can use energy conservation\footnote{Avoiding this assumption and the time reversal invariance of the Hamiltonian together with the assumption of homogeneous temperature, $\beta_i = \beta$ $\forall i$, would lead to the work fluctuation theorems. In this case, energy conservation leads to $\sum_{\alpha, i} \Delta \epsilon_{\alpha i} = W$, where $W$ denotes the work performed on the quantum system.} together with particle number conservation between the two quantum measurements
\bea
\sum_{\alpha , i} \Delta \epsilon_{\alpha i }  = 0 ,\\
\sum_{\alpha , i} \Delta n_{\alpha i }  = 0.
\eea
These constraints allow us to eliminate the functional dependence on the fluctuations of one of the reservoirs in Eqs. (\ref{enmatfluct}) and (\ref{revenmatfluct}). We thus introduce the index $j \in \{ 1L, 1R, 2L \}$,  and define the distribution
\be 
p(\Delta \epsilon_{j}, \Delta n_{j}) \equiv \left.  p(\Delta \epsilon_{\alpha i}, \Delta n_{\alpha i})\right|_{\Delta x_{2R } = -\Delta x_{1L }-\Delta x_{1R} - \Delta x_{2L} }  
\ee 
for the probability of energy and charge fluctuations in reservoirs $j$ and where the notation $\Delta x_{j}$ is used to denote both $\Delta \epsilon_{j}$ and $\Delta n_{j}$.

Using these relations and definitions, we write down the \textbf{quantum exchange fluctuation theorem} in the circuit
\be \label{QEFT} \boxed{
\frac{p(\Delta \epsilon_{j}, \Delta n_{j})} {p(-\Delta \epsilon_{j}, -\Delta n_{j})} = \mbox{e}^{\sum_{j' }\left[ A^{\epsilon}_{j'} \Delta \epsilon_{j'} + A^{n}_{j'} \Delta n_{j'}  \right] } }
\ee
in terms of the macroscopic thermodynamic affinities driving the nonequilibrium fluxes defined as
\bea \label{affen}
A^{\epsilon}_{j} & \equiv & \beta_{j} - \beta_{2R}, \\ \label{affpart}
A^{n}_{j} & \equiv & -\beta_{j} \mu_{j} + \beta_{2R} \mu_{2R}.
\eea

In the case of homogeneous initial and final conditions, i.e. $\beta_{j} = \beta$ and $\mu_{j} = \mu$, $\forall j$, all the thermodynamic affinities are zero
\be \label{equ}
A^{\epsilon}_{j}  = A^{n}_{j} = 0  \qquad \forall j .
\ee
In this case, the probabilities of observing energy and matter fluctuations in the reservoirs are even
\be
p(\Delta \epsilon_{j}, \Delta n_{j})=  p(-\Delta \epsilon_{j}, -\Delta n_{j}),
\ee
which implies the vanishing of the first moments, hence demonstrating the absence of macroscopic currents of energy and particles at thermodynamic equilibrium.

The quantum exchange fluctuation theorem can be alternatively expressed in terms of the generating function defined as the Laplace transform of the probability distribution
\bea \label{245}
G(\lambda_{j} , \xi_{j}) & \equiv & \langle \mbox{e}^{-\sum_{j} \left( \lambda_{j} \Delta \epsilon_{j} + \xi_{j} \Delta n_{j} \right) } \rangle\\
& = & \int \prod_{j} d(\Delta \epsilon_{j}) d( \Delta n_{j}) \, \mbox{e}^{-\sum_{j} \left( \lambda_{j} \Delta \epsilon_{j} + \xi_{j} \Delta n_{j} \right)} p(\Delta \epsilon_{j}, \Delta n_{j}),
\eea
where the parameters $\{ \lambda_{j}, \xi_{j} \}$ are often called the counting parameters. The moments of the distribution $p(\Delta \epsilon_{j}, \Delta n_{j})$ are obtained by applying multiple derivatives to this generating function evaluated at the origin $\lambda_{j} = \xi_{j} = 0$. As an example, the first moments of the ditribution are given by
\bea
\langle \Delta \epsilon_{j'}\rangle & = & -\left. \partial_{\lambda_{j'}} G(\lambda_{j} , \xi_{j}) \right|_{\lambda_{j} = \xi_{j} = 0}, \\
\langle \Delta n_{j'}\rangle & = &- \left. \partial_{\xi_{j'}} G(\lambda_{j} , \xi_{j}) \right|_{\lambda_{j} = \xi_{j} = 0}.
\eea

From this definition, we see that (\ref{QEFT}) is equivalent to a symmetry property of the generating function
\be \label{QEFTgen} \boxed{
G(\lambda_{j} , \xi_{j}) = G( A^{\epsilon}_{j} -\lambda_{j} , A^{n}_{j} - \xi_{j})
}
\ee
in terms of the thermodynamic affinities $A^{\epsilon}_{j} $ and $A^{n}_{j} $.

In the several models considered in this thesis, the two parallel channels will not exchange particles so that the total particle number in each conduction channel is conserved during the whole evolution
\be
\left[ N_{\alpha} , H(t) \right] = 0 \quad \forall t,
\ee
where $N_{\alpha} = N_{\alpha L}+N_{\alpha R} $ denotes the particle number operator in channel $\alpha$. As a consequence, there are only two independent matter currents by virtue of
\bea
\Delta n_{\alpha L} = - \Delta n_{\alpha R} 
\eea
for $\alpha =1$ and $2$. By further assuming a homogeneous temperature across the two circuits $\beta_{\alpha i} = \beta$, we get a bivariate fluctuation theorem for the particle fluxes in each circuit
\be \label{bivariateft} 
\frac{P(\Delta n_{1L},\Delta n_{2L}  )} {P(-\Delta n_{1L}, -\Delta n_{2L}  )} = \mbox{e}^{\left[  A^{n}_{1L} \Delta n_{1L} +  A^{n}_{2L} \Delta n_{2L}  \right] } 
\ee
in terms of the thermodynamic affinities $A^{n}_{\alpha L} =- \beta (\mu_{\alpha L} - \mu_{\alpha R})$. Again this relation leads to a symmetry property for the generating function of the moments
\be \label{253}
G(\xi_{1L} , \xi_{2L}) =  G(A^{n}_{1L} - \xi_{1L} , A^{n}_{2L} - \xi_{2L}) ,
\ee
where $\xi_{\alpha L}$ is the counting parameter for the number of particles that flow out of reservoir $\alpha L$.


\section{Long-time limit} \label{longtimelimit}


In order to obtain the fluctuation theorem (\ref{QEFT}) or  alternatively (\ref{QEFTgen}), we considered the intermediate quantum systems $H_{1C}$ and $H_{2C}$ to be part of the macroscopic left-hand reservoir. By doing so, we implicitly assumed them to be at equilibrium with these reservoirs at initial and final times. In typical counting experiments however \cite{PhysRevLett.96.076605, fujisawa2006, PhysRevLett.99.206804, gustavsson2009, PhysRevB.81.125331, PhysRevX.2.011001}, the statistical ensemble of these quantum subsystems is not controlled. As a result, this assumption is not necessarily valid, and we may observe deviations to the fluctuation relation (\ref{QEFT}) for measurements performed during a finite amount of time \cite{PhysRevB.81.125331}. 

Nevertheless these finite-time effects should decrease for increasing measurement times and we expect a steady state fluctuation theorem to hold for the statistics of the currents in the long-time limit \cite{RevModPhys.81.1665}. With this in mind, we introduce the cumulant generating function for the time averaged fluctuations of energy and matter fluxes in the long-time limit and make the connection to the large deviation function. These functions will show to be independent of the statistical properties of the finite quantum systems in latter applications. Here below, we give a steady state fluctuation theorem for the currents in terms of the cumulant generating function and the large deviation function \cite{andrieux2009monnai, RevModPhys.81.1665}.


\subsection{Cumulant generating function of the currents}


In the long time limit, the energy and number of particles flowing out of the reservoir $j \in \left\{1L, 1R, 2L\right\}$ is expected to grow linearly in time\footnote{This is true for all the systems considered in this thesis, though exceptions do exist \cite{esposito2008continuous}.}\cite{RevModPhys.81.1665}. We are thus interested in describing the statistical properties of the currents defined as
\bea \label{energycurrent}
J_{\epsilon j} & = & \frac{\Delta \epsilon_{j}}{T} , \\ \label{particlecurrent}
J_{n j} & = &  \frac{\Delta n_{j}}{T} ,
\eea
where $T$ denotes the time during which the interaction between the reservoirs is turned on. This can be done through the definition of the \textbf{cumulant generating function}
\bea \label{CGF}
\mathcal{G} (\lambda_{j} , \xi_{j}) & \equiv & \lim_{T \rightarrow \infty} - \frac{1}{T} \ln \langle \mbox{e}^{- \sum_{j} T \left( \lambda_{j} J_{\epsilon j}+ \xi_{j} J_{n j}\right)} \rangle_T \\
& = & \lim_{T \rightarrow \infty} - \frac{1}{T} \ln G(\lambda_{j} , \xi_{j})
\eea
in terms of the average (\ref{average2}) and provided the limit exists. The cumulant generating function generates all the statistical cumulants of the distribution corresponding to the random variables (\ref{energycurrent}) and (\ref{particlecurrent}) by applying multiple derivatives evaluated at zero counting parameters. As an example, the $k^{\mbox{th}}$ cumulants of the energy and particle currents out of reservoir $l$ are given by
\bea
K_{\epsilon l}^{k} & = & (-1)^{k+1} \partial_{\lambda_{l}}^{k}  \left. \mathcal{G} (\lambda_{j} , \xi_{j}) \right|_{\lambda_{j} = \xi_{j} = 0}, \\
K_{n l}^{k} & = & (-1)^{k+1} \partial_{\xi_{l}}^{k}  \left. \mathcal{G} (\lambda_{j} , \xi_{j}) \right|_{\lambda_{j} = \xi_{j} = 0}.
\eea

In typical experimental situations, one controls the statistical properties of the macroscopic reservoirs through the adjustment of temperatures and chemical potentials. However, it is in general not possible to control the initial and final conditions of the intermediate quantum systems $H_{\alpha C}$. As a consequence, the identification (\ref{systemresabs}) together with the initial equilibrium condition (\ref{initialdensity}) may not be valid and we expect deviations to relations (\ref{QEFT}) and (\ref{QEFTgen}) for short times. Nevertheless, these effects should become negligible when the interaction time $T$ increases as a consequence of the finiteness of the energy and matter content in the intermediate quantum systems implying
\be
 \lim_{T \rightarrow \infty} \frac{\Delta \epsilon_{\alpha C}}{T} =  \lim_{T \rightarrow \infty} \frac{\Delta n_{\alpha C}}{T} = 0,
\ee
where $\Delta \epsilon_{\alpha C}$ and $\Delta n_{\alpha C}$ denote, respectively, the energy and matter changes within the intermediate system $\alpha C$ during the time $T$ of interaction.

In the long-time limit, we thus expect that the cumulant generating function still satisfies the symmetry relation
\be \label{longtimeFT} \boxed{
\mathcal{G} (\lambda_{j} , \xi_{j}) =\mathcal{G} ( A^{\epsilon}_{j} -\lambda_{j} ,  A^{n}_{j} - \xi_{j})
}
\ee
in terms of the thermodynamic forces (\ref{affen}) and (\ref{affpart}), and regardless of the initial and final conditions of the intermediate quantum systems in both channels. This last point has been studied to some extent and it was shown that the initial correlations between two macroscopic reservoirs do not affect the long-time cumulants of the energy and particle transfer \cite{RevModPhys.81.1665}. In this thesis, we will directly evaluate the cumulant generating function for the specific models considered, and demonstrate the symmetry relation (\ref{longtimeFT}) in the long-time limit.

It is also worthy to mention that a rigorous proof of the symmetry relation (\ref{longtimeFT}) for the long time limit cumulant generating function from the transient fluctuation relation (\ref{QEFT}) was given by making the substitution (\ref{systemresabs}) and provided the limit (\ref{CGF}) exists without imposing the identical initial and final conditions \cite{andrieux2009monnai}.


\subsection{Large deviation theory and steady state fluctuation relations}\label{largedevth1}


The large deviation function of the currents $\mathcal{I} (J_{\epsilon j},J_{n j}) $ is defined in terms of the probability  $P(J_{\epsilon j},J_{nj})$ that the currents (\ref{energycurrent}) and (\ref{particlecurrent}) take the values $J_{\epsilon j}$ and $J_{n j}$ as
\be
\mathcal{I} (J_{  \epsilon j},J_{n j}) = \lim_{T \rightarrow \infty} -\frac{1}{T} \ln P(J_{\epsilon j},J_{n j})
\ee
provided the limit exists. Random variables fulfilling this last requirement are said to satisfy a large deviation principle. The existence of this limit relies on the exponential dominant behavior of the probability distribution $P(J_{\epsilon j },J_{n j})$ in the long time limit, that is 
\be \label{263}
P(J_{\epsilon j},J_{n j}) \approx C(J_{\epsilon j},J_{n j},T) \mbox{e}^{ -T  \, \mathcal{I} (J_{\epsilon j},J_{n j}) }
\ee
for large $T$, with
\be
\lim_{T \rightarrow \infty} \frac{1}{T} \ln C(J_{\epsilon j},J_{n j},T) = 0.
\ee

A fundamental result is the \textbf{G\"artner-Ellis theorem} \cite{gartner1977, ellis1984, touchette2009} relating the cumulant generating function to the large deviation function which can be stated as follows. Suppose the cumulant generating function $\mathcal{G} (\lambda_{j}, \xi_{j})$ defined in (\ref{CGF}) exists and is differentiable for all $\lambda_{ j}, \xi_{j} in \mathbb{R}$, then one is ensured that the stochastic variables $J_{\epsilon j}, J_{n j}$ do satisfy a large deviation principle with a large deviation function given as the Legendre-Fenchel transform\footnote{Note that the sign difference appearing in equation (\ref{gartner}) with respect to the usual definition of the Legendre-Fenchel transform is due to a sign difference in our definition of the cumulant generating function (\ref{CGF}) with respect to the usual convention used in mathematics. By instead defining $\mathcal{G} (\lambda_{j} , \xi_{j}) \equiv  \lim_{T \rightarrow \infty} \frac{1}{T} \ln \langle \mbox{e}^{ \sum_{j} T \left( \lambda_{j} J_{\epsilon j}+ \xi_{j} J_{n j}\right)} \rangle$ one gets the same result as in \cite{touchette2009}.} of the cumulant generating function defined by
\be \label{gartner}
\mathcal{I} (J_{\epsilon j},J_{n j}) = \sup\limits_{\lambda_{j}, \xi_{j}}  \left[   \mathcal{G}(\lambda_{j},\xi_{j})  -\sum_{j} (\lambda_{j} J_{\epsilon j} + \xi_{j} J_{n j}) \right].
\ee
This theorem does not prove that the random variables $J_{\epsilon j}$ and $J_{n j}$ do satisfy a large deviation principle since it assumes the existence of the cumulant generating function, which in itself is a non trivial point. However, it is in general much easier to calculate the cumulant generating function than the large deviation function. In the applications we consider, the direct calculation of the cumulant generating function will be enough to ensure the existence of a large deviation function as a consequence of the G\"artner-Ellis theorem. This large deviation function can be easily obtained from (\ref{gartner}).

Moreover, if the cumulant generating function calculated satisfies relation (\ref{longtimeFT}), one is ensured that the probability distribution $P(J_{\epsilon j} , J_{ n j})$ of the currents satisfies a steady state fluctuation theorem in the long-time limit. Indeed, the property (\ref{gartner}) can be used together with the symmetry relation (\ref{longtimeFT}) of the cumulant generating function to show that
\be
\mathcal{I} (J_{ j \epsilon},J_{n j}) -\mathcal{I} (-J_{ j \epsilon}, -J_{n j}) =- \sum_{j} \left[A^{\epsilon}_{j} J_{ j \epsilon} +A^{n}_{j} J_{n j}\right].
\ee
By using the definition of the large deviation function, this last equation can be translated into a \textbf{steady state fluctuation theorem} for the currents $J_{ j \epsilon}$ and $J_{n j}$ in the long-time limit \cite{andrieux2009monnai, RevModPhys.81.1665}
\be \boxed{
\lim_{T \rightarrow \infty} \frac{1}{T} \ln \frac{P(J_{ j \epsilon},J_{n j})}{P(-J_{ j \epsilon},-J_{n j})} =  \sum_{j} \left[A^{\epsilon}_{j} J_{ j \epsilon} +A^{n}_{j} J_{n j}\right] }.
\ee

An important point to note is that these fluctuation relations as well as the results discussed above are universal and independent of the detailed microscopic description composing the reservoirs of the system and the interaction protocol. The results derived here are directly generalized to an arbitrary number of reservoirs with several particle species in each reservoir. We decided however to directly consider the case of a four terminal circuit to settle notations in future discussions.

%% file: Chapter3.tex

\setcounter{chapter}{2}

\chapter{Counting statistics in open quantum systems} 

\label{Chapter3} 

\lhead{Chapter 3. \emph{Counting statistics in open quantum systems}} 


\section{Density matrix formalism}


In the previous chapter, we introduced the Laplace transform of the probability distribution of energy and matter transfers between grand-canonical baths. Here, we show how this generating function can be evaluated by constructing a modified evolution operator keeping track of the energy and particle flows out of the reservoirs \cite{2004PhRvB70k5327R, 2008PhRvL100o0601F, 2005EL69475F, 2005PhRvB72p5347W, 2006PhRvL96b6805B, 2006PhRvE73d6129E, 2006PhRvB73c3312K, 2006RvMaP18619D, 2007PhRvB76p1404E, 2007PhRvB75o5316E, PhysRevB.76.085408, PhysRevB.77.195315, RevModPhys.81.1665}. This modified evolution operator depends on the counting parameters associated with the fluctuating variables.

We first write the generating function in terms of a modified density matrix whose evolution is obtained by using the modified evolution operator. From this, we define a reduced density which will be used to describe the effective dynamics and counting statistics in open quantum systems.


\subsection{Modified evolution operator} \label{meo}


In this Section, we are going to rewrite the counting statistics for the fluctuations of energy and matter in each part of our system. In order to lighten notation, we will denote by $X$ the energy or particle number operator in one of the reservoirs composing our system. The generalization to the simultaneous counting of several operators is straightforward.

We denote by $x_{k}$ the eigenvalue of $X$ corresponding to the state vector $ | \Psi _{k} \rangle $ so that (\ref{eigenbasis1}) and (\ref{eigenbasis2}) are rewritten as
\be
X | \Psi _{k} \rangle=  x_{k} | \Psi _{k} \rangle .
\ee

With these notations, the probability $p(\Delta x , t)$ to observe a change $\Delta x = x_{k'} -x_{k}$ over two successive measurements of the operator $X$ at times $0$ and $t$ is given by
\be \label{32}
p(\Delta x , t) = \sum_{kk'} \delta (\Delta x - (x_{k'} -x_{k})) p_{kk'}
\ee
in terms of the probability $p_{kk'}$, introduced in Eq. (\ref{227}), to measure the system in states $ | \Psi _{k} \rangle$ at the initial time $0$ and $ | \Psi _{k'} \rangle$ at the final time $t$. We define the projection operator over the state vector $ | \Psi _{k} \rangle $ as
\be
P_{k} \equiv  | \Psi _{k} \rangle \langle \Psi _{k}|
\ee
satisfying $P_{k}^{2} = P_{k}$, and use this definition to express the probability $p_{kk'}$  as
\be \label{projectorsprob}
p_{kk' } = \mbox{Tr} \{ P_{k'} U (t ,0) P_{k} \,  \rho_{0} \, P_{k} U^{\dagger} (t,0) P_{k'} \}
\ee
in terms of the initial density matrix $\rho_{0}$. 

By further using the relation $f(X) = \sum_{k} f (x_{k}) P_{k}$, we can write the generating function of the distribution $p(\Delta x , t)$ as \cite{2007PhRvB75o5316E,PhysRevB.76.085408}
\bea \label{35}
G( -i \lambda ,t ) & \equiv & \int d(\Delta x) e^{i \lambda \Delta x} p(\Delta x , t) \\
& = &  \sum_{kk'} e^{i \lambda (x_{k'} -x_{k})} \mbox{Tr} \left\{ P_{k'} U (t ,0) P_{k} \,  \rho (0) \, P_{k} U^{\dagger} (t,0) P_{k'}  \right\} \\
& = &\mbox{Tr} \left\{ P_{k'} e^{i (\lambda/2) X } U (t ,0) e^{-i (\lambda/2) X } \tilde{\rho}_{0} \, e^{-i (\lambda/2) X } U^{\dagger} (t,0) e^{i (\lambda/2) X }P_{k'}  \right\} \label{lastgen}
\eea
in terms of $\tilde{\rho}_{0} = \sum_{k} P_{k} \,  \rho_{0} \, P_{k}$. As mentioned earlier, the operator $X$ stands for the energy or particle number operators of the macroscopic reservoirs composing our system. The reservoirs over which we perform the counting statistics will always be considered to be initially in a grand-canonical equilibrium such as (\ref{initialdensity}). As a consequence, the initial density matrix is always taken diagonal in the eigenbasis $ \{| \Psi _{k} \rangle \}$ so that
\be \label{38}
\tilde{\rho}_{0} =   \rho_{0}
\ee
for the purpose of this thesis\footnote{It is however possible to deal with initial coherences in the basis $\{ | \Psi _{k} \rangle \}$ as shown in \cite{RevModPhys.81.1665}.}.

Equation (\ref{lastgen}) for the generating function can be written in a more compact and convenient form for future calculations by introducing the modified density matrix $\rho (\lambda , t)$ defined by 
\be
\rho (\lambda , t) = U^{\lambda} (t ,0)  \rho_{0} U^{-\lambda \, \dagger} (t,0),
\ee
where the evolution is performed by means of the modified evolution operator
\be
U^{\lambda} (t ,0) \equiv e^{i (\lambda/2) X } U (t ,0) e^{-i (\lambda/2) X } 
\ee
in terms of the evolution operator $U (t,0)$, the measured operator $X$, and the counting parameter $\lambda$. With these definitions and the use of (\ref{lastgen}), the generating function simply reads
\be
G(-i \lambda , t) = \mbox{Tr} \{ \rho (\lambda , t) \},
\ee
the trace being performed over the whole Hilbert space.

The modified density matrix reduces to the usual density matrix for $\lambda = 0$, i.e. $\rho (0 , t) = \rho (t)$, so that the generating function is correctly normalized according to $G(0) = 1$. By introducing the modified Hamiltonian
\be \label{modifiedHam3}
H^{\lambda} \equiv e^{i (\lambda /2) X } H e^{-i (\lambda /2) X } ,
\ee
the modified evolution operator is shown to be solution of the differential equation
\be
\frac{d}{dt} U^{\lambda} (t ,0) = - \frac{i}{\hbar} H^{\lambda} U^{\lambda} (t ,0)
\ee
with the initial condition $U^{\lambda} (0 ,0)=I$. Similarly, the modified density operator obeys the equation of motion
\be \label{genvonneuman}
\frac{d}{dt} \rho (\lambda , t) = -\frac{i}{\hbar} \left[ H^{\lambda} , \rho (\lambda , t) \right]_{\lambda}
\ee
with initial condition $\rho (\lambda , 0) = \rho_{0}$ and in terms of the modified commutator 
\be \label{gencom}
\left[ O^{\lambda} , \, . \, \right]_{\lambda} \equiv O^{\lambda} \, . \, - \, . \, O^{-\lambda}
\ee
for any operator $O^{\lambda}$ depending on the counting parameter $\lambda$. Equation (\ref{genvonneuman}) is the starting point to derive a quantum master equation for the reduced density matrix introduced in next section.


\subsection{Reduced density matrix of an open quantum system}


The systems we will consider in this work are generically composed of a finite size open quantum subsystem $S$ exchanging energy and matter with infinite size reservoirs - which are also referred to as macroscopic baths - composing the environment $R$. The Hilbert space of the total system is given by the direct product $\mathcal{H} = \mathcal{H}_{S} \otimes \mathcal{H}_{R}$ of the subsystem Hilbert space $\mathcal{H}_{S}$ and the environment Hilbert space $\mathcal{H}_{R}$. This system is conveniently described by the Hamiltonian
\be
H = H_{S} +  H_{R} + V,
\ee
where $H_{S}$ is the Hamiltonian of subsystem $S$, $H_{R}$ is the Hamiltonian of the environment $R$, and $V$ is the interaction Hamiltonian describing in our case the exchange of energy and matter between these two parts. In the following, the interaction Hamiltonian is assumed to take the form
\be \label{interactionexpl}
V = \sum_{\kappa} S_{\kappa} R_{\kappa},
\ee
where $S_{\kappa}$ and $R_{\kappa}$ are operators acting on $\mathcal{H}_{S}$ and $\mathcal{H}_{R}$, respectively.

For our purpose, it is also useful to write the system operators in a basis $\left\{ | s\rangle \right\}$ composed of the system Hamiltonian eigenstates
\be \label{eigenbasis}
H_{S} = \sum_{s} E_s |s \rangle \langle s |,
\ee
where $E_{s}$ is the energy eigenvalue corresponding to the eigenstate $|s \rangle$. The system operators $S_{\kappa}$ can thus be expanded in this basis according to
\be
S_{\kappa} = \sum_{s s'} \langle s | S_{\kappa} | s' \rangle  |s \rangle \langle s' |
\ee
in terms of the matrix elements $\langle s | S_{\kappa} | s' \rangle $ of operator $S_{\kappa}$.

As mentioned earlier, the operator $X$ introduced in section \ref{meo} denotes the energy or particle number operator in the environment and, as a consequence, is an operator acting on $\mathcal{H}_{R}$. Bearing this in mind, the modified Hamiltonian accounting for the fluctuations of $X$ reads\footnote{The energy and particle number operators both commute with $H_{R}$ in the models we study. An exception is the model considered in Chapters \ref{Chapter5} where a tunneling term in the quantum point contact detector Hamiltonian is id in the free evolution $H_{R}$. However, no counting is performed on the fluxes in the quantum point contact in the mentioned Chapter.}
\be \label{modham}
H^{\lambda} = H_{S} + H_{R} + V^{\lambda},
\ee
where the modified interaction Hamiltonian has been defined as 
\bea \label{interactioncounting} 
V^{\lambda} & \equiv & \mbox{e}^{i (\lambda /2) X} V \mbox{e}^{-i ( \lambda /2) X} \\
& = & \sum_{\kappa} S_{\kappa}  \left[ \mbox{e}^{i (\lambda /2) X} R_{\kappa} \mbox{e}^{-i ( \lambda /2) X} \right]  \\
& \equiv & \sum_{\kappa} S_{\kappa} R_{\kappa}^{\lambda} \label{interactionexplicit}.
\eea

In order to calculate the statistics of operator $X$, we need to solve equation (\ref{genvonneuman}) or, at least, find a steady state solution when considering the long-time limit. However, such procedure is not well suited when considering an environment with an infinite number of degrees of freedom. Indeed, the time evolution of individual modes within the environment is not relevant when considering the statistics of nonequilibrium fluxes through an open quantum system, not to mention the technical difficulties that may arise when solving the dynamics of an infinite number of variables. Our interest thus lies in describing the effective dynamics in the subsystem $S$ obtained by tracing all the degrees of freedom in the environment. This can be done formally by introducing the reduced density matrix of subsystem $S$ as
\be \label{reducedens}
\rho_{S} (\lambda , t) \equiv \mbox{Tr}_{R} \{ \rho (\lambda , t) \},
\ee
where the trace $\mbox{Tr}_{R}$ is taken over the environment Hilbert space $\mathcal{H}_{R}$. One readily verifies from this definition that
\be \label{genfunred}
G(-i \lambda , t) = \mbox{Tr}_{S} \left\{ \rho_{S} (\lambda , t) \right\},
\ee
where the trace $\mbox{Tr}_{S}$ is taken over the subsystem Hilbert space $\mathcal{H}_{S}$. The definition (\ref{reducedens}) reduces to the usual reduced density matrix in the absence of counting parameter so that $\rho_{S} (t) = \rho_{S} (0, t)$, where $\rho_{S} (t) \equiv \mbox{Tr}_{R}\{  \rho (t) \}$, $\rho (t)$ being the total density matrix. 

As a consequence of the evolution equation (\ref{genvonneuman}) for $\rho (\lambda , t)$, we directly infer an evolution equation for the reduced density matrix as
\be
\dot{\rho}_{S} (\lambda , t) = -\frac{i}{\hbar}\mbox{Tr}_{R} \left\{ \left[ H_{\lambda} , \rho (\lambda , t) \right]_{\lambda} \right\}.
\ee
However, this equation is tractable only if it can be closed in the quantity of interest $\rho_{S} (\lambda , t)$. This last condition is satisfied only by the introduction of additional assumptions over the environment properties and its interaction with the quantum subsystem as explained in the following section.


\section{Master equation for a weakly coupled open quantum system}


In this Section, we write a closed equation for the reduced density matrix of the subsystem $S$, which describes its effective dynamics under the influence of the environment, and accounts for the energy and particle transfers between the macroscopic baths through the counting parameters. We treat the system-environment interaction to second order in perturbation theory and consider a factorized form for the initial density matrix. This is known as the \textbf{Born approximation} and is justified in the weak coupling regime if the reservoirs are large enough to remain unaffected by the small quantum system and the interaction \cite{breuer2007theory}.

Moreover, we perform the \textbf{Markovian approximation} on the resulting equation which amounts to neglect the effect of the transient dynamics and memory effects of the environment on the subsystem dynamics \cite{JChemphys1992, PhysRevLett.73.1060, JChemphys1997, JChemphys1999, JPhysChemB2005, PhysRevA.77.032104, breuer2007theory}. This is reasonable provided we consider macroscopic baths with short correlation times compared to the time scales of the system evolution. This gives us a Markovian quantum master equation for the effective dynamics of the modified density matrix in subsystem $S$.

We finally write this equation as a \textbf{Lindblad master equation} \cite{Linblad, RevModPhys.52.569, breuer2007theory} by averaging over the fast oscillations due to the subsystem free dynamics. This procedure is known in literature as the \textbf{rotating wave approximation} \cite{gardinerzoller, breuer2007theory, tannoudjiintproc, PhysRevA.78.022106, RevModPhys.52.569, vankampen}. A consequence of this approximation is the dynamical decoupling between the sets of diagonal and off-diagonal components of the density matrix, which corresponds at zero counting parameter to the populations and coherences in subsystem $S$, respectively.


\subsection{Perturbative expansion and the Born-Markov approximation} \label{pertexpandbmapprox}


The derivation of a master equation for the reduced density matrix by perturbation theory in the interaction Hamiltonian $V$ is most easily accomplished in the so-called interaction picture. With this in mind, we rewrite the total Hamiltonian (\ref{modham}) as 
\be
H^{\lambda} = H_{0} + V^{\lambda}
\ee
with 
\be \label{freehamiltonian}
H_{0} \equiv  H_{S}   +  H_{R}
\ee
describing the free evolution of both the subsystem and the environment. The interaction representation $O_{I}$ of any operator $O$ is defined by
\bea
O_{I} (t) & = & \mbox{e}^{\frac{i}{\hbar} H_{0} t } \, O \, \mbox{e}^{-\frac{i}{\hbar} H_{0} t} \\
& \equiv & \mbox{e}^{-\mathcal{L}_{0} t} \, O ,
\eea
where we introduced the superoperator $\mathcal{L}_{0}$ acting on operators as 
\be
\mathcal{L}_{0} \, O = -\frac{i}{\hbar}\left[ H_{0} , O \right].
\ee
Accordingly, the density matrix in the interaction representation $\rho_{I} (\lambda , t)$ satisfies the equation of motion 
\bea 
\dot{\rho}_{I} (\lambda ,t) & = & -\frac{i}{\hbar} \left[ V_{I}^{\lambda} (t), \rho_{I} (\lambda , t) \right]_{\lambda} \\
& \equiv & \mathcal{L}_{I}^{\lambda} (t) \rho_{I} (\lambda , t) \label{intpictevol}
\eea
obtained from the dynamical equation for the modified density matrix (\ref{genvonneuman}), together with the definition of the interaction representation and Eq. (\ref{gencom}).

The last equation (\ref{intpictevol}) can be integrated over time and solved up to second order in the interaction Hamiltonian $V$ which yields
\be \label{rhoI}
\rho_{I} (\lambda , t) \approx \rho_{I} (0) + \int_{0}^{t} dt_{1} \, \mathcal{L}_{I}^{\lambda} (t_{1}) \rho_{I} (0) + \int_{0}^{t} dt_{1} \, \int_{0}^{t_{1}} dt_{2}  \, \mathcal{L}_{I}^{\lambda} (t_{1}) \mathcal{L}_{I}^{\lambda} (t_{2}) \rho_{I} (0)
\ee
in terms of the initial density matrix $\rho_{I} (0) = \rho_{0}$. Our aim is to trace this equation over the environment degrees of freedom in order to write an effective evolution operator for the reduced density matrix $\rho_{S} (\lambda , t)$. However, simply tracing equation (\ref{rhoI}) does not give a closed expression in terms of $\rho_{S} (\lambda , t)$. In order to get the desired result, we must make the assumption that the initial density matrix of the total system is initially in a factorized form
\be \label{factorization}
\rho_{0} = \rho_{S} (0) \otimes \rho_{R},
\ee
where $\rho_{S} (0)$ is the initial reduced density matrix of subsystem $S$ and $\rho_{R}$ is the initial density matrix of the environment. This assumption, together with the second order approximation (\ref{rhoI}) are known as the \textbf{Born approximation} which amounts to neglect the influence of the small quantum subsystem on the weakly coupled and macroscopic baths.

Back to the Schr\"odinger picture and defining the superoperator $\mathcal{L}_{S} O_{S} \equiv -\frac{i}{\hbar} \left[ H_{S}, O_{S} \right]$ for any operator $O_{S}$ of subsystem $S$, we can trace equation (\ref{rhoI}) to get
\be
\rho_{S} (\lambda , t) = \mathcal{U} (\lambda , t) \rho_{S} (\lambda , 0),
\ee
where the effective evolution superoperator $ \mathcal{U} (\lambda , t) $ performs the evolution on the subsystem density matrix and is given by
\be \label{uevolution}
 \mathcal{U} (\lambda , t)  = \mbox{e}^{\mathcal{L}_{S} t} \mbox{Tr}_{R} \left\{ \left[  I + \int_{0}^{t} dt_{1} \,\mathcal{L}^{\lambda}_{I} (t_{1}) + \int_{0}^{t} dt_{1} \, \int_{0}^{t_{1}} dt_{2} \, \mathcal{L}^{\lambda}_{I} (t_{1})  \mathcal{L}^{\lambda}_{I} (t_{2})  \right] \rho_{R} \right\}.
\ee
We note that this operator is generally not unitary due to the irreversible character of the effective dynamics of the subsystem $S$ influenced by the environment $R$.

We are now in position to write down a dynamical equation for the subsystem density matrix by noting that 
\be
\dot{\rho}_{S} (\lambda , t) = \dot{\mathcal{U}} (\lambda , t) \mathcal{U}^{-1} (\lambda , t) \rho_{S} (\lambda , t),
\ee
where $\mathcal{U}^{-1} (\lambda , t)$ is the inverse of the effective evolution operator calculated to second order in the interaction Hamiltonian $V$. By inserting (\ref{uevolution}) into this last equation, we get a modified master equation describing the effective evolution of the open quantum system
\be \label{effevol}
\dot{\rho}_{S} (\lambda , t) = \left[ \mathcal{W}^{\lambda}_{0} + \mathcal{W}^{\lambda}_{1} + \left( \mathcal{W}_{2}^{\lambda} - (\mathcal{W}_{1}^{\lambda} )^{2} \right) \right] \rho_{S} (\lambda , t),
\ee
where we defined the super operators
\bea
 \mathcal{W}^{\lambda}_{0} & \equiv & \mathcal{L}_{S} ,\\
 \mathcal{W}^{\lambda}_{1} & \equiv & \mbox{Tr}_{R} \left\{ \mathcal{L}^{\lambda}_{I} (0) \rho_{R} (t) \right\}, \\
 \mathcal{W}^{\lambda}_{2} & \equiv & \int_{0}^{t} d\tau \, \mbox{Tr}_{R} \left\{ \mathcal{L}^{\lambda}_{I} (0) \mathcal{L}^{\lambda}_{I} (-\tau) \rho_{R} (t) \right\}  \label{corfunc1},
\eea
together with the environment density matrix at time $t$ given by 
\be \label{envevol}
\rho_{R} (t) \equiv \mbox{e}^{-\frac{i}{\hbar} H_{R} t} \rho_{R} \, \mbox{e}^{ \frac{i}{\hbar} H_{R} t}.
\ee
For our purpose, we may assume the environment density matrix $\rho_{R}$ to be stationary with respect to the environment evolution satisfying
\be \label{stationarycondition}
\left[ \rho_{R} , H_R \right] = 0.
\ee

The equation (\ref{effevol}) is closed in the quantity $\rho_{S} (\lambda ,t)$ as announced and reduces to a perturbative dynamical equation for the quantity $\rho_{S} (t) = \rho_{S} (0 , t)$ when setting the counting parameter $\lambda$ to $0$. It is also non-Markovian as a consequence of the time integral over past history from times $0$ to $t$ in the term (\ref{corfunc1}) of its generator.

However, as we consider an environment of infinite size, we may assume its time correlation functions
\be
\mbox{Tr} \left\{ R_{\kappa}  R_{\kappa'} (-\tau)\, \rho_{R} \right\}
\ee
to decay fast enough on a short time scale $\tau_C$ which is such that 
\be \label{markovhyp}
\tau_{C} \ll \tau_R,
\ee
where $\tau_R$ is the time scale over which the state of the subsystem $S$ varies appreciably \cite{breuer2007theory,JChemphys1999}.

As a consequence, we may approximate the integral in (\ref{corfunc1}) by sending its upper bound to $\infty$. Equation (\ref{effevol}) thus becomes a homogeneous Markovian master equation with
\bea
 \mathcal{W}^{\lambda}_{0} & = & \mathcal{L}_{S}  \label{zerothorder} , \\
 \mathcal{W}^{\lambda}_{1} & = & \mbox{Tr}_{R} \left\{ \mathcal{L}^{\lambda}_{I} (0) \rho_{R}  \right\} \label{linorder} , \\ 
 \mathcal{W}^{\lambda}_{2} & \approx & \int_{0}^{\infty} d\tau \, \mbox{Tr}_{R} \left\{ \mathcal{L}^{\lambda}_{I} (0) \mathcal{L}^{\lambda}_{I} (-\tau) \rho_{R}  \right\} \label{corfunc2},
\eea
and where the environment stationary density matrix satisfyies Eq. (\ref{stationarycondition}). The hypothesis (\ref{markovhyp}) and its consequences (\ref{linorder}) and (\ref{corfunc2}) on the quantum master equation are usually called the \textbf{Markovian approximation} \cite{JChemphys1992, PhysRevLett.73.1060, JChemphys1997, JChemphys1999, JPhysChemB2005, PhysRevA.77.032104, breuer2007theory}. Together with the second order approximation in the system-environment interaction (\ref{rhoI}) and the initial factorization of the total density matrix (\ref{factorization}), it is also referred to as the \textbf{Born-Markov approximation}.

Considering the above results and using the expression (\ref{interactionexpl}) for the interaction between $S$ and $R$, we can explicitly write down the action of superoperators (\ref{zerothorder})-(\ref{corfunc2}) on the reduced density matrix as\footnote{We assume the measured operator $X$ to commute with the initial density matrix of the reservoirs in obtaining the following results, i.e. $\left[ X , \rho_R \right]$. This assumption is satisfied in all the cases considered in this thesis. However it is by no means fundamental and is used only to lighten notations. Otherwise, generalization is straightforward by introducing the additional correlation functions $\langle R^{\pm\lambda}_{\kappa} (\tau)R^{\pm \lambda}_{\kappa'}  \rangle$.}
\bea
 \mathcal{W}^{\lambda}_{0} \rho_{S}(\lambda , t)& = & -\frac{i}{\hbar}  \left[ H_{S}, \rho_{S} (\lambda , t) \right] \label{zerothorder2} , \\
 \mathcal{W}^{\lambda}_{1} \rho_{S}(\lambda , t)& = &- \frac{i}{\hbar}  \sum_{\kappa} \left( \langle R^{\lambda}_{\kappa} \rangle S_{\kappa} \rho_ {S} (t) - \langle R^{-\lambda}_{\kappa} \rangle  \rho_ {S} (t) S_{\kappa} \right) , \\
 \mathcal{W}^{\lambda}_{2}\rho_{S}(\lambda , t) & = &\frac{1}{\hbar^{2}} \int_{0}^{\infty}d\tau \sum_{\kappa , \kappa '}  \Big(   \alpha_{\kappa \kappa '} (\lambda , \tau )  S_{\kappa '} (-\tau) \rho_{S}(t) S_{\kappa} + \alpha_{\kappa ' \kappa } (\lambda ,-\tau)  S_{\kappa}\rho_{S}(t)S_{\kappa '} (-\tau) 
 \nonumber \\
&-&   \alpha_{\kappa \kappa '} (0 ,\tau )  S_{\kappa} S_{\kappa '} (-\tau) \rho_{S}(t)   - \alpha_{\kappa ' \kappa } (0 ,-\tau) \rho_{S}(t) S_{\kappa '} (-\tau)S_{\kappa}   \Big)  \label{secondordermarkov},
\eea
where we defined the average of a bath operator on the initial density matrix of the environment as
\be \label{mean}
 \langle R^{\lambda}_{\kappa} \rangle = \mbox{Tr}_{R} \{ R^{\lambda}_{\kappa}  \rho_{R} \}
\ee
and the modified correlation functions in the environment by\footnote{Equation (\ref{stationarycondition}) can be used to show that the correlation functions calculated by averaging over $\rho_{R}$ only depend on the time difference of their arguments, so that $\langle R^{\lambda}_{\kappa} (t_{1}) R^{\lambda}_{\kappa '} (t_{2})  \rangle = \langle R^{\lambda}_{\kappa} (t_{1}-t_{2}) R^{\lambda}_{\kappa '}  \rangle $.}
\begin{equation} \label{corfunc}
\alpha_{\kappa \kappa '} (\lambda ,\tau ) =  \langle R^{-\lambda}_{\kappa} (\tau)R^{\lambda}_{\kappa'}  \rangle
\end{equation}
with bath operators evolving according to the environment Hamiltonian $H_{R}$
\be \label{bathevol}
 R^{\lambda}_{\kappa}(\tau) = e^{i H_{R} \tau} R^{\lambda}_{\kappa} e^{-i H_{R} \tau}.
\ee
The correlation functions of the environment $\alpha_{\kappa \kappa '}(\tau)$ are given in terms of the modified correlation functions evaluated at zero counting parameter
\be
\alpha_{\kappa \kappa '}(\tau) = \alpha_{\kappa \kappa '}(0, \tau ).
\ee
From the above results, it is clear that the influence of the environment on the effective dynamics of the subsystem $S$ is encoded in these correlation functions.


\subsection{Lindblad form and rotating wave approximation}\label{linbrot}


The Markovian master equation (\ref{effevol}) with (\ref{zerothorder2})-(\ref{bathevol}) describing the effective evolution of the reduced density matrix of the open system $S$ is known to break the positivity condition on the density matrix for initial conditions chosen near the border of its domain of positivity. This problem can be dealt with by applying a slippage on the initial condition describing the non-Markovian dynamics of the quantum system $S$ during the short relaxation time of the bath coupling operators \cite{JChemphys1999}.

However, in the specific cases we will consider, the master equation can be written as a Lindblad master equation which is known to be the most general quantum master equation preserving the positivity of the density matrix. This is done by performing the \textbf{rotating wave approximation} (RWA) which consists in averaging out the fast oscillations due to the free subsystem evolution by coarse graining the dynamics over a time scale $\Delta t$ corresponding to the sampling time of the observation. This approximation is justified provided the time scale $\Delta t$ is intermediate between the time scale $\tau_S$ of the free oscillations and the relaxation time $\tau_{R}$ of the subsystem
\be
\tau_{S} \equiv \omega_S^{-1} \ll \Delta t \ll \tau_R,
\ee
where $\omega_{S} = \min_{ss'} \hbar^{-1}( E_{s}-E_{s'} ) $ is the smallest Bohr frequency of the system \cite{gardinerzoller, breuer2007theory, tannoudjiintproc}.

In order to perform the RWA, it is useful to write the master equation (\ref{effevol}) in the system eigenbasis (\ref{eigenbasis}). By introducing the matrix elements $\rho_{ss'} (\lambda ,t) \equiv \langle s | \rho_S (\lambda , t) | s' \rangle$, we can write
\be
\dot{\rho}_{ss'} (\lambda, t) = \sum_{\tilde{s} \tilde{s}'} \mathcal{W}^{\lambda}_{ss' | \tilde{s}\tilde{s}' } \rho_{\tilde{s}\tilde{s}'} (\lambda, t),
\ee
where we introduced the evolution superoperator
\be
 \mathcal{W}^{\lambda} =  \left[ \mathcal{W}^{\lambda}_{0} + \mathcal{W}^{\lambda}_{1} + \left( \mathcal{W}_{2}^{\lambda} - (\mathcal{W}_{1}^{\lambda} )^{2} \right) \right] 
\ee
with (\ref{zerothorder2}) - (\ref{secondordermarkov}). The matrix elements of this superoperator are defined through its action on the projectors $|\tilde{s} \rangle  \langle \tilde{s}' |$ by
\be \label{superoperatorRWA}
\mathcal{W}^{\lambda}_{ss' |\tilde{s}\tilde{s}' } = \langle s |\left( \mathcal{W}^{\lambda} \left(|\tilde{s} \rangle  \langle \tilde{s}' | \right) \right) |s' \rangle.
\ee

This equation is conveniently expressed in the interaction picture with respect to the free Hamiltonian $H_{S}$ as
\be
\dot{\rho}^{I}_ {ss'} (\lambda, t) =  \sum_{\tilde{s}\tilde{s}'} \left[ i \omega_{\tilde{s}\tilde{s}'} \delta_{s \tilde{s}} \delta_{\tilde{s}'s'} +  \mathcal{W}^{\lambda} _{ss' | \tilde{s}\tilde{s}'}\right] e^{i (\omega_{\tilde{s}\tilde{s}'} + \omega_{s's} )t} \rho^{I}_{\tilde{s}\tilde{s}'} (\lambda, t),
\ee
where
\be
\rho^{I}_ {ss'} (\lambda, t) = e^{i \omega_{ss'} t} \rho_{ss'} (\lambda, t)
\ee
and with the Bohr frequencies of the system defined by $\omega_{ss'} \equiv \frac{E_{s}-E_{s'}}{\hbar}$.

The RWA is performed by applying the average $\lim_{\Delta t \rightarrow \infty} \frac{1}{\Delta t} \int_{-\Delta t/2}^{\Delta t/2} dt$ to the superoperator (\ref{superoperatorRWA}) itself, and using relation
\be
\lim_{\Delta t \rightarrow \infty} \frac{1}{\Delta t} \int_{-\Delta t/2}^{\Delta t/2} dt \, e^{i (\omega_{\tilde{s}\tilde{s}'} + \omega_{s's} )t} 
= (1- \delta_{\tilde{s} s }) \delta_{\tilde{s}\tilde{s}'} \delta_{s's} + \delta_{\tilde{s}s} \delta_{\tilde{s}' s'} .
\ee
Back to the Schr\"odinger picture, this treatment results in dynamical equations for the diagonal $\rho_{ss} (\lambda , t)$ and off-diagonal $\rho_{ss'} (\lambda ,t)$ elements of the reduced density matrix with $s\neq s'$. At zero counting parameter, $\lambda = 0$, the diagonal elements become the occupation probabilities in the subsystem $S$, also called populations, while the off-diagonal elements give the coherences in the quantum subsystem. It is straightforward to show that the diagonal elements evolve according to
\be \label{stochmod}
\dot{\rho}_{ss} (\lambda ,t) = \hbar^{-2}\sum_{\kappa \kappa '} \sum_{\tilde{s}} \langle s | S_{\kappa} |\tilde{s} \rangle \langle \tilde{s}| S_{\kappa'} | s \rangle \left[  \hat{\alpha}_{\kappa' \kappa} ( \lambda, \omega_{  \tilde{s} s}) - \hat{\alpha}_{\kappa \kappa'} (0, - \omega_{\tilde{s}s} ) \delta_{s\tilde{s}}  \right] \rho_{\tilde{s}\tilde{s}} (\lambda ,t)
\ee
in terms of the Fourier transforms of the modified correlation functions of the environment
\be \label{fouriertransform}
\hat{\alpha}_{\kappa \kappa'} (\lambda , \omega) = \int_{-\infty}^{\infty}d\omega e^{i \omega \tau} \alpha_{ \kappa \kappa'} (\lambda, \tau).
\ee

The off-diagonal matrix elements $\rho_{ss'} (\lambda , t)$, for $s\neq s'$, evolve independently of each other according to
\be \label{rwacoherences}
\dot{\rho}_ {ss'} (\lambda , t) =(- \Upsilon_{ss'} - i \Omega_{ss'} ) \rho_{ss'} ( \lambda, t),
\ee
with the damping rates given by
\begin{multline} \label{damping}
\Upsilon_{ss'} = \hbar^{-2} \sum_{\kappa \kappa'} \left[ - \langle s' | S_{\kappa} |s' \rangle  \langle s |S_{\kappa'} | s \rangle \hat{\alpha}_{\kappa \kappa'} (0,0) \right. \\ \left.
 + \frac{1}{4 \pi}\sum_{\tilde{s}} \left( \langle s | S_{\kappa} | \tilde{s} \rangle \langle \tilde{s} | S_{\kappa'} | s \rangle \hat{\alpha}_{\kappa \kappa'} (0, \omega_{s \tilde{s} } ) + \langle s' | S_{\kappa} | \tilde{s} \rangle \langle s | S_{\kappa'} | s' \rangle \hat{\alpha}_{\kappa \kappa'} (0, \omega_{\tilde{s} s'} )  \right) \right] 
\end{multline}
and the frequencies by
\begin{multline}  \label{oscillations}
\Omega_{ss'} = \omega_{ss'} + \lambda \sum_{\kappa} \left(\langle s |S_{\kappa} | s \rangle  - \langle s' | S_{\kappa} | s' \rangle \right) \langle R_{\kappa} \rangle \\+ \frac{\lambda^{2}}{2 \pi} \sum_{\tilde{ s}} \sum_{\kappa \kappa'} \left[   \mbox{p.v.}\int_{-\infty}^{\infty} dx \,  \langle s | S_{\kappa} | \tilde{s} \rangle \langle \tilde{s} |  S_{\kappa'} | s \rangle \frac{\hat{\alpha}_{\kappa \kappa'} (0, x)}{\omega_{s\tilde{s}} - x}  + \right. \\
 \left. \mbox{p.v.}\int_{-\infty}^{\infty} dx \, \langle s' | S_{\kappa} | \tilde{s} \rangle \langle \tilde{s} | S_{\kappa'} | s' \rangle  \frac{\hat{\alpha}_{\kappa \kappa'} (0, x)}{\omega_{ \tilde{s}s'} + x}  \right],
\end{multline}
where $\mbox{p.v.}$ denotes the Cauchy principal value of the integrals\footnote{The Cauchy principal value arises from the Fourier transform of the Heaviside theta function which is a combination of a Dirac delta distribution and a Cauchy principal value distribution. Indeed, one gets
\bea
\int_{0}^{\infty} d\tau \, f(\tau) e^{i \omega \tau} = \frac{1}{2} \hat{f} (\omega) -\frac{1}{2 \pi i} \mbox{p.v.} \int_{-\infty}^{\infty} dx \, \frac{\hat{f} (x)}{\omega - x},
\eea 
where $\hat{f} (\omega)$ is the Fourier transform of $f(\tau)$.}.

Most remarkably, these equations do not couple the sets of diagonal and off-diagonal matrix elements so that we are able to study the stochastic dynamics of the modified occupation probabilities 
\be \label{370}
g_{s} (\lambda , t) \equiv \rho_{ss} (i \lambda  , t)
\ee
independently of the quantum coherences. 

Moreover, since the generating function of the fluctuations of operator $X$ is expressed through the trace (\ref{genfunred}), we get the important result that the counting statistics of $X$ is completely determined by the diagonal elements of the modified density matrix $g_{s} (\lambda ,t)$ and their closed dynamics.


\section{Stochastic dynamics}


In this section, we give a formal solution to the modified rate equation (\ref{stochmod}) which enables us, in the long-time limit, to write the cumulant generating function in terms of the dominant eigenvalue of the modified rate matrix \cite{RevModPhys.81.1665}. This is an important practical result since it will be used in several applications to be found in this thesis.

We end up by converting the modified rate equation into a rate equation for the occupation probabilities in the subsystem $S$ conditional to the observation of a change $\Delta x$ of $X$ since initial time. The resulting generator is a convolution operator with an integration over all the possible changes in $\Delta x$ caused by stochastic jumps in $S$. The corresponding rate matrix will be used in order to characterize the nonequilibrium thermodynamics associated with the energy and particle fluxes in Chapter \ref{Chapter6}.


\subsection{General solution of the modified master equation} \label{generalsolutionof}


The dynamical equation (\ref{stochmod}) for the quantities $g_{s} (\lambda , t)$ can be written in a more compact form by introducing the vector
\be
{\bf g} (\lambda , t) = \left[ 
\begin{array}{c}
g_{1} (\lambda , t) \\
g_{2} (\lambda , t) \\
\vdots \\
g_{n_{S}} (\lambda ,t )
\end{array}
 \right],
\ee
where $n_{S}$ is the number of eigenstates in the quantum system $S$. The modified master equation (\ref{stochmod}) thus reads in matrix form 
\be \label{stochcompact}
\dot{{\bf g}}(\lambda , t) = {\bf W}( \lambda) \cdot {\bf g}(\lambda , t),
\ee
where the dot $\cdot$ denotes the matrix product and ${\bf W} (\lambda)$ is a matrix whose matrix elements are given by
\be \label{transitionratesgeneral3}
\left[ {\bf W} (\lambda) \right]_{ss'} = \hbar^{-2} \sum_{\kappa \kappa '}  \langle s | S_{\kappa} | s' \rangle \langle s' | S_{\kappa'} | s \rangle \left[  \hat{\alpha}_{\kappa' \kappa} ( i \lambda, \omega_{s' s}) - \hat{\alpha}_{\kappa \kappa'} (0, -\omega_{s' s} ) \delta_{ss'}  \right].
\ee

In this regard, it is straightforward to write a formal solution of equation (\ref{stochcompact}) as
\be \label{generalsolution}
{\bf g}(\lambda , t) =\mbox{e}^{ t {\bf W} (\lambda) } \cdot {\bf p}_0
\ee
in terms of the matrix exponential of the rate matrix ${\bf W} (\lambda )$ and where ${\bf p}_0 $ is the initial probability distribution over the quantum states of subsystem $S$.


\subsection{Full counting statistics in the long-time limit}\label{fcslongtime}


As shown in section \ref{meo}, the generating function of the statistical distribution describing the time fluctuations of operator $X$ is given in terms of the reduced density matrix of the subsystem $S$ by
\bea
G (\lambda , t) & = & \mbox{Tr} \left\{ \rho_{S} ( -i \lambda , t) \right\} \nonumber \\
&  = & {\bf 1}^{\top} \cdot {\bf g} (\lambda , t) \nonumber \\
& = & {\bf 1 }^{\top} \cdot \mbox{e}^{ t {\bf W} ( \lambda ) }  \cdot {\bf p }_0
 \label{genfuncvect},
\eea
where we denoted the trace in terms of the vectors ${\bf g} (\lambda , t)$ and ${\bf 1}^{\top}$ which is the matrix transposition of
\be
{\bf 1 }= \left[ 
\begin{array}{c}
1 \\
1 \\
\vdots \\
1
\end{array}
 \right].
\ee 
In the last line of equation (\ref{genfuncvect}), we used the general solution (\ref{generalsolution}) of the modified master equation. By further introducing the right eigenvectors ${\bf u }_{s}$ of the modified rate matrix
\be
{\bf W} (\lambda ) \cdot {\bf u}_{s} = - w_{s}(\lambda) {\bf u}_{s}
\ee
the generating function be written as
\be \label{geneigen}
G (\lambda , t) = \sum_{\tilde{s}} \mbox{e}^{-w_{\tilde{s}} (\lambda) t} \left( {\bf 1} \cdot {\bf u}_{\tilde{s}} \right) \left( {\bf u}_{\tilde{s}} \, \cdot {\bf p}_0 \right).
\ee
This last relation enables us to study the statistics of the current fluctuations
\be \label{379}
J_{x} = \frac{\Delta x}{t}
\ee
of operator $X$ in the long-time limit. Indeed the cumulant generating function describing the fluctuations of the currents has been defined in Chapter \ref{Chapter1} as
\be \label{longtimecgfvect}
\mathcal{G} (\lambda) \equiv \lim_{t \rightarrow \infty} -\frac{1}{t} \ln G(\lambda,t).
\ee
By introducing (\ref{geneigen}) into this definition and keeping the dominant terms in the long time-limit we simply get \cite{RevModPhys.81.1665}
\be
\mathcal{G} (\lambda) =  w (\lambda),
\ee
where $w(\lambda)$ is the dominant eigenvalue of the modified rate matrix, that is
\be \label{382}
 w(\lambda) = -\max_{s}  \mbox{Re} \left\{- w_{s} (\lambda) \right\}
\ee
the maximum being taken over the whole set of eigenvalues of ${\bf W} (\lambda )$ and $\mbox{Re} \left\{ - w_{s} (\lambda) \right\}$ denoting the real part of $w_{s} (\lambda)$. This remarkable result will be used in order to calculate the cumulant generating function in all the applications we consider. In practice, the evaluation of the largest eigenvalue ${\bf W} (\lambda )$ is easily performed.


\subsection{Back to the probability distribution} \label{backtoprob}


From the results of section \ref{generalsolutionof}, a rate equation can be derived for the occupation probabilities $p_{s} (\Delta x , t)$ that the system $S$ is in quantum state $|s \rangle $ at time $t$ while observing a change $\Delta x$ in operator $X$ between the times $0$ and $t$. This quantity is simply given as the inverse Fourier transform of $g_{s} (i \lambda , t)$, so that we have in matrix notation
\be \label{383}
{\bf p} (\Delta x , t) = \frac{1}{2 \pi} \int_{-\infty}^{\infty}d\lambda \, e^{-i \lambda \Delta x} \, {\bf g} (-i \lambda , t).
\ee

By applying an inverse Fourier transform to each member of equation (\ref{stochcompact}) we obtain the rate equation
\be \label{rateequpart}
\dot{{\bf p}} (\Delta x , t) = \int d\delta x\,   \hat{{\bf W}} (\delta x) \, \cdot {\bf p} (\Delta x - \delta x , t) ,
\ee
where we introduced the inverse Fourier transform of the matrix ${\bf W} (\lambda )$
\be \label{rateparticspace}
\hat{{\bf W}} (\delta x)  =\frac{1}{2 \pi} \int_{-\infty}^{\infty}d\lambda \, e^{-i \lambda \delta x}  \, {\bf W} (-i  \lambda ).
\ee

The integral over $\delta x$ in (\ref{rateequpart}) is an integral over all the possible changes in $\delta x$ which may occur when the system undergoes a transition between two quantum states. In some cases this integral will reduce to a discrete sum, but not necessarily as we will see for example in Chapter \ref{Chapter6}, where the energy transfers between a double quantum dot and a quantum point contact can take a continuous set of values.

The convolution in equation (\ref{rateequpart}) is a direct consequence of the convolution theorem for the Fourier transform: the product in the left-hand side of (\ref{stochcompact}) becomes a convolution in (\ref{rateequpart}). This makes the rate matrix of the populations more difficult to deal with. However, its rate matrix $(\ref{rateparticspace})$ will be useful in discussing the resulting nonequilibrium thermodynamics associated with the nonequilibrium fluxes of energy and matter.

%% file: Chapter4.tex
\chapter{Fluctuation theorems for capacitively coupled electronic currents} 

\label{Chapter4} 

\lhead{Chapter 4. \emph{Fluctuation theorems for capacitively coupled electronic currents}} 

Current fluctuations in mesoscopic circuits have generated a large amount of works in recent years, both theoretical and experimental. The study of these fluctuations gives us access to additional information about the transport processes beyond the mean currents. Fluctuations in such circuits are nowadays accessible through the real time tracking of the quantum dots electronic occupation by use of an auxiliary detection circuit. The transitions between charge states on the quantum dots induce measurable changes in the detectors current through Coulomb interaction. However, this interaction also affects the measured circuit, and we expect the statistical properties of its fluctuations to be affected by the detector which is itself an out of equilibrium circuit. Recalling the fluctuation relations established for coupled circuits in Chapter \ref{Chapter2}, this last point suggests that the detection setup should be included in a theoretical description of the transport processes in such systems.

In the present Chapter, we address this issue by performing the full counting statistics in a system composed of two capacitively coupled parallel transport channels, each containing a single quantum dot in contact with two electron reservoirs.  The nonequilibrium steady state of the system is controlled by two affinities or thermodynamic forces, each one determined by the two reservoirs of each channel. We investigate the status of a single-current fluctuation theorem starting from the fundamental two-current fluctuation theorem established in Chapter \ref{Chapter2}, which is a consequence of microreversibility. We show that the single-current fluctuation theorem holds in the limit of a large Coulomb repulsion between the two parallel quantum dots, as well as in the limit of a large current ratio between the parallel channels. In this latter limit, the symmetry relation of the single-current fluctuation theorem is satisfied with respect to an effective affinity that is much lower than the affinity determined by the reservoirs. This back-action effect is quantitatively characterized. The analogy with the experimental situation is made by considering the circuits with large and low currents as the detector and measured circuits respectively.

Furthermore, our analysis leads to the evaluation of the entropy production in the electronic device.
The fluctuation theorem has for consequence the non-negativity of the entropy production and is thus compatible with the second law of thermodynamics.  The directionality due to the nonequilibrium driving of the device is characterized by the probability distributions of the current fluctuations, by the mean values of the currents, and also by the entropy production.  The analysis based on the fluctuation theorem allows us to understand the connections between these complementary and fundamental aspects of such nonequilibrium electronic devices.

\begin{figure}
	\centering
		\includegraphics[width=11cm]{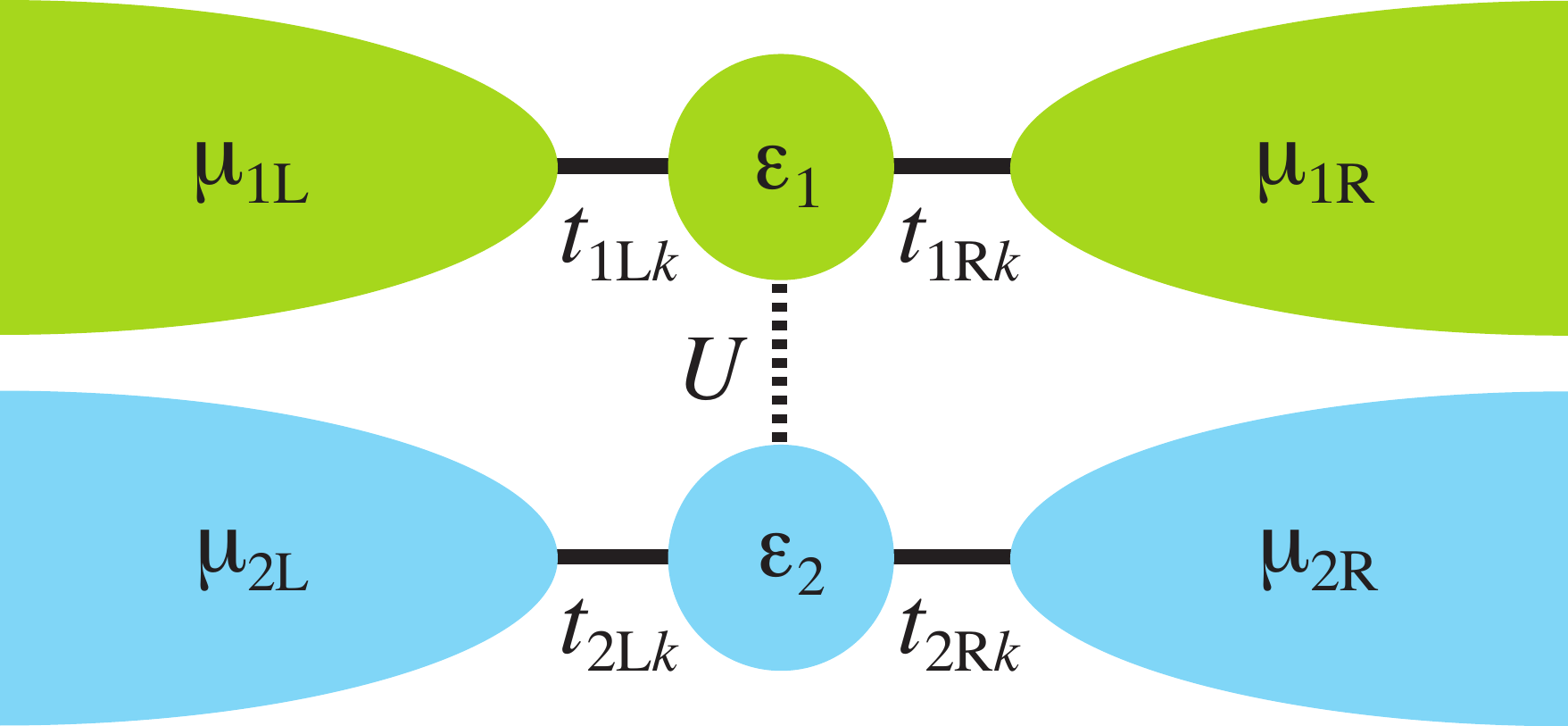}
	\caption{Schematic representation of two quantum dots in parallel.  Each quantum dot is coupled
to two reservoirs of electrons.  Moreover, both quantum dots influence each other by the Coulomb electrostatic interaction.}
	\label{fig1}
\end{figure}


\section{Capacitively coupled parallel transport channels}
\label{Model}


In this Section, we present the Hamiltonian model of the two channel circuits discussed above. By using the results of Chapter \ref{Chapter3}, we derive a quantum master equation ruling the occupancies of the quantum dots and the electron currents for quantum dots weakly coupled to the reservoirs within the Markovian and rotating wave approximations. This equation is first expressed in terms of the counting parameters introduced in Chapter \ref{Chapter3}, allowing us to determine the characteristic function and its time evolution. We alternatively write it in Fourier conjugated space with respect to the counting parameters, which yields a stochastic equation that can be used to simulate trajectories of quantum dots occupation and electronic currents in each conduction channel.


\subsection{The Hamiltonian}


The vehicle of our study is the Hamiltonian model considered in Ref.~\cite{SKB10}.
Each transport channel ($\alpha=1$ or $\alpha  =2$) is composed of one quantum dot with a single energy level $\epsilon_\alpha$ for the electron.  
This level is either occupied or empty and the spin degree of freedom is ignored.
Moreover, the quantum dots are capacitively coupled by electrostatic repulsion if both are occupied.  This electrostatic repulsion is taken into account by an Anderson-type term with the parameter $U$.
The parameter $U$ is thus the energy contribution of the Coulomb repulsion when both quantum dots are occupied by an electron.  The system Hamiltonian is therefore given in second quantization by
\be
H_{\rm S} = \epsilon_1 \, d_1^{\dagger} d_1 +\epsilon_2 \, d_2^{\dagger} d_2 +U d_1^{\dagger} d_1 d_2^{\dagger} d_2,
\label{H_S}
\ee
where $d_\alpha$ and $d_\alpha^{\dagger}$ denote the fermionic annihilation and creation operators of an electron on the quantum dot labeled by $\alpha=1,2$.  This Hamiltonian is diagonalized in the four-state basis $\{\vert 00\rangle,\vert 10\rangle,\vert 01\rangle,\vert 11\rangle\}$ with the corresponding energy eigenvalues $\{0,\epsilon_1,\epsilon_2,\epsilon_1+\epsilon_2+U\}$.

Each quantum dot is in tunneling contact with two reservoirs on its left- and right-hand sides (see \textsc{Figure}~\ref{fig1}).
The system has thus four reservoirs $j=1{\rm L},1{\rm R},2{\rm L},2{\rm R}$, which are denoted as $j= \alpha i$
by the label $\alpha=1,2$ of the channel and the side $i={\rm L},{\rm R}$ where the reservoir stands.
The Hamiltonian of all the reservoirs can be expressed as
\be 
H_{\rm B} = \sum_j H_j 
\label{H_B}
\ee
in terms of the Hamiltonians of the individual reservoirs, which are defined as
\be 
H_j = \sum_k \epsilon_{j k}\, c_{jk}^{\dagger} c_{jk},
\label{H_j4}
\ee
where $c_{jk}$ and $c_{jk}^{\dagger}$ denote the annihilation and creation operators of electrons in the corresponding states.  The reservoirs are supposed to be much larger than the system itself so that the eigenvalues $\{\epsilon_{jk}\}$ of each reservoir form a very dense spectrum which is quasi continuous and characterized by a density of states $D_j(\epsilon)=\sum_k \delta(\epsilon-\epsilon_{jk})$. We also define the electron number operator in reservoir $j$ as $N_j=\sum_k c_{jk}^{\dagger} c_{jk}$.

The tunneling Hamiltonian establishing the interaction between the quantum dots and the reservoirs has the form
\be
V= \sum_{\alpha=1,2}\sum_{i ={\rm L}, {\rm R}}\sum_k t_{ \alpha i k} \, d_{\alpha}^{\dagger} c_{ \alpha i k} + {\rm H.\,c.},
\label{H_SR}
\ee
where we have here specified the channels and the reservoirs by writing $j= \alpha i$. The effect of the electrostatic interaction on the energy barriers between the quantum dots and the reservoirs could be taken into account by including corresponding capacitances, as considered in Ref.~\cite{PhysRevLett.104.076801}. As a result, the positions of the energy levels in the quantum dots are in general shifted when a voltage is changed in the reservoirs. However, our primary interest is here focused essentially on the rate processes taking place in the quantum dots.  For this purpose, we may already use the Hamiltonian model of Ref.~\cite{SKB10} where the capacitances between the quantum dots and the reservoirs are absent.

This interaction Hamiltonian can be written in the form (\ref{interactionexpl}) by defining the system operators operators
\bea
A^{+}_{1} \equiv d_1^{\dagger} & \mbox{and} & A^{-}_{1} \equiv d_1 , \\
A^{+}_{2} \equiv d_2^{\dagger} & \mbox{and} & A^{-}_{2} \equiv d_2 ,
\eea
and their associated environment operators as
\bea \label{B1parQD}
B^{+}_{1} \equiv \sum_{i= {\rm L}, {\rm R}} \sum_{k} t_{1 i k} \, c_{1 i k} & \mbox{and} & B^{-}_{1} \equiv - \sum_{i= {\rm L}, {\rm R}} \sum_{k} t_{1 i k} \, c_{1 i k}^{\dagger} , \\ \label{B2parQD}
B^{+}_{2} \equiv \sum_{i= {\rm L}, {\rm R}} \sum_{k} t_{2 i k} \, c_{2 i k} & \mbox{and} & B^{-}_{2} \equiv - \sum_{i= {\rm L}, {\rm R}} \sum_{k} t_{2 i k} \,  c_{2 i k}^{\dagger},
\eea
the minus sign arising due to the anti-commutation properties of fermionic creation and annihilation operators.

Finally, the total Hamiltonian is defined as the sum:
\be
H=H_{\rm S}+H_{\rm B}+V.
\label{H}
\ee

We notice that the electron number operators of each transport channel
\be
N_\alpha = d_\alpha^{\dagger} d_\alpha +\sum_{i={\rm L}, {\rm R}}\sum_k c_{\alpha i k}^{\dagger} c_{ \alpha i k} \qquad \alpha=1,2
\ee
separately commutes with the total Hamiltonian
\be
[ H,N_1 ] = [H, N_2 ] = 0
\label{c=2}
\ee
so that the electron number is conserved in each transport channel and there is no electron exchange between the channels.  In contrast, the number operators of the reservoirs $N_j=N_{\alpha i }$ with $\alpha=1,2$ and $i={\rm L},{\rm R}$, do not commute with the total Hamiltonian unless the tunneling amplitudes $\left\{ t_{\alpha i k} \right\}$ are equal to zero.

We are going to perform the counting statistics of the electron flow out of the left reservoirs of each channel. The number of particles flowing out of the reservoir $\alpha L$ for $\alpha =1$ or $2$ is minus the change of particle number in the same reservoir. As a consequence, the modified Hamiltonian introduced in (\ref{interactioncounting}) takes the form
\be
H = H_{{\rm S}} + H_{{\rm B}} + V^{\lambda_1  \lambda_2}
\ee 
with 
\bea \label{413}
V^{\lambda_1  \lambda_2} & = &  \mbox{e}^{i (\lambda_1 /2) N_{1}}   \mbox{e}^{i (\lambda_2 /2) N_{2}}  \, V \,  \mbox{e}^{-i (\lambda_1 /2) N_{1}} \mbox{e}^{-i (\lambda_2 /2) N_{2}} \\
& = &  \sum_{\alpha=1,2} \mbox{e}^{i \lambda_{\alpha}} \, A^{+}_{\alpha} B^{+}_\alpha +  \mbox{e}^{-i \lambda_{\alpha}} \, A^{-}_{\alpha } B^{-}_\alpha \label{414},
\eea
where the counting parameters $\lambda_{1}$ and $\lambda_{2}$ account for the fluctuations of the particle number operators $N_{1 {\rm L}}$ and $N_{2{\rm L}}$, respectively.


\subsection{The modified density matrix}


In Chapter \ref{Chapter3}, we derived a modified master equation for the reduced density matrix of an intermediate quantum system accounting for the particle and energy exchanges with the macroscopic reservoirs. Here, we apply these results to the Hamiltonian described above and give a modified master equation for the system composed of the parallel quantum dots by tracing out the environment degrees of freedom.

In the following, we suppose that the two quantum dots are weakly coupled to the reservoirs by small enough tunneling amplitudes $\{ t_{jk}\}$ so that we may carry out the Born perturbative approximation up to second order in the tunneling amplitudes.

The reservoirs are initially in grand-canonical equilibrium states characterized by the chemical potentials $\mu_j$ with $j \in\{ 1 {\rm L},1 {\rm R},2 {\rm L},2 {\rm R}\}$ and a uniform temperature $T$.  We denote by $\beta=(k_{\rm B}T)^{-1}$ the inverse temperature with the Boltzmann constant $k_{\rm B}$.  On the other hand, the quantum dots are in an arbitrary statistical mixture $\rho_{\rm S} (0)$. As a consequence, the initial density matrix of the total system has the factorized form
\be
\rho(0)= \rho_{\rm S} (0) \prod_{j} \otimes \, \rho_j ,
\label{rho_0}
\ee
where 
\be \label{grandcan4}
\rho_j = \, {\rm 
e}^{-\beta(H_j-\mu_{j} N_j - \phi_{j})}
\ee
denotes the grand-canonical density operator in the reservoir $j$ and
\be
\phi_j = -\beta^{-1} \ln \left[ \mbox{Tr} \left\{ {\rm 
e}^{-\beta(H_j-\mu_{j} N_j )} \right\} \right]
\ee
is the thermodynamic grand-canonical potential for the reservoir $j$.

In order to study the statistical fluctuations of the number of electron transfers between the left and right reservoirs in channels $1$ and $2$, we introduce the modified density matrix
\be \label{418}
\rho (\lambda_1 , \lambda_2 , t ) \equiv \mbox{e}^{i (\lambda_1 /2) N_{1}}   \mbox{e}^{i (\lambda_2 /2) N_{2}} \rho (t) \mbox{e}^{-i (\lambda_1 /2) N_{1}}   \mbox{e}^{-i (\lambda_2 /2) N_{2}},
\ee
where $\rho (t)$ is the total density matrix of the system at time $t$. The corresponding reduced density matrix thus reads
\be
\rho_S (\lambda_1 , \lambda_2 ,t ) = \mbox{Tr}_{R} \left\{ \rho (\lambda_1 , \lambda_2 , t ) \right\},
\ee
where the trace is taken over the environment degrees of freedom.

As shown in Chapter \ref{Chapter3}, the influence of the environment on the effective dynamics of $\rho_S (\lambda_1 , \lambda_2 ,t ) $ is characterized by the correlation functions of the operators (\ref{B1parQD}) and (\ref{B2parQD}). These correlation functions are evaluated for these operators and the reservoir Hamiltonians (\ref{H_j4}) in the next Section. 


\subsection{The environment correlation function}


The charging transition rates $a_j$ and $\overline{a}_j$ of the quantum dot $i$ from reservoir $j=i \alpha$ are defined by
\bea \label{420}
&& a_j \equiv  \int d\tau \mbox{e}^{i \epsilon_i \tau} \langle B^{-}_{j} (\tau) B^{+}_{j} \rangle , \\
&& \bar{a}_j \equiv  \int d\tau \mbox{e}^{i(  \epsilon_i + U) \tau} \langle B^{-}_{j} (\tau) B^{+}_{j} \rangle , \label{421}
\eea
where $\langle \, \cdot \, \rangle \equiv \mbox{Tr} \{ \, \cdot \, \rho_j \}$ denotes an average with the grand-canonical density matrix (\ref{grandcan4}) in the reservoir $j$.

The reservoir Hamiltonians (\ref{H_j4}) being quadratic, these correlation functions can be evaluated analytically (see Appendix \ref{AppendixC}) to yield
\bea
&& a_j =  \Gamma_j f_j  \label{aj} , \\
&& \bar{a}_j = \bar\Gamma_j \bar{f}_j \label{aUj},
\eea
in terms of the Fermi-Dirac distributions
\bea
&& f_j = \frac{1}{1+{\rm e}^{\beta(\epsilon_j-\mu_j)}} \label{fj} ,\\
&& \bar{f}_j = \frac{1}{1+{\rm e}^{\beta(\epsilon_j+U-\mu_j)}} \label{fUj},
\eea
where $\epsilon_j=\epsilon_\alpha$ for $j= \alpha i$.
The rate constants are given by
\bea
 \Gamma_j & = & \frac{2\pi}{ \hbar^{2}} \sum_k \vert t_{jk}\vert^2 \delta(\epsilon_j-\epsilon_{jk}) \\
& = & \frac{2\pi}{ \hbar^{2}} \vert t_j(\epsilon_j)\vert^2 D_j(\epsilon_j) ,\\
 \bar\Gamma_j & = & \frac{2\pi}{ \hbar^{2}} \sum_k \vert t_{jk}\vert^2 \delta(\epsilon_j+U-\epsilon_{jk}) \\
& = & \frac{2\pi}{ \hbar^{2}} \vert t_j(\epsilon_j+U)\vert^2 D_j(\epsilon_j+U),
\eea
where the quantities $t_j(\epsilon)$ are the tunneling amplitudes as a function of energy and $D_j(\epsilon) \equiv \sum_{k} \delta (\epsilon - \epsilon_{jk}) $ the density of states of the reservoir $j$.

Similarly, we define the discharging transition rates $b_j$ and $\overline{b}_j$ as
\bea \label{430}
&& b_j \equiv  \int d\tau\mbox{e}^{i \epsilon_i \tau} \langle B^{+}_{j} (\tau) B^{-}_{j} \rangle , \\ \label{431}
&& \bar{b}_j \equiv  \int d\tau \mbox{e}^{i(  \epsilon_i + U) \tau} \langle B^{+}_{j} (\tau) B^{-}_{j} \rangle.
\eea
By using the results of Appendix \ref{AppendixC} we get
\bea
&& b_j = \Gamma_j (1-f_j) \label{bj} , \\
&& \bar{b}_j = \bar\Gamma_j (1-\bar{f}_j) \label{bUj}.
\eea

The transition rates (\ref{aj}), (\ref{aUj}) and (\ref{bj}), (\ref{bUj}) are proportional to the equilibrium bath correlation functions, and as such do satisfy the Kubo-Martin-Schwinger condition \cite{Kubo1998} so that
\bea \label{KMS1}
\frac{a_{j}}{b_{j}} & = & \mbox{e}^{-\beta (\epsilon_{j} - \mu_{j})} , \\
\frac{\overline{a}_{j}}{\overline{b}_{j}} & = & \mbox{e}^{-\beta (\epsilon_{j} + U - \mu_{j})} ,  \label{KMS2}
\eea
in terms of the quantum dot energies $\epsilon_{1}$, $\epsilon_{2}$ and the inter-dot interaction energy $U$. This property is a key point in proving the fluctuation theorem for the stochastic dynamics of a system interacting with large equilibrium baths. We will later use this property in order to establish the fluctuation theorem for the electron currents in the present model.


\subsection{The modified master equation}\label{414sec}


The total system is characterized by two sets of time scales:

(1) The correlation times of the reservoirs:  The correlation time of the reservoir $j$ can be estimated as $\tau^{j}_{C}\sim \Delta\epsilon_j^{-1}$ in terms of the width $\Delta\epsilon_j$ of the function giving the charging rate $a_j(\epsilon)=2\pi \hbar^{-2} \vert t_j(\epsilon)\vert^2 D_j(\epsilon) f_j(\epsilon)$ versus the energy $\epsilon$.

(2) The relaxation times induced by the electron exchanges with the reservoirs: $\tau^j_{\rm R}\sim \Gamma_j^{-1}$.

In consistency with the assumption of weak coupling, we suppose that the correlation times are much shorter than the relaxation times and that the secular approximation is performed by averaging the equation of motion over a time scale $\Delta t$ which is intermediate between both
\be
\tau_{C}^{j} \ll \Delta t \ll \tau^{j}_{\rm R}
\label{Dt_begin}
\ee
justifying the use of the Markovian approximation \cite{tannoudjiintproc,PhysRevA.78.022106}. Moreover, since the perturbative expansion is limited to second order, resonance effects are neglected.  Consequently, the thermal energy should be supposed to be larger than the natural width of the quantum dot energy levels \cite{B91}
\be
\Gamma_{\alpha{\rm L}}+ \Gamma_{\alpha{\rm R}} \ll k_{\rm B} T \qquad \alpha=1,2.
\label{hG<kT}
\ee

As a consequence of these approximations and as shown in Chapter \ref{Chapter3}, we obtain a closed set of dynamical equations in terms of the vector
\be \label{stateresgen}
{\bf g}(\lambda_{1}, \lambda_{2} ,t)=
\left(
\begin{array}{c}
g_{00}(\lambda_{1}, \lambda_{2},t) \\
g_{10}(\lambda_{1}, \lambda_{2},t) \\
g_{01}(\lambda_{1}, \lambda_{2},t) \\
g_{11}(\lambda_{1}, \lambda_{2},t)
\end{array}
\right)
\ee
composed of the diagonal elements of the modified reduced density matrix $g_{ \nu_1 \nu_2} (\lambda_{1},\lambda_{2} ,t) = \langle \nu_1 \nu_2 | \rho_{ {\rm S}} ( i \lambda_{1}, i \lambda_{2} , t) | \nu_1 \nu_2 \rangle$ where the indices $\nu_1 = 0,1$ and $\nu_2 = 0,1$ denote the occupancies in the corresponding quantum dot. This modified rate equation reads
\be \label{modifiedrateQDpar}
\dot{{\bf g}} (\lambda_{1}, \lambda_{2} ,t) = {\bf W} (\lambda_{1} , \lambda_{2}) \cdot {\bf g} (\lambda_{1}, \lambda_{2} ,t),
\ee
where the rate matrix ${\bf W} (\lambda_{1} , \lambda_{2})$ can be written as a sum
\be \label{modratematqdpar}
{\bf W} (\lambda_{1} , \lambda_{2}) = {\bf W}_{1} (\lambda_{1} ) + {\bf W}_{2} ( \lambda_{2})
\ee
with 
\be 
\mbox{\bf W}_1 (\lambda_{1})
=
\left(
\begin{array}{cccc}
-a_{1L}-a_{1 {\rm R}} & b_{1 {\rm L}}\, \mbox{ e }^{\lambda_{1}} +b_{1 {\rm R}} & 0 & 0 \\
a_{1 {\rm R}}\, \mbox{ e }^{- \lambda_{1}}  +a_{1 {\rm R}} & -b_{1{\rm L}}-b_{1 {\rm R}} & 0 & 0 \\
0 & 0 & -\bar{a}_{1{\rm L}}-\bar{a}_{1{\rm R}} & \bar{b}_{1{\rm L}}\,\mbox{ e }^{\lambda_{1}} 
+\bar{b}_{1{\rm R}}  \\
0 & 0 & \bar{a}_{1{\rm L}}\, \mbox{ e }^{-\lambda_{1}}  +\bar{a}_{1{\rm R}} & -\bar{b}_{1{\rm L}}-\bar{b}_{1{\rm R}}  \\
\end{array}
\right)
\label{L1}
\ee
accounting for the transitions due to the tunneling events between the quantum dot ${\rm No.} \,1$ and the reservoirs $1{\rm L}$ and $1{\rm R}$, while
\be 
\mbox{\bf W}_2 (\lambda_{2})
=
\left(
\begin{array}{cccc}
-a_{2{\rm L}}-a_{2{\rm R}} & 0 &b_{2{\rm L}}\, \mbox{ e }^{ \lambda_{2}}  +b_{2{\rm R}} &  0 \\
0 & -\bar{a}_{2{\rm L}}-\bar{a}_{2{\rm R}} 
& 0 &\bar{b}_{2{\rm L}} \,\mbox{ e }^{ \lambda_{2}} +\bar{b}_{2{\rm R}}\\
a_{2{\rm L}} \, \mbox{ e }^{- \lambda_{2}} +a_{2{\rm R}} & 0 &-b_{2{\rm L}}-b_{2{\rm R}} &  0 \\
0 & \bar{a}_{2{\rm L}} \, \mbox{ e }^{-\lambda_{2}} +\bar{a}_{2{\rm R}} & 0 &-\bar{b}_{2{\rm L}}-\bar{b}_{2{\rm R}}\\
\end{array}
\right)
\label{L2}
\ee
accounts for the transitions due to the tunneling events between the quantum dot ${\rm No.} \,2$ and the reservoirs $2 {\rm L}$ and $2 {\rm R}$.

A rate equation can also be written for the probabilities $p_{\nu_1 \nu_2} (n_{1}, n_{2},t) $ that the system is in the state $| \nu_1 \nu_2\rangle$ while having observed $n_{1}$ and $n_{2}$ electrons flowing out of, respectively, the reservoir $1{\rm L}$ and $2{\rm L}$ during time $t$\footnote{Note that these stochastic variables are time-dependent, although we did not explicit this time dependence in order to simplify notations.}. As we pointed out in section \ref{backtoprob}, this is done by applying a Fourier transform to each member of the modified rate equation (\ref{modifiedrateQDpar}). The resulting equation can be written in a fairly simple form by introducing the operators
\be
\hat{E}_\alpha^{\pm} \equiv \exp\left( \pm \frac{\partial}{\partial n_\alpha} \right),
\label{Eop}
\ee
which increase or decrease the numbers $n_{\alpha}$ of electrons transferred in channel ${\rm No. } \,\alpha$ so that
\be
\hat{E}_\alpha^{\pm} f(n_\alpha) = f(n_\alpha \pm 1)
\ee
when acting on any function $f(n_{\alpha})$. The rate equation for the probability vector
\be \label{probabvect}
{\bf p}(n_{1}, n_{2} ,t)=
\left(
\begin{array}{c}
p_{00}(n_{1}, n_{2},t) \\
p_{10}(n_{1}, n_{2},t) \\
p_{01}(n_{1}, n_{2},t) \\
p_{11}(n_{1}, n_{2},t)
\end{array}
\right)
\ee
thus reads
\be
\partial_t\,{\bf p}(n_1,n_2, t) = \left(\hat{\mbox{\bf W}}_1 +\hat{\mbox{\bf W}}_2 \right) \cdot {\bf p}(n_1,n_2, t) ,
\label{probastoch}
\ee
where the matrices $\hat{\mbox{\bf W}}_1$ and $\hat{\mbox{\bf W}}_1$ are directly obtained from, respectively, $\mbox{\bf W}_1 (\lambda_{1})$ and $\mbox{\bf W}_1(\lambda_{2})$ by making the substitution
\be
{\rm e}^{\pm\lambda_\alpha} \to \hat{E}_\alpha^{\pm}.
\ee

Equations (\ref{modifiedrateQDpar}) and (\ref{probastoch}) both reduce to an equation for the stochastic jumps in the quantum dots obtained by whether setting $\lambda_1 = \lambda_2 = 0$ in the first case or summing over all the possible electrons transfers $n_1$ and $n_2$ in the second one. This results in a stochastic equation for the occupation probabilities on the quantum dots
\be \label{probabocc}
\left(
\begin{array}{c}
p_{00}(t) \\
p_{10}(t) \\
p_{01}(t) \\
p_{11}(t)
\end{array}
\right)
=
\left(
\begin{array}{c}
g_{00}(0,0,t) \\
g_{10}(0,0,t) \\
g_{01}(0,0,t) \\
g_{11}(0,0,t)
\end{array}
\right)
=
\sum_{n_1 = -\infty}^{\infty}
\sum_{n_2 =- \infty}^{\infty}
\left(
\begin{array}{c}
p_{00}(n_{1}, n_{2},t) \\
p_{10}(n_{1}, n_{2},t) \\
p_{01}(n_{1}, n_{2},t) \\
p_{11}(n_{1}, n_{2},t)
\end{array}
\right).
\ee
In the following, we write the steady state probabilities by using an upper case and dropping the time index $t$. As an example
\be \label{ssprobabocc}
P_{\nu_1 \nu_2} = \lim_{t \rightarrow \infty} p_{\nu_1 \nu_2} (t)
\ee
is the occupation probability of the quantum dots at steady state.


\section{The two-current fluctuation theorem and its consequences}
\label{FCS-FT}


This section is devoted to the full counting statistics of the two interacting currents, for which the fundamental fluctuation theorem is established. We also consider the finite-time statistics, and show the convergence to the steady state fluctuation theorem presented in Chapter \ref{Chapter2}, regardless of the initial probability distribution on the quantum dots. Moreover, the consequences of the fluctuation theorem on the transport coefficients and the connection with the entropy production of the device are discussed.


\subsection{The cumulant generating function and the affinities}


In order to perform the counting statistics of the electrons transferred from the left reservoirs to the quantum dots, we introduce the cumulant generating function of the currents
\be 
\mathcal{G}(\lambda_1,\lambda_2)  \equiv \lim_{t\to \infty} - \frac{1}{t} \ln \left\langle \exp(-\lambda_1 n_1 - \lambda_2 n_2 ) \right\rangle_t
\label{Q}
\ee
in terms of the counting parameter $\lambda_\alpha$ of the corresponding transport channel and where $n_\alpha$ denotes the number of particles transferred out of reservoir $\alpha = 1 {\rm L}$ or $  2 {\rm L}$ during time $t$. As shown in section \ref{fcslongtime}, the cumulant generating function is given as the leading eigenvalue of the eigenvalue problem
\be
\mbox{\bf W} (\lambda_{1}, \lambda_{2}) \cdot {\bf v} = - \mathcal{G} (\lambda_1 , \lambda_2)  \, {\bf v} 
\label{eigenvalue_problem}
\ee
in terms of the modified rate matrix (\ref{modratematqdpar}) - (\ref{L2}).

In order to obtain a fluctuation relation of the form (\ref{longtimeFT}) for the cumulant generating function, we note that the four-by-four matrix $\mbox{\bf W}(\lambda_1 , \lambda_2)$ obeys the symmetry
\be
\mbox{\bf M}^{-1} \cdot \mbox{\bf W}(\lambda_1 , \lambda_2) \cdot \mbox{\bf M}
= \mbox{\bf W}(A_{1} - \lambda_1 , A_{2}  - \lambda_2)^{\top}
\label{symmetry}
\ee
with
\be \label{Msym}
\mbox{\bf M} = 
\left(
\begin{array}{cccc}
1 & 0 & 0 &  0 \\
0 & {\rm e}^{-\beta(\epsilon_1-\mu_{1{\rm R}})} & 0 & 0 \\
0 & 0 & {\rm e}^{-\beta(\epsilon_2-\mu_{2{\rm R}})} & 0 \\
0 & 0 & 0 & {\rm e}^{-\beta(\epsilon_1+\epsilon_2+U-\mu_{1{\rm R}}-\mu_{2{\rm R}})} \\
\end{array}
\right)
\ee
and the affinities ${\bf A}=(A_1,A_2)$ defined by
\bea 
&& A_1 = \ln\frac{a_{1{\rm L}}b_{1{\rm R}}}{b_{1{\rm L}}a_{1{\rm R}}} 
= \ln\frac{\bar{a}_{1{\rm L}}\bar{b}_{1{\rm R}}}{\bar{b}_{1{\rm L}}\bar{a}_{1{\rm R}}} 
= \beta \left( \mu_{1{\rm L}} -\mu_{1{\rm R}} \right) \label{A1}\\
&& A_2 = \ln\frac{a_{2{\rm L}}b_{2{\rm R}}}{b_{2{\rm L}}a_{2{\rm R}}} 
= \ln\frac{\bar{a}_{2{\rm L}}\bar{b}_{2{\rm R}}}{\bar{b}_{2{\rm L}}\bar{a}_{2{\rm R}}} 
= \beta \left( \mu_{2{\rm L}} -\mu_{2{\rm R}} \right). \label{A2}  
\eea
These relations are a direct consequence of the local detailed balance conditions (\ref{KMS1}) and (\ref{KMS2}) satisfied by the transition rates. We notice that the affinities (\ref{A1}) and (\ref{A2}) can also be obtained by using Schnakenberg graph analysis \cite{Andrieux2007,S76}.
These quantities are the two independent thermodynamic forces able to drive the system away from equilibrium.
The fact that there exists only two independent affinities although the system contains four reservoirs is due to the existence of the two constants of motion (\ref{c=2}) given by the particle numbers in the two transport channels.

If the system was fully connected, only the total particle number would be a constant of motion and there would exist three independent affinities.  More generally, a system composed of $r$ reservoirs and partitioned into $c$ disconnected but interacting transport channels has $c$ constant particle numbers and can be driven away from equilibrium by $r-c$ independent affinities.  Here, $r=4$ and $c=2$ so that there is only $r-c=2$ independent affinities.

As mentioned earlier, the cumulant generating function is given by the leading eigenvalue of Eq. (\ref{eigenvalue_problem}), i.e., by the smallest root of the quartic characteristic polynomial
\be
\det\left\{ \mbox{\bf W}(\lambda_1 , \lambda_2) + \mathcal{G} (\lambda_1 , \lambda_2) \, \mbox{\bf 1}\right\}=0
\label{det}
\ee
of the four-by-four matrix (\ref{modratematqdpar}) - (\ref{L2}).
Therefore, the symmetry (\ref{symmetry}) implies that the cumulant generating function obeys a corresponding symmetry \cite{Kurchan1998,andrieuxthesis}. Indeed, given a solution $\mathcal{G} (\lambda_1 , \lambda_2)$ of the eigenvalue problem (\ref{det}), it is easily shown that $\mathcal{G} (A_{1} - \lambda_1 , A_{2}-\lambda_2)$ is also a solution by the following sequence of equalities
\bea \nonumber
& & \mbox{det}  \left\{ {\bf W} (\lambda_1 , \lambda_2) + \mathcal{G} (A_1 - \lambda_1 ,A_2  - \lambda_2 ) {\bf 1} \right\}  \\ \nonumber
& &=  \mbox{det} \left\{ {\bf M}  \, \cdot {\bf W} (A_1 - \lambda_1 , A_2 - \lambda_2)^{\top} \cdot {\bf M}^{-1} + \mathcal{G} (A_1 - \lambda_1 ,A_2  - \lambda_2 ) {\bf M}  \, \cdot {\bf M}^{-1} \right\}  \\ \nonumber
& &=  \mbox{det} \left\{  {\bf W} (A_1 - \lambda_1 , A_2 - \lambda_2)+ \mathcal{G} (A_1 - \lambda_1 ,A_2  - \lambda_2 ) {\bf 1} \right\}  \\ \nonumber
&& = 0.
\eea

In this way, the fundamental result is proved that the cumulant generating function satisfies the \textbf{steady state fluctuation theorem}
\be
\mathcal{G}(\lambda_1,\lambda_2)=\mathcal{G}(A_1-\lambda_1,A_2-\lambda_2)
\label{FT4}
\ee
in terms of the affinities $A_1$ and $A_2$ given by Eqs. (\ref{A1}) and (\ref{A2}).

As mentioned in Section \ref{largedevth1}, this last relation is equivalent to a fluctuation relation in the long-time limit for the probability distribution $p(J_1 , J_2, t)$ of the currents defined by
\bea
J_1 = \frac{n_1}{t} \label{I1} , \\
J_2 = \frac{n_2}{t} \label{I2},
\eea
in terms of the numbers of electrons $n_1$ and $n_2$ transferred in, respectively, channels $1$ and $2$ during time $t$\footnote{We did not explicitly write the time dependence of these stochastic variables in order to simplify notations, just as we did for $n_1$ and $n_2$.}. The probability distribution $P (J_1 , J_2, t)$ of the currents thus satisfies
\be
\lim_{t \rightarrow \infty} \frac{1}{t} \ln \frac{p(J_1,J_2 , t)}{p(-J_1,-J_2 , t)} = A_1 J_1+A_2 J_2 
\label{FT-p}
\ee
with the thermodynamic affinities given by (\ref{A1}) and (\ref{A2}).


\subsection{Finite-time counting statistics}


As shown in Section \ref{fcslongtime}, the characteristic function of the probability distribution $p (n_{1} , n_2 ,t)$ for the number of particles $n_1$ and $n_2$ transferred in channels $1$ and $2$ during time $t$ can be expressed in terms of the vector (\ref{stateresgen}) as
\bea \label{finitetimefcs}
G(\lambda_1 , \lambda_2 ,t) & = & {\bf 1}^{\top} \cdot {\bf g} (\lambda_1 , \lambda_2 , t) \\
& = & {\bf 1}^{\top} \cdot \mbox{e}^{t {\bf W} (\lambda_1 , \lambda_2) } \cdot {\bf p}_0, \label{finitetimefcs2}
\eea
where ${\bf p}_0$ denotes the initial probability distribution over the quantum dot occupancies. The probability distribution $p (n_1 , n_2 ,t)$ for observing $n_1$ and $n_2$ electrons transferred in channels $1$ and $2$ during time $t$ is given by the Fourier coefficients of (\ref{finitetimefcs})
\be \label{probadistrtot}
p (n_1 , n_2 ,t) = \int \int \, \frac{d \lambda_1}{2 \pi}  \, \frac{d \lambda_2}{2 \pi} \, G( -i \lambda_1 , -i \lambda_2 , t) \mbox{e}^{-i \lambda_1 n_1 - i \lambda_2 n_2}.
\ee

\begin{figure}
\centering
		\includegraphics[width=10cm]{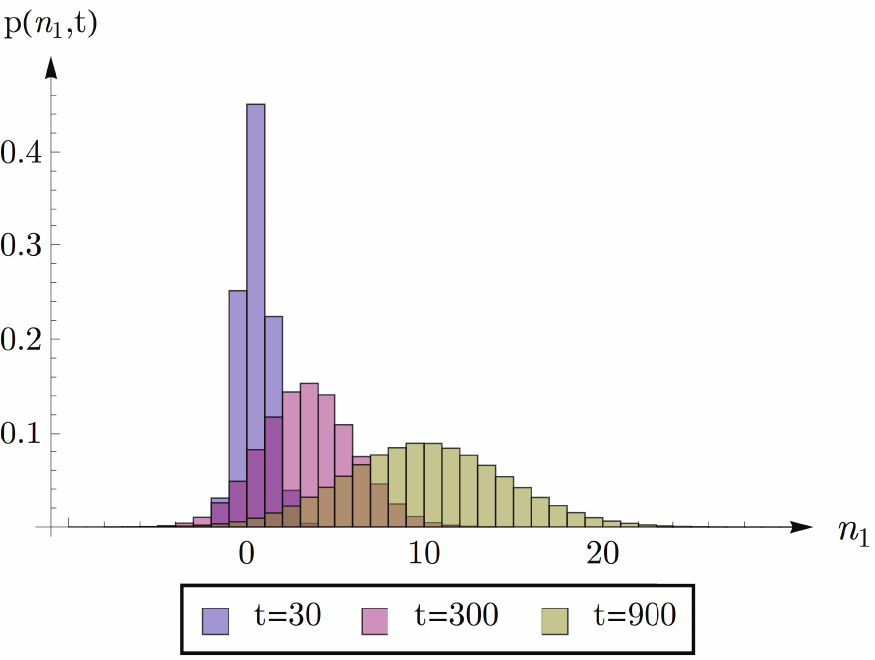}
	\caption{Time evolution of the marginal probability distribution $p_{1} (n_1 ,t)$ of the number of electrons $n_1$ transferred accross channel $\mbox{No.} \,1$ during time $t$. Parameters are chosen as $\beta =1$, $\epsilon_1  = 0.4$, $\epsilon_2 = 0.1$, $U = 0.5$, $\mu_{1{\rm L}}=-\mu_{1{ \rm R}}=0.5$, $\mu_{2 {\rm L}} = \mu_{2{\rm R}}=0$ and $\Gamma_{\alpha i} = 0.1$ for $\alpha = 1,2$ and $i={\rm L}, {\rm R}$.}
	\label{histogram1}
\end{figure}

\begin{figure}
\centering
		\includegraphics[width=10cm]{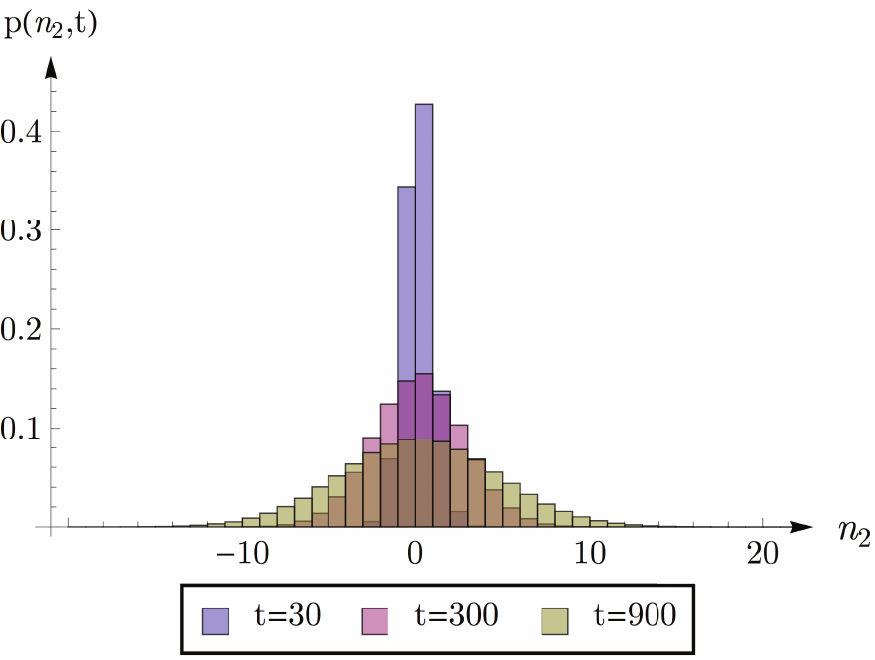}
	\caption{Time evolution of the marginal probability distribution $p_{2} (n_2 ,t)$ of the number of electrons $n_2$ transferred accross channel $\mbox{No.} \,2$ during time $t$. Parameters are chosen as in \textsc{Figure} \ref{histogram1}.}
	\label{histogram2}
\end{figure}

In \textsc{Figures} \ref{histogram1} and \ref{histogram2}, we illustrate the time evolution of the marginal distributions
\bea \label{marginaldis}
p (n_1 , t) & = & \sum_{n_2 =0}^{\infty} p (n_1 , n_2 ,t), \\
p (n_2 , t) & = & \sum_{n_1 =0}^{\infty} p (n_1 , n_2 ,t),
\eea
by plotting them as a function of, respectively, $n_1$ and $n_2$ for different values of time $t$. These probability distributions have their mean value\footnote{In the situation illustrated in \textsc{Figure} \ref{histogram2}, the mean of the probability distribution $p(n_2 , t)$ of the number of electrons $n_2$ transferred in circuit $\mbox{No.}\,2$ is zero as a consequence of the absence of bias in this conduction channel, i.e. $\mu_{2L} = \mu_{2R}$.} and variance linearly growing with time in the long time limit. In particular, this implies that the currents (\ref{I1}) and (\ref{I2}) are asymptotically constant in time with variance decreasing as $t^{-1}$.

Though the cumulant generating function (\ref{Q}) satisfies the steady state fluctuation theorem (\ref{FT4}), the finite-time generating function (\ref{finitetimefcs}) does not in general satisfy a fluctuation theorem at all times and for arbitrary initial condition. As mentioned in Chapter \ref{Chapter2}, this is due to the fact that we need to assume that the intermediate quantum systems is initially equilibrated with respect to one of the reservoirs in order to show the finite-time fluctuation theorem (\ref{QEFTgen}). In the present case, a fluctuation theorem for the characteristic function (\ref{finitetimefcs}) holds at all times provided that the initial condition on the quantum dots is chosen as the stationary distribution with respect to both right leads\footnote{Indeed, as the two quantum dots exchange energy via the interaction energy $U$, each quantum dot does not equilibrate separately with its right lead. Instead, the quantum dots may relax to an stationary state with respect to both right leads as a consequence of the homogeneous temperature between the leads and the absence of tunneling between the two quantum dots. This last assumption forbids the setting up of a steady state and additional particle current due to an eventual difference between $\mu_{1R}$ and $\mu_{2R}$.}. This distribution is shown to be given by the probability vector
\bea \label{idealinitial}
{\bf p}^{st}_{0} &  = & \left( \mbox{Tr} \left\{ {\bf M} \right\} \right)^{-1} {\bf M} \cdot {\bf 1} \\
& = & \frac{1}{\zeta}\left[
\begin{array}{c}
1 \\
\mbox{e}^{-\beta(\epsilon_1 - \mu_{1 {\rm R}})}\\
\mbox{e}^{-\beta(\epsilon_2 - \mu_{2  {\rm R}})} \\
\mbox{e}^{-\beta (\epsilon_1 + \epsilon_2 + U - \mu_{1  {\rm R}} - \mu_{2  {\rm R}})}
\end{array}
  \right],
\eea
where
\be
\zeta \equiv\frac{1}{1+\mbox{e}^{-\beta(\epsilon_1 - \mu_{1 {\rm R}})}+\mbox{e}^{-\beta(\epsilon_2 - \mu_{2  {\rm R}})} + \mbox{e}^{-\beta (\epsilon_1 + \epsilon_2 + U - \mu_{1  {\rm R}} - \mu_{2  {\rm R}})}}
\ee
is a normalization factor. Indeed, by using the symmetry (\ref{symmetry}) and writing the rate matrix as 
\be \label{ratedec4app}
{\bf W} (\lambda_{1} , \lambda_{2})\equiv {\bf W}_{{\rm L}} (\lambda_{1} , \lambda_{2}) +  {\bf W}_{\rm R} 
\ee
with 
\be \label{ratedec4app1}
\mbox{\bf W}_{\rm L} (\lambda_{1},\lambda_2)
=
\left(
\begin{array}{cccc}
-a_{1{\rm L}} - a_{2 {\rm L}}& b_{1{\rm L}}\, \mbox{ e }^{ \lambda_{1}}  & b_{2L} \mbox{e}^{\lambda_{2}} & 0 \\
a_{1{\rm L}}\, \mbox{ e }^{- \lambda_{1}}  & -b_{1{\rm L}}- \overline{a}_{2 {\rm L}}& 0 & \overline{b}_{2 {\rm L}} \mbox{e}^{ \lambda_{2}} \\
a_{2L} \mbox{e}^{- \lambda_{2}} & 0 & -\bar{a}_{1{\rm L}}-b_{2{\rm L}} & \bar{b}_{1{\rm L}}\,\mbox{ e }^{ \lambda_{1}}  \\
0 & \overline{a}_{2 {\rm L}} \mbox{e}^{- \lambda_2} & \bar{a}_{1{\rm L}}\, \mbox{ e }^{- \lambda_{1}}  & -\bar{b}_{1{\rm L}}-\bar{b}_{2{\rm L}}  \\
\end{array}
\right)
\ee
and
\be \label{ratedec4app2}
\mbox{\bf W}_{\rm R}
=
\left(
\begin{array}{cccc}
-a_{1{\rm R}} - a_{2R}& b_{1{\rm R}}\, \mbox{ e }^{ \lambda_{1}}  & b_{2 {\rm R}} \mbox{e}^{ \lambda_{2}} & 0 \\
a_{1{\rm R}}\, \mbox{ e }^{- \lambda_{1}}  & -b_{1{\rm R}}- \overline{a}_{2 {\rm R}}& 0 & \overline{b}_{2 {\rm R}} \mbox{e}^{ \lambda_{2}} \\
a_{2R} \mbox{e}^{- \lambda_{2}} & 0 & -\bar{a}_{1{\rm R}}-b_{2{\rm R}} & \bar{b}_{1{\rm R}}\,\mbox{ e }^{ \lambda_{1}}  \\
0 & \overline{a}_{2 {\rm R}} \mbox{e}^{- \lambda_2} & \bar{a}_{1{\rm R}}\, \mbox{ e }^{- \lambda_{1}}  & -\bar{b}_{1{\rm R}}-\bar{b}_{2{\rm R}}  \\
\end{array}
\right).
\ee
It is possible to show (see Appendix \ref{AppendixB}) by using the symmetry (\ref{symmetry}) that the vector (\ref{idealinitial}) is a steady state solution of the dynamics induced by the right leads, i.e.
\be
{\bf W}_{{\rm R}} \cdot {\bf p}^{st}_{0} = 0.
\ee

Now, by using this initial condition in (\ref{finitetimefcs2}) and the symmetry relation (\ref{symmetry}), we show that 
\bea \label{symmetryfinitetcalc}
G(\lambda_{1} , \lambda_2 ,t) & = &  {\bf 1}^{\top} \cdot \mbox{e}^{t \,  {\bf M} \cdot {\bf W} (A_1 -\lambda_1 ,A_2 - \lambda_2)^{\top} \cdot {\bf M}^{-1} } \cdot {\bf p}^{st}_0  \nonumber \\
& = & {\bf 1}^{\top}  \cdot {\bf M} \cdot  \mbox{e}^{t \,  {\bf W} (A_1 -\lambda_1 ,A_2 - \lambda_2)^{\top} } \cdot {\bf M}^{-1} \cdot {\bf p}^{st}_0 \nonumber \\
& = & {\bf 1}^{\top}  \cdot  \mbox{e}^{t \,  {\bf W} (A_1 -\lambda_1 ,A_2 - \lambda_2)}  \cdot {\bf p}^{st}_0 \nonumber \\
& = &G(A_1 -\lambda_1 , A_2 - \lambda_2 ,t) \label{symmetryfinitet}
\eea
thus proving the fluctuation theorem at finite-times provided that the initial distribution on the quantum dots is chosen as the stationary distribution with respect to the right leads $1 {\rm R}$ and $2{\rm R}$ given in (\ref{idealinitial}). Again, the symmetry relation (\ref{symmetryfinitet}) is equivalent to a fluctuation relation for the probability distribution (\ref{probadistrtot}), that is
\be \label{finitetFT}
 \ln \frac{p(n_1 , n_2 ,t)}{p(-n_1, -n_2 ,t)} = A_1 n_1 + A_2 n_2
\ee
in terms of the thermodynamic affinities (\ref{A1}) and (\ref{A2}).

\begin{figure}
\centering
		\includegraphics[width=12cm]{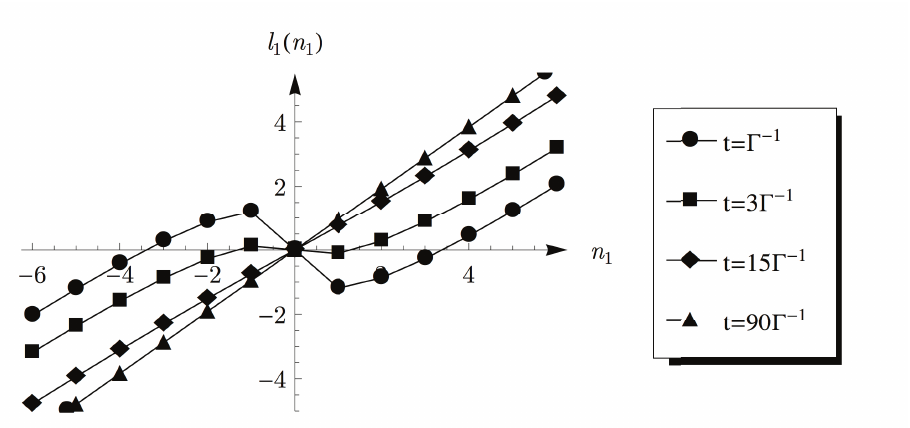}
	\caption{Test of the fluctuation relation (\ref{finitetFT}) at finite times by plotting the quantity $l_t (n_1)$ defined in (\ref{deftest}) as a function of the number of electrons $n_1$ transferred in the conduction channel $\mbox{No.} \,1$ during time $t$. The initial condition was chosen as $p_{00}=p_{10}=p_{01}=0$ and $p_{11}=1$. Parameters were chosen as $\beta = 1$, $\epsilon_1 = 0.4$, $\epsilon_2 = 0.1$, $U=0.5$, $\mu_{1{\rm L}}=-\mu_{1 {\rm R}}=0.5$, $\mu_{2 {\rm L}}= \mu_{2 {\rm R}}=0$, $\Gamma \equiv \Gamma_{1 {\rm L}} = \Gamma_{1 {\rm R}} = 0.1$ and $ \Gamma_{2 {\rm L}} = \Gamma_{2 {\rm R}} = 0.1$. We observe deviations to a linear behavior for short times, but the fluctuation theorem is satisfied for larger time values ($t \gg \Gamma_{1 \alpha}^{-1}$ in this case for $\alpha=\mbox{L},\mbox{R}$).}
	\label{fluctufinitet}
\end{figure}

Deviations to this symmetry relation are expected whenever we consider arbitrary initial conditions. However, as mentioned in Chapter \ref{Chapter2}, we expect these effects to decrease in time and eventually vanish in the long-time limit so that equation (\ref{finitetFT}) is strictly valid in this limit. This is illustrated in \textsc{Figure} \ref{fluctufinitet} where we plotted the quantity 
\be \label{deftest}
l_t (n_{1}) =   \ln \frac{p(n_1 ,t)}{p (-n_1, t)}
\ee
in terms of the marginal distribution (\ref{marginaldis}) for different time values. The affinity in channel $2$ was set to $0$ so that we expect this logarithm to become linear in $n_1$ with slope $A_1$ for large time values, as deduced from the steady state fluctuation theorem (\ref{FT-p}). For initial conditions chosen far from (\ref{idealinitial}), deviations to this linear behavior are observed at short times.

In typical counting statistics experiments, chemical potentials of the reservoirs as well as the temperature over the whole setup are controlled contrary to the initial state of the intermediate quantum dots of the circuit considered. As a result, deviations to the finite-time fluctuation theorem (\ref{symmetryfinitet}) for times such that $t \lesssim \Gamma^{-1}_{j}$ is consistent with the results illustrated in \textsc{Figure} \ref{fluctufinitet}. The symmetry relation (\ref{symmetryfinitet}) is however recovered for longer measurement times.

In the following, we concentrate our study to the case of long enough measurement times. As a consequence, we will consider the long-time limit cumulant generating function (\ref{Q}) for the characterization of the current fluctuations.


\subsection{The average currents and the response coefficients}


The average values of the particle currents in the long time limit are given in terms of the cumulant generating function according to
\bea
\langle J_{\alpha} \rangle & \equiv & \lim_{t \rightarrow \infty} \frac{\langle n_\alpha \rangle}{t} \\
& = & \left.\frac{\partial \mathcal{G}}{\partial\lambda_{\alpha}}\right\vert_{\lambda_{1} = \lambda_{2}=0} 
\label{J_a}
\eea
for $\alpha=1,2$. These currents can be expressed in terms of the probabilities (\ref{ssprobabocc}) of the four quantum dot states in the non-equilibrium steady state corresponding to the affinities (\ref{A1}) and (\ref{A2}). In order to obtain these currents, we first write the characteristic polynomial (\ref{det}) as
\be \label{charactpol}
\mathcal{G}^4 + C_3 \mathcal{G}^3 + C_2 \mathcal{G}^2 + C_1 \mathcal{G} + C_0 = 0,
\ee
which is quartic in\footnote{We do not write explicitly the dependence on the counting parameters of the cumulant generating function $\mathcal{G} (\lambda_1 , \lambda_2) $, unless necessary, in order to simplify notation.} $\mathcal{G}$ as a consequence of the dimensionality of the rate matrix ${\bf W} (\lambda_1, \lambda_2)$ of our model. We note that the last term is simply given by $C_0 = \det \left\{ {\bf W} (\lambda_1 , \lambda_2) \right\}$. Since the matrix ${\bf W} (\lambda_1, \lambda_2)$ reduces to the matrix of a jump stochastic process conserving probability if $\lambda_1=\lambda_2=0$, the leading eigenvalue vanishes in this limit
\be
\mathcal{G}(0,0)=0.
\label{Q=0}
\ee
We also take the partial derivative $\partial_\alpha$ of the characteristic determinant with respect to the counting parameter $\lambda_\alpha$ to get 
\be \label{derivationchar}
\left(4\mathcal{G}^3 + 3C_3 \mathcal{G}^2 + 2C_2 \mathcal{G}+ C_1\right)\partial_\alpha \mathcal{G} 
+ \partial_\alpha C_3 \mathcal{G}^3 + \partial_\alpha C_2 \mathcal{G}^2 + \partial_\alpha C_1 \mathcal{G} + \partial_\alpha C_0=0.
\ee

We can use this last expression together with the normalization condition (\ref{Q=0}) to write the average current (\ref{J_a}) as
\be 
\langle J_\alpha \rangle = - \left.\frac{\partial_\alpha C_0}{C_1}\right\vert_{\lambda_{1} = \lambda_{2} = 0}.
\ee
By calculating the coefficients $C_{0}$ and $C_1$ of the characteristic polynomial (\ref{charactpol}) corresponding to the rate matrix (\ref{modratematqdpar}) - (\ref{L2}) we finally get
\bea
&& \langle J_1 \rangle = a_{\rm 1L} P_{00} - b_{\rm 1L} P_{10} + \bar{a}_{\rm 1L} P_{01} - \bar{b}_{\rm 1L} P_{11} \label{J1}, \\
&& \langle J_2 \rangle = a_{\rm 2L} P_{00} - b_{\rm 2L} P_{01} + \bar{a}_{\rm 2L} P_{10} - \bar{b}_{\rm 2L} P_{11} \label{J2},
\eea
in terms of the steady state occupation probabilities $P_{\nu_1 \nu_2}$ of the quantum dots at steady state.

These currents are nonlinear functions of the affinities, which can be expanded in powers of the affinities in order to identify the linear and nonlinear response coefficients
\be
\langle J_\alpha \rangle =\langle  J_{\alpha} \rangle (A_1,A_2) = \sum_\beta L_{\alpha,\beta}A_\beta
+\frac{1}{2}\sum_{\beta,\gamma} M_{\alpha,\beta\gamma}A_\beta A_\gamma 
+\frac{1}{6}\sum_{\beta,\gamma,\delta} N_{\alpha,\beta\gamma\delta}A_\beta A_\gamma A_\delta + \cdots.
\ee
As a consequence of the fluctuation theorem (\ref{FT4}), the linear response coefficients $L_{\alpha,\beta}$ are given in terms of the second derivatives of the cumulant generating function with respect to the counting parameters and they  obey the Onsager reciprocity relations
\be
L_{\alpha,\beta}=L_{\beta,\alpha}=- \frac{1}{2} \frac{\partial^2\mathcal{G}}{\partial\lambda_\alpha\partial\lambda_\beta}\Big\vert_{\lambda_1= \lambda_2=0,A_1 = A_2=0}.
\label{L12}
\ee
Similar relationships have been established for the nonlinear response coefficients \cite{AG07JSM}.

Again, the linear response coefficients can be calculated in terms of the characteristic determinant (\ref{det}) of the matrix ${\bf W} (\lambda_1, \lambda_2)$ by further differentiating (\ref{derivationchar}) with respect to $\lambda_{\beta}$ and using the fact that the average currents vanishe at zero bias. The Onsager coefficients (\ref{L12}) are thus given by
\be
L_{\alpha,\beta}=L_{\beta,\alpha}= \left.\frac{\partial_\alpha \partial_\beta C_0}{2C_1}\right\vert_{\lambda_1 = \lambda_2 = 0, A_1 = A_2 = 0}.
\label{L12_det}
\ee

In the present model, the Onsager coefficients can be calculated and turn out to be proportional to
\be
L_{1,2}\propto \left(\Gamma_{\rm 1L} \bar{\Gamma}_{\rm 1R}-\bar{\Gamma}_{\rm 1L} \Gamma_{\rm 1R}\right)
\left(\Gamma_{\rm 2L} \bar{\Gamma}_{\rm 2R}-\bar{\Gamma}_{\rm 2L} \Gamma_{\rm 2R}\right)
\label{L12_prop}.
\ee
In general, this Onsager coefficient is thus non vanishing and there is a phenomenon of Coulomb drag according to which a current may be induced in a circuit at equilibrium if the other circuit is out of equilibrium, as shown in Ref.~\cite{PhysRevLett.104.076801}.  

However, the Onsager coefficient vanishes under the condition that the rate constants of one circuit do not depend on the Coulomb repulsion parameter $U$.  In this particular case, there is no Coulomb drag because
\be
\langle J_1 \rangle (0,A_2)=0 \qquad \mbox{and} \qquad J_2(A_1,0)=0 \qquad \mbox{if}\quad \Gamma_j = \bar{\Gamma}_j
\label{J_a=0}
\ee
as can be obtained by evaluating (\ref{J1}) and (\ref{J2}) for the particular values of the affinities in this last relation.  Equation (\ref{J_a=0}) implies the vanishing of the Onsager coefficient as well as the nonlinear response coefficients allowing the coupling of one current to the affinity of the other circuit:
\be
L_{1,2}=M_{1,22}=N_{1,222}=\cdots =0  \quad \mbox{and} \quad
L_{2,1}=M_{2,11}=N_{2,111}=\cdots =0  \quad \mbox{if} \quad \Gamma_j = \bar{\Gamma}_j.
\label{LMN}
\ee
Nevertheless, these coefficients do not vanish in general.


\subsection{The entropy production and the energy dissipation}


A further consequence of the fluctuation theorem (\ref{FT4}) is that the average currents (\ref{J_a}) obeys the second law of thermodynamics according to which the entropy production is always non-negative
\be
\frac{1}{k_{\rm B}} \frac{d_{\rm i}S}{dt} = A_1 \langle J_1 \rangle + A_2 \langle J_2 \rangle \geq 0 ,
\ee
where $k_{\rm B}$ is Boltzmann's constant \cite{Andrieux2007,RevModPhys.81.1665}. Indeed, by using the symmetry relation (\ref{FT-p}) on the generating function (\ref{Q}) and its normalization condition (\ref{Q=0}), we get
\bea
0 & = & \lim_{t \rightarrow \infty} - \frac{1}{t} \ln \langle \mbox{e}^{-A_1 n_1 - A_2 n_2 } \rangle \\
& \leq & \lim_{t \rightarrow \infty} - \frac{1}{t} \ln  \mbox{e}^{-A_1 \langle n_1 \rangle - A_2 \langle n_2 \rangle } \\
& = &  A_1 \left( \lim_{t \rightarrow \infty}\frac{\langle n_1 \rangle}{t}  \right) + A_2 \left(  \lim_{t \rightarrow \infty}\frac{\langle n_2 \rangle}{t} \right)  \\
& = & A_1 \langle J_1 \rangle +A_2 \langle  J_2 \rangle,
\eea
where we used Jensen's inequality for the exponential $\langle \mbox{e}^{X} \rangle \geq \mbox{e}^{\langle X \rangle}$ in the second line.

The power dissipated in each circuit is defined as the product of the voltage $V_\alpha$ by the electric current $I_\alpha=eJ_\alpha$ where $e$ is the electric charge of the particle: $\Pi_\alpha=V_\alpha I_\alpha$, with $\alpha=1,2$.  Since the affinities are related to the voltages by
\be
A_\alpha = \frac{eV_\alpha}{k_{\rm B} T},
\ee
we have that the dissipated power in the circuit $\alpha$ is given by
\be
\Pi_\alpha = k_{\rm B} T \, A_\alpha J_\alpha 
\ee
and the entropy production is thus proportional to the total dissipated power:
\be
\frac{d_{\rm i}S}{dt} = \frac{1}{T} \left( \Pi_1 + \Pi_2 \right) \geq 0 .
\ee
Therefore, the entropy production of the system characterizes the energy dissipation of the quantum measurement performed on one quantum dot by the current flowing in the other circuit playing the role of the detector in the experiments of Refs.~\cite{fujisawa2006,PhysRevLett.96.076605}.  We shall evaluate these quantities under such specific conditions in the following sections.


\section{Single-current fluctuation theorem} \label{scft}


In previous section, we showed that a bivariate fluctuation theorem (\ref{FT4}) and (\ref{FT-p}) holds for both current in channels $\mbox{No.} \, 1$ and $2$ with respect to the affinities (\ref{A1}) and (\ref{A2}). However, in typical electron counting experiments the statistics of the detection circuit is not accessible, and the experimentally accessible statistics is the one of the measured channel. In the present model, we assume that the channel $\mbox{No.} \, 2$ plays the role of detector and the statistics of channel $\mbox{No.} \, 1$ is thus obtained by considering the two current generating function (\ref{Q}) for $\lambda_2 = 0$. Regarding the fluctuation relations, the bivariate fluctuation theorem does not imply a fluctuation theorem for the sole current observed in channel $1$, that is, in general
\be
\mathcal{G} (\lambda_1 ,0) \neq \mathcal{G} (\tilde{A}_1 -\lambda_1 ,0)
\ee
for any $\tilde{A}_1$. Nevertheless, conditions can be found for which a univariate fluctuation theorem holds as observed experimentally, with a measurable effective affinity $\tilde{A}_1$.

In this section, we first consider the limit of a large Coulomb repulsion between the quantum dots, in which case a single-current fluctuation theorem is obtained but without modification of the effective affinity. We next consider the limit where the current in one circuit is much larger than the one in the other circuit. It is in this limit that the single-current fluctuation theorem is obtained with an important modification of the effective affinity with respect to which the symmetry relation of the single-current fluctuation theorem holds.


\subsection{The large Coulomb repulsion limit}
\label{Uinfty}


In the limit where the Coulomb repulsion between both quantum dots is large, the coupling parameter $U$ takes large values so that the charging rates of a second electron on the two quantum dots vanish:
\be
\bar{a}_j=0 \qquad \mbox{for} \quad j=1{\rm L}, 1{\rm R}, 2{\rm L}, 2{\rm R} \qquad \mbox{if} \quad U=\infty
\label{bara}.
\ee
As a consequence, the probabilities $p_{11} (n_1, n_2 ,t)$ in (\ref{probabvect}) and $P_{11} $ in (\ref{ssprobabocc}) 
that the system is in the fourth state $\vert 11\rangle$ also vanish:
\be
p_{11}(n_1,n_2 ,t)=0 \qquad \mbox{and} \qquad  P_{11}=0 \qquad \mbox{if} \quad U=\infty.
\label{P11=0}
\ee

In this limit, the occupancy of one quantum dot is stopping the current in the other quantum dot.  For instance, the average current in the secondary circuit (\ref{J2}) has two contributions depending on the occupancy of the quantum dot No.\,1:
\be
\langle J_2 \rangle = \left.  \langle J_2\rangle \right\vert_{\nu_1=0} + \left. \langle J_2\rangle \right\vert_{\nu_1=1} .
\ee
However, the contribution when the quantum dot No.\,1 is occupied is vanishing
\be
\left. \langle J_2\rangle \right\vert_{\nu_1=1} = \bar{a}_{\rm 2L} P_{10} - \bar{b}_{\rm 2L} P_{11} = 0 \qquad\mbox{if} \quad U=\infty
\ee
since $\bar{a}_{\rm 2L}=0$ according to Eq. (\ref{bara}) and $P_{11} = 0$ because of Eq. (\ref{P11=0}).
Therefore, the secondary circuit has a non-vanishing current only when the quantum dot No.\,1 is empty and vice versa.

The cumulant generating function can be obtained by considering the three-by-three matrix obtained by removing the fourth row and column from the matrix (\ref{modratematqdpar}) - (\ref{L2}).  In this case, the characteristic determinant (\ref{det}) depends on the counting parameters only in the following combinations:
\bea
&& a_{\rm 1R} b_{\rm 1L} {\rm e}^{\lambda_1} + a_{\rm 1L} b_{\rm 1R} {\rm e}^{-\lambda_1} ,\\
&& a_{\rm 2R} b_{\rm 2L} {\rm e}^{\lambda_2} + a_{\rm 2L} b_{\rm 2R} {\rm e}^{-\lambda_2},
\eea
which remain invariant under the independent substitutions $\lambda_1 \to A_1-\lambda_1$ and/or $\lambda_2 \to A_2-\lambda_2$ with the affinities (\ref{A1}) and (\ref{A2}).  Consequently, we obtain the symmetry relations:
\be
\mathcal{G}(\lambda_1,\lambda_2)=\mathcal{G}(A_1-\lambda_1,\lambda_2)=\mathcal{G}(\lambda_1,A_2-\lambda_2)=\mathcal{G}(A_1-\lambda_1,A_2-\lambda_2)
\qquad \mbox{if} \quad U=\infty ,
\ee
which implies the single-current fluctuation theorem:
\be
\mathcal{G}(\lambda_1,0)=\mathcal{G}(A_1-\lambda_1,0)
\qquad \mbox{if} \quad U=\infty
\ee
but with the unmodified affinity (\ref{A1}). Therefore, this limit cannot explain the modification of the affinity observed in the experiments reported in Ref.~\cite{fujisawa2006}.


\subsection{The large current ratio limit}
\label{Limit}


In typical counting statistics experiments \cite{fujisawa2006,PhysRevLett.96.076605}, the detector circuit which is used to observe the occupancy of the quantum dot carries a current which is typically much larger than the current in the measured circuit by a huge factor $10^7$-$10^8$. If the detector is taken as the circuit No.\,2 in the present model, the rate constants of that circuit are much larger than the ones of the circuit No.\,1:
\be
\Gamma_{\rm 1L}, \Gamma_{\rm 1R}, \bar{\Gamma}_{\rm 1L}, \bar{\Gamma}_{\rm 1R} \ll \Gamma_{\rm 2L}, \Gamma_{\rm 2R}, \bar{\Gamma}_{\rm 2L}, \bar{\Gamma}_{\rm 2R} .
\label{conditions}
\ee
Under such circumstances, the relaxation times $\tau^{1i}_{R}\sim \Gamma_{1i}^{-1}$ of the circuit No.\,1 are much longer than the relaxation times $\tau^{2i}_{R}\sim \Gamma_{2i}^{-1}$ of the circuit No.\,2 and the monitoring of the slow circuit by the fast one is performed over a time scale $\overline{\Delta   t}$ such that
\be
\tau^{2i}_{R} \ll \overline{\Delta   t }\ll \tau^{1i}_{R}
\label{Dt_real}
\ee
instead of the time scale (\ref{Dt_begin}). This basically means that the excitations induced by the slow circuit on the detector relax quasi-instantaneously. As a result, we may assume that the detector is in a stationary state during the whole period where the state of quantum dot $\mbox{No.} \, 1$ remains unchanged, while transitions in the quantum dot $\mbox{No.} \, 1$ induce sudden jumps between nonequilibrium stationary states in the detector. By observing the current in circuit $\mbox{No.  2}$, we thus observe trajectories like the one presented in \textsc{Figure}~\ref{fig24} where the sudden jumps are caused by tunneling events in channel $\mbox{No. 1}$. 

Our aim is here to obtain the cumulant generating function for the counting statistics in the sole circuit No.\,1 without measuring the current in the fast circuit No.\,2, as it is the case in Refs.~\cite{fujisawa2006,PhysRevLett.96.076605}. 
This amounts to consider the two-current generating function (\ref{Q}) for $\lambda_2=0$.
Accordingly, we focus on the time evolution of the probabilities defined by
\be
p_{\nu_1}(n_1 , t)=\sum_{\nu_2=0,1}\sum_{n_2=-\infty}^{+\infty} p_{\nu_1\nu_2}(n_1,n_2 ,t)
\label{p_nu1_n1}.
\ee

As mentioned above, since the electron transfers in the circuit No.\,2 are much faster than in the circuit No.\,1, the circuit No.\,2 can be supposed to be in a stationary state during the whole period when the circuit No.\,1 is in a given state.
Such stationary states conditional to the state $\nu_1$ of the quantum dot No.\,1 are obtained by finding the zero eigenvectors of the transition matrix (\ref{L2}) with $\hat{E}_2^{\pm}=1$.  The conditional probabilities $P_{\nu_2\vert\nu_1}$ that the quantum dot No.\,2 has the occupancy $\nu_2$ provided that the quantum dot No.\,1 is in the state $\nu_1$ are given by
\bea
&& P_{0\vert 0} = \frac{b_2}{a_2+b_2} \label{P00} ,\\
&& P_{1\vert 0} = \frac{a_2}{a_2+b_2} \label{P10} , \\
&& P_{0\vert 1} = \frac{\bar{b}_2}{\bar{a}_2+\bar{b}_2} \label{P01} , \\
&& P_{1\vert 1} = \frac{\bar{a}_2}{\bar{a}_2+\bar{b}_2} \label{P11},
\eea
with
\bea
&& a_2 = a_{2{\rm L}} + a_{2{\rm R}} ,\\
&& b_2 = b_{2{\rm L}} + b_{2{\rm R}} ,\\
&& \bar{a}_2 = \bar{a}_{2{\rm L}} + \bar{a}_{2{\rm R}} ,\\
&& \bar{b}_2 = \bar{b}_{2{\rm L}} + \bar{b}_{2{\rm R}}.
\eea

Under the conditions (\ref{conditions}), the probability that the system is in the state $\vert\nu_1\nu_2\rangle$ and that $n_1$ electrons have been transferred in the circuit No.\,1 factorizes into the probability (\ref{p_nu1_n1}) and the probability of the occupancy $\nu_2$ of the quantum dot No.\,2 conditioned to the occupancy $\nu_1$:
\be
p_{\nu_1\nu_2}(n_1 ,t)=p_{\nu_1}(n_1 ,t) P_{\nu_2\vert\nu_1}.
\ee

Substituting these relations into the master equation (\ref{p_nu1_n1}) and summing over $n_2$ and $\nu_2$, we get  the master equations for the probabilities $p_{\nu_1}(n_1)$ as follows:
\bea 
\partial_t \, p_0(n_1 ,t) &=& -\Big(a_{\rm L}+a_{\rm R}\Big)\, p_0(n_1 ,t) + \Big(b_{\rm L}\hat{E}_1^+ +b_{\rm R}\Big)\, p_1(n_1 ,t)  ,  \label{p0t} \\
\partial_t \, p_1(n_1 ,t) &=&  \Big(a_{\rm L}\hat{E}_1^- +a_{\rm R}\Big)\, p_0(n_1 ,t) - \Big(b_{\rm L}+b_{\rm R}\Big)\, p_1(n_1 ,t), \label{p1t}
\eea
where
\bea
&& a_{\rm L} =a_{1{\rm L}} P_{0\vert 0}+ \bar{a}_{1{\rm L}} P_{1\vert 0} \label{aL} ,\\
&& a_{\rm R} =a_{1{\rm R}} P_{0\vert 0} + \bar{a}_{1{\rm R}} P_{1\vert 0} \label{aR} , \\
&& b_{\rm L} =b_{1{\rm L}} P_{0\vert 1} + \bar{b}_{1{\rm L}} P_{1\vert 1} \label{bL} , \\
&& b_{\rm R} =b_{1{\rm R}} P_{0\vert 1} + \bar{b}_{1{\rm R}} P_{1\vert 1} \label{bR}
\eea
are the charging and discharging rates of the first quantum dot
averaged over the conditional stationary probabilities of the second quantum dot.  The master equations (\ref{p0t})-(\ref{p1t}) rule the process in the slow circuit No.\,1 as monitored by the fast circuit No.\,2 over the time scale (\ref{Dt_real}).

Taking a solution of the form $p_{\nu_1}(n_1)\sim \exp(\lambda_1n_1-\mathcal{G} t)$ for Eqs. (\ref{p0t})-(\ref{p1t}), 
the cumulant generating function (\ref{Q}) with $\lambda_2=0$ has thus for approximation the leading eigenvalue of the matrix
\be
\tilde{ {\bf W}} =
\left(
\begin{array}{cc}
-a_{\rm L}-a_{\rm R} &  b_{\rm L}{\rm e}^{+\lambda_1}+b_{\rm R}  \\
a_{\rm L}{\rm e}^{-\lambda_1}+a_{\rm R} & -b_{\rm L}-b_{\rm R} \\
\end{array}
\right),
\label{tilde_L_0}
\ee
which is given by
\begin{multline}
\mathcal{G}(\lambda_1,0) \\
\simeq  \frac{1}{2} \left[ 
a_{\rm L}+a_{\rm R} +b_{\rm L}+b_{\rm R} -
\sqrt{\left(a_{\rm L}+a_{\rm R} -b_{\rm L}-b_{\rm R}\right)^2
+ 4 \left( a_{\rm L}{\rm e}^{-\lambda_1}+a_{\rm R}\right)\left(b_{\rm L}{\rm e}^{+\lambda_1}+b_{\rm R}\right)}\right]
\label{tilde_Q_0}
\end{multline}
in the limit (\ref{conditions}) where the current in the second quantum dot is much larger than in the first one.
In this limit, the generating function (\ref{tilde_Q_0}) obeys the {it single-current fluctuation theorem}:
\be
\boxed{
\mathcal{G}(\lambda_1,0) = \mathcal{G}(\tilde A_1- \lambda_1,0) }
\label{FT14}
\ee
with the effective affinity for the first quantum dot obtained as
\be
\boxed{
\tilde A_1 \equiv \ln \frac{a_{\rm L} b_{\rm R}}{a_{\rm R} b_{\rm L}} }
\label{eff_A_1}
\ee
in terms of the averaged rates (\ref{aL})-(\ref{bR}).  This constitutes the main result of the present Chapter.

We notice that similar results hold in the other limit where the circuit No.\,1 is much faster than the circuit No.\,2 because both circuits have the same structure and are symmetrically coupled together through the Coulomb repulsion of parameter $U$ in equation~(\ref{H_S}).

The result (\ref{FT14}) shows that the generating function of the counting statistics in the slow quantum dot No.\,1 has the symmetry of a single-current fluctuation theorem under the experimental conditions (\ref{conditions}) but with respect to the effective affinity (\ref{eff_A_1}).  This latter may differ by orders of magnitude with respect to the affinity (\ref{A1}) driving the circuit out of equilibrium.  The reason for this modification is the back-action of the other circuit to which the quantum dot is capacitively coupled.  Indeed, the charging and discharging rates of the quantum dot No.\,1 are averaged over the two possible states of the quantum dot No.\,2 according to Eqs. (\ref{aL})-(\ref{bR}) so that their effective values are modified by the back-action of the circuit No.\,2.  This modification of the transition rates is reminiscent of the influence of environmental noises as described by the $P(E)$ theory \cite{IN92}.

In the following section, the dependence of the effective affinity (\ref{eff_A_1}) on the applied voltages and other parameters is numerically investigated under specific conditions, showing the importance of the back-action effect.


\subsection{Single-current fluctuation theorem at finite times}


Equations (\ref{p0t}) and (\ref{p1t}) for the occupation and transfer probabilities (\ref{p_nu1_n1}) can be cast into an equation for the vector
\be \label{4129}
{\bf g} (\lambda_1 , t) = \left(  
\begin{array}{c}
g_{0} (\lambda_1 ,t) \\
g_{1} (\lambda_1 , t )
\end{array}
\right)
\equiv 
\left(
\begin{array}{c}
g_{00} (\lambda_1,0 ,t)+g_{01} (\lambda_1,0 ,t) \\
g_{10} (\lambda_1,0 ,t)+g_{11} (\lambda_1,0 ,t)
\end{array}
\right)
\ee
so that
\be
\dot{{\bf g}} (\lambda_1 , t) = \tilde{{\bf W}} (\lambda_1  , t) \cdot  {\bf g } (\lambda_1 ,t)
\ee
with the coarse-grained modified rate matrix (\ref{tilde_L_0}).

The characteristic function at time $t$ of the electron number variable $n_1$ is thus given by
\bea
G(\lambda_1 , 0 ,t) & = & \langle \mbox{e}^{-\lambda_1 n_1} \rangle_t \label{charfunconevar} \\
& = & {\bf 1}^{\top} \cdot {\bf g} (\lambda_1 ,t) \\
& = & {\bf 1}^{\top} \cdot \mbox{e}^{t \, \tilde{{\bf W}} (\lambda_1 ,t)} \cdot {\bf p}_0,
\eea
where ${\bf p}_0$ is the initial probability distribution over the quantum dot No. $1$ regardless of the state of the quantum dot No. $2$, while ${\bf 1} $ is the vector 
\be
{\bf 1} = \left( \begin{array}{c}
1 \\
1
\end{array} \right).
\ee

The question arises whether the finite-time characteristic function (\ref{charfunconevar}) satisfies an effective single-current fluctuation theorem as does the cumulant generating function in the large current ratio limit. Here below, we show that this is indeed the case, the symmetry holding with an effective affinity given by (\ref{eff_A_1}).

In order to show this, we first note that the modified rate matrix $\tilde{{\bf W}} (\lambda_1)$ obeys the symmetry relation
\be
\tilde{{\bf M}}^{-1} \cdot \tilde{{\bf W}} (\lambda_1) \cdot \tilde{{\bf M}} = \tilde{{\bf W}} (\tilde{A}_1 -\lambda_1) ^{\top},
\ee
where the matrix $\tilde{{\bf M}} $ is given by
\be
\tilde{{\bf M}} =\left(
\begin{array}{cc}
\frac{b_{{\rm R}}}{a_{{\rm R}}+b_{{\rm R}}} & 0 \\
0 & \frac{a_{{\rm R}}}{a_{{\rm R}}+b_{{\rm R}}} 
\end{array}
\right)
\ee
and the effective affinity (\ref{eff_A_1}). The existence of such a matrix is the basic element in showing a finite-time fluctuation theorem for particular initial condition on the probability distribution over the quantum dot No. $1$.

Indeed, by introducing the probability distribution
\be \label{incondcoarse}
{\bf p}^{st}_0 = (\mbox{Tr} \tilde{{\bf M}})^{-1} \, \tilde{{\bf M}} \cdot {\bf 1},
\ee
one shows by the same steps as followed in (\ref{symmetryfinitetcalc}) - (\ref{symmetryfinitet}) that the generating function at time $t$ with initial condition chosen as ${\bf p}^{st}_{0}$ obeys the symmetry relation
\be \label{finitetefffluct}
G(\lambda_1 , 0 , t ) = G (\tilde{A}_1 -\lambda_1 , 0, t) \quad \forall t
\ee
thus proving the finite-time single-current fluctuation theorem by choosing the particular initial condition (\ref{incondcoarse}).

The interpretation of the probability vector (\ref{incondcoarse}) is given by considering the rate matrix $\tilde{{\bf W}} (\lambda_1)$ as the sum
\be
\tilde{{\bf W} }(\lambda_1) = \tilde{{\bf W} }_{{\rm L}}(\lambda_1) + \tilde{{\bf W} }_{{\rm R}}
\ee
with
\be
\tilde{{\bf W} }_{{\rm L}}(\lambda_1)  = \left(
\begin{array}{cc}
-a_{\rm L} &  b_{\rm L}{\rm e}^{+\lambda_1}  \\
a_{\rm L}{\rm e}^{-\lambda_1}& -b_{\rm L} \\
\end{array}
\right)
\quad
\mbox{and}
\quad
\tilde{{\bf W} }_{{\rm R}}
=
\left(
\begin{array}{cc}
-a_{\rm R} &  b_{\rm R}  \\
a_{\rm R} & -b_{\rm R} \\
\end{array}
\right),
\ee
where $\tilde{{\bf W} }_{{\rm L}}(\lambda_1)  $ accounts for the tunneling events between the quantum dot No. $1$ and reservoir the $1 {\rm L}$, while $\tilde{{\bf W} }_{{\rm R}}$ accounts for the processes induced by channel $2$ and reservoir $1 {\rm R}$ on the quantum dot No. $1$. 
By using the symmetry relation $\tilde{{\bf M}}^{-1} \cdot \tilde{{\bf W}}_{{\rm R}}  \cdot \tilde{{\bf M}} = \tilde{{\bf W}}_{{\rm R}}  ^{\top} $, one thus shows that the probability vector (\ref{incondcoarse}) is the stationary distribution with respect to the effective dynamics of quantum dot No. $1$ induced by reservoir $1 {\rm R}$, as well as the out of equilibrium channel $2$, that is (see Appendix \ref{AppendixB})
\be \label{eqap4}
\tilde{{\bf W}}_{{\rm R}} \cdot {\bf p}_{0}^{st} = 0.
\ee

As a consequence, the reservoir over which we count the electron flow - the reservoir $1 {\rm L}$ in our case - should be connected to the device after the other parts of the system have reached a stationary state in order to test the finite-time single-current fluctuation theorem (\ref{finitetefffluct}).

The symmetry of the fluctuation theorem (\ref{finitetefffluct}) is equivalent to a fluctuation relation for the probability distribution of the particles transferred through channel No. $1$, $p(n_1 , t) ={\bf 1}^{\top} \cdot {\bf p} (n_1, t)$, so that
\be \label{fluctrelcoarse}
 \ln \frac{p (n_1  , t)}{p (-n_1 , t)} = \tilde{A} n_1,
\ee
which is valid at any time $t$ provided the initial probability distribution on quantum dot No. $1$ is chosen as (\ref{incondcoarse}).

\begin{figure}
\centering
		\includegraphics[width=13cm]{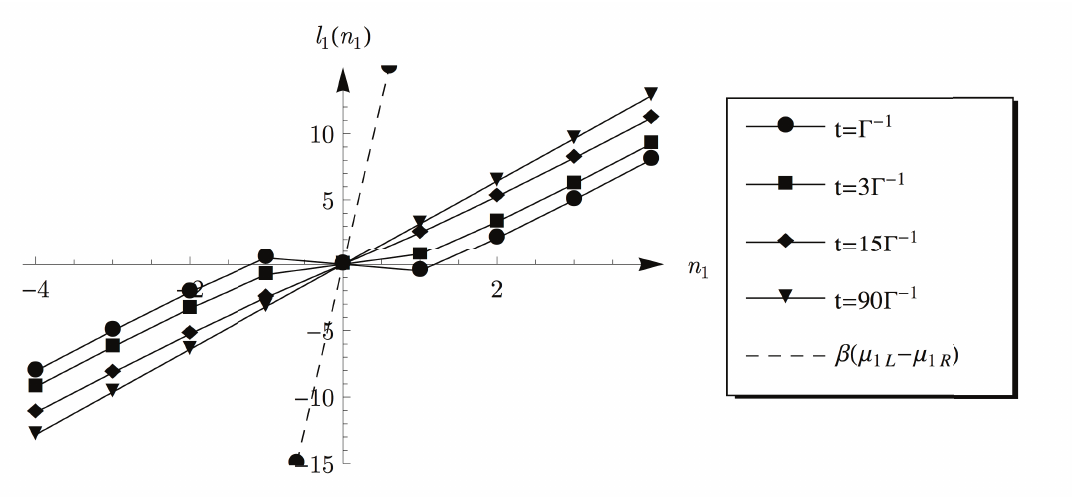}
	\caption{Test of the effective fluctuation relation (\ref{fluctrelcoarse}) at finite times by plotting the quantity $l_t (n_1)$ defined in (\ref{l1coarse}) as a function of the number of electrons $n_1$ transferred in the conduction channel $\mbox{No.} \,1$ during time $t$. The initial condition on the probability distribution over the quantum dot No. $1$ was chosen as $p_{0}=0$ and $p_{1}=1$. Parameters were chosen as in (\ref{U})-(\ref{e2}). We observe deviations to a linear behavior for short times, but the single-current fluctuation theorem is recovered in the  long time limit.}
	\label{fluctufinitecoarse}
\end{figure}

Deviations to relation (\ref{fluctrelcoarse}) are expected though for initial distributions far from (\ref{incondcoarse}). To illustrate this point, we show in \textsc{Figure} \ref{fluctufinitecoarse} the time evolution of the quantity
\be \label{l1coarse}
l_{t} (n_1 ,t ) =  \ln \frac{ p(n_1 , t)}{p(-n_1 , t)}
\ee
as a function of the number of electrons transferred out of reservoir $1 {\rm L}$. We observe strong deviations to (\ref{fluctrelcoarse}) at finite times, though this relation is recovered in the long-time limit, in consistency with (\ref{FT14}). It is worthy to point out the difference between the effective affinity and the ideal one (\ref{A1}) which is made manifest by the difference in slopes of the straight line $l_{t} (n_1)$ for large times $t$ and the dashed straight line with slope $\beta (\mu_{1 {\rm L}} - \mu_{1 {\rm R}})$.

\section{Numerical results}
\label{Numerics}


In this section, the effects exposed in Section \ref{scft} are numerically demonstrated with the model for parameter values corresponding to typical experimental conditions. We analyze the dependence of the effective affinity on the parameters of the Hamiltonian model and, especially, on the electrostatic interaction between both circuits.

Parameters are estimated by making an analogy with the experiment reported in \cite{fujisawa2006}. In this experiment, the detector is made of a quantum point contact (QPC) circuit sensitive to the electronic occupation in the double quantum dot (DQD) of the measured circuit. Though the model studied in this chapter is not identical to the experimental setup, it enables us to show how an effective single-current fluctuation theorem emerges for the slower circuit in the limit of large current ratio. This was the first theoretical work to put this point in evidence by use of a coarse-grained description of the slower degrees of freedom.


\subsection{Parameter values}


In typical counting statistics experiments \cite{fujisawa2006,PhysRevB.81.125331}, the affinities take quite large values because the voltages are large with respect to the temperature.  With the voltages $V_{\rm DQD}=300\; \mu$V, $V_{\rm QPC}=800\; \mu$V, and the electronic temperature $T=130$ mK reported in Ref.~\cite{fujisawa2006}, we can estimate the affinities as follows:
\bea
&& A_1 = A_{\rm DQD}=\frac{eV_{\rm DQD}}{k_{\rm B}T} = 25 \label{A1_act} , \\
&& A_2 = A_{\rm QPC}=\frac{eV_{\rm QPC}}{k_{\rm B}T}= 70 \label{A2_act} .
\eea
Since the QPC current is reduced by about 10\% if the QD is occupied, the parameter $U$ of the Coulomb repulsion between both quantum dots can be taken as
\be
\beta U = 32.8.
\label{U}
\ee
Moreover, the QPC current is about $10^7$-$10^8$ larger than the quantum dot current.

As mentioned earlier, the role of the detector is played by the circuit No.\,2, while the circuit No.\, 1 in our model will be the measured circuit.  The energy level of the second quantum dot is supposed to be in the middle between the reservoirs electrochemical potentials and the couplings to the reservoirs are chosen symmetric and independent of the energy.  Under such assumptions, possible parameter values are given by 
\bea
&& \beta\mu_{1{\rm L}} = 25 \label{m1L} , \\
&& \beta\mu_{1{\rm R}} = 0 \label{m1R} , \\
&& \Gamma_{1{\rm L}}= \Gamma_{1{\rm R}} =\bar{\Gamma}_{1{\rm L}}= \bar{\Gamma}_{1{\rm R}} = 1 \label{G1} , \\
&& \beta\mu_{2{\rm L}} = 70 \label{m2L} , \\
&& \beta\mu_{2{\rm R}} = 0 \label{m2R} , \\
&& \Gamma_{2{\rm L}}= \Gamma_{2{\rm R}} = \bar{\Gamma}_{2{\rm L}}= \bar{\Gamma}_{2{\rm R}} = 10^4 \label{G2} , \\
&& \beta\epsilon_2 = 35 \label{e2},
\eea
while the level of the quantum dot No.\,1 has the energy $\epsilon_1$, which may take different values in the following numerical calculations. We suppose that the correlation times of the reservoirs are short enough for the conditions (\ref{Dt_begin}) to hold in consistency with the perturbative approximation and that the thermal energy is sufficiently large to satisfy the conditions (\ref{hG<kT}).  Here, we use the rates of the quantum dot No.\,1 in order to fix the unit of time.

A remark is that there is no Coulomb drag for the conditions chosen in the present section because we have here taken rate constants such that $\Gamma_j=\bar{\Gamma}_j$ in equations (\ref{G1}) and (\ref{G2}).  Therefore, the Onsager coefficient (\ref{L12_prop}) vanishes together with higher-order coefficients according to equations (\ref{J_a=0}) and (\ref{LMN}) and the Coulomb drag does not manifest itself for the conditions we here consider.

\begin{figure}
	\centering
		\includegraphics[width=10cm]{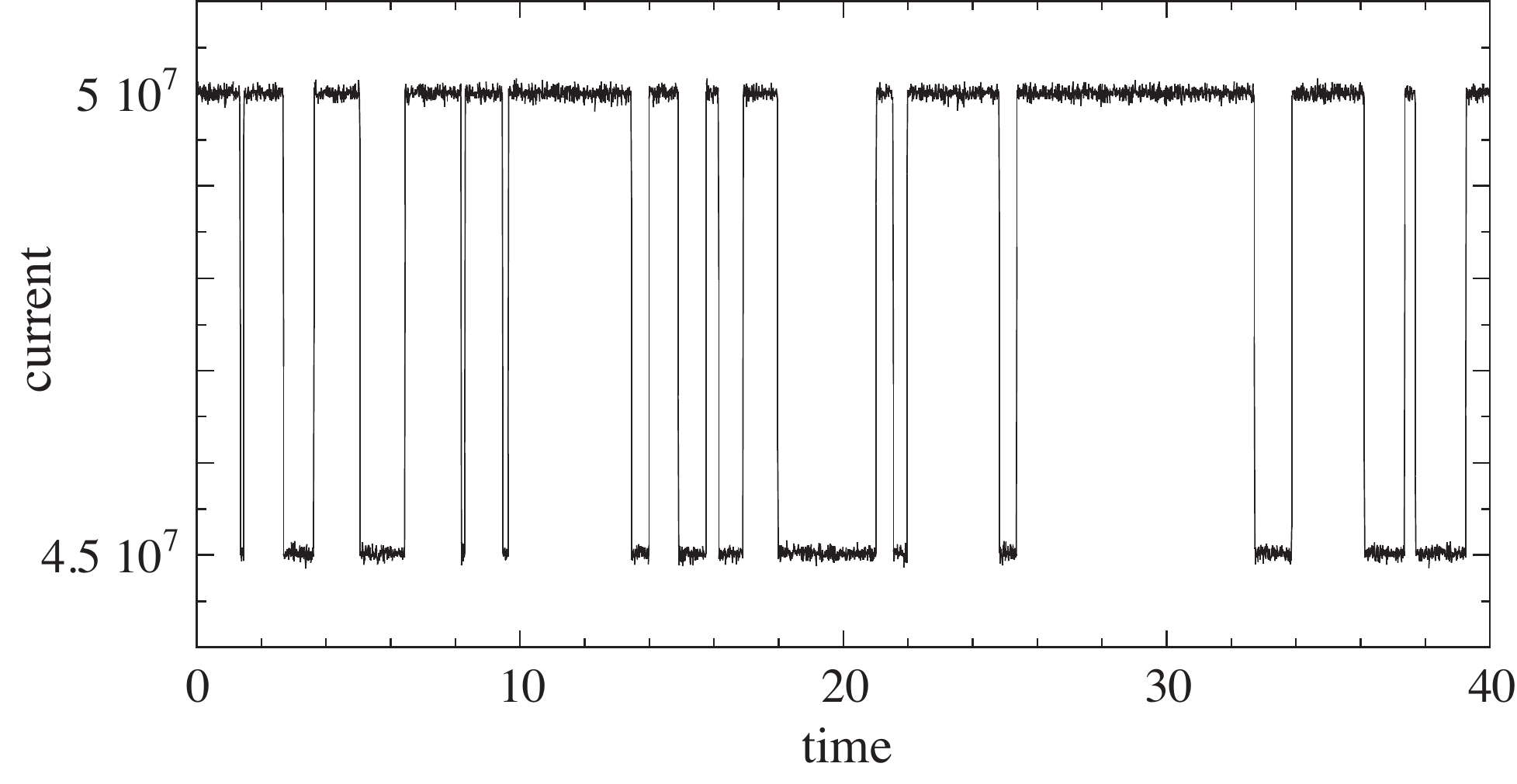}
	\caption[Simulation of current trajectory in the detector]{Simulation with Gillespie's algorithm of the current in circuit No. 2 measuring the quantum dot occupancy.  The parameter values are given by equations (\ref{U})-(\ref{e2}) and $\beta\epsilon_1=0$.  The effective affinity of the circuit No. 1 is $\tilde{A}_1=1.17$.  The mean value of the current in channel $\mbox{No.} \ ,1$ is $J_1\simeq 0.17$ electrons per unit time.  The mean value of the current in channel $\mbox{No.}\, 2$ is $J_2\simeq 4.8\times10^7$ electrons per unit time.  The quantum dot $\mbox{No.}\, 1$ is empty (resp. occupied) when the current takes the value $5\times 10^7$ (resp. $4.5\times 10^7$).}
	\label{fig24}
\end{figure}


\subsection{Stochastic simulations}


The random time evolution of the system can be generated by simulating the stochastic jump process of the master equation (\ref{probastoch}) with Gillespie's algorithm \cite{G76,G77}.  Four possible transitions may occur from each of the four states.  The transition rates are given by equations (\ref{aj})-(\ref{bUj}) with the Fermi-Dirac distributions (\ref{fj})-(\ref{fUj}) and the rate constants (\ref{G1})-(\ref{G2}). 

\begin{figure}
\centering
		\includegraphics[width=14cm]{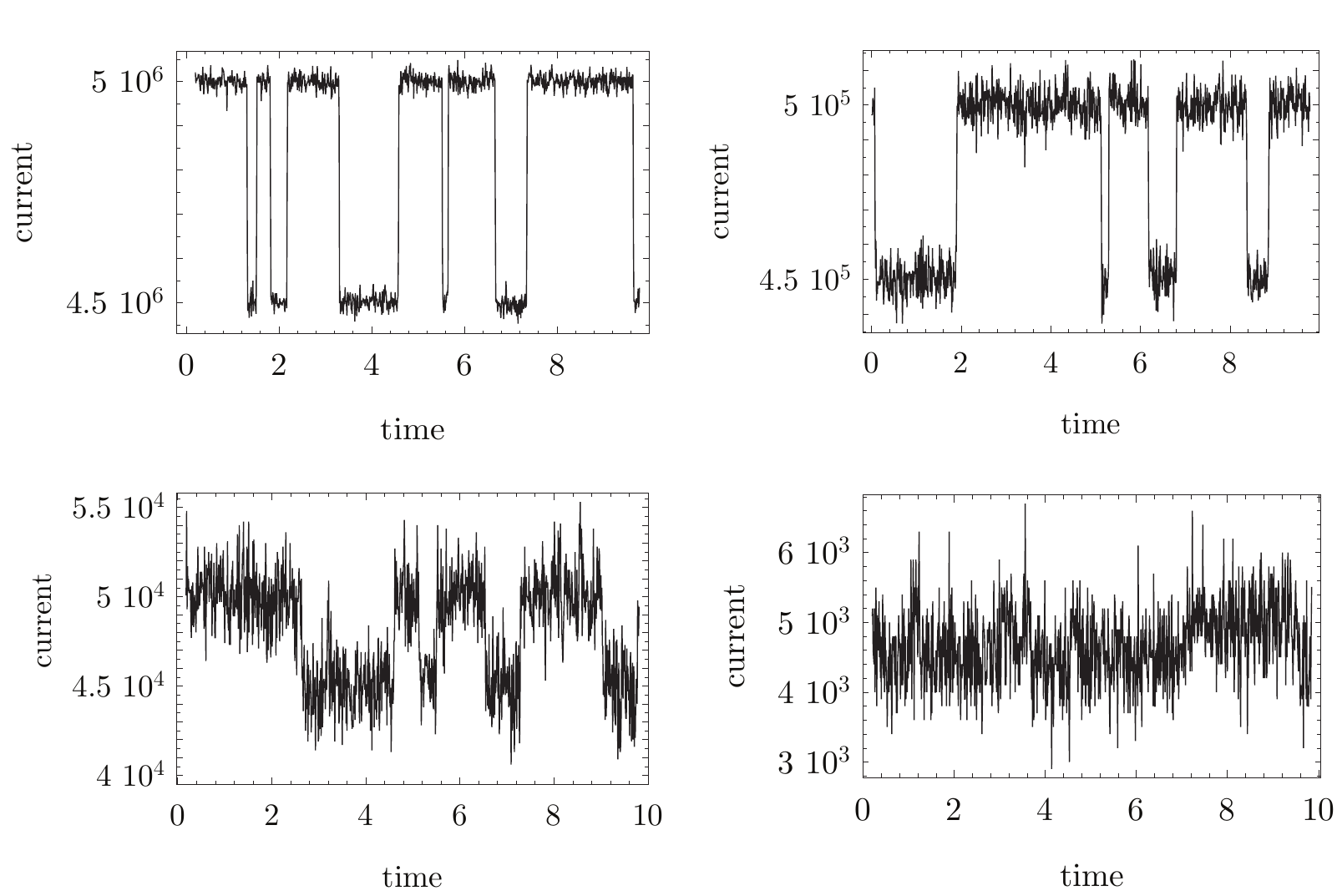}
	\caption{Current trajectories in circuit No. 2 measuring the quantum dot occupancy.  The parameter values are chosen as in \textsc{Figure} \ref{fig24} except for the tunneling amplitudes $\Gamma_{2\alpha}$ for $\alpha={\rm L, {\rm R}}$ which takes the values $\Gamma_{2\alpha} = 10^{3}$, $10^{2}$, $10^1$ and $10^{0}$ from left to right and top to bottom. The fluctuations of the detector current increase as its tunneling amplitudes $\Gamma_{2\alpha}$ decreases eventually forbidding the identification of the electronic occupation in quantum dot $\mbox{No.} 1$. }
	\label{currenttrajs}
\end{figure}

\textsc{Figure} \ref{fig24} depicts the current in the circuit No.\,2 averaged over a time interval $\Delta t = 0.01$, which is shorter than the typical dwell time of the quantum dot No.\,1, as required by equation (\ref{Dt_real}).  We see that the current is reduced by about 10\% when the quantum dot No.\,1 is occupied, which is in agreement with the choice for the parameter (\ref{U}).  The ratio between the mean values of the currents is here given by $J_2/J_1=2.8\times 10^8$, while the ratio of the dissipated powers takes the value $\Pi_2/\Pi_1=(A_2J_2)/(A_1J_1)=7.9\times 10^8$.  Such very large ratios are required in order for the secondary current to distinguish between the two states of the QD in the primary circuit.  Indeed, simulations show that the fluctuations of the secondary current would be larger for smaller values of the current ratio as illustrated in \textsc{Figure} \ref{currenttrajs}.  Thanks to the large ratio, the instantaneous occupancy in the circuit No.\,1 can be monitored by the current in the circuit No.\,2 over the time scale (\ref{Dt_real}), which is longer than the time scale of the fast circuit No.\,2 but shorter than the one of the circuit No.\,1.

\begin{figure}
\centering
		\includegraphics[width=13cm]{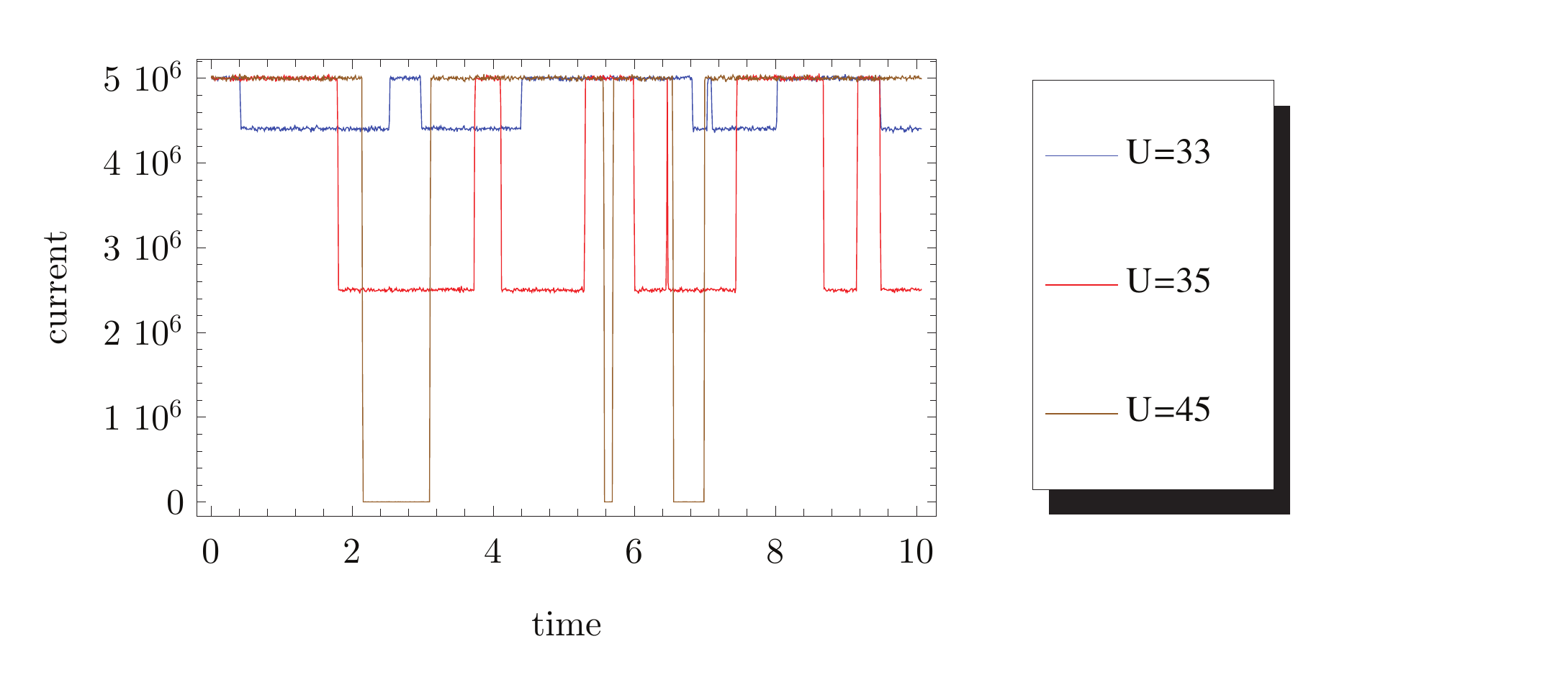}
	\caption{Current trajectories in circuit No. 2 for different values of the inter-dot interaction parameter $U$. An increase of the interaction strength leads to higher suppression of the detector current when quantum dot $\mbox{No.} \, 1$ is occupied.}
	\label{currenttrajU}
\end{figure}

Finally, we show the influence of the Coulomb interaction parameter $U$ on the current trajectories in the detection circuit in \textsc{Figure} \ref{currenttrajU}. As can be seen, the current when quantum dot $\mbox{No.} \, 1$ is occupied eventually vanishes when $U$ is sufficiently large confirming the results of section \ref{Uinfty}. A sufficiently large value of $U$ is needed in order to distinguish between the two charge states in the quantum dot $\mbox{No.} \, 1$.


\subsection{The cumulant generating function and its properties}


\begin{figure}
\centerline{
\includegraphics[width=8.5cm]{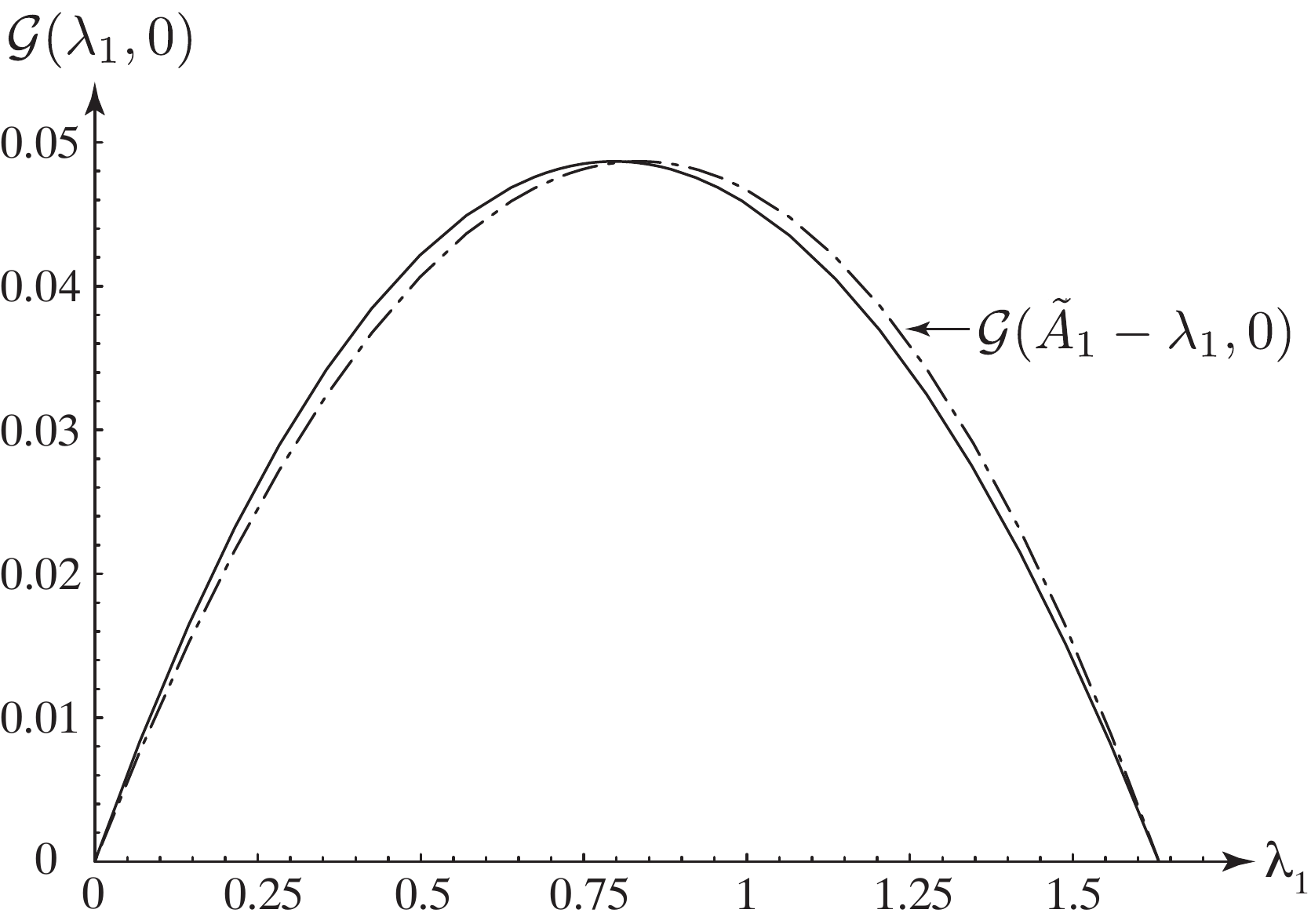}}
\caption{The cumulant generating function versus the counting parameter $\lambda_1$ at $\lambda_2=0$ and the symmetric function with respect to the effective affinity $\tilde{A}_1=1.6319$ (dotted-dashed line) for the parameter values $\beta U=30$, $\beta\epsilon_1=0$, $\beta\epsilon_2=35$, $\beta\mu_{1{\rm L}}=25$, $\beta\mu_{1{\rm R}}=0$, $\beta\mu_{2{\rm L}}=70$, $\beta\mu_{2{\rm R}}=0$, $\Gamma_{1{\rm L}}=\Gamma_{1{\rm R}}=\bar{\Gamma}_{1{\rm L}}= \bar{\Gamma}_{1{\rm R}} =1$, $\Gamma_{2{\rm L}}=\Gamma_{2{\rm R}}= \bar{\Gamma}_{2{\rm L}}= \bar{\Gamma}_{2{\rm R}}=1$.}
\label{fig34}
\end{figure}

The cumulant generating function $\mathcal{G} (\lambda_1 , 0)$ of the current in the circuit No.\,1 is calculated by the leading root of the characteristic polynomial (\ref{det}) of the four-by-four matrix (\ref{modratematqdpar}) - (\ref{L2}) with $\lambda_2=0$.

The lack of symmetry of the single-current generating function $\mathcal{G}(\lambda_1,0)$ is manifest if the rate constants of both circuits are of the same order of magnitude.  The generating function and its symmetric with respect to the effective affinity is depicted in \textsc{Figure}~\ref{fig34} for $\Gamma_{2\alpha}/\Gamma_{1\alpha}=1$ (with $\alpha={\rm L, R}$) and $\beta U=30$.  Here, the effective affinity is taken as the non-trivial root of the generating function such that $\mathcal{G}(\tilde{A}_1,0)=0$.  We clearly see that the generating function is not symmetric with respect to the effective affinity $\mathcal{G}(\lambda_1,0)\neq \mathcal{G}(\tilde{A}_1-\lambda_1,0)$ so that the single-current fluctuation theorem does not hold in general although the two-current fluctuation theorem always does.  Furthermore, we notice that the effective affinity $\tilde{A}_1=1.6319$ is much smaller than the affinity determined by the reservoirs: $A_1=\beta(\mu_{\rm 1L}-\mu_{\rm 1R})=25$.

\begin{figure}
\centerline{\includegraphics[width=8.5cm]{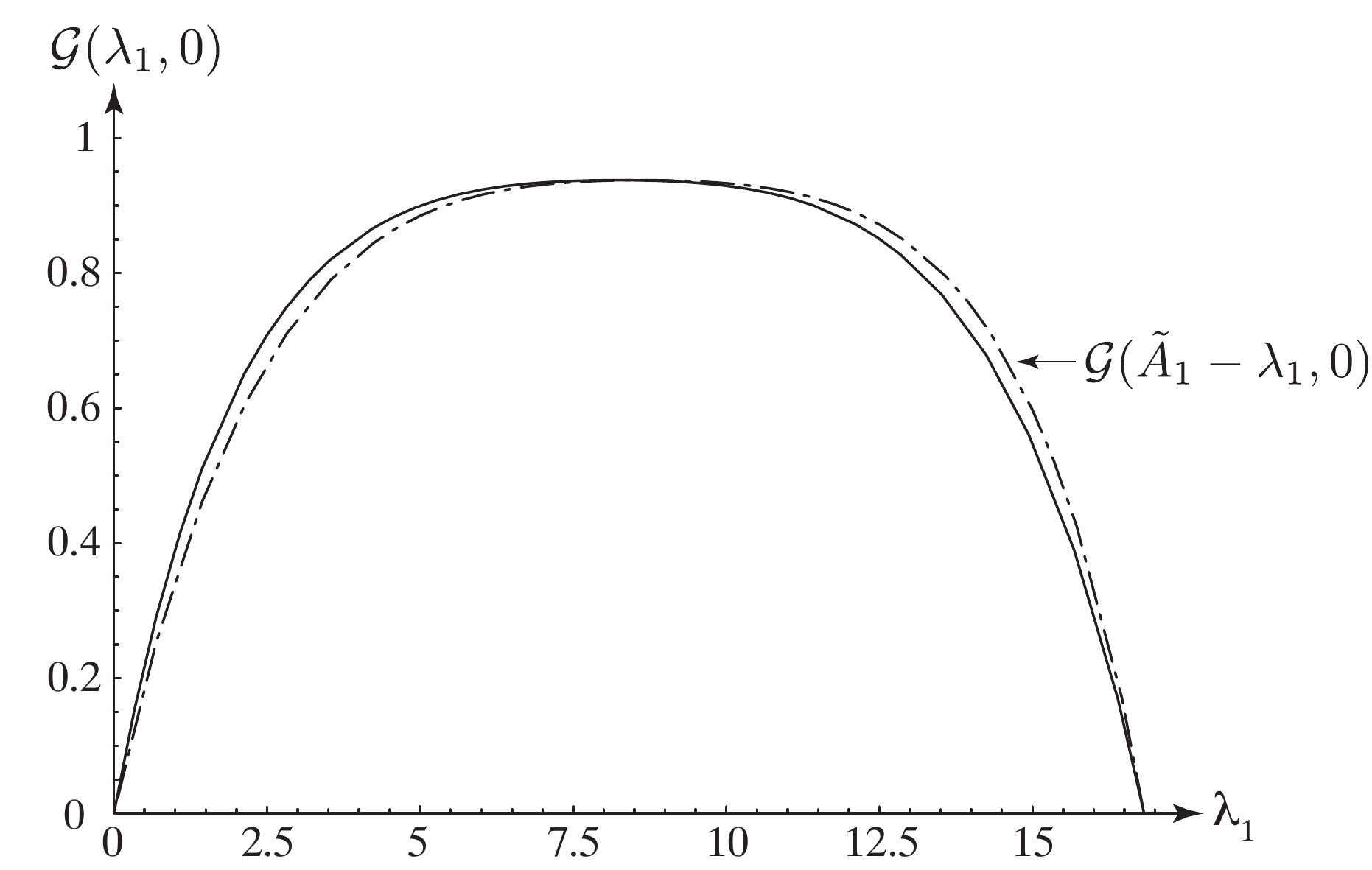}}
\caption{The cumulant generating function versus the counting parameter $\lambda_1$ at $\lambda_2=0$ and the symmetric function with respect to the effective affinity $\tilde{A}_1=16.8356$ (dotted-dashed line) for the parameter values $\beta U=10$, $\beta\epsilon_1=10$, $\beta\epsilon_2=35$, $\beta\mu_{1{\rm L}}=25$, $\beta\mu_{1{\rm R}}=0$, $\beta\mu_{2{\rm L}}=70$, $\beta\mu_{2{\rm R}}=0$, $\Gamma_{1{\rm L}}=\Gamma_{1{\rm R}}=\bar{\Gamma}_{1{\rm L}}= \bar{\Gamma}_{1{\rm R}} =1$, $\Gamma_{2{\rm L}}=\Gamma_{2{\rm R}}= \bar{\Gamma}_{2{\rm L}}= \bar{\Gamma}_{2{\rm R}}=2$.}
\label{fig44}
\end{figure}

In \textsc{Figure} \ref{fig44}, the single-current generating function is also depicted for the smaller value of the Coulomb repulsion $\beta U=10$ and $\Gamma_{2\alpha}/\Gamma_{1\alpha}=2$.  Here, the effective affinity takes a larger value, but again the asymmetry of the generating function is still manifest.  We note that the shape of the generating function now deviates from the parabolic shape seen in \textsc{Figure} \ref{fig34} as its maximum approaches the unity value.

Although the ratio of the rate constants is of order unity in both \textsc{Figure} \ref{fig34} and \ref{fig44}, the difference between the generating function and its symmetric is smaller than 5\% and could remain unobservable if the counting statistics was not precise enough.

\begin{figure}
\centerline{\includegraphics[width=8cm]{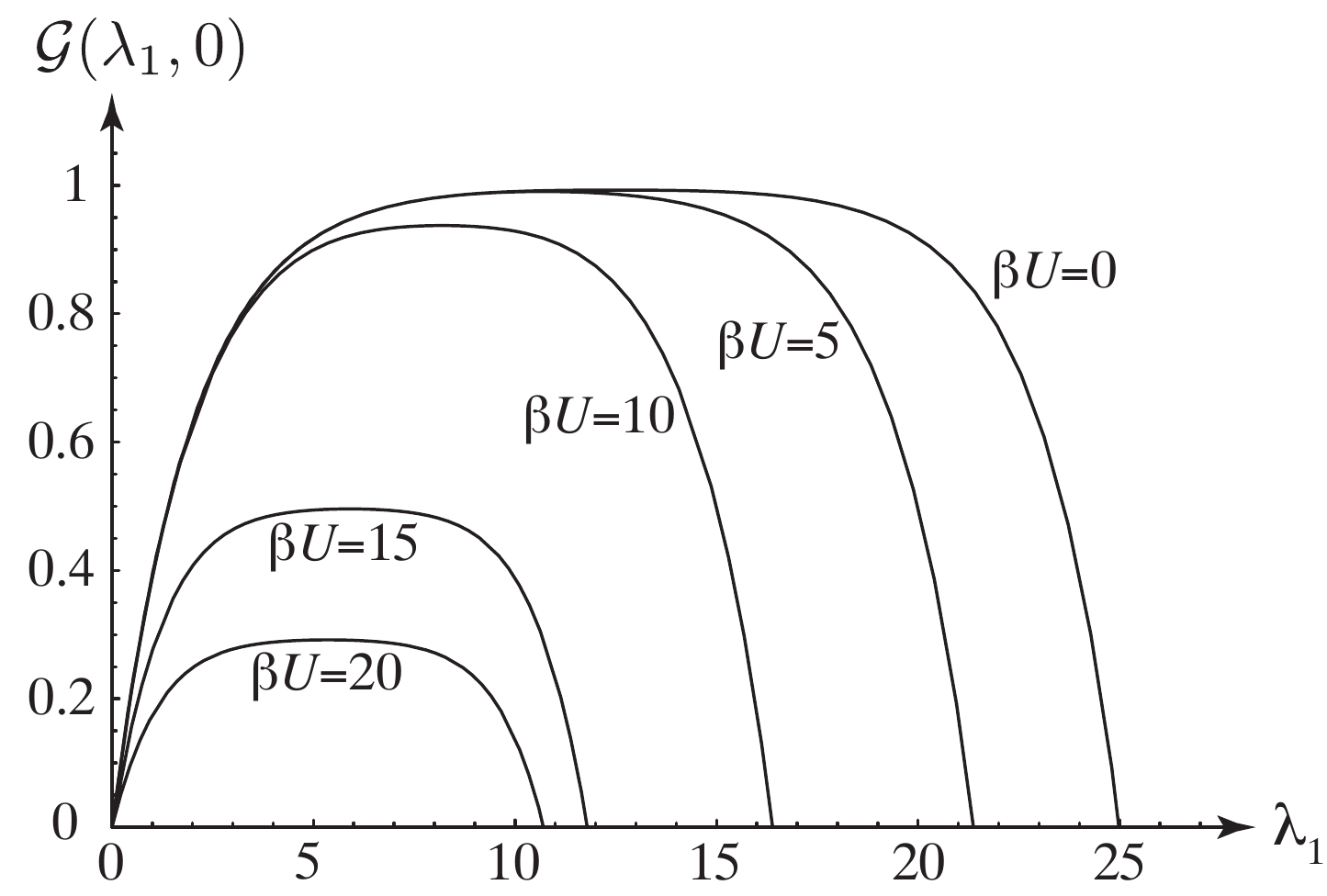}}
\caption{The cumulant generating function versus the counting parameter $\lambda_1$ at $\lambda_2=0$ for different values of the electrostatic coupling parameter $\beta U$. The other parameters take the values $\beta\epsilon_1=10$, $\beta\epsilon_2=35$, $\beta\mu_{1{\rm L}}=25$, $\beta\mu_{1{\rm R}}=0$, $\beta\mu_{2{\rm L}}=70$, $\beta\mu_{2{\rm R}}=0$, $\Gamma_{1{\rm L}}=\Gamma_{1{\rm R}}=\bar{\Gamma}_{1{\rm L}}= \bar{\Gamma}_{1{\rm R}} =1$, $\Gamma_{2{\rm L}}=\Gamma_{2{\rm R}}=100$.}
\label{fig54}
\end{figure}

\textsc{Figure} \ref{fig54} shows the deformation of the generating function $\mathcal{G}(\lambda_1,0)$ as the electrostatic coupling parameter $U$ varies from zero to $\beta U=20$ for $\Gamma_{2i}/\Gamma_{1i}=100$.  In the absence of electrostatic coupling, the single-current fluctuation theorem holds in the circuit No.\,1 since it is decoupled from the rest of the system.  In this case, the affinity takes the value $A_1=25$ determined by the two reservoirs of this circuit, as seen in \textsc{Figure} \ref{fig54}.  However, the non-trivial root $\tilde{A}_1$ of the generating function decreases as the Coulomb repulsion $U$ increases, showing the back-action effect of the secondary circuit due to the capacitive coupling.  In the same progression, the maximum of the generating function is also reduced.

For the ratio of rate constants taken in \textsc{Figure} \ref{fig54}, the generating function is already practically indistinguishable from its symmetric $\mathcal{G}(\tilde{A}_1-\lambda_1,0)$ so that the single-current fluctuation theorem is already effective and the considerations of Section \ref{Limit} apply.  In particular, the effective affinity is now very well approximated by Eq. (\ref{eff_A_1}).  

\subsection{The large current ratio limit and the effective affinity}


\begin{figure}
\centerline{\includegraphics[width=9cm]{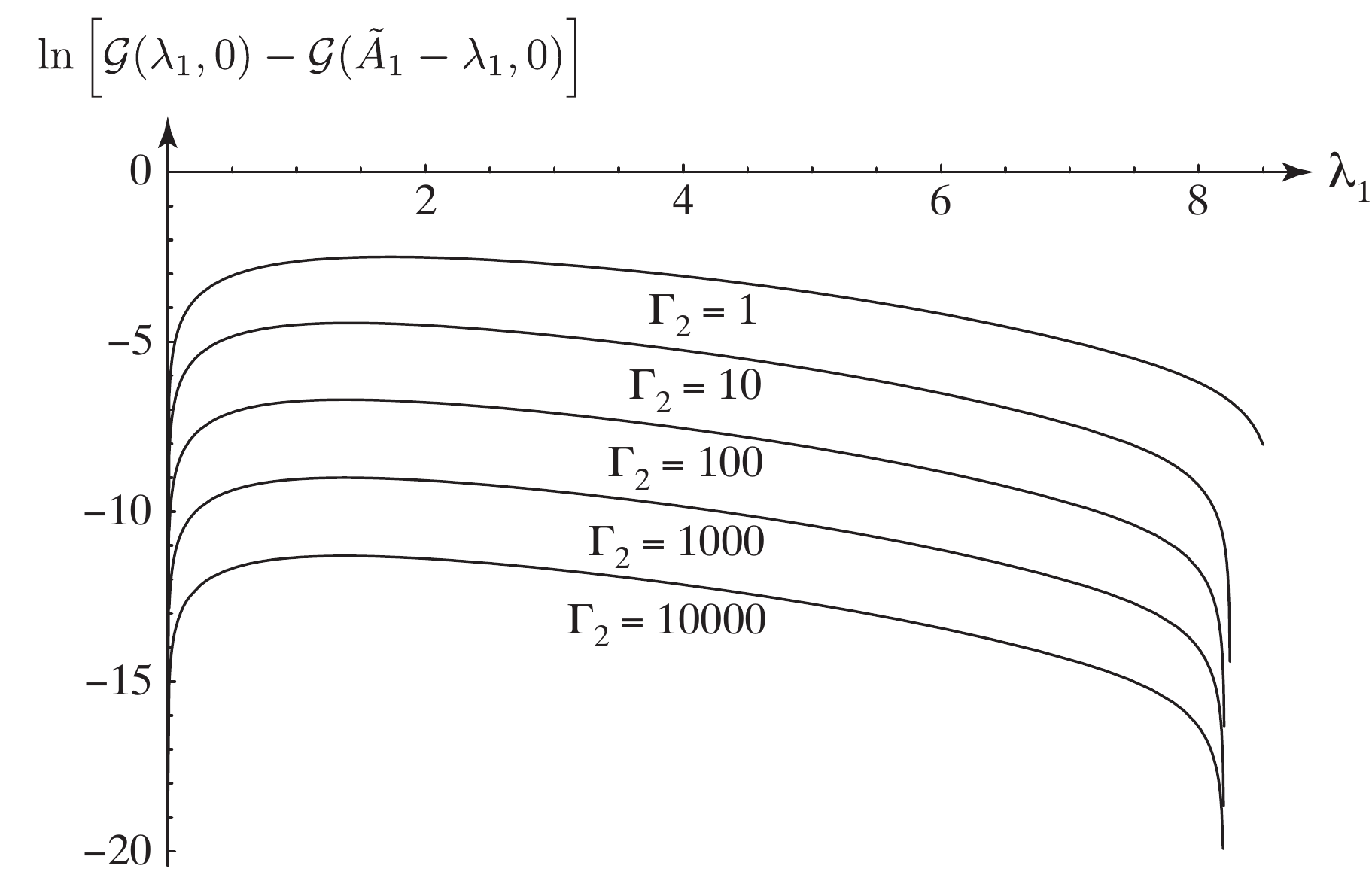}}
\caption{The difference between the cumulant generating function and its symmetric with respect to the effective affinity $\tilde{A}_1$ versus the counting parameter $\lambda_1$ at $\lambda_2=0$ for the parameter values $\beta U=10$, $\beta\epsilon_1=10$, $\beta\epsilon_2=35$, $\beta\mu_{1{\rm L}}=25$, $\beta\mu_{1{\rm R}}=0$, $\beta\mu_{2{\rm L}}=70$, $\beta\mu_{2{\rm R}}=0$, $\Gamma_{1{\rm L}}=\Gamma_{1{\rm R}}=\bar{\Gamma}_{1{\rm L}}= \bar{\Gamma}_{1{\rm R}} =1$, and $\Gamma_2\equiv\Gamma_{2{\rm L}}=\Gamma_{2{\rm R}}= \bar{\Gamma}_{2{\rm L}}= \bar{\Gamma}_{2{\rm R}}=1, 10, 100, 1000, 10000$.  As observed in \textsc{Figures} \ref{fig34} and \ref{fig44}, the difference $\mathcal{G}(\lambda_1,0)-\mathcal{G}(\tilde{A}_1-\lambda_1,0)$ is positive for $\lambda_1<\tilde{A}_1/2$ and negative for $\lambda_1>\tilde{A}_1/2$.  Here, we only depict the difference for $\lambda_1<\tilde{A}_1/2$.  The other half has a similar structure if the absolute value of the difference is taken before the logarithm.}
\label{fig64}
\end{figure}

In the limit where the ratio of rate constants tends to infinity, the generating function becomes identical with its symmetric, as argued in Section \ref{Limit}.  In order to verify this prediction, we depict in \textsc{Figure} \ref{fig64} the difference between both functions versus the counting parameter $\lambda_1$.  We observe in this figure that the difference is reduced by one order of magnitude each time the ratio of rate constants $\Gamma_2/\Gamma_1$ is increased by the same factor.  Consequently, the single-current fluctuation theorem is well established in the large ratio limit $\Gamma_2/\Gamma_1\to \infty$. In this limit, the effective affinity is given by Eq. (\ref{eff_A_1}).

\begin{figure}
\centerline{\includegraphics[width=9cm]{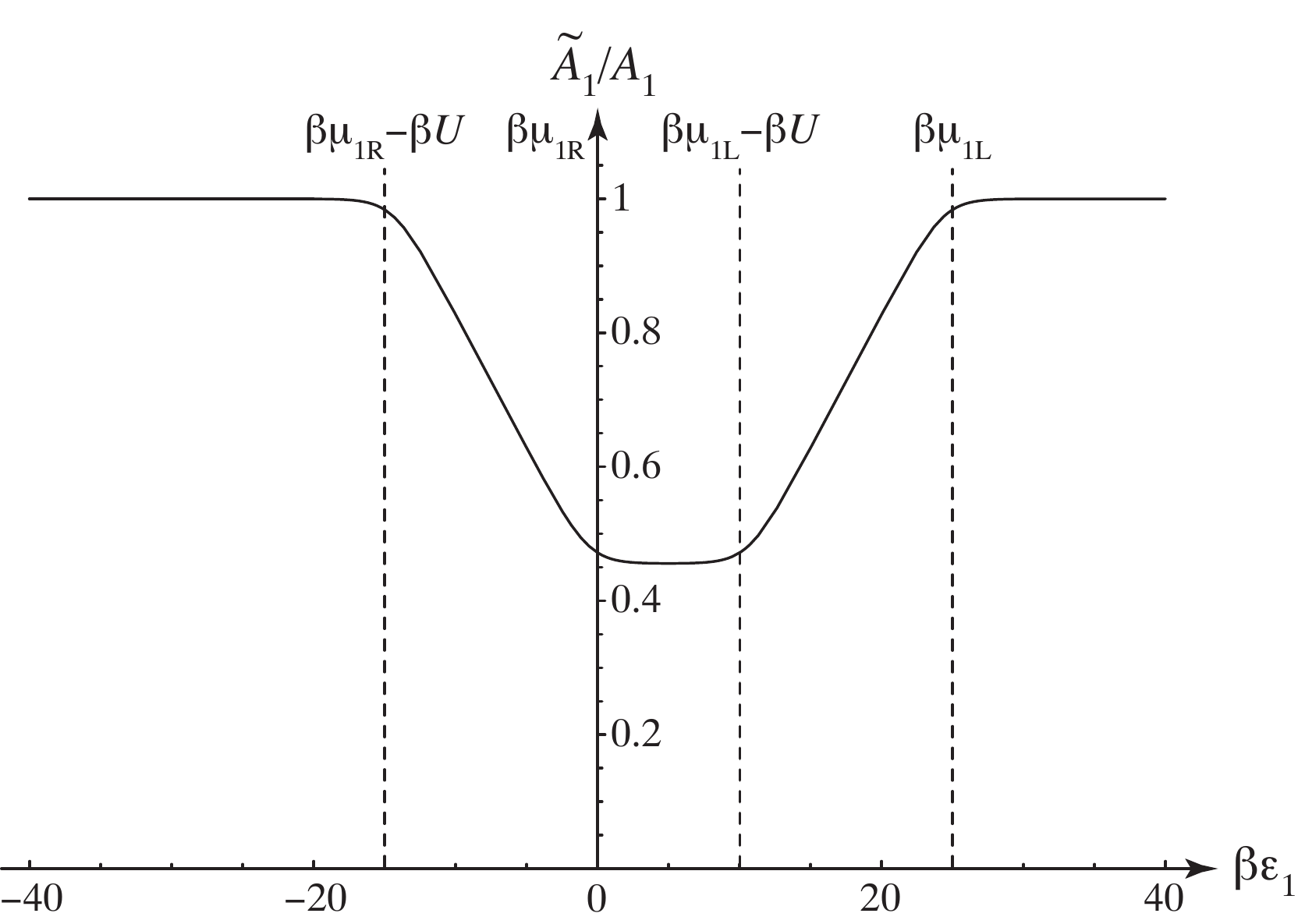}}
\caption{The effective affinity (\ref{eff_A_1}) of the quantum dot versus the dimensionless energy $\beta\epsilon_1$ of its level for the parameter values $\beta U=15$ and (\ref{m1L})-(\ref{e2}).}
\label{fig74}
\end{figure}

The effective affinity is depicted in \textsc{Figure} \ref{fig74} as a function of the energy $\beta\epsilon_1$ of the quantum dot No.\,1 for $\beta U=15$.  We observe that the effective affinity takes the actual value (\ref{A1_act}) determined by the reservoirs for either low or large values of the energy $\beta\epsilon_1$.  However, the effective affinity undergoes a significant reduction in between, down to a minimum of about $\tilde{A}_1\simeq 0.45 \times A_1$.  The function has a characteristic shape, which can be explained in terms of the Fermi-Dirac distributions (\ref{fj})-(\ref{fUj}) entering in the expression (\ref{eff_A_1}) of the effective affinity.  Away from their critical energy $\epsilon_j=\mu_j$ or $\epsilon_j=\mu_j-U$, these Fermi-Dirac distributions approximately behave as constants or Maxwell-Boltzmann exponential distributions.  As the consequence of the logarithm defining the effective affinity (\ref{eff_A_1}), this latter switches between either constant or linear dependences on the energies or chemical potentials.  Supposing that $\mu_{\rm 2R} < \epsilon_2 < \mu_{2L}-U$ and $0<U<\mu_{\rm 1L}-\mu_{\rm 1R}$, we find that the effective affinity is approximately given by
\be
\tilde{A}_1 \simeq
\left\{
\begin{array}{ll}
\beta(\mu_{\rm 1L}-\mu_{\rm 1R}) & \mbox{for} \quad \epsilon_1<\mu_{\rm 1R}-U \\ 
\beta(-\epsilon_1-U+\mu_{\rm 1L}) & \mbox{for} \quad \mu_{\rm 1R}-U < \epsilon_1< \mu_{\rm 1R} \\ 
\beta(\mu_{\rm 1L}-\mu_{\rm 1R}-U) & \mbox{for} \quad \mu_{\rm 1R}<\epsilon_1<\mu_{\rm 1L}-U \\ 
\beta(\epsilon_1-\mu_{\rm 1R}) & \mbox{for} \quad \mu_{\rm 1L}-U < \epsilon_1< \mu_{\rm 1L} \\ 
\beta(\mu_{\rm 1L}-\mu_{\rm 1R}) & \mbox{for} \quad \mu_{\rm 1L} <\epsilon_1 \\ 
\end{array}
\right.
\label{eff_A_1_shape1}
\ee
up to corrections that are smaller than $\beta=(k_{\rm B}T)^{-1}$ in the zero temperature limit $T\to 0$.
Crossovers happen where the energy $\epsilon_1$ coincides with the values of the chemical potentials of the left- and right-hand reservoirs and the chemical potentials reduced by the Coulomb repulsion $U$.  The slope of the effective affinity versus $\beta\epsilon_1$ is successively $\{0,-1,0,+1,0\}$, as seen in \textsc{Figure} \ref{fig74}.  According to Eq. (\ref{eff_A_1_shape1}), the minimum value of the effective affinity is approximately given by $\tilde{A}_1\simeq A_1-\beta U = 10$ in the middle interval $\beta\mu_{\rm 1R}=0<\beta\epsilon_1<\beta\mu_{\rm 1L}-\beta U=10$. The affinity $A_1=25$ of the reservoirs is recovered for $\beta\epsilon_1<\beta\mu_{\rm 1R}-\beta U=-15$ and for $\beta\epsilon_1>\beta \mu_{\rm 1L}=25$, which explains the features observed in \textsc{Figure} \ref{fig74}.

\begin{figure}
\centerline{\includegraphics[width=9cm]{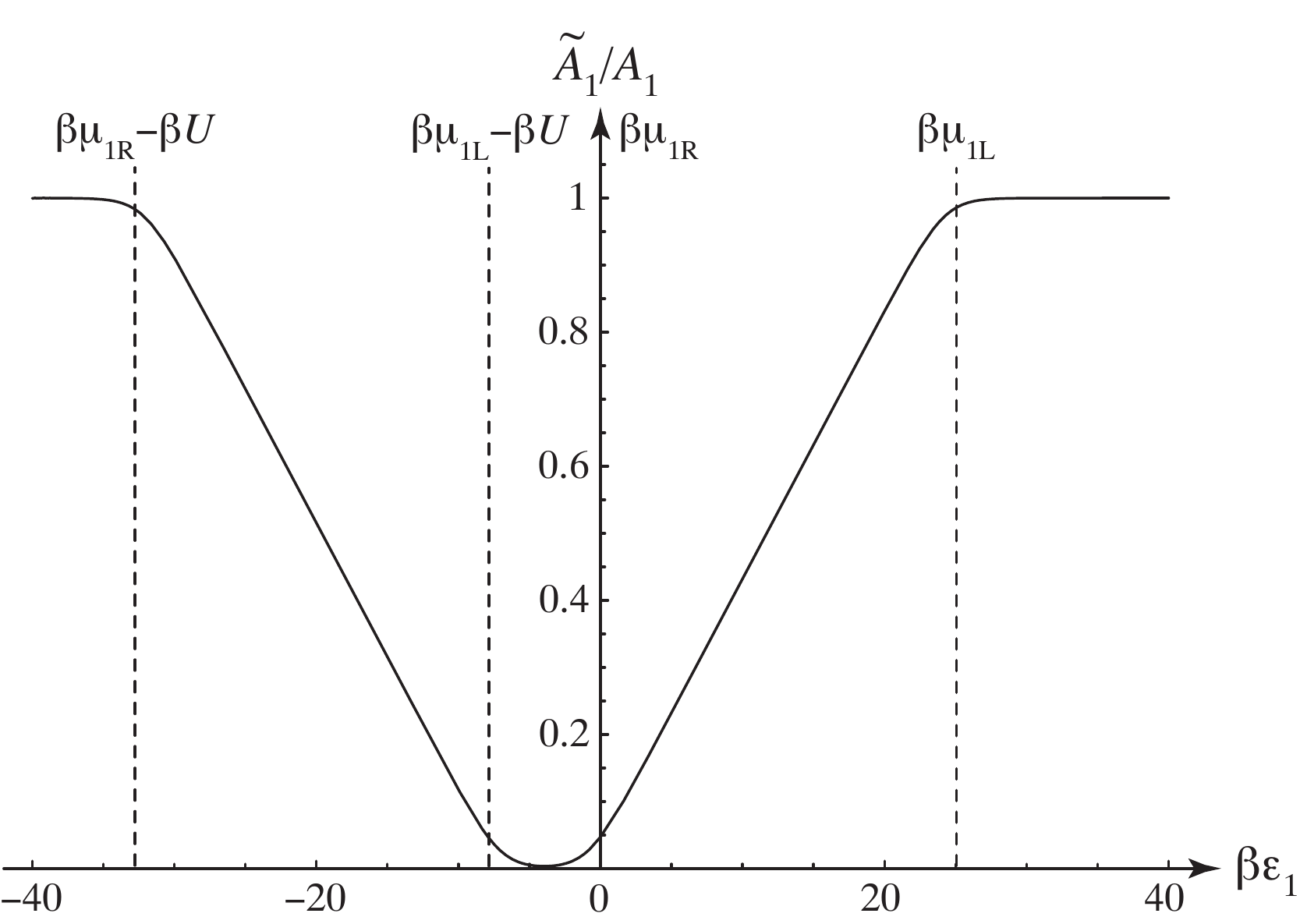}}
\caption{The effective affinity (\ref{eff_A_1}) of the quantum dot versus the dimensionless energy $\beta\epsilon_1$ of its level for the parameter values (\ref{U})-(\ref{e2}).}
\label{fig84}
\end{figure}

Equation (\ref{eff_A_1_shape1}) predicts that the minimum value of the effective affinity could be further decreased by increasing the Coulomb repulsion $U$.  This is indeed the case as observed in \textsc{Figure} \ref{fig84}, which depicts the effective affinity versus the energy $\epsilon_1$ now for the value (\ref{U}).
Here, we see that the effective affinity may vary from the maximum value given by the affinity $A_1 = 25$ imposed by the reservoirs down to the very small minimum value $\tilde A_1 \simeq 0.083565$ at $\beta\epsilon_1\simeq-3.9525$, i.e., a drop by a factor $300$.  

If the condition $\mu_{\rm 2R} < \epsilon_2 < \mu_{2L}-U$ is still satisfied for the parameter values of \textsc{Figure} \ref{fig84}, the Coulomb repulsion is now larger than the difference of chemical potentials:  $U>\mu_{\rm 1L}-\mu_{\rm 1R}$.  In this other regime, the effective affinity is approximately given by
\be
\tilde{A}_1 \simeq
\left\{
\begin{array}{ll}
\beta(\mu_{\rm 1L}-\mu_{\rm 1R}) & \mbox{for} \quad \epsilon_1<\mu_{\rm 1R}-U \\ 
\beta(-\epsilon_1-U+\mu_{\rm 1L}) & \mbox{for} \quad \mu_{\rm 1R}-U < \epsilon_1< \mu_{\rm 1L}-U \\ 
0 & \mbox{for} \quad \mu_{\rm 1L}-U<\epsilon_1<\mu_{\rm 1R} \\ 
\beta(\epsilon_1-\mu_{\rm 1R}) & \mbox{for} \quad \mu_{\rm 1R} < \epsilon_1< \mu_{\rm 1L} \\ 
\beta(\mu_{\rm 1L}-\mu_{\rm 1R}) & \mbox{for} \quad \mu_{\rm 1L} <\epsilon_1 \\ 
\end{array}
\right.
\label{eff_A_1_shape2}
\ee
up to corrections that are smaller than $\beta=(k_{\rm B}T)^{-1}$ in the zero temperature limit $T\to 0$.
In the middle interval $\beta\mu_{\rm 1L}-\beta U=-7.8<\beta\epsilon_1<\beta\mu_{\rm 1R}=0$, the minimum effective affinity reaches a value that vanishes in the low temperature limit $T\to 0$.  The actual value of the affinity $A_1=25$ is recovered for $\beta\epsilon_1<\beta\mu_{\rm 1R}-\beta U=-32.8$ or $\beta\epsilon_1>\beta \mu_{\rm 1L}=25$.

\begin{figure}
\centerline{\includegraphics[width=9cm]{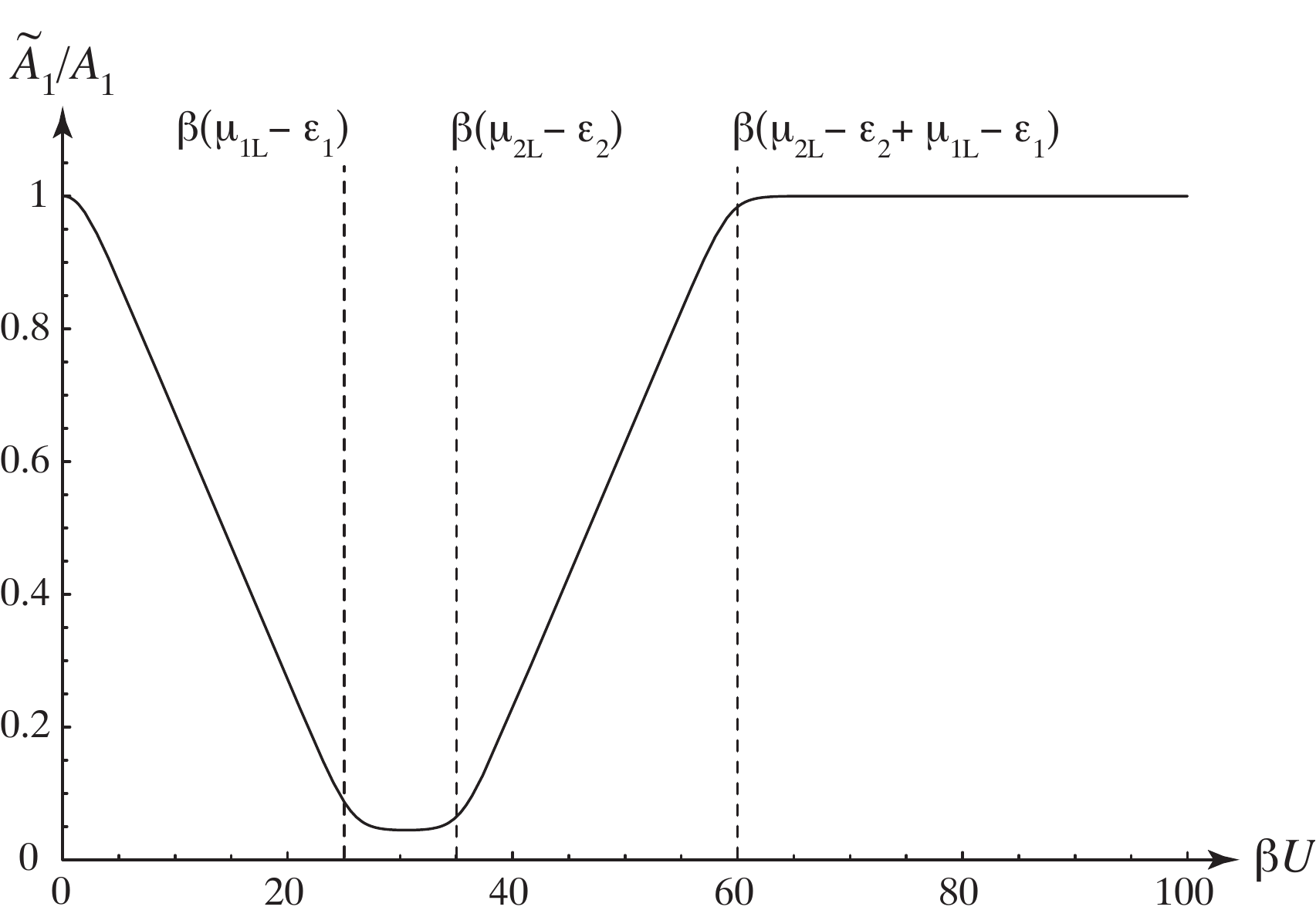}}
\caption{The effective affinity (\ref{eff_A_1}) of the quantum dot versus the dimensionless electrostatic coupling constant $\beta U$ for the parameter values $\beta\epsilon_1=0$ and (\ref{m1L})-(\ref{e2}).}
\label{fig94}
\end{figure}

The dependence of the effective affinity (\ref{eff_A_1}) on the Coulomb repulsion is shown in \textsc{Figure} \ref{fig94} for a given value of the energy $\beta\epsilon_1=0$.  Here also, the effective affinity can be reduced down to a much lower value than the one determined by the reservoirs.  By a reasoning similar to the one used to get equations (\ref{eff_A_1_shape1}) and (\ref{eff_A_1_shape2}), we can obtain the approximate dependence of the effective affinity on the parameter $U$ under the conditions $\mu_{\rm 1R}-\epsilon_1 <0<\mu_{\rm 1L}-\epsilon_1 < \mu_{\rm 2L}-\epsilon_2$ as follows:
\be
\tilde{A}_1 \simeq
\left\{
\begin{array}{ll}
\beta(-U+\mu_{\rm 1L}-\mu_{\rm 1R})  & \mbox{for} \quad 0 < U < \mu_{\rm 1L}-\epsilon_1 \\ 
\beta(\epsilon_1-\mu_{\rm 1R})  &\mbox{for} \quad \mu_{\rm 1L}-\epsilon_1 < U < \mu_{\rm 2L}-\epsilon_2 \\ 
\beta(U+\epsilon_2-\mu_{\rm 2L}+\epsilon_1-\mu_{\rm 1R}) & \mbox{for} \quad \mu_{\rm 2L}-\epsilon_2 < U < \mu_{\rm 2L}-\epsilon_2+\mu_{\rm 1L}-\epsilon_1 \\ 
\beta(\mu_{\rm 1L}-\mu_{\rm 1R}) & \mbox{for} \quad \mu_{\rm 2L}-\epsilon_2+\mu_{\rm 1L}-\epsilon_1 < U \\ 
\end{array}
\right.
\label{eff_A_1_shapeU}
\ee
up to corrections smaller than $\beta=(k_{\rm B}T)^{-1}$ as $T\to 0$.
The piecewise linear approximation obtained from the Fermi-Dirac distributions here also explains the successive slopes $-1$, $0$, $+1$, and $0$, observed in the plot of the effective affinity versus $\beta U$.
We notice that the different linear pieces of the approximation match together at the crossover values of the variable $\beta U$.  The minimum value is reached in the interval $\beta(\mu_{\rm 1L}-\epsilon_1)=25 < \beta U < \beta(\mu_{\rm 2L}-\epsilon_2)=35$ while the affinity $A_1=25$ of the reservoirs is recovered for $\beta U > 
\beta(\mu_{\rm 2L}-\epsilon_2+\mu_{\rm 1L}-\epsilon_1)=60$, as indeed confirmed by \textsc{Figure} \ref{fig94}.

\begin{figure}
\centerline{\includegraphics[width=9cm]{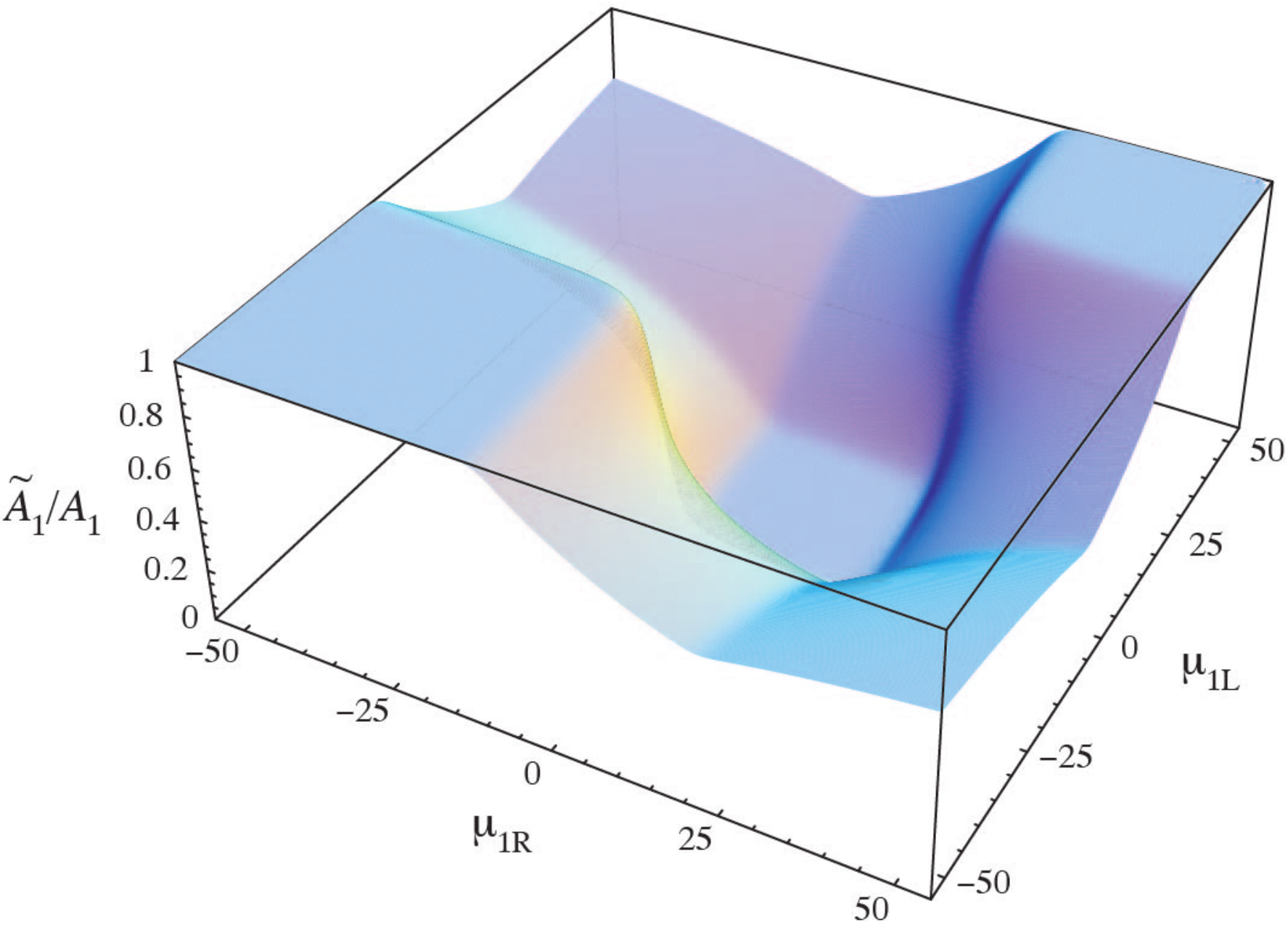}}
\caption{The effective affinity (\ref{eff_A_1}) divided by the actual affinity $A_1=\beta(\mu_{\rm 1L}-\mu_{\rm 1R})$ of the quantum dot No. 1 versus the chemical potentials $\mu_{\rm 1R}$ and $\mu_{\rm 1L}$ for the parameter values $\epsilon_1=-10$, $\epsilon_2=35$, $U=32.8$, $\Gamma_{\rm 1L}=\Gamma_{\rm 1R}=1$, $\Gamma_{\rm 2L}=\Gamma_{\rm 2R}=10^8$, $\mu_{\rm 2L}=70$, $\mu_{\rm 2L}=0$, and the inverse temperature $\beta=1$.  We notice that the affinities themselves change their sign along the diagonal line $\mu_{\rm 1L}=\mu_{\rm 1R}$.}
\label{fig104}
\end{figure}

\begin{figure}
\centerline{\includegraphics[width=9cm]{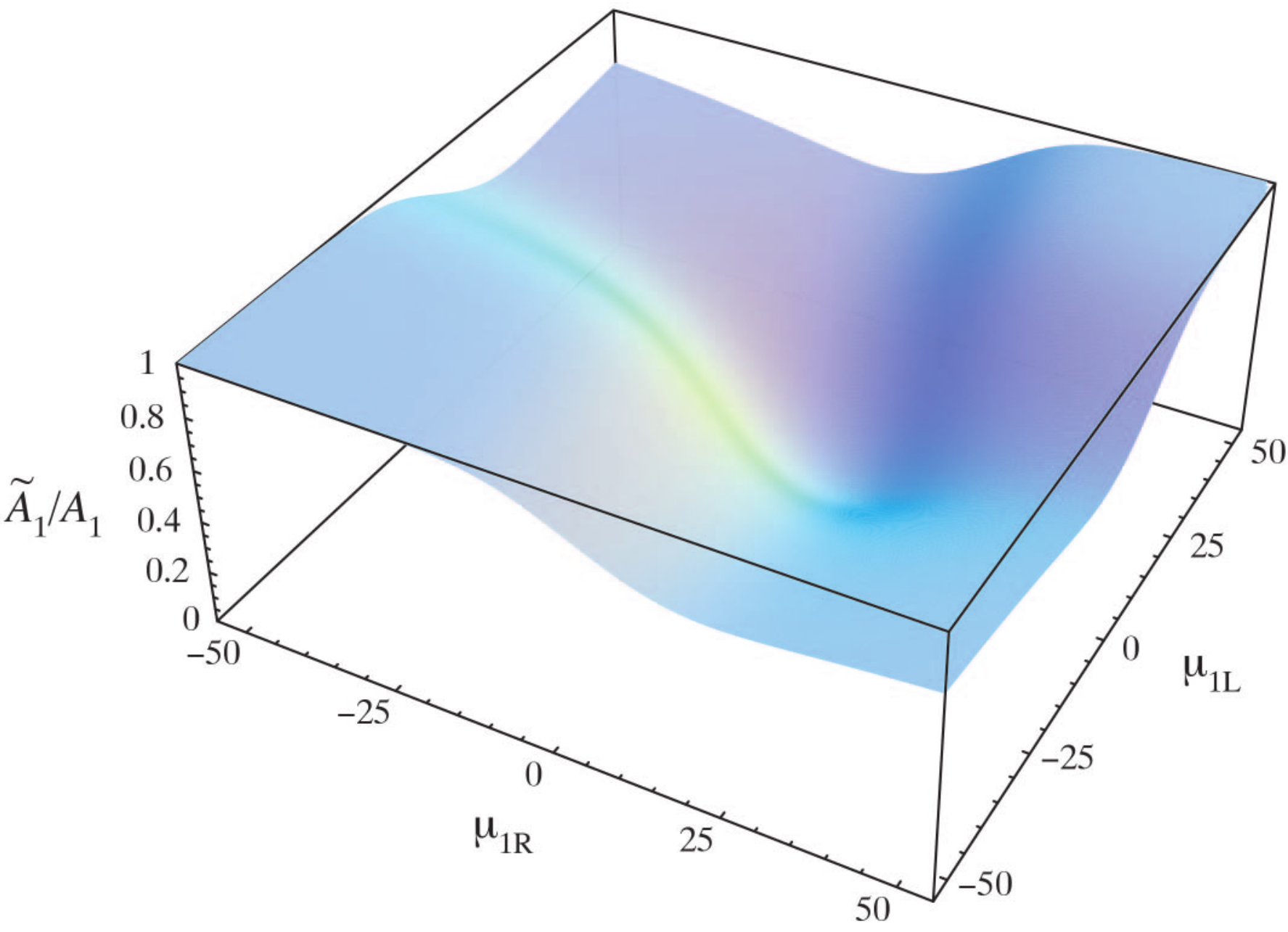}}
\caption{The effective affinity (\ref{eff_A_1}) divided by the actual affinity $A_1=\beta(\mu_{\rm 1L}-\mu_{\rm 1R})$ of the quantum dot No. 1 versus the chemical potentials $\mu_{\rm 1R}$ and $\mu_{\rm 1L}$ for the same parameter values as in \textsc{Figure} \ref{fig104} but the inverse temperature $\beta=0.2$.  Here also, the affinities themselves change their sign along the diagonal line $\mu_{\rm 1L}=\mu_{\rm 1R}$.}
\label{fig114}
\end{figure}

\textsc{Figure} \ref{fig104} shows how the effective affinity (\ref{eff_A_1}) behaves as a function of the chemical potentials $\mu_{\rm 1R}$ and $\mu_{\rm 1L}$ of the reservoirs connected to the quantum dot No. 1.  This figure confirms that the effective affinity undergoes crossovers if the chemical potentials take the values $\epsilon_1$ and $\epsilon_1+U$.  On the one hand, the effective affinity reaches its lower values in the domain where $\epsilon_1 < \mu_{\rm 1L} < \epsilon_1+U$ and $\epsilon_1 < \mu_{\rm 1R} < \epsilon_1+U$.  On the other hand, the actual value of the affinity is recovered in the domains $\mu_{\rm 1L},\mu_{\rm 1R}<\epsilon_1$ and $\epsilon_1+U<\mu_{\rm 1L},\mu_{\rm 1R}$.  As the temperature increases, the effective affinity becomes smoother as we observe in \textsc{Figure} \ref{fig114} for a temperature five times higher. 

The lowering of the effective affinity under specific conditions can be explained in the present model as the effect of the back-action of the secondary circuit interacting with the observed quantum dot.  The charging and discharging rates of the quantum dot can be drastically modified by the coupling to the secondary circuit.  In this way, the effective affinity can be much reduced in some regimes which are determined by the value of the energy $\epsilon_1$ of the quantum dot with respect to the values of the chemical potentials and the electrostatic coupling parameter $U$. This back-action effect tends to disappear as the temperature increases at constant voltages.  

\section{Summary of the results}
\label{Summary4}

%
In the present Chapter, we have reported the study of the single-current fluctuation theorem in a Hamiltonian model of quantum electron transport in two capacitively coupled channels, each containing a quantum dot \cite{SKB10}. Such a system is similar to the electronic devices used in typical counting statistics experiments \cite{fujisawa2006,PhysRevLett.96.076605} where the current in one circuit can continuously monitor the state of the quantum dot in the other circuit thanks to the capacitive coupling. The model allows us to investigate the effects of the back action of the monitoring circuit on the counting statistics in the light of the so-called fluctuation theorems.

Since both circuits are capacitively coupled and microreversibility holds for the total Hamiltonian (\ref{H}), a fluctuation theorem is satisfied for the two currents flowing across the system. This two-current fluctuation theorem (\ref{FT4}) or (\ref{FT-p}) relates the counting statistics of opposite random electron transfers in both circuits to the affinities or thermodynamic forces (\ref{A1})-(\ref{A2}) driving the system away from equilibrium.
The fluctuation theorem is valid far from equilibrium in the strongly nonlinear regimes encountered in electronic circuits composed of quantum dots and quantum point contacts.

However, in counting statistics experiments, one circuit is used to monitor the current fluctuations in the other circuit so that the counting statistics cannot be carried out on both currents together and is thus restricted to a single current. Accordingly, such experiments can only test a single-current fluctuation theorem.
In general, the two-current fluctuation theorem does not imply the single-current fluctuation theorem except under certain conditions \cite{AG07} or in some limits as we have demonstrated in the present Chapter.

In Section \ref{Uinfty}, we have studied the limit of large capacitive coupling between both circuits. In this limit, the state of simultaneous occupancy of both quantum dots in the two parallel channels is at a so high energy that it is energetically forbidden. The consequence is that the two single-occupancy states are separately accessible only from the empty state and the single-current fluctuation theorem holds with respect to the affinity determined by the chemical potentials of the reservoirs.

In Section \ref{Limit}, we have instead considered the limit where the current in one circuit is much larger than in the other circuit.  Indeed, a large current ratio is a key feature of typical counting statistics experiments \cite{fujisawa2006,PhysRevLett.96.076605} where the current ratio reaches values as high as $10^7$-$10^8$. The circuit with the very large current performs the continuous-time monitoring of the quantum state of the quantum dot in the other circuit.  In such a limit, the charging and discharging rates of the slow quantum dot take values averaged over the very fast fluctuations of the monitoring circuit. This is the essence of the back action of the monitoring circuit onto the quantum dot circuit. As a consequence of the large current ratio limit, the single-current fluctuation theorem holds but with respect to the effective affinity (\ref{eff_A_1}), which can be significantly reduced with respect to the actual value of the affinity determined by the reservoirs of the corresponding circuit.  

As shown in Section \ref{Numerics}, e.g. by equation (\ref{eff_A_1_shape1}), the reduction of the affinity is due to the capacitive coupling between both circuits and occurs when the transition energies $\{\epsilon_1,\epsilon_1+U\}$ of the quantum dot lie within the bias window $[\mu_{\rm 1R},\mu_{\rm 1L}]$ where the dynamics of the system is sensitive to the fluctuations of the detector.
In terms of the parameter $U$ of the Coulomb electrostatic interaction appearing in the Hamiltonian (\ref{H_S}), the affinity is lowered according to $\tilde{A}_1 \simeq A_1 - \beta U$ under the conditions specified around equation (\ref{eff_A_1_shape1}). This result explicitly expresses the effect of the back action between both circuits on the single-current fluctuation theorem.  This back-action effect can be reduced if the Coulomb repulsion $U$ is decreased, but the monitoring circuit can no longer resolve the two states of the quantum dot as in \textsc{Figure} \ref{fig24} if $U$ is too small.  On the other hand, the back-action effect is also reduced for large values of the Coulomb repulsion as shown by equation (\ref{eff_A_1_shapeU}) and in \textsc{Figure} \ref{fig94}.  Indeed, for a large Coulomb repulsion, the affinity recovers the value determined by the reservoirs and the back-action effect disappears.  This case corresponds to the situation considered in reference~\cite{CTH10} where a quantum fluctuation theorem has been obtained in a multiple measurements scheme.

From a general viewpoint, the two-current fluctuation theorem implies the non-negativity of the entropy production in agreement with the second law of thermodynamics. The dissipation of energy can thus be evaluated in the electron transport process used to perform quantum measurement in the experiments of references~\cite{fujisawa2006,PhysRevLett.96.076605}.  This dissipation of energy accompanying quantum measurement is expected on fundamental ground \cite{vN55}.  The necessity of resolving the quantum dot state in real time has for direct consequence that the dissipation in the monitoring circuit is much higher than in the quantum dot by a factor $\Pi_2/\Pi_1=(A_2/A_1)\times(J_2/J_1)$ of the same order of magnitude as the current ratio $J_2/J_1$. If the quantum dot state is monitored with a sampling time $\Delta t$, the secondary circuit playing the role of the detector should have transitions on equal or shorter time scales according to equation (\ref{Dt_real}).  Since the secondary circuit is driven out of equilibrium by the affinity $A_2$, its electron current should satisfy $J_2 \gtrsim (\Delta t)^{-1}$, so that the dissipated power should be bounded by $\Pi_2=k_{\rm B}T A_2J_2 \gtrsim k_{\rm B}T A_2(\Delta t)^{-1}$.  The higher the time resolution, the higher the dissipation rate.

In summary, we have shown that the single-current fluctuation theorem is valid under different limiting conditions and provided a fundamental understanding of the back-action effect of the monitoring circuit on the affinity of the monitored circuit, as observed in reference~\cite{fujisawa2006}.  The present study extends the analysis of reference~\cite{PhysRevB.81.125331,UGMSFS10bis} in showing how the effective affinity of the single-current fluctuation theorem can be directly expressed in terms of the parameters entering the Hamiltonian of the system. 

%% file: Chapter5.tex

\chapter{Single-current fluctuation theorem in a double quantum dot coupled to a quantum point contact} 

\label{Chapter5} 

\lhead{Chapter 5. \emph{Single-current fluctuation theorem in a DQD coupled to a QPC}} 

As mentionned in the introduction and the previous Chapter, typical experiments on full counting statistics are carried out with quantum dots capacitively coupled to an auxiliary circuit playing the role of charge detector and often taken as a quantum point contact (QPC) \cite{PhysRevLett.96.076605,gustavsson2009}. Due to the electrostatic Coulomb interaction, the current in the QPC is sensitive to the electronic occupancy of the quantum dots, thus allowing the measurement of single-electron transitions. Moreover, by coupling the QPC asymmetrically to the quantum dots, it is possible to infer the directionality of the flow of charges across the quantum dot system, providing the full counting statistics of single-electron transfers \cite{fujisawa2006,PhysRevLett.99.206804,ubbelohde2012measurement}.

With these devices, several experiments have established that the full counting statistics obeys the symmetry predicted by the fluctuation theorem \cite{PhysRevB.81.125331,PhysRevX.2.011001}.  Fundamentally, the fluctuation theorem is bivariate and holds for the two currents in the quantum dot and detector circuits and it is remarkable that the symmetry of the fluctuation theorem is observed for the sole current in the quantum dot circuit.  However, the back-action of the detector onto the quantum dot circuit modifies the symmetry by shifting the value of the voltage across the quantum dot circuit to an effective value.  Since this effective value is experimentally accessible, a key issue is to understand how this value depends on the capacitive coupling between the detector and the quantum dot circuit, as well as on the non-equilibrium driving forces.

In this Chapter, we address this issue in the case of a double quantum dot weakly coupled to two electrodes and probed by a QPC detector sensitive to the electronic occupation of the double quantum dot via Coulomb interaction.  The currents are driven by the two electric potential differences applied to both conduction channels.  We use a non-perturbative analysis for the QPC circuit that is considered in fully nonequilibrium regimes.

First, we show that, at finite and homogeneous temperature, the QPC behaves as a source of Bose-like fluctuations driving transitions between the charge eigenstates of the double quantum dot.  Our result is consistent, in the low-temperature limit, with the experimental observation of a threshold for the current induced in the double quantum dot channel as a function of the bias across the QPC \cite{PhysRevLett.99.206804}. This effect is directly related to the Coulomb drag exerted by the QPC onto the double quantum dot current \cite{PhysRevLett.104.076801,levchenko2008coulomb,andrieux2009monnai}.

Secondly, we demonstrate the emergence of single-current fluctuation theorems for the current in the sole double quantum dot under different experimentally relevant conditions.  These conditions suppose that the QPC is faster than the double quantum dot.  A single-current fluctuation theorem holds if the tunneling rate between the two quantum dots composing the double quantum dot is smaller than their tunneling rates with the electrodes.  Another single-current fluctuation theorem is obtained if the QPC induces transitions between the double quantum dot internal states at a rate faster than the double quantum dot charging and discharging rates.  In every case, we investigate how the single-current fluctuation theorem can characterize the double quantum dot and its capacitive coupling to the QPC.


\section{Double quantum dot coupled to a quantum point contact}
\label{Model}

\begin{figure}[htbp]
\centerline{\includegraphics[width=10cm]{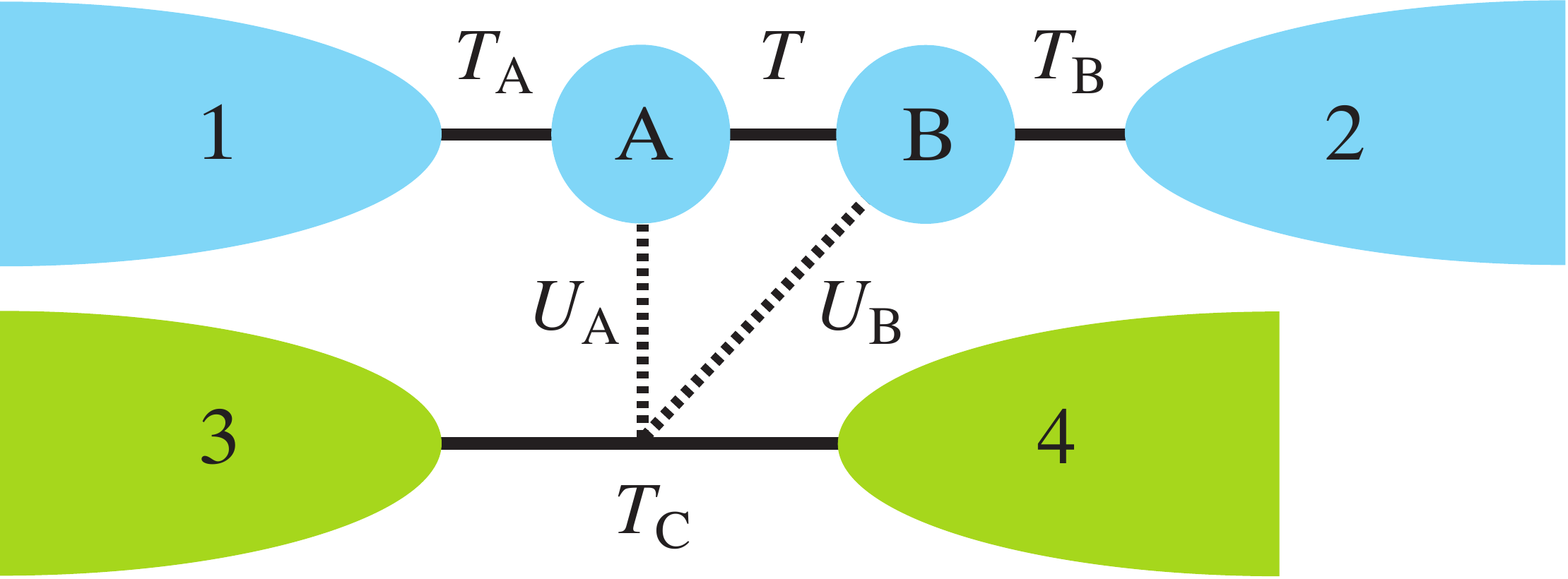}}
\caption{Schematic representation of a double quantum dot capacitively coupled to a QPC.  The double quantum dot is composed of the two quantum dots A and B that are coupled together as well as to the electrodes 1 and 2.  The QPC is coupling the electrodes 3 and 4.  The solid lines depict the couplings by tunneling and the dashed lines the capacitive couplings between the QPC and each QD.  The symbols are explained in the text.}
\label{fig15}
\end{figure}

In this section, we give a theoretical description for a double quantum dot capacitively coupled to a 	quantum point contact (QPC) starting from the Hamiltonian of the system, which is schematically depicted in \textsc{Figure}~\ref{fig15} \cite{GUMS11,OLY10}.  Without the capacitive coupling to the double quantum dot, the Hamiltonian model of the QPC is solved non perturbatively, leading to the Landauer-B\"uttiker formula for its average current and allowing us to calculate the correlation functions of its properties when it is in an arbitrary nonequilibrium steady state.  On the other hand, the capacitive coupling of the double quantum dot to the QPC, as well as the coupling of the double quantum dot to its reservoirs by direct tunneling are treated perturbatively at second order in the corresponding coupling parameters and with the rotating-wave approximation.  In this way, a modified master equation is obtained for the transitions between the internal states of the double quantum dot accounting for the current fluctuations in the double quantum dot channel.  The state of double occupancy in the double quantum dot is supposed to lie high enough in energy to play a negligible role.  In order to obtain the transition rates as explicitly as possible in terms of the parameters of the Hamiltonian operator, we consider tight-binding models for the reservoirs and the QPC \cite{T01,S07}. Moreover, the wide-band approximation is used for the QPC.  As in our previous work \cite{cuetara2011fluctuation}, the capacitances of the tunneling junctions between the QDs and the reservoirs are absent since our interest is here focused on the nonequilibrium conditions influencing the transport process.

\subsection{The Hamiltonian}


The double quantum dot is supposed to be composed of two quantum dots in series that can exchange electrons by direct tunneling.  This system is modeled in the local basis by the Hamiltonian 
\be
H_{\rm AB} = \epsilon_{\rm A} \, d_{\rm A}^{\dagger} d_{\rm A} + \epsilon_{\rm B} \, d_{\rm B}^{\dagger} d_{\rm B} + T \left( d_{\rm A}^{\dagger} d_{\rm B} +  d_{\rm B}^{\dagger} d_{\rm A}\right)
\label{H_DQD},
\ee
where $d_{\rm A}$ and $d_{\rm B}$ are the fermionic annihilation operators of an electron in the corresponding dot. The same Hamiltonian holds for both spin orientations\footnote{We do not consider here the effects due to an eventual external magnetic field.}, which are treated similarly and thus implicitly in our notations.  The energy of one electron in the dot A (respectively the dot B) is equal to $\epsilon_{\rm A}$ (respectively $\epsilon_{\rm B}$).  The tunneling amplitude between both dots is denoted $T$.

The double quantum dot is connected to the reservoirs $j=1,2$ and the QPC to the reservoirs $j=3,4$.
The reservoirs can be modeled by tight-binding Hamiltonians such as
\be
H_j = -\gamma \sum_{l=0}^{\infty} \left(d_{j,l}^{\dagger} d_{j,l+1} +  d_{j,l+1}^{\dagger} d_{j,l}\right),
\label{H_j}
\ee
where $d_{j,l}$ denotes the fermionic annihilation operator for an electron on the site of index $l\in{\mathbb N}$ in the $j^{\rm th}$ reservoir.  The advantage of such models is that these Hamiltonian operators
are exactly diagonalizable as shown below in Section \ref{AppB}.  The parameter $\gamma>0$ determines the width of the allowed energy band according to the dispersion relation $\epsilon_k=-2\gamma \cos k$ with the wavenumber $0\leq k \leq \pi$.  The band width is thus equal to $\Delta\epsilon=4\gamma$.
For simplicity, the parameter $\gamma$ is supposed to be common to every reservoir.

The double quantum dot is coupled by tunneling to the reservoirs $j=1,2$ with the following interaction operators:
\bea
&& V_{1{\rm A}} = T_{\rm A} \left(d_{1,0}^{\dagger} d_{\rm A} +  d_{\rm A}^{\dagger} d_{1,0}\right) \label{V1A} ,  \\
&& V_{2{\rm B}} = T_{\rm B} \left(d_{2,0}^{\dagger} d_{\rm B} +  d_{\rm B}^{\dagger} d_{2,0}\right) \label{V2B},
\eea
where $T_{\rm A}$ denotes the tunneling amplitude between the dot A and the reservoir $j=1$, while $T_{\rm B}$ is the tunneling amplitude between the dot B and the reservoir $j=2$.

The Hamiltonian of the QPC is taken as
\be
H_{\rm C} = H_3 + H_4 + T_{\rm C} \left(d_{3,0}^{\dagger} d_{4,0} +  d_{4,0}^{\dagger} d_{3,0}\right)
\label{H_C},
\ee
where $T_C$ denotes the amplitude for electron tunneling between the leads $3$ and $4$. Since this Hamiltonian is also quadratic in the annihilation-creation operators, it is exactly diagonalizable as shown in Section \ref{AppC}, which provides the electronic scattering properties of the QPC.

The capacitive coupling between the double quantum dot and the QPC is described by the interaction
\be
V_{\rm ABC} = \left(U_{\rm A} \, d_{\rm A}^{\dagger} d_{\rm A} + U_{\rm B} \, d_{\rm B}^{\dagger} d_{\rm B}\right)\left(d_{3,0}^{\dagger} d_{4,0} +  d_{4,0}^{\dagger} d_{3,0}\right)
\label{VABC},
\ee
where $U_{\rm A}$ and $U_{\rm B}$ are the parameters characterizing the electrostatic Coulomb interaction between the electrons in the quantum dots and the ones at the edges of the reservoirs $j=3,4$ \cite{OLY10}. Because of the interaction (\ref{VABC}), the tunneling amplitude in the QPC depends on the occupation of the double quantum dot.

Finally, the total Hamiltonian of the double quantum dot and QPC channels can be written as the sum of system and environment Hamiltonians, respectively $H_S$ and $H_R$, interacting through an interaction $V$ as
\be \label{totdqdqpc}
H= H_{\rm S} + H_{\rm R} + V.
\ee
In the present case, the system is made of the double quantum dot
\be
H_{\rm S} =H_{\rm AB}
\ee
while the environment is made of the reservoirs $j=1$, $2$ and QPC Hamiltonians
\be
H_{\rm R} = H_1 + H_2 + H_{\rm C}.
\ee
The double quantum dot and  its environment are coupled together by the interaction Hamiltonian
\be
V = V_{1{\rm A}} + V_{2{\rm B}} + V_{\rm ABC},
\label{V}
\ee
which is treated perturbatively at second order in the coupling parameters $T_{\rm A}$, $T_{\rm B}$, $U_{\rm A}$, and $ U_{\rm B}$.

In the present chapter, we are going to investigate the emergence of a single-current fluctuation theorem in the double quantum dot channel, taking into account the back-action of the out of equilibrium QPC detector. In order to do so, we perform the counting statistics of the electron current through the double quantum dot. As explained in chapter \ref{Chapter3}, this can be done through the introduction of the modified Hamiltonian
\be \label{511}
H^{\lambda} = \mbox{e}^{i \frac{\lambda}{2} N_1} H \mbox{e}^{-i \frac{\lambda}{2} N_1},
\ee
where $N_1$ is the particle number operator in reservoir $1$ of the double quantum dot channel, and $\lambda$ its associated counting parameter. This transformation only affects the interaction potential $V_{1A}$ which is consequently redefined as
\be 
V_{1A}^{\lambda} = \sum_{s=+ , -} T_{\rm A} \left( \mbox{e}^{i \lambda } d_{1,0}^{\dagger} d_{\rm A} +  \mbox{e}^{-i \lambda } d_{\rm A}^{\dagger} d_{1,0}\right) \label{V1A}.
\ee


\subsection{The Hamiltonian in the double quantum dot eigenbasis}


The diagonalization of the double quantum dot Hamiltonian (\ref{H_DQD}) can be performed analytically.  In the following, we assume for simplicity that the only states entering the dynamics are the empty and the single-charge eigenstates. The states of the local basis with one charge are defined as
\bea
&&\vert 1_{\rm A}0_{\rm B}\rangle \equiv d_{\rm A}^{\dagger} \vert 0_{\rm A}0_{\rm B}\rangle \nonumber  ,\\
&&\vert 0_{\rm A}1_{\rm B}\rangle \equiv d_{\rm B}^{\dagger} \vert 0_{\rm A}0_{\rm B}\rangle,
\eea
where $\vert 0_{\rm A}0_{\rm B}\rangle$ is the ground state of the double quantum dot.  Discarding the double-occupancy state, the eigenstates are thus expressed as
\bea
\vert 0 \rangle & = & \vert 0_{\rm A}0_{\rm B}\rangle \nonumber , \\
\vert + \rangle & = & \cos\frac{\theta}{2} \; \vert 1_{\rm A}0_{\rm B}\rangle + \sin\frac{\theta}{2} \; \vert 0_{\rm A}1_{\rm B}\rangle \nonumber , \\
\vert - \rangle & = &  \sin\frac{\theta}{2} \; \vert 1_{\rm A}0_{\rm B}\rangle - \cos\frac{\theta}{2} \; \vert 0_{\rm A}1_{\rm B}\rangle \label{eigenstates5} ,
\eea
in terms of the mixing angle
\be \label{theta5}
\tan\theta = \frac{2T}{\epsilon_{\rm A}-\epsilon_{\rm B}}.
\ee
Accordingly, the Hamiltonian of the double quantum dot can be written in the basis of its eigenstates $\{ \vert s \rangle  \}$ as
\be \label{eigenbas5}
H_{\rm S}=\sum_s \epsilon_s \, \vert s \rangle  \langle s \vert ,
\ee
where the corresponding eigenvalues are given by
\bea 
\epsilon_0 & = & 0 \nonumber , \\
\epsilon_{\pm} & = & \frac{\epsilon_{\rm A}+\epsilon_{\rm B}}{2} \pm \sqrt{\left(\frac{\epsilon_{\rm A}-\epsilon_{\rm B}}{2}\right)^2 + T^2}  \label{HSeigen}.
\eea
If $T$ vanishes, the mixing angle goes to $\theta=0$ for $\epsilon_{\rm A}>\epsilon_{\rm B}$ and to $\theta=\pi$ for $\epsilon_{\rm B}>\epsilon_{\rm A}$.

In order to write the interaction Hamiltonians $V_{1A}$ and $V_{2B}$ in the form (\ref{interactionexpl}), we introduce the system and reservoir operators respectively as
\be
\left\{
\begin{array}{l}
 S^{+}_{s}  \equiv  |s \rangle \langle 0 |  \\ 
 S^{-}_{s}  \equiv  | 0 \rangle \langle s | 
\end{array} \right.
\quad \mbox{and} \quad 
\left\{
\begin{array}{l}
 R^{+}_{js}  \equiv  T_{js} \, d_{j,0}   \\
 R^{-}_{js}  \equiv  - T_{js} \, d^{\dagger}_{j,0} 
\end{array} \right.
\ee
for $s = +$ and $-$  and $j=1$ and $2$.

With these definitions, the interaction operators (\ref{V1A}) and (\ref{V2B}) can expressed in the double quantum dot eigenbasis as
\bea
 V_{1{\rm A}} & = & \sum_{s=\pm}  \left( \mbox{e}^{-i \lambda} S^{+}_{s} R^{+}_{1s} +  \mbox{e}^{i \lambda} \, S^{-}_{s} R^{-}_{1 s}  \right) \label{V1A_bis} , \\
 V_{2{\rm B}} & = &  \sum_{s=\pm} \left( S^{+}_{s} R^{+}_{2s} +  \, S^{-}_{s} R^{-}_{2s}  \right)  \label{V2B_bis} ,
\eea
where $T_{js}$ are the tunneling amplitudes to reservoirs $j=1$ and $2$ in the eigenbasis, given by
\be \label{T1+} 
\begin{array}{l}
 T_{1+}   =   T_{\rm A} \, \cos\frac{\theta}{2} \\
 T_{1-}  =  T_{\rm A} \, \sin\frac{\theta}{2}
\end{array}
\quad \mbox{and} \quad
\begin{array}{l}
 T_{2+} =  T_{\rm B} \, \sin\frac{\theta}{2} \\ 
 T_{2-}  =  -T_{\rm B} \, \cos\frac{\theta}{2} 
\end{array}
\ee
in terms of the parameters of the local basis. 

The interaction with the QPC is similarly expressed as
\be
V_{\rm ABC} =\left(\sum_{s,s'=\pm} U_{ss'} \, \vert s \rangle \langle s' \vert\right)\left(d_{3,0}^{\dagger} d_{4,0} +  d_{4,0}^{\dagger} d_{3,0}\right) ,
\label{VABC_bis} 
\ee
showing that the QPC can induce transitions between the equal-charge eigenstates $\vert + \rangle$ and $\vert - \rangle$ of the double quantum dot. The coupling coefficients $U_{ss'}$ characterize the strength of the Coulomb interaction between the double quantum dot and the QPC and are given by
\bea
&&U_{++} = \frac{1}{2}(U_{\rm A}+U_{\rm B}) + \frac{1}{2}(U_{\rm A}-U_{\rm B}) \cos\theta \label{U++} , \\
&&U_{+-} = U_{-+}=  \frac{1}{2}(U_{\rm A}-U_{\rm B}) \sin\theta \label{U+-} , \\
&&U_{--} = \frac{1}{2}(U_{\rm A}+U_{\rm B}) - \frac{1}{2}(U_{\rm A}-U_{\rm B}) \cos\theta \label{U--} ,
\eea
with the parameters of the local basis. We notice that $U_{+-} = U_{-+}=0$ at $T=0$, i.e. when the double quantum dot channel is open.

We also point out that the coupling parameters in the capacitive interaction (\ref{VABC_bis}) do satisfy
\be
U_{+-}=U_{-+}.
\ee
Moreover, these coupling parameters are proportional to $U_{\rm A}-U_{\rm B}$, which characterizes the degree of asymmetry in the capacitive coupling between the QPC and the double quantum dot (see \textsc{Figure}~\ref{fig15}).  A direct consequence of this fact is the vanishing of the back-action if the QPC is symmetrically coupled to the double quantum dot.


\subsection{Diagonalization of the reservoir Hamiltonians}\label{AppB}


In this section, we perform the analytical diagonalization of the tight-biding Hamiltonians (\ref{H_j}) describing the reservoirs coupled to the double quantum dot. These being quadratic in the creation-annihilation operators, they can be analytically diagonalized into \cite{T01,S07}
\be
H_j = \int_0^{\pi} dk \, \epsilon_k \, c_{j,k}^{\dagger} c_{j,k} \qquad (j=1,2,3,4)
\label{H_j_diag}
\ee
with the energy eigenvalues
\be
\epsilon_k = - 2 \gamma \, \cos k \qquad (0\leq k \leq \pi)
\label{E_k}.
\ee
The new annihilation operators $\left\{ c_{jk} \right\}$ are related to the previous ones by
\bea
&& c_{j,k}= \sum_{l=0}^{\infty}  \phi_k(l) \, d_{j,l}  \label{loc5} , \\
&& d_{j,l}=\int_0^{\pi} dk \, \phi_k(l) \, c_{j,k} \label{dc},
\eea
in terms of the real eigenfunctions
\be
\phi_k(l) = \sqrt{\frac{2}{\pi}} \, \sin k (l+1) \qquad (l=0,1,2,3,...) .
\label{phi_k}
\ee
This set of eigenfunctions forms a complete orthonormal basis
\bea
&& \int_0^{\pi} dk\, \phi_k(l)\, \phi_k(l') = \delta_{ll'} , \\
&& \sum_{l=0}^{\infty}  \phi_k(l) \, \phi_{k'}(l) = \delta(k-k').
\eea

The creation-annihilation operators $\{ c_{j,k} \}$ and $\{ c_{j,k}^{\dagger} \}$ anticommute as a consequence of the fermionic character of the electrons. The annihilation operators have the following free time evolution:
\be
c_{j,k}(t) = {\rm e}^{iH_jt} c_{j,k} {\rm e}^{-iH_jt} =c_{j,k} \, {\rm e}^{-i\epsilon_kt} .
\label{c_t}
\ee
Similarly, the electron number operator is diagonalized into
\bea
N_j & \equiv & \sum_{l=0}^{\infty} d^{\dagger}_{j,l} d_{j,l} \nonumber \\
 & = &\int_0^{\pi} dk \, c_{j,k}^{\dagger} c_{j,k}
\label{N_j_diag}.
\eea

If a reservoir is composed of $L$ sites of indices $0\leq l \leq L-1$, the wavenumber takes the discrete values $k=n\pi/L$ with $n=1,2,3,...,L$ separated by $\Delta k =\pi/L$ and the density of states is given by
\be
D(\epsilon)=\sum_k \delta(\epsilon-\epsilon_k) =\frac{L}{\pi\sqrt{4\gamma^2-\epsilon^2}} .
\ee
The energy band extends over the interval $-2\gamma \leq \epsilon\leq +2\gamma$ and the average local density of states in the middle of the band is given by
\be
g\equiv \frac{D(0)}{L} = \frac{1}{2\pi\gamma} .
\label{LDOS}
\ee

This ends the spectral analysis of the tight-binding Hamiltonians of reservoirs $j=1$ and $2$ of the double quantum dot channel. These results enable us to calculate the equilibrium averages of quadratic combinations of creation-annihilation operators. The correlation functions in these reservoirs are evaluated in Section \ref{eqcorrfunctions}.

A large reservoir in the grand-canonical equilibrium ensemble at the inverse temperature $\beta$ and the chemical potential $\mu_j$ is described by the density operator
\be
\rho_j = {\rm e}^{-\beta(H_j-\mu_jN_j - \phi_j)} ,
\label{rho_j}
\ee
where the thermodynamic grand potential is defined by $\phi_j \equiv -\beta^{-1} \ln \left[ \mbox{Tr} \left\{ \mbox{e}^{- \beta(H_j-\mu_jN_j) } \right\} \right]$.
As shown in Appendix \ref{AppendixC}, the averages of quadratic combinations of creation-annihilation operators  with respect to the grand-canonical statistical ensemble (\ref{rho_j}) are given by
\bea
&&\langle c_{j,k}^{\dagger} \, c_{j,k'}\rangle = f_{jk} \; \delta(k-k') \label{<c+c>} , \\
&&\langle c_{j,k} \, c_{j,k'}^{\dagger}\rangle = \left(1-f_{jk}\right) \, \delta(k-k') \label{<cc+>}.
\eea
where $f_{jk} = f_j(\epsilon_k)$ is the Fermi-Dirac 
\be
f_j(\epsilon) = \frac{1}{{\rm e}^{\beta(\epsilon-\mu_j)}+1}.
\label{FD}
\ee


\subsection{Diagonalization of the QPC Hamiltonian}\label{AppC}


The diagonalization of the QPC Hamiltonian (\ref{H_C}) is solved as a scattering problem in which the point contact between the reservoirs $j=3$ and $j=4$ is the scatterer \cite{T01,S07}.  Since the Hamiltonian and the particle number are quadratic, they can be transformed into
\bea
&& H_{\rm C} = \int_{-\pi}^{+\pi} dq \, \epsilon_q \, c_{q}^{\dagger} c_{q}
\label{H_C_diag}, \\
&& N_{\rm C} = \int_{-\pi}^{+\pi} dq \, c_{q}^{\dagger} c_{q}.
\label{N_C_diag} 
\eea
with the energy eigenvalues
\be
\epsilon_q = - 2 \gamma \, \cos q \qquad (-\pi\leq q \leq +\pi)
\label{E_q}
\ee
with $\gamma >0$.
The annihilation operators are transformed according to
\bea
&& c_{q}= \sum_{j=3,4} \sum_{l=0}^{\infty}  \psi_q^*(j,l) \, d_{j,l} , \\
&& d_{j,l}=\int_{-\pi}^{+\pi} dq \, \psi_q(j,l) \, c_{q},
\eea
in terms of the scattering eigenfunctions
\bea
\left\{
\begin{array}{l}
\psi_q(3,l) = \frac{1}{\sqrt{2\pi}}\left({\rm e}^{-iql}+r_q\, {\rm e}^{iql}\right) \\
\psi_q(4,l) = \frac{1}{\sqrt{2\pi}}\; t_q\, {\rm e}^{iql}
\end{array}
\right. \label{psi+}\\
\left\{
\begin{array}{l}
\psi_{-q}(3,l) = \frac{1}{\sqrt{2\pi}}\; t_q\, {\rm e}^{iql}\\
\psi_{-q}(4,l) = \frac{1}{\sqrt{2\pi}}\left({\rm e}^{-iql}+r_q\, {\rm e}^{iql}\right) 
\end{array}
\right. \label{psi-}
\eea
for $q>0$ and $l=0,1,2,3,...$.  The transmission amplitude is given by
\be \label{tq}
t_q = - T_{\rm C} \, \gamma\, \frac{{\rm e}^{iq}-{\rm e}^{-iq}}{T_{\rm C}^2-\gamma^2{\rm e}^{-2iq}}
\ee
and it is related to the reflection amplitude by
\be \label{rq}
1+r_q= -\frac{\gamma}{T_{\rm C}} \, t_q \, {\rm e}^{-iq}.
\ee
The transmission probability ${\cal T}_{\epsilon} $ is given by
\be
{\cal T}_{\epsilon} = \vert t_{q(\epsilon)}\vert^2 = \frac{T_{\rm C}^2\, (4\gamma^2-\epsilon^2)}{(T_{\rm C}^2+\gamma^2)^2-T_{\rm C}^2\epsilon^2}
\label{T_E}
\ee
characterizing the probability flux of the transmitted wave relative to that of the incident wave with energy $\epsilon$.
The transmission probability is maximal in the middle of the band and vanishes at the edges of the energy band.  Since the band width $\Delta\epsilon=4\gamma$ is related to the local density of states in the middle of the band by Eq.~(\ref{LDOS}), the transmission probability can be written at its maximal value as
\be
{\cal T}_0 = \frac{4\kappa}{(1+\kappa)^2}
\ee
in terms of the dimensionless contact transparency \cite{GK06}
\be
\kappa = (2\pi \, g \, T_{\rm C})^2 = (T_{\rm C}/\gamma)^2 .
\label{kappa}
\ee
We notice that the transparency satisfies $\kappa < 1$ because of the condition $\vert T_{\rm C}\vert <\gamma$, which is required for the absence of bound state. Under this last condition, the scattering eigenfunctions (\ref{psi+})-(\ref{psi-}) form a complete orthonormal basis
\bea
&& \int_{-\pi}^{+\pi} dq\, \psi_q(j,l)\, \psi_q^*(j',l') = \delta_{jj'}\, \delta_{ll'}  ,\\
&& \sum_{j=3,4}\sum_{l=0}^{\infty}  \psi_q(j,l) \, \psi_{q'}^*(j,l) = \delta(q-q').
\eea

Here, the annihilation operators have the free time evolution:
\be
c_{q}(t) = {\rm e}^{iH_{\rm C}t} c_{q} {\rm e}^{-iH_{\rm C}t} =c_{q} \, {\rm e}^{-i\epsilon_qt} .
\ee

After diagonalization, the Hamiltonian operator (\ref{H_C_diag}) splits as
\be
H_{\rm C}= H_{\rm C}^{(+)} + H_{\rm C}^{(-)} 
\ee
 into the operators
\bea
&& H_{\rm C}^{(+)} = \int_{0}^{+\pi} dq \, \epsilon_q \, c_{q}^{\dagger} c_{q} \label{HC+} ,\\
&& H_{\rm C}^{(-)} = \int_{-\pi}^{0} dq \, \epsilon_q \, c_{q}^{\dagger} c_{q} \label{HC-}.
\eea
A similar decomposition holds for the particle number: $N_{\rm C}= N_{\rm C}^{(+)} + N_{\rm C}^{(-)}$.

If both reservoirs coupled by the QPC extended over $L$ sites of indices $0\leq l \leq L-1$, the wavenumber $q$ would take discrete values separated by $\Delta q =\pi/L$.
Accordingly, a nonequilibrium steady state for the QPC could be defined with the density operator \cite{T01,LS11} 
\be
\rho_{\rm C} = {\rm e}^{-\beta(H_{\rm C}^{(+)}-\mu_3N_{\rm C}^{(+)} - \phi_{C}^{(+)}) }
\;  {\rm e}^{-\beta(H_{\rm C}^{(-)}-\mu_4N_{\rm C}^{(-)}- \phi_{C}^{(-)})}
\label{rho_C} ,
\ee
where the potentials are defined by $\phi_{C}^{(\pm)} =- \beta^{-1} \ln \left[ \mbox{Tr} \left\{ \mbox{e}^{- \beta(H_{C}^{(\pm)}-\mu_j N_{C}^{(\pm)}) } \right\} \right]$ for $j=3$ and $4$.
In this statistical ensemble, the quadratic combinations of the creation-annihilation operators have the statistical averages (see Appendix \ref{AppendixC})
\bea
&&\langle c_{q}^{\dagger} \, c_{q'}\rangle = f_{jq} \; \delta(q-q') , \\
&&\langle c_{q} \, c_{q'}^{\dagger}\rangle = \left(1-f_{jq}\right) \, \delta(q-q') ,
\eea
with $j=3$ for $q>0$, $j=4$ for $q<0$, and the notation $f_{jq}=f_j(\epsilon_q)$ for the Fermi-Dirac distribution (\ref{FD}) at the inverse temperature $\beta$, the chemical potential $\mu_j$, and the wavenumber $q$.


\section{Derivation of the modified master equation}


Here, we apply the results of Chapter \ref{Chapter3} in order to obtain a modified master equation for the reduced density matrix of the double quantum dot $\rho_{S} (\lambda , t)$ accounting for the electrons flowing out of reservoir $j=1$ by means of the counting parameter $\lambda$. In the following, we assume the double quantum dot to be weakly coupled to its reservoirs as well as to the quantum point contact detector. This enables us to perform the Born perturbative approximation to second order in the tunneling amplitudes $\{ T_{js} \}$ and the interaction parameters $\{ U_{ss'} \}$.

The density matrix of the total system is assumed to be initially in the factorized form 
\be
\rho (0) = \rho_S(0) \otimes \rho_{1} \otimes \rho_{2} \otimes \rho_{C} ,
\ee
where $\rho_{S} (0)$ denotes an arbitrary initial statistical mixture on the double quantum dot, $\rho_j$ denotes the grand-canonical ensembles (\ref{rho_j}) over reservoirs $j=1$ and $2$, while $\rho_{C}$ denotes the nonequilibrium stationary state of the quantum point contact (\ref{rho_C}).

We consider the dynamics of the double quantum dot system over a time scale which is intermediate between the correlation time of the interaction operators in the environment and the relaxation time induced by the environment on the double quantum dot. We also assume the typical time scale of the free oscillations in the double quantum dot to be much shorter than the sampling time of the observations. As a consequence, we perform the Markovian and rotating wave approximations resulting in a modified rate equation of the form (\ref{stochmod}) with the correlation functions of the environment defined in (\ref{corfunc}).

These correlation functions completely encode the effects of the environment on the double quantum dot dynamics. In the following sections, we evaluate these correlations functions for the double quantum dot channel reservoirs as well as for the quantum point contact. The double quantum reservoirs being at equilibrium, their correlation functions are evaluated over the equilibrium ensembles (\ref{rho_j}). In contrast, the correlation functions of the quantum point contact are evaluated at the nonequilibrium steady state (\ref{rho_C}).

\subsection{Electron tunneling between the double quantum dot and its reservoirs}\label{eqcorrfunctions}


The charging rate into the eigenstate $\vert s\rangle$ of the double quantum dot from the reservoir $j$ is given by
\bea
a_{js} & \equiv & \int_{-\infty}^{\infty} dt \, \mbox{e}^{-i \epsilon_s t}\langle R^{-}_{js } (t) \, R^{+}_{js} \rangle \nonumber  \\
& = & T_{js}^2 \int_{-\infty}^{+\infty} dt\; {\rm e}^{-i\epsilon_s t} \left\langle {\rm e}^{iH_jt} \, d_{j,0}^{\dagger}\, {\rm e}^{-iH_jt} \, d_{j,0}\right\rangle \label{Gjs2},
\eea
where $j=1,2$, $s=\pm$, and the average is carried out over the equilibrium ensemble (\ref{rho_j}) of the $j^{\rm th}$~reservoir: $\langle\cdot\rangle={\rm Tr}\rho_j(\cdot)$.  With Eq.~(\ref{dc}) for $l=0$ and Eq.~(\ref{c_t}), we find
\bea
&&\left\langle {\rm e}^{iH_jt} \, d_{j,0}^{\dagger}\, {\rm e}^{-iH_jt} \, d_{j,0}\right\rangle
 = \int_0^{\pi} dk \, \phi_k(0) \; {\rm e}^{i\epsilon_kt} \nonumber\\
&& \qquad\qquad\times \int_0^{\pi} dk' \, \phi_{k'}(0)  \left\langle c_{j,k}^{\dagger}\, c_{j,k'}\right\rangle.
\eea
Using the average (\ref{<c+c>}) together with the explicit form (\ref{phi_k}) of the eigenfunction at $l=0$ and the corresponding energy eigenvalue (\ref{E_k}), the rate (\ref{Gjs2}) is obtained as
\be
a_{js} =  \Gamma_{js} \, f_j(\epsilon_s) \label{ajs}
\ee
in terms of the Fermi-Dirac distribution (\ref{FD}) and the rates
\be
\Gamma_{js} =T_{js}^2 \, \frac{2}{\gamma} \, \sqrt{1-\left(\frac{\epsilon_s}{2\gamma}\right)^2}
\label{Gjs_bis} .
\ee
In the wide-band approximation for which $\vert\epsilon_s\vert\ll2\gamma$, the local density of states is evaluated by Eq.~(\ref{LDOS}) in the middle of the band so that the rates become
\be
\Gamma_{js} = 4\pi g \, T_{js}^2 .
\label{Gjs_wb}
\ee

The discharging rate into the eigenstate $\vert s\rangle$ of the double quantum dot from the reservoir $j$ is given by
\bea
b_{js} & \equiv &\int_{-\infty}^{\infty} dt \, \mbox{e}^{i \epsilon_s t}\langle R^{+}_{j s } (t) \, R^{-}_{j s} \rangle  \\
& = & T_{js}^2 \int_{-\infty}^{+\infty} dt\; {\rm e}^{i\epsilon_s t} \left\langle {\rm e}^{iH_jt} \, d_{j,0}\, {\rm e}^{-iH_jt} \, d_{j,0}^{\dagger}\right\rangle
\eea
with $j=1,2$ and $s=\pm$.  The calculation is similar as in the previous one, using instead the average (\ref{<cc+>}) to get the discharging rate (\ref{bjs}) as
\be
 b_{js} = \Gamma_{js} \, \left[1-f_j(\epsilon_s)\right] \label{bjs}
\ee 
with (\ref{Gjs_wb}) in the wide-band approximation. 

We notice that the charging and discharging rates obey the local detailed balance conditions:
\be
\frac{a_{js}}{b_{js}}={\rm e}^{-\beta(\epsilon_s-\mu_j)}.
\ee
The thermal energy is assumed to be larger than the natural width of the double quantum dot energy levels, $\beta\hbar(\Gamma_{1s}+\Gamma_{2s})\ll 1$, in consistency with the neglect of resonance effects by second-order perturbation theory \cite{B91}.


\subsection{Calculation of the non-equilibrium correlation functions}


The capacitive coupling of the double quantum dot with the QPC is again treated perturbatively at second order and in the rotating wave approximation, but the QPC is supposed to be in the nonequilibrium steady state (\ref{rho_C}).  At the Hamiltonian level of description, the capacitive coupling is expressed with the interaction operator (\ref{VABC_bis}), which has the form
\be
V_{\rm ABC} = S \, R
\ee
with the subsystem operator $S=\sum_{s,s'=\pm} U_{ss'} \, \vert s \rangle \langle s' \vert$ and the QPC operator $R=d_{3,0}^{\dagger} d_{4,0} +  d_{4,0}^{\dagger} d_{3,0}$. Recalling (\ref{transitionratesgeneral3}), the transition rates associated with this interaction are given by
\be
c_{ss'} =L_{ss'}^{(0)}= \hbar^{-2} \hat \alpha(\omega_{ss'}) \vert\langle s\vert S\vert s'\rangle\vert^2
\label{css}
\ee
with $s=-s'=\pm$, $\omega_{ss'}=\epsilon_s-\epsilon_{s'}$,
\be
\vert\langle +\vert S\vert -\rangle\vert^2 =\vert\langle -\vert S\vert +\rangle\vert^2 = U_{+-}^2 = U_{-+}^2 = \frac{1}{4}\,(U_{\rm A}-U_{\rm B})^2 \sin^2\theta
\label{S+-}
\ee
and the spectral function
\be \label{576}
\hat \alpha(\omega) =\int_{-\infty}^{+\infty} dt \, {\rm e}^{-i\omega t} \langle \overline R(t) \, \overline R\rangle,
\ee
where $\overline R = R-\langle R \rangle$, $\langle\cdot\rangle={\rm tr}\rho_{\rm C}(\cdot)$, and
\be
\overline{R}(t) \equiv {\rm e}^{iH_{\rm C} t} \overline{R}\,  {\rm e}^{-iH_{\rm C} t}.
\ee
Using the expression of the operator $R$ and Wick's lemma, the spectral function becomes
\begin{multline} \label{578}
\hat \alpha(\omega) = \int_{-\infty}^{+\infty} dt \, {\rm e}^{-i\omega t} \left( \langle d_{3,0}^{\dagger}(t) \, d_{4,0}\rangle \langle d_{4,0}(t) \, d_{3,0}^{\dagger}\rangle + \langle d_{3,0}^{\dagger}(t) \, d_{3,0}\rangle \langle d_{4,0}(t) \, d_{4,0}^{\dagger}\rangle \right. \\
\left. \qquad\qquad\quad\quad + \langle d_{4,0}^{\dagger}(t) \, d_{4,0}\rangle \langle d_{3,0}(t) \, d_{3,0}^{\dagger}\rangle+ \langle d_{4,0}^{\dagger}(t) \, d_{3,0}\rangle \langle d_{3,0}(t) \, d_{4,0}^{\dagger}\rangle\right).
\end{multline}
The correlation functions of the creation-annihilation operators are obtained as
\bea
&&\langle d_{3,0}^{\dagger}(t) \, d_{4,0}\rangle = -\frac{1}{2\pi} \int_0^{\pi} dk \, \vert t_k\vert^2 \, {\rm e}^{i\epsilon_kt} \frac{\gamma}{T_{\rm C}} \left( {\rm e}^{ik} f_{3k} + {\rm e}^{-ik} f_{4k}\right) ,\\
&&\langle d_{4,0}(t) \, d_{3,0}^{\dagger}\rangle = -\frac{1}{2\pi} \int_0^{\pi} dk \, \vert t_k\vert^2 \, {\rm e}^{-i\epsilon_kt} \frac{\gamma}{T_{\rm C}} \left( {\rm e}^{ik} (1-f_{3k}) + {\rm e}^{-ik} (1-f_{4k})\right) , \\
&&\langle d_{3,0}^{\dagger}(t) \, d_{3,0}\rangle = \frac{1}{2\pi} \int_0^{\pi} dk \, \vert t_k\vert^2 \, {\rm e}^{i\epsilon_kt}  \left( \frac{\gamma^2}{T_{\rm C}^2}\, f_{3k} + f_{4k}\right) , \\
&&\langle d_{4,0}(t) \, d_{4,0}^{\dagger}\rangle = \frac{1}{2\pi} \int_0^{\pi} dk \, \vert t_k\vert^2 \, {\rm e}^{-i\epsilon_kt} \left( 1-f_{3k} + \frac{\gamma^2}{T_{\rm C}^2}\, (1-f_{4k})\right) ,
\eea
and similar expressions with transposed indices $3$ and $4$.  As a consequence, we have that
\begin{multline} \label{583}
\hat \alpha(\omega) = \frac{1}{(2\pi)}\int_0^{\pi}dk\int_0^{\pi}dq \, \vert t_k\vert^2 \vert t_q\vert^2 \, \delta(\epsilon_k-\epsilon_q-\omega)  \\
 \left\{ \frac{2\,\gamma^2}{T_{\rm C}^2}\left(\cos(k+q)+1\right) \left(f_{3k}\, (1-f_{3q})+f_{4k}\, (1-f_{4q})\right) \right. \\
\left. +\left(\frac{\gamma^4}{T_{\rm C}^4} +\frac{2\,\gamma^2}{T_{\rm C}^2}\cos(k-q)+1\right) \left(f_{3k}\, (1-f_{4q})+f_{4k}\, (1-f_{3q})\right) \right\}.
\end{multline}

In the wide-band approximation, the transmission coefficients as well as the functions $\cos(k\pm q)$  should be evaluated at the values of the wavenumbers $0\leq k,q\leq\pi$ corresponding to the middle of the energy band.  Given the dispersion relation (\ref{E_q}), the only possibility is $k=q=\pi/2$.  Therefore, $\cos(k+q)=-1$, so that the first term is negligible in the wide-band approximation.
On the other hand, $\cos(k-q)=1$, and, using expressions (\ref{tq}) and (\ref{rq}), we find
\be \label{584}
\hat \alpha(\omega) \simeq \frac{1}{(2\pi)}\, \frac{4\, \gamma^2}{(T_{\rm C}^2+\gamma^2)^2}  \int_{-\infty}^{+\infty} d\epsilon \, \left\{ f_{3}(\epsilon)\left[1-f_{4}(\epsilon-\omega)\right]+f_{4}(\epsilon)\left[1-f_{3}(\epsilon-\omega)\right]\right\} .
\ee
The integral of the first term is evaluated as follows:
\be
\int_{-\infty}^{+\infty} d\epsilon \, f_{3}(\epsilon)\left[1-f_{4}(\epsilon-\omega)\right] = \frac{\omega-\Delta\mu_{\rm C}}{{\rm e}^{\beta(\omega-\Delta\mu_{\rm C})}-1}
\ee
with $\Delta\mu_{\rm C}=\mu_3-\mu_4$ and the other similarly.  Using the local density of states in the middle of the band given by Eq.~(\ref{LDOS}), the dimensionless contact transparency (\ref{kappa}) together with the interaction parameters (\ref{S+-}), we finally get
\be
c_{ss'} =\frac{8\pi g^2 U_{ss'}^2}{(1+\kappa)^2} \ \left[ \frac{\omega_{ss'}-\Delta\mu_{\rm C}}{{\rm e}^{\beta(\omega_{ss'}-\Delta\mu_{\rm C})}-1} + \frac{\omega_{ss'}+\Delta\mu_{\rm C}}{{\rm e}^{\beta(\omega_{ss'}+\Delta\mu_{\rm C})}-1}\right]
\label{c+-}
\ee
for the transition rates (\ref{css}).

These transition rates are proportional to the intensity of the capacitive coupling: $U_{+-}^2=U_{-+}^2=(U_{\rm A}-U_{\rm B})^2\sin^2\theta/4$.  The expression (\ref{c+-}) shows the Bose-like character of the random transitions due to the back-action of the QPC onto the double quantum dot circuit.  If the QPC is at equilibrium with $\Delta\mu_{\rm C}=0$, these transition rates satisfy the condition of local detailed balance:
\be
\frac{c_{+-}}{c_{-+}}={\rm e}^{-\beta\omega_{+-}} \qquad\mbox{for} \qquad \Delta\mu_{\rm C}=0.
\label{c_ratio_eq}
\ee
However, this condition is not satisfied under general nonequilibrium conditions $\Delta\mu_{\rm C}\neq 0$ for the QPC.

We notice that, if the QPC is at a uniform temperature different from the double quantum dot temperature, the inverse temperature $\beta$ in the rates~(\ref{c+-}) should be replaced by the inverse temperature $\beta_{\rm C}$ of the QPC.


\subsection{The modified master equation}\label{modmastereq5}

We are now in position to write down a dynamical equation for the vector
\be 
{\bf g}(\lambda,t)=
\left(
\begin{array}{c}
g_{0}(\lambda,t) \\
g_{+}(\lambda,t) \\
g_{-}(\lambda,t) 
\end{array}
\right)
\ee
composed of the diagonal elements of the modified reduced density matrix of the double quantum dot $g_{s} (\lambda , t) = \langle s | \rho_{S} (  i \lambda , t) | s\rangle$ for $s=0, \, \pm$. Indeed, recalling equations (\ref{transitionratesgeneral3}) together with the results obtained in the two previous sections, we obtain the modified master equation
\be \label{mastereqmod}
\dot{{\bf g}} (\lambda ,t ) = {\bf W}(\lambda) \cdot {\bf g} (\lambda , t)
\ee
in terms of the rate matrix ${\bf W}(\lambda) $ given by
\be 
\mbox{\bf W} (\lambda)
=
\left(
\begin{array}{ccc}
-a_{1+}-a_{2+}-a_{1-}-a_{2-} & b_{1+}\,\mbox{e}^{\lambda}+ +b_{2+} & b_{1-}\,\mbox{e}^{\lambda} +b_{2-} \\
a_{1+}\,\mbox{e}^{-\lambda} +a_{2+} & -b_{1+}-b_{2+}-c_{-+} & c_{+-} \\
a_{1-}\,\mbox{e}^{-\lambda} +a_{2-} & c_{-+} & -b_{1-}-b_{2-}-c_{+-} 
\end{array}
\right).
\label{Wdqdqpc}
\ee
This rate matrix is expressed in terms of the charging and discharging rates (\ref{ajs}) and (\ref{bjs}) describing the tunneling of electrons between the double quantum dot and its electrodes, as well as the rate (\ref{c+-}) describing the exchange of energy between the quantum point contact and the double quantum dot resulting from tunneling events within the quantum point contact.

Alternatively, we can apply an inverse Fourier transform to each member of the modified rate equation to get a dynamical equation for the probabilities $p_s (n, t)$ of observing the system in the state $|s \rangle$ at time $t$ and $n$ electrons having the left reservoir $1$ since the initial time $t=0$. If the probabilities $\{p_s(n,t)\}_{s=0,\pm}$ are gathered in the array
\be
{\bf p}(n,t)=
\left(
\begin{array}{c}
p_{0}(n,t) \\
p_{+}(n,t) \\
p_{-}(n,t) 
\end{array}
\right),
\ee
we obtain the master equation
\be
\partial_t\,{\bf p}(n,t) = \hat{\mbox{\bf W}} \cdot {\bf p}(n,t) 
\label{master-m}
\ee
with the matrix
\be
\hat{\mbox{\bf W}}
=
\left(
\begin{array}{ccc}
-a_{1+}-a_{2+}-a_{1-}-a_{2-} & b_{1+}\,\hat{E}^- +b_{2+} & b_{1-}\,\hat{E}^- +b_{2-} \\
a_{1+}\,\hat{E}^- +a_{2+} & -b_{1+}-b_{2+}-c_{-+} & c_{+-} \\
a_{1+}\,\hat{E}^- +a_{2-} & c_{-+} & -b_{1-}-b_{2-}-c_{+-} 
\end{array}
\right)
\label{Ldqdqpc}
\ee
in terms of the operators
\be
\hat E^{\nu} \equiv \exp\left(\nu \frac{\partial}{\partial n}\right),
\label{E}
\ee
which change the number $n$ of transferred electrons according to $\hat E^{\pm}p_s(n,t)=p_s(n\pm 1,t)$.

A stochastic master equation is obtained for the occupation probabilities on the quantum dot $p_{s}(t)$ by setting $\lambda = 0$ in equations (\ref{mastereqmod}) with (\ref{Wdqdqpc}) or, equivalently, by integrating (\ref{master-m}) with (\ref{Ldqdqpc}) over the number $n$ of transferred electrons.


\section{The mean currents}


In this Section we define the mean currents in the double quantum dot and QPC channels as the average of the time derivative of the particle number operators in the reservoirs $1$ and $3$. These currents are driven by two thermodynamic forces resulting from the electric biases applied to each channel. We also explain how the current in the QPC channel is modulated by the changes of electronic occupation in the double quantum dot.


\subsection{Definitions}


The total Hamiltonian (\ref{totdqdqpc}) - (\ref{V}) commutes separately with the total numbers of electrons in each conduction channel:
\bea
&& \left[ H, N_1+N_2+N_{\rm AB}\right] = 0  ,\\
&& \left[ H, N_3+N_4\right] = 0 ,
\eea
where
\be
N_j = \sum_{l=0}^{\infty} d_{j,l}^{\dagger} d_{j,l} 
\label{N_j}
\ee
is the number of electrons in the $j^{\rm th}$ reservoir and
\be
N_{\rm AB} = d_{\rm A}^{\dagger} d_{\rm A} + d_{\rm B}^{\dagger} d_{\rm B}
\ee
the number of electrons in the double quantum dot.  Therefore, the electron current is conserved separately in the double quantum dot circuit, as well as in the QPC circuit.

The current in the double quantum dot circuit can be defined as the rate of decrease of the electron number in the reservoir $j=1$ by
\be
J_{\rm D} \equiv -\frac{dN_1}{dt} = i\left[ N_1,H\right] = i \, T_{\rm A} \left(d_{1,0}^{\dagger} d_{\rm A}-d_{\rm A}^{\dagger} d_{1,0}\right)
\ee
in units where Planck's constant is equal to $\hbar=1$.
The current in the QPC is similarly defined as the rate of decrease of the electron number in the reservoir $j=3$ by
\be
J_{\rm C} \equiv -\frac{dN_3}{dt} = i\left[ N_3,H\right] .
\ee
The average values of the electric currents are thus given by
\be
I_{\rm D} = e \langle J_{\rm D}\rangle \qquad\mbox{and}\qquad I_{\rm C} = e \langle J_{\rm C}\rangle,
\ee
where $e$ is the electron charge.

As mentioned earlier, the reservoirs are assumed to be initially in grand-canonical statistical ensembles at homogeneous temperature so that the inverse temperature $\beta$ is uniform across the whole system.  However, the reservoirs have different chemical potentials given by $\mu_j$ with $j=1,2,3,4$.  Because of the separate charge conservation in both circuits, the non-equilibrium conditions of this isothermal system are specified by two dimensionless affinities
\bea
&& A_{\rm D} = \beta(\mu_1-\mu_2) = \beta \Delta\mu_{\rm D}= \beta e V_{\rm D} \label{A_D} ,\\
&& A_{\rm C} = \beta(\mu_3-\mu_4) = \beta \Delta\mu_{\rm C}= \beta e V_{\rm C} \label{A_C},
\eea
corresponding to the differences of electric potentials in the two circuits, respectively $V_{\rm D}=\Delta\mu_{\rm D}/e$ and $V_{\rm C}=\Delta\mu_{\rm C}/e$.


\subsection{The fluctuating current in the QPC}


As it is the case in typical experiments on full counting statistics \cite{PhysRevLett.96.076605,gustavsson2009,fujisawa2006,PhysRevLett.99.206804}, the correlation time of the QPC is shorter than the dwell time of the electrons in the double quantum dot.  Therefore, the current in the QPC rapidly jumps to a value that is fixed by the state $\vert s_t\rangle$ of the double quantum dot.  This latter has the slow time evolution of the stochastic process ruled by the master equation (\ref{mastereqmod})-(\ref{Wdqdqpc}).  In this respect, electron transport in the double quantum dot can be simulated by Monte Carlo algorithms to obtain random histories $\{\vert s_t\rangle,n_t\}_{t\in{\mathbb R}}$ for the system \cite{gardinerzoller,GMWS01}. Due to the capacitive coupling, the occupancy of the quantum dots by electrons modulates the current in the QPC.  Indeed, if the double quantum dot is in the instantaneous state $\vert s_t\rangle$, the capacitive interaction (\ref{VABC}) modifies the tunneling amplitude $T_{\rm C}$ of the QPC into the time-dependent effective amplitude
\be
\tilde T_{\rm C}(t) = T_{\rm C} + \langle s_t\vert U_{\rm A} \, d_{\rm A}^{\dagger} d_{\rm A} + U_{\rm B} \, d_{\rm B}^{\dagger} d_{\rm B}\vert s_t\rangle.
\label{T_C(t)}
\ee

Now, the fast current in the QPC can be obtained thanks to the Landauer-B\"uttiker formula with a transmission probability given in terms of the time-dependent tunneling amplitude (\ref{T_C(t)}).
Over time scales longer than the QPC correlation time, $\Delta t\gg \tau_C$, the current in the QPC is thus given for each spin orientation by
\be
\langle J_{\rm C} \rangle_t = \frac{1}{2\pi} \int_{-2\gamma}^{+2\gamma} d\epsilon \; \tilde{\cal T}_{\epsilon}(t) \left[ f_3(\epsilon)-f_4(\epsilon)\right]
\ee
in terms of the transmission probability~(\ref{T_E}) but with the tunneling amplitude $T_{\rm C}$ replaced by the time-dependent expression (\ref{T_C(t)}).  

If the double quantum dot is occupied, Coulomb repulsion raises the barrier in the QPC, thus, lowering its current.  If the double quantum dot is temporarily in the state $\vert +\rangle$, the tunneling amplitude takes the value $\tilde T_{\rm C}(t) = T_{\rm C} +U_{++}$ with the Coulomb repulsion (\ref{U++}).  Instead, the tunneling amplitude is equal to $\tilde T_{\rm C}(t) = T_{\rm C} +U_{--}$ with (\ref{U--}) if the double quantum dot state is $\vert -\rangle$.  The QPC current is thus sensitive to the directionality of the electron jumps in the double quantum dot if the Coulomb repulsion is asymmetric between the two states $\vert \pm \rangle$, which requires that the mixing angle is not close to $\theta=\pi/2$ otherwise $U_{++}=U_{--}$.



\section{Full counting statistics of electron transport in the double quantum dot}
\label{FCS}


Here below, we introduce the cumulant generating function of the electron number transfers in the double quantum dot channel. This function is used in order to express the mean current in the double quantum dot as a function of the rate matrix elements introduced in Section \ref{modmastereq5}.

Subsequently, we investigate the behavior of the current in the double quantum dot as a function of both affinities (\ref{A_D}) and (\ref{A_C}). We observe that the QPC may drag a current in the double quantum dot when the eigenstates of the latter are well localized in consistency with experimental observations.


\subsection{Cumulant generating function}


The counting statistics of electron transfers in the double quantum dot is fully characterized in terms of the generating function of the statistical cumulants for the random number $n$ of electrons transferred from the reservoir $j=1$:
\be \label{cgfdqdqpc}
\mathcal{G}(\lambda) \equiv \lim_{t\to\infty} - \frac{1}{t} \ln \left\langle \mbox{e}^{-\lambda n} \right\rangle_t,
\ee
where the average $\langle\cdot\rangle_t$ is carried out over the probability distribution $p(n,t)=\sum_s p_s(n,t)$ that $n$ electrons have been transferred during the time interval $[0,t]$.
The probabilities $p_s(n,t)$ denote the solutions of the master equation (\ref{master-m}) - (\ref{Ldqdqpc}).  We showed in Section \ref{fcslongtime} that the cumulant generating function is given by the leading eigenvalue of the modified rate matrix (\ref{Wdqdqpc}).  The generating function is thus obtained by solving the eigenvalue problem
\be
\mbox{\bf W}(\lambda)\cdot {\bf v} = - \mathcal{G}(\lambda) \, {\bf v},
\label{eigenvalue_eq}
\ee
where ${\bf v}={\bf v}(\lambda)$ is the associated eigenvector.  Since the matrix elements depend on the chemical potentials of the reservoir, the generating function characterizes the current fluctuations in a stationary state of the double quantum dot that is generally out of equilibrium.  The complete equilibrium state is reached if both affinities (\ref{A_D}) and (\ref{A_C}) are vanishing.

The average current in the double quantum dot as well as the higher cumulants are given by taking successive derivatives of the generating function with respect to the counting parameter $\lambda$.  In particular, the average current is obtained as
\be
\langle J_{\rm D}\rangle = \frac{\partial \mathcal{G}}{\partial\lambda}\Big\vert_{\lambda=0}
\label{J_D-Q}
\ee
in a stationary state of the double quantum dot.  The second derivative gives the diffusivity of the current fluctuations around its average value.


\subsection{The average current in the double quantum dot}


In order to calculate explicitly the average current through the double quantum dot channel, it is useful to rewrite the cumulant generating function in terms of the rate matrix by use of equations (\ref{genfuncvect}) and (\ref{longtimecgfvect}) as
\bea
\mathcal{G} (\lambda) & = &  \lim_{t \rightarrow \infty} -\frac{1}{t}  \ln G(\lambda,t) \\
& = & \lim_{t \rightarrow \infty} -\frac{1}{t}  \ln \left( {\bf 1 }^{\top} \cdotp \mbox{e}^{ t {\bf W}(\lambda) }  \cdotp {\bf p }_0 \right),
\eea
where $ {\bf p }_0$ denotes the initial probability distribution over the double quantum dot. We further introduce the steady state probability vector
\be
{\bf P} = \lim_{t \rightarrow \infty}\mbox{e}^{ t {\bf W}(0) }  \cdotp {\bf p }_0
\ee
whose components $\left[ {\bf P} \right]_{s} = P_s$ give the steady state occupation probabilities over the double quantum dot.
Therefore, the average current through the double quantum dot can be calculated from (\ref{J_D-Q}) as
\be
\langle J_{\rm D}\rangle = (a_{1+}+a_{1-}) \, P_0 - b_{1+}\, P_+ - b_{1-} \, P_- .
\label{J_D}
\ee
The net average current from the reservoir $j=1$ has thus two positive contributions due to the charging transitions $\vert 0 \rangle \to \vert\pm\rangle$ from the reservoir $j=1$ and two negative contributions due to the discharging transitions $\vert\pm\rangle \to \vert 0 \rangle$ back to the reservoir $j=1$.  

\begin{figure}[h]
\centerline{\includegraphics[width=8cm]{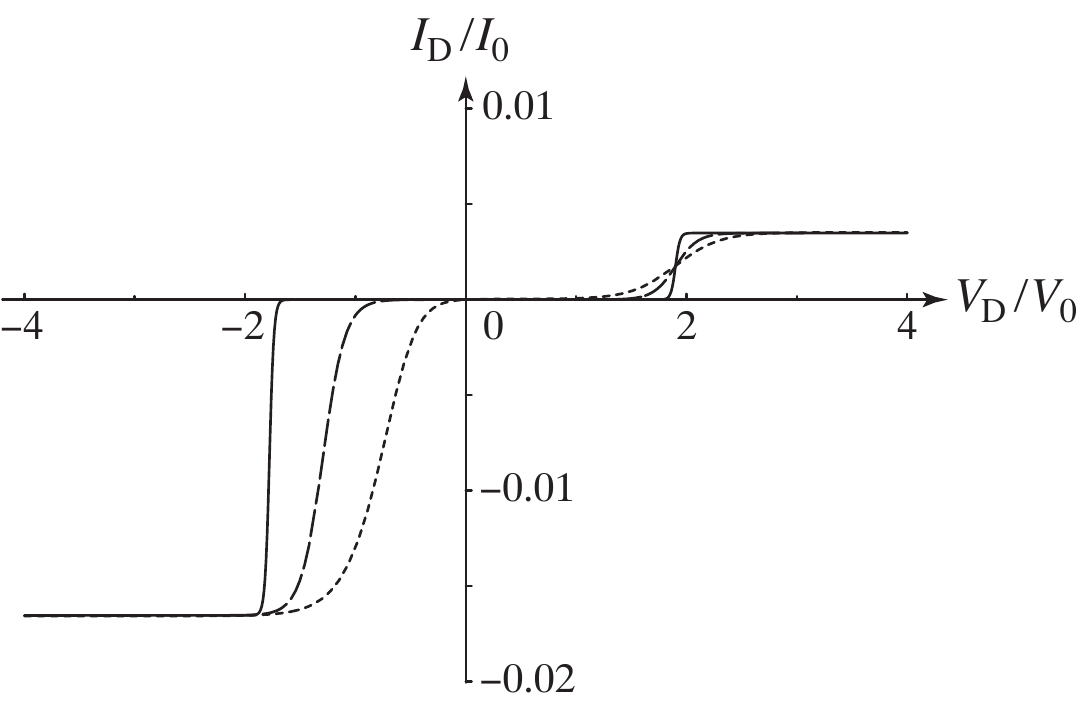}}
\caption{The average current $I_{\rm D}=e\langle J_{\rm D}\rangle$ in the double quantum dot versus the applied potential difference, $V_{\rm D}=\Delta\mu_{\rm D}/e=(\mu_1-\mu_2)/e$, if the QPC is at equilibrium with $\Delta\mu_{\rm C} = 0$.  The inverse temperature is $\beta = 10, 20, 100$ for respectively the dotted, dashed, and continuous lines.  The other parameters are $\mu_1+\mu_2= 3.3$, $\epsilon_{\rm A} = 0.7$, $\epsilon_{\rm B}  = 1.3$, $T = 0.01$, $T_{\rm A} = T_{\rm B} = 1$, $U_{\rm A} = -2.1$, $U_{\rm B} =- 0.6$, $\kappa = 0.2$, and $g = 1$.  The units are $I_0=eT_{\rm A}^2$ and $V_0=(\epsilon_{\rm A}+\epsilon_{\rm B})/(2e)$.}
\label{fig25}
\end{figure}

\textsc{Figure}~\ref{fig25} shows several $I$-$V$ characteristic curves of the double quantum dot circuit in the absence of bias in the QPC at different temperatures.  Since the QPC is at equilibrium $\Delta\mu_{\rm C}=0$, the average current $I_{\rm D}$ in the double quantum dot vanishes with the applied potential $V_{\rm D}$.  In \textsc{Figure}.~\ref{fig25}, the eigenstates of the double quantum dot have the energies $\epsilon_+\simeq 1.3$, $\epsilon_-\simeq 0.7$, and $\epsilon_0=0$.  The steps in the $I$-$V$ curves arise because every Fermi-Dirac distribution $f_j(\epsilon_s)$ undergo a similar step at the thresholds $\epsilon_s=\mu_j$ with $j=1,2$ and $s=\pm$.  {\it A priori}, thresholds are thus expected at the values $V_{\rm D}\simeq \pm 0.7$ if $\epsilon_+=\mu_j$, and $V_{\rm D}\simeq \pm 1.9$ if $\epsilon_-=\mu_j$ ($j=1,2$).  Nevertheless, only the latter ones appear in \textsc{Figure}~\ref{fig25} under the condition $\Delta\mu_{\rm C}=0$.  The reason is that the capacitive coupling to the QPC favors the transitions $\vert + \rangle\to\vert - \rangle$ because of Eq.~(\ref{c_ratio_eq}) and thus depopulates the level $\vert + \rangle$.  However, the level $\vert - \rangle$ remains below the Fermi energies of both reservoirs if $0<V_{\rm D}< 1.9$ so that the current is essentially stopped in this range.

\begin{figure}[h]
\centerline{\includegraphics[width=8cm]{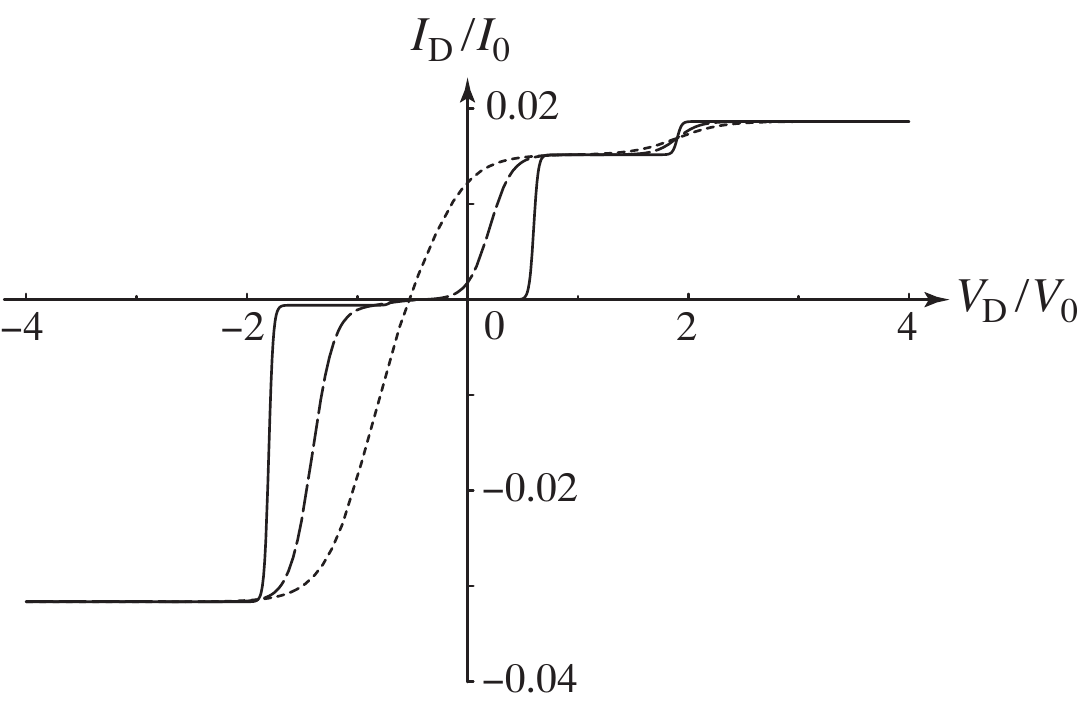}}
\caption{The average current $I_{\rm D}=e\langle J_{\rm D}\rangle$ in the double quantum dot versus the applied potential difference, $V_{\rm D}=\Delta\mu_{\rm D}/e$, if the QPC is out of equilibrium with $\Delta\mu_{\rm C} = -2$.  The inverse temperature is $\beta = 10, 20, 100$ for respectively the dotted, dashed, and continuous lines.  The other parameters and the units are the same as in \textsc{Figure}~\ref{fig25}.}
\label{fig35}
\end{figure}

In contrast, \textsc{Figure}~\ref{fig35} shows that, for $\Delta\mu_{\rm C}=-2$, the current no longer vanishes with the voltage bias applied to the double quantum dot and may even go against the bias if $V_{\rm D}\lesssim 0$.
This remarkable effect is due to the Coulomb drag that manifests itself because the QPC is out of equilibrium and capacitively coupled to the double quantum dot circuit.  In the linear regime for small enough values of the applied voltages, the average currents are related to the affinities (\ref{A_D})-(\ref{A_C}) according to $\langle J_{m}\rangle \simeq \sum_{m'}{\cal L}_{mm'} A_{m'}$ in terms of the Onsager coefficients ${\cal L}_{mm'}$ with $m,m'={\rm C,D}$.  Because of the asymmetric capacitive coupling the coefficients ${\cal L}_{\rm CD}={\cal L}_{\rm DC}$ are non vanishing, allowing the Coulomb drag effect observed in \textsc{Figure}~\ref{fig35} specially at high temperature for $\beta=10$.
If $V_{\rm D}=0$, the average current (\ref{J_D}) is proportional to
\be
\langle J_{\rm D}\rangle \propto \left(\Gamma_{1+}\Gamma_{2-}-\Gamma_{1-}\Gamma_{2+}\right)\left( c_{-+}\, {\rm e}^{\beta\epsilon_-}-c_{+-}\, {\rm e}^{\beta\epsilon_+}\right)
\label{J_D-V_D=0}
\ee
with $\Gamma_{1+}\Gamma_{2-}-\Gamma_{1-}\Gamma_{2+}=(4\pi g T_{\rm A}T_{\rm B})^2\cos\theta$.  This current vanishes if the QPC is at equilibrium when the local detailed balance condition (\ref{c_ratio_eq}) holds, which is no longer the case out of equilibrium.  We notice that the current (\ref{J_D-V_D=0}) also vanishes if the mixing angle reaches the value $\theta=\pi/2$.  The drag effect thus requires a good localization of the eigenstates in either one or the other of both dots, a condition which is met if the tunneling amplitude $T$ is not too large.  

\begin{figure}[htbp]
\centerline{\includegraphics[width=7cm]{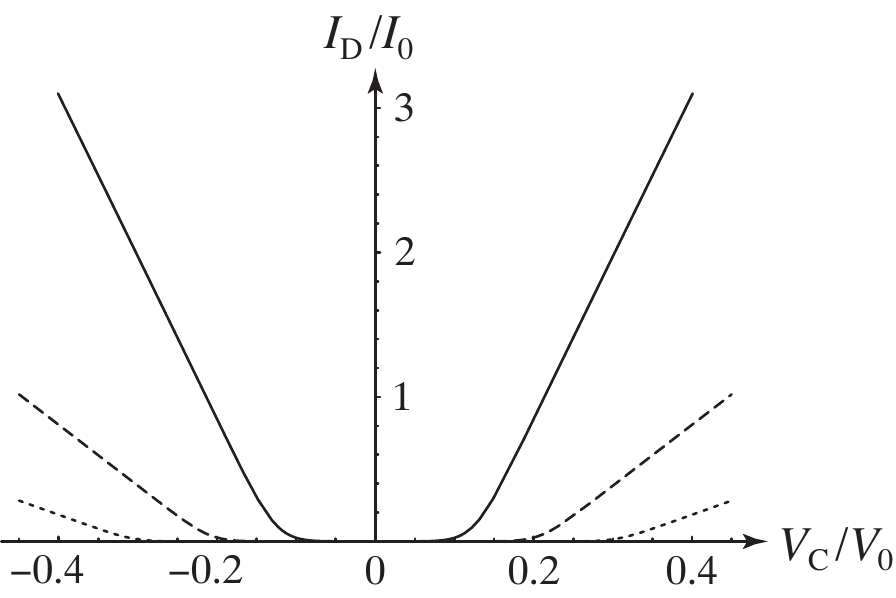}}
\caption{The average current $I_{\rm D}=e\langle J_{\rm D}\rangle$ in the double quantum dot without applied potential, $\mu_1=\mu_2=0.11$,  versus the potential difference in the QPC, $V_{\rm C}=\Delta\mu_{\rm C}/e=(\mu_3-\mu_4)/e$.  The energy of the quantum dot B is $\epsilon_{\rm B}  = 0.21, 0.3, 0.4$ for respectively the continuous, dashed, and dotted lines.  The other parameters are $\beta = 100$, $\epsilon_{\rm A} = 0.1$, $T = 0.03$, $\sqrt{g}\,T_{\rm A} = \sqrt{g}\,T_{\rm B} = 10$, $g\,U_{\rm A} =- 5$, $g\,U_{\rm B} =- 1.2$, and $\kappa = 0.17$.  The values of the parameters $\beta$, $\epsilon_{\rm B}-\epsilon_{\rm A}$, $T$, and $\kappa$ are estimated from experimental conditions \cite{PhysRevLett.99.206804}.  For the given values of $\mu_1=\mu_2$, $\sqrt{g}\,T_{\rm A}=\sqrt{g}\,T_{\rm B}$, and $g\,U_{\rm A}-g\,U_{\rm B}$, the two largest rates are $a_{1-}\simeq 1.0$\,kHz and $b_{2+}\simeq 1.2$\,kHz in accordance with experimental data \cite{PhysRevLett.99.206804}.  The units are $I_0=e/{\rm s}$ and $V_0=1$~mV with energies counted in meV.}
\label{fig55}
\end{figure}

\begin{figure}[htbp]
\centerline{\includegraphics[width=7cm]{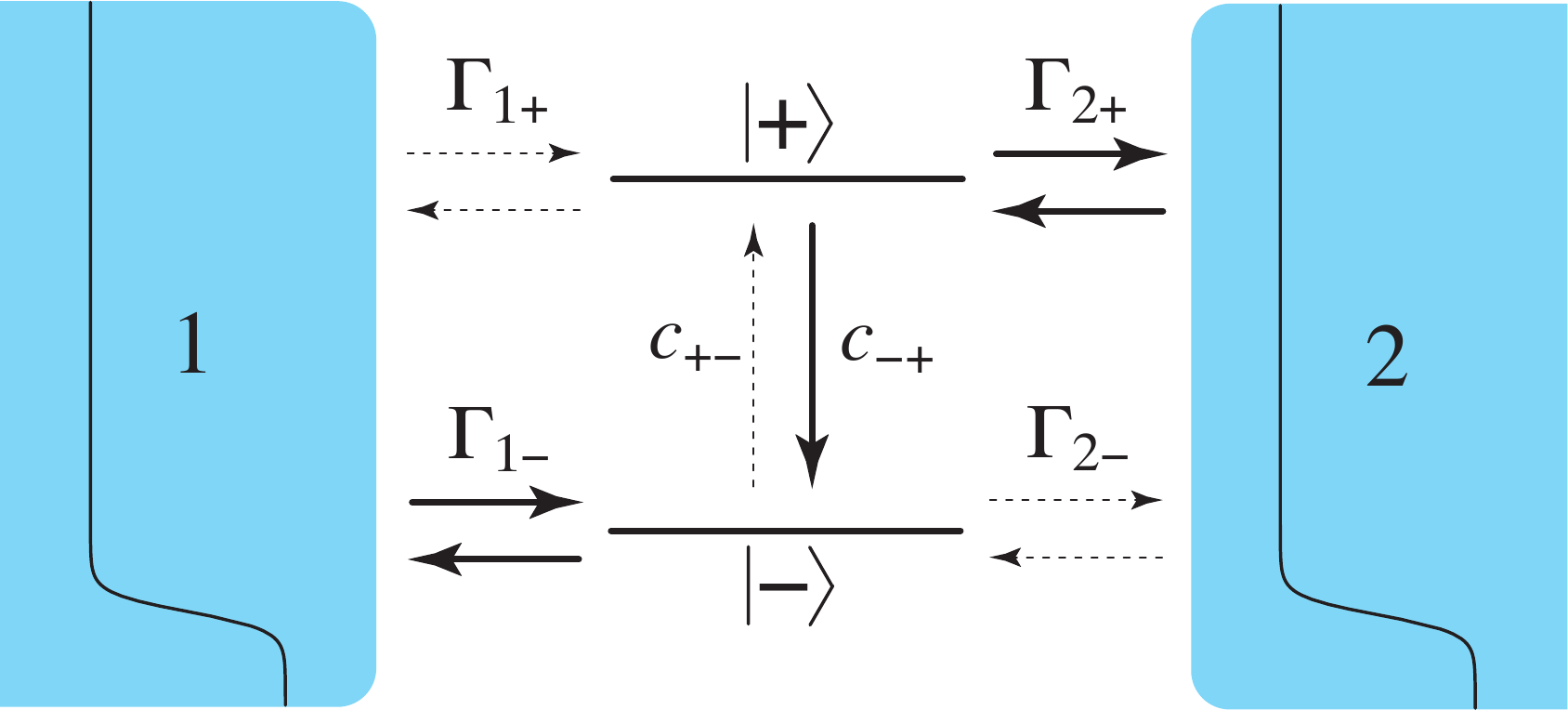}}
\caption{Schematic representation of the transitions in the double quantum dot circuit if the QPC fulfills the condition $\vert\Delta\mu_{\rm C}\vert < \epsilon_+-\epsilon_-$ for which Eqs.~(\ref{c+-0})-(\ref{c-+0}) hold and, moreover, if $\Gamma_{1-},\Gamma_{2+}\gg\Gamma_{1+},\Gamma_{2-}$, as it is the case for $\theta\simeq\pi$.  The solid line in each reservoir depicts the corresponding Fermi-Dirac distribution.}
\label{fig45}
\end{figure}

\textsc{Figure}~\ref{fig35} also shows that, for $\Delta\mu_{\rm C}=-2$, the four thresholds $V_{\rm D}\simeq \pm 0.7$ and $V_{\rm D}\simeq \pm 1.9$ appear in the $I$-$V$ curves of the double quantum dot.  This is explained by the behavior of the transition rates (\ref{c+-}) as a function of the bias in the QPC.  At low enough temperature, under the condition that
\be
\vert\Delta\mu_{\rm C}\vert < \omega_{+-}=\sqrt{(\epsilon_{\rm A}-\epsilon_{\rm B})^2+4T^2} ,
\label{threshold}
\ee
the transition rates (\ref{c+-}) are given by
\bea
&& c_{+-} \simeq 0 \label{c+-0} , \\
&& c_{-+} \simeq \frac{16\pi g^2 U_{ss'}^2}{(1+\kappa)^2} \, \omega_{+-} \label{c-+0}.
\eea
Therefore, the rate $c_{+-}$ of the transition $\vert -\rangle\to \vert +\rangle$ vanishes if $\vert\Delta\mu_{\rm C}\vert < \epsilon_+-\epsilon_-$ so that the upper energy level $\epsilon_+$ remains depopulated, as shown in \textsc{Figure}~\ref{fig45}.

However, both transition rates $c_{\pm\mp}$ are positive if $\vert\Delta\mu_{\rm C}\vert > \epsilon_+-\epsilon_-$, allowing the upper energy level $\epsilon_+$ to become populated.  As mentioned earlier, the thresholds at $V_{\rm D}\simeq \pm 0.7$ correspond to the condition $\epsilon_+=\mu_j$.  Therefore, these thresholds also appear in the $I_{\rm D}$-$V_{\rm D}$ curves for $\vert\Delta\mu_{\rm C}\vert > \epsilon_+-\epsilon_-\simeq 0.6$.  This is the case in \textsc{Figure}~\ref{fig35} where $\Delta\mu_{\rm C}=-2$ so that the four thresholds $V_{\rm D}\simeq \pm 0.7$ and $V_{\rm D}\simeq \pm 1.9$ are visible under such conditions.

The Coulomb drag effect is also illustrated in \textsc{Figure}~\ref{fig45}, which shows the current in the unbiased double quantum dot circuit versus the potential difference in the QPC for different values of the detuning $\epsilon_{\rm B}-\epsilon_{\rm A}$ between the energy levels of the quantum dots.  The double quantum dot current remains vanishing as long as the QPC potential difference satisfies the condition~(\ref{threshold}).  Again, under this condition, the upper energy level $\epsilon_+$ is not populated because the transition $\vert -\rangle\to \vert +\rangle$ does not occur according to Eq.~(\ref{c+-0}).  Since the lower level is charged from the left-hand reservoir $j=1$ and the upper level is discharged to the right-hand reservoir $j=2$, the double quantum dog current remains switched off.  Instead, beyond the threshold for $\vert\Delta\mu_{\rm C}\vert >\epsilon_+ -\epsilon_-=\omega_{+-}$, the upper energy level $\epsilon_+$ becomes populated thanks to the transitions induced by the QPC, which can thus exert Coulomb drag on the double quantum dot circuit.  The double quantum dot current is thus mainly determined by the rate $c_{+-}$.  In \textsc{Figure}~\ref{fig45}, the parameter values are taken to compare with the experimental observations reported in Ref.~\cite{PhysRevLett.99.206804}.  We see the remarkable agreement with \textsc{Figure}~\ref{fig45} of that reference.

These results show that the back-action of the QPC may strongly affect the transport properties of the double quantum dot circuit.  By enabling transitions between the states $\vert+\rangle$ and $\vert-\rangle$, the current across the double quantum dot circuit is enhanced if the QPC is driven out of equilibrium.  The capacitive coupling between the QPC and the double quantum dot should be asymmetric to allow the Coulomb drag to manifest itself.  Furthermore, we notice that the back-action tends to decrease if the dimensionless contact transparency $\kappa$ of the QPC increases.  A perturbative treatment of conductance in the QPC would thus overestimate the effects of back-action.


\subsection{The effective affinity}

\begin{figure}[b]
\centerline{\includegraphics[width=7cm]{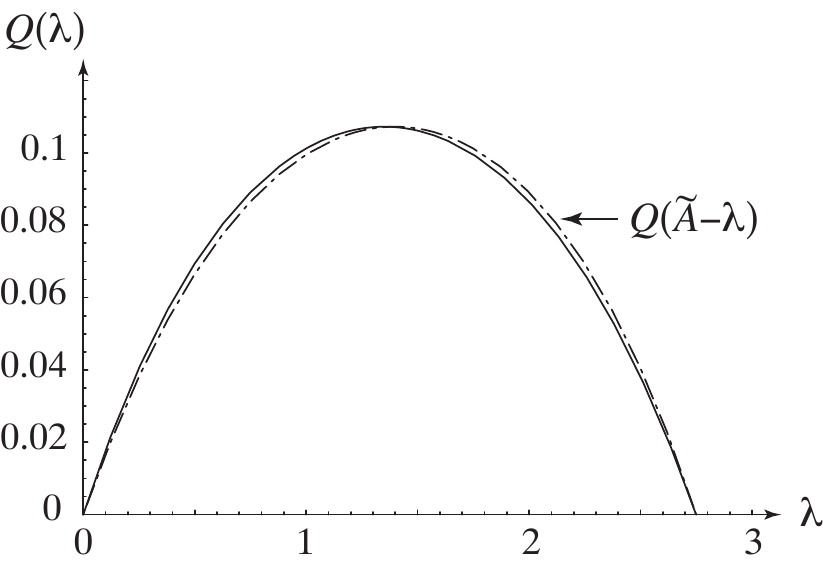}}
\caption{The cumulant generating function (\ref{cgfdqdqpc}) versus the counting parameter $\lambda$ compared to the function transformed by the reflection $\lambda\to\tilde A-\lambda$ with the effective affinity $\tilde A=2.7495$ (dotted-dashed line) for the parameter values $\mu_1=1.5$, $\mu_2=1$, $\Delta\mu_{\rm C}=5$, $\beta = 10$, $\epsilon_{\rm A} = 0.7$, $\epsilon_{\rm B}  = 1.2$, $T = 0.5$, $\sqrt{g}\,T_{\rm A} = \sqrt{g}\,T_{\rm B} = 1$, $g\,U_{\rm A} = -0.21$, $g\,U_{\rm B} =- 0.06$, and $\kappa = 0.2$.  The generating function has the units of $gT_{\rm A}^2$.}
\label{fig65}
\end{figure}
\begin{figure}[b]
\centerline{\includegraphics[width=7cm]{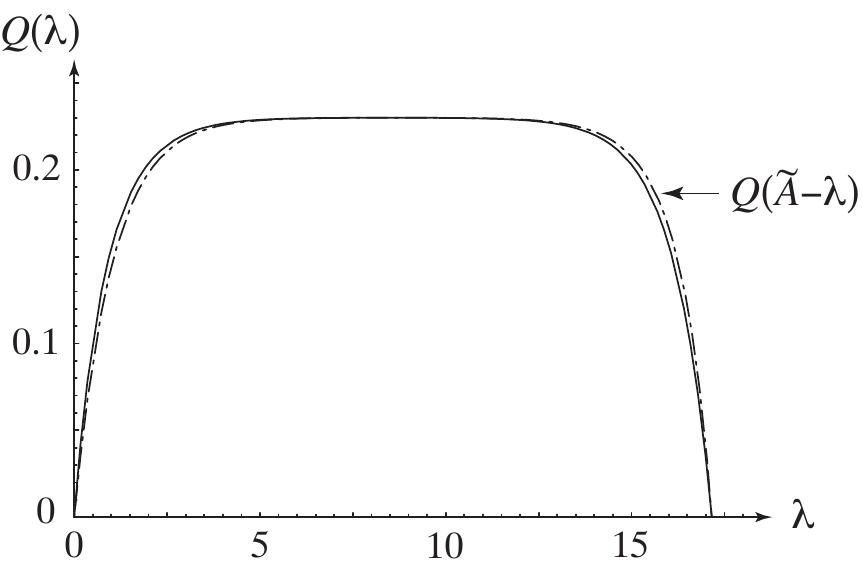}}
\caption{The cumulant generating function (\ref{cgfdqdqpc}) versus the counting parameter $\lambda$ compared to the function transformed by the reflection $\lambda\to\tilde A-\lambda$ with the effective affinity $\tilde A=17.1636$ (dotted-dashed line) for the parameter values $\mu_1=3$, $\mu_2=1$, $\Delta\mu_{\rm C}=5$, $\beta = 10$, $\epsilon_{\rm A} = 0.7$, $\epsilon_{\rm B}  = 1.2$, $T = 0.5$, $\sqrt{g}\,T_{\rm A} = \sqrt{g}\,T_{\rm B} = 1$, $g\,U_{\rm A} =- 0.21$, $g\,U_{\rm B} = -0.06$, and $\kappa = 0.2$. The units are the same as in \textsc{Figure}~\ref{fig65}.}
\label{fig75}
\end{figure}

The cumulant generating function (\ref{cgfdqdqpc}) is shown in Figures ~\ref{fig65} and \ref{fig75} for low and high potential differences in the double quantum dot.  As expected by its definition, the generating function vanishes at $\lambda=0$ where its slope gives the average current by Eq.~(\ref{J_D-Q}).  Moreover, the generating function also vanishes at a non-zero value of the counting parameter $\lambda=\tilde A$, which defines the {\it effective affinity}.    Since the generating function is given by the smallest root of the eigenvalue polynomial
\be
\det\left[ \mbox{\bf W}(\lambda)+\mathcal{G}(\lambda)\mbox{\bf 1}\right]=0
\ee
and $\mathcal{G}(\tilde A)=0$, the effective affinity can be obtained by solving
\be
\det\mbox{\bf W}(\tilde A)= D_+\left({\rm e}^{-\tilde A}-1\right)+D_-\left({\rm e}^{\tilde A}-1\right)=0,
\ee
where $D_{\pm}$ are quantities given in terms of the coefficients of the matrix (\ref{Wdqdqpc}).
The non-trivial root is equal to $\tilde A = \ln(D_+/D_-)$, which gives the {\it effective affinity}
\be
\tilde A =\ln \frac{a_{1+}\left[b_{2+}(b_{1-}+b_{2-}+c_{+-})+b_{2-}c_{-+}\right]+a_{1-}\left[b_{2-}(b_{1+}+b_{2+}+c_{-+})+b_{2+}c_{+-}\right]}{a_{2+}\left[b_{1+}(b_{1-}+b_{2-}+c_{+-})+b_{1-}c_{-+}\right]+a_{2-}\left[b_{1-}(b_{1+}+b_{2+}+c_{-+})+b_{1+}c_{+-}\right]}.
\label{Aff_eff}
\ee


In Figures~\ref{fig65} and \ref{fig75}, the generating function $\mathcal{G}(\lambda)$ is compared to $\mathcal{G}(\tilde A-\lambda)$.  For typical values of the parameters, the difference between $\mathcal{G}(\lambda)$ and $\mathcal{G}(\tilde A-\lambda)$ turns out to remain small in this system.  However, the symmetry under the transformation $\lambda\to\tilde A-\lambda$ is not valid in general.  Therefore, it is remarkable that there exist special conditions under which this symmetry nevertheless holds, as demonstrated in the following section.


\section{Single-current fluctuation theorems}
\label{FT}


As we showed in Chapter \ref{Chapter2}, a bivariate fluctuation theorem holds for both currents in the double quantum dot and QPC circuits with respect to the basic affinities (\ref{A_D})-(\ref{A_C}) \cite{Andrieux2006,PhysRevB.78.115429,andrieux2009monnai,RevModPhys.81.1665}. However, the conditions of observation are such that the full counting statistics of a circuit requires its coupling to another circuit so that bivariate statistics is not available.  Besides, the bivariate fluctuation theorem does not generally imply a fluctuation theorem for the sole current that is observed.  Nevertheless, conditions can be found for which univariate fluctuation theorems hold \cite{cuetara2011fluctuation,RJ07,GS11,MLBBS12,SSBE13}.

In the present system, several such conditions exist: 

(1) If the QPC is at equilibrium, it has no other influence than an equilibrium environment so that the only source of nonequilibrium driving comes from the voltage applied to the double quantum dot.

(2) If the tunneling amplitude between the two dots composing the double quantum dot is small enough $\vert T\vert\ll\vert\epsilon_{\rm A}-\epsilon_{\rm B}\vert$, each dot equilibrates with its next-neighboring reservoir on a time scale that is shorter than electron transfer time scale.

(3) If the QPC induces fast transitions $\vert + \rangle \leftrightharpoons \vert - \rangle$ in the limit $c_{\pm\mp}\gg a_{js},b_{js}$, the two states $\vert +\rangle$ and $\vert -\rangle$ can be lumped together. Accordingly, the three-state process reduces by coarse graining to a two-state process, in which the double quantum dot is either empty or singly occupied, and a fluctuation theorem always holds for two-state processes \cite{PhysRevB.72.235328,Andrieux2006}.

In these limiting cases, the difference between the cumulant generating function $\mathcal{G}(\lambda)$ and the transformed function $\mathcal{G}(\tilde A-\lambda)$ goes to zero so that the symmetry relation
\be
\mathcal{G}(\lambda)=\mathcal{G}(\tilde A-\lambda)
\label{FT-Q}
\ee
is obtained.  As a corollary, the probability $p(n,t)=\sum_s p_s(n,t)$ that $n$ electrons are transferred across the double quantum dot during the time interval $[0,t]$ obeys the {\it fluctuation theorem}
\be
\frac{p(n,t)}{p(-n,t)}\simeq {\rm e}^{\tilde A \, n} \qquad\mbox{for}\quad t\to\infty
\label{FT1}
\ee
as proved using the theory of large deviations \cite{touchette2009}. The symmetry of the fluctuation theorem is established with respect to the effective affinity (\ref{Aff_eff}) taken in the limit where the equality (\ref{FT-Q}) is valid.  An important result is that the effective affinity may differ from the value (\ref{A_D}) fixed by the reservoirs alone, because of the back-action of the capacitively coupled circuit.  Here below, the effective affinity is given for the different limits where the univariate fluctuation theorem holds.


\subsection{The QPC is at equilibrium}


In this case, the rates of the transitions induced by the QPC obey the local detailed balance condition (\ref{c_ratio_eq}) so that the matrix (\ref{Wdqdqpc}) obeys the symmetry relation
\be
\mbox{\bf M}^{-1}\cdot\mbox{\bf W}(\lambda)\cdot \mbox{\bf M}= \mbox{\bf W}(A_{\rm D}-\lambda)^{\rm T}
\ee
with the matrix
\be
\mbox{\bf M} = 
\left(
\begin{array}{ccc}
1 & 0 & 0 \\
0 & {\rm e}^{-\beta(\epsilon_+-\mu_2)} & 0  \\
0 & 0 & {\rm e}^{-\beta(\epsilon_--\mu_2)}
\end{array}
\right)
\label{M}
\ee
and the affinity (\ref{A_D}).  Consequently, the leading eigenvalue giving the cumulant generating function by Eq.~(\ref{eigenvalue_eq}) has the symmetry $\lambda\to A_{\rm D}-\lambda$.
Therefore, the effective affinity reduces to the standard one
\be
\tilde A=A_{\rm D}=\beta (\mu_1-\mu_2)
\ee
if the QPC is at equilibrium.  This result can also be obtained from the definition (\ref{Aff_eff}) of the effective affinity.


\subsection{The limit $\vert T\vert\ll\vert\epsilon_{\rm A}-\epsilon_{\rm B}\vert$}


In this limit, the tunneling amplitude between the two dots composing the double quantum dot is smaller than the difference between the energy levels of the dots, $\vert T\vert\ll\vert\epsilon_{\rm A}-\epsilon_{\rm B}\vert$.  Therefore, each dot is more strongly coupled to the next-neighboring reservoir than to the other dot.  Therefore, each dot equilibrates with the nearby reservoir on a time scale faster than for electron transfers.  There exist two subcases whether $\epsilon_{\rm A}-\epsilon_{\rm B}$ is positive or negative.

If $\epsilon_{\rm A}>\epsilon_{\rm B}$, the mixing angle goes to zero $\theta\to 0$ so that $\vert + \rangle \simeq \vert 1_{\rm A}0_{\rm B}\rangle$ and $\vert - \rangle \simeq -\vert 0_{\rm A}1_{\rm B}\rangle$.  In this subcase, the transition rates separate in the two groups:
\be
a_{1+},b_{1+},a_{2-},b_{2-} \gg a_{1-},b_{1-},a_{2+},b_{2+} ,c_{+-},c_{-+} = O(\theta^2).
\ee
Accordingly, the matrix (\ref{Wdqdqpc}) splits in two as $\mbox{\bf W}(\lambda)=\mbox{\bf W}_0(\lambda)+\mbox{\bf W}_1(\lambda)$ where $\mbox{\bf W}_0=O(\theta^0)$ and $\mbox{\bf W}_1=O(\theta^2)$ as $\theta\to 0$.  In Eq.~(\ref{eigenvalue_eq}), the eigenvector and the eigenvalue can be expanded similarly as ${\bf v}={\bf v}_0 + {\bf v}_1+\cdots$ and $\mathcal{G}=\mathcal{G}_0+\mathcal{G}_1+\cdots$.  At zeroth order, the eigenvalue vanishes $\mathcal{G}_0=0$, the right-hand eigenvector is given by
\be
{\bf v}_0=
\left(
\begin{array}{c}
1 \\
{\rm e}^{-\beta(\epsilon_+-\mu_1)}{\rm e}^{-\lambda} \\
{\rm e}^{-\beta(\epsilon_--\mu_2)} 
\end{array}
\right)
\ee
and the left-hand eigenvector such that ${\bf u}_0^{\rm T}\cdot\mbox{\bf W}_0=0$ by
\be
{\bf u}_0^{\rm T} = \left( \ 1 \quad {\rm e}^{\lambda} \quad 1 \ \right).
\ee
At first order in $\theta^2$, we find that
\be
\mathcal{G}_1 \simeq - \frac{{\bf u}_0^{\rm T}\cdot\mbox{\bf W}_1\cdot{\bf v}_0}{{\bf u}_0^{\rm T}\cdot{\bf v}_0},
\ee
which gives the leading approximation to the cumulant generating function $Q\simeq Q_1$:
\be
\mathcal{G}(\lambda)\simeq J_+\left(1-{\rm e}^{-\lambda}\right) + J_-\left(1-{\rm e}^{\lambda}\right)  
\label{Q_theta=0}
\ee
with
\bea
&&J_+= \frac{a_{1-}+(b_{2+}+c_{-+}){\rm e}^{-\beta(\epsilon_+-\mu_1)}}{1+{\rm e}^{-\beta(\epsilon_+-\mu_1)}+{\rm e}^{-\beta(\epsilon_--\mu_2)}}\\
&& J_-= \frac{a_{2+}+(b_{1-}+c_{+-}){\rm e}^{-\beta(\epsilon_--\mu_2)}}{1+{\rm e}^{-\beta(\epsilon_+-\mu_1)}+{\rm e}^{-\beta(\epsilon_--\mu_2)}}.
\eea
The cumulant generating function (\ref{Q_theta=0}) has the symmetry (\ref{FT-Q}) of the fluctuation theorem with the effective affinity:
\be
\tilde A = \beta(\mu_1-\mu_2) + \ln\frac{a_{1-}{\rm e}^{-\beta\mu_1}+a_{2+}{\rm e}^{-\beta\mu_2}+c_{-+}{\rm e}^{-\beta\epsilon_+}}{a_{1-}{\rm e}^{-\beta\mu_1}+a_{2+}{\rm e}^{-\beta\mu_2}+c_{+-}{\rm e}^{-\beta\epsilon_-}}.
\ee
We notice that, in the logarithm, the numerator and the denominator are both vanishing proportionally to $\theta^2$ so that their ratio is non vanishing in the limit $\theta\to 0$ and thus gives a contribution to the effective affinity beyond its standard value (\ref{A_D}).  This expression is equivalently obtained from Eq.~(\ref{Aff_eff}) in the limit $\theta\to 0$.

If $\epsilon_{\rm A}<\epsilon_{\rm B}$, the mixing angle has the limit $\theta\to \pi$ so that $\vert -\rangle \simeq \vert 1_{\rm A}0_{\rm B}\rangle$ and $\vert + \rangle \simeq \vert 0_{\rm A}1_{\rm B}\rangle$.
In this other subcase, the transition rates separate as
\be
a_{1-},b_{1-},a_{2+},b_{2+} \gg  a_{1+},b_{1+},a_{2-},b_{2-},c_{+-},c_{-+} = O(\delta^2)
\ee
with $\delta=\pi-\theta$.  Hence, the cumulant generating function can be obtained with a similar method as in the previous limit to get
\be
\mathcal{G}(\lambda)\simeq J'_+\left(1-{\rm e}^{-\lambda}\right) + J'_-\left(1-{\rm e}^{\lambda}\right)  
\label{Q_theta=pi}
\ee
with
\bea
&&J'_+= \frac{a_{1+}+(b_{2-}+c_{+-}){\rm e}^{-\beta(\epsilon_--\mu_1)}}{1+{\rm e}^{-\beta(\epsilon_--\mu_1)}+{\rm e}^{-\beta(\epsilon_+-\mu_2)}}\\
&& J'_-= \frac{a_{2-}+(b_{1+}+c_{-+}){\rm e}^{-\beta(\epsilon_+-\mu_2)}}{1+{\rm e}^{-\beta(\epsilon_--\mu_1)}+{\rm e}^{-\beta(\epsilon_+-\mu_2)}}.
\eea
Here also, the symmetry (\ref{FT-Q}) of the fluctuation theorem is satisfied by the generating function (\ref{Q_theta=pi}) but with the effective affinity:
\be
\tilde A = \beta(\mu_1-\mu_2) + \ln\frac{a_{1+}{\rm e}^{-\beta\mu_1}+a_{2-}{\rm e}^{-\beta\mu_2}+c_{+-}{\rm e}^{-\beta\epsilon_-}}{a_{1+}{\rm e}^{-\beta\mu_1}+a_{2-}{\rm e}^{-\beta\mu_2}+c_{-+}{\rm e}^{-\beta\epsilon_+}},
\label{Aff_eff_theta=pi}
\ee
which is equivalently given by Eq.~(\ref{Aff_eff}) in the limit $\theta\to \pi$.
In the logarithm, the numerator and the denominator are both vanishing proportionally to $\delta^2=(\pi-\theta)^2$ so that their ratio is non vanishing in the limit $\theta\to \pi$ and here also modifies the effective affinity with respect to its standard value (\ref{A_D}).

\begin{figure}[htbp]
\centerline{\includegraphics[width=9cm]{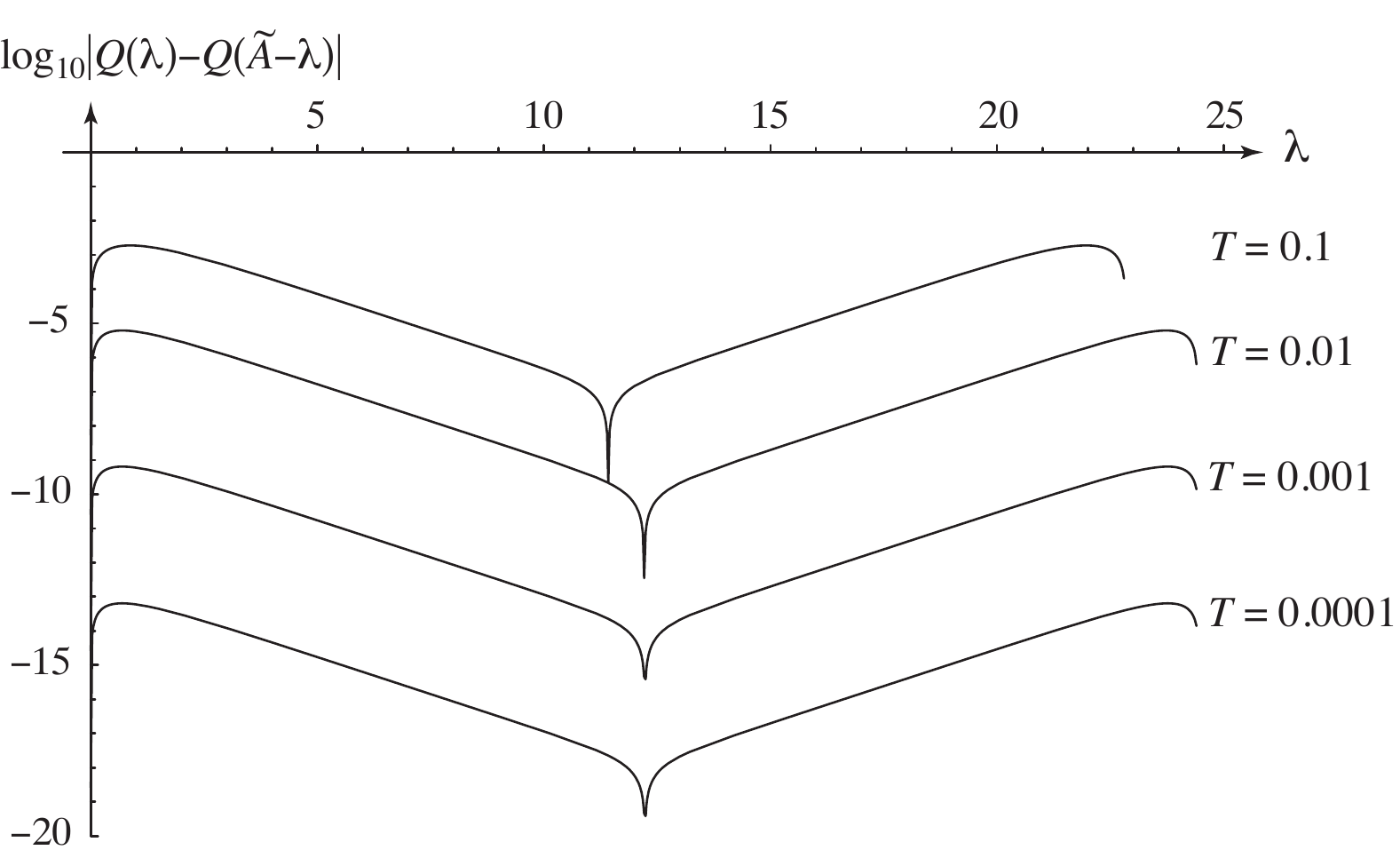}}
\caption{The difference between both sides of Eq.~(\ref{FT-Q}) in absolute value versus the counting parameter $\lambda$ for smaller and smaller values of the tunneling amplitude: $T=0.1$, $T=0.01$, $T=0.001$, and $T=0.0001$.  The other parameters are $\mu_1=3$, $\mu_2=1$, $\Delta\mu_{\rm C}=5$, $\beta =10$, $\epsilon_{\rm A} = 0.7$, $\epsilon_{\rm B}  = 1.2$, $\sqrt{g}\,T_{\rm A} = \sqrt{g}\,T_{\rm B} = 0.1$, $g\,U_{\rm A} =-0.21$, $g\,U_{\rm B} = -0.06$, and $\kappa = 0.2$.  For $T=0.0001$, the effective affinity is equal to $\tilde A=24.4594$ and the mixing angle to $\theta=\pi-0.0004$.  In the limit $T=0$, the effective affinity (\ref{Aff_eff_theta=pi}) is equal to $\tilde A=\beta(\mu_1-\mu_2)+\Delta\tilde A$ with $A_{\rm D}=\beta(\mu_1-\mu_2)=20$ and $\Delta\tilde A=4.45945$. The generating function $\mathcal{G}(\lambda)$ has the units of $gT_{\rm A}^2$.  A dip appears in the middle of each curve because the generating function $\mathcal{G}(\lambda)$ is always equal to $\mathcal{G}(\tilde A-\lambda)$ at $\lambda=\tilde A/2$, as seen in Figures~\ref{fig65} and~\ref{fig75}.}
\label{fig85}
\end{figure}

\textsc{Figure}~\ref{fig85} shows that the difference between both sides of Eq.~(\ref{FT-Q}) is indeed vanishing as the tunneling amplitude $T$ gets smaller and smaller so that  the symmetry (\ref{FT-Q}) is indeed satisfied in the limit $\theta\to\pi$.


\subsection{The QPC induces fast transitions $\vert + \rangle \leftrightharpoons \vert - \rangle$}\label{QPCinducedfast}


Here, we consider the limit
\be
\vert U_{\rm A}-U_{\rm B}\vert \gg \vert T_{\rm A}\vert, \vert T_{\rm B}\vert
\ee
so that the rates of the transitions $\vert + \rangle \leftrightharpoons \vert - \rangle$ are larger than the other rates:
\be
c_{+-}, c_{-+} \gg a_{js}, b_{js}
\label{limit_red}
\ee
with $j=1,2$ and $s=\pm$.  In this case, there is no distinction between the states $\vert\pm\rangle$ on the intermediate time scale $\Delta t$ between the short time of the transitions $\vert + \rangle \leftrightharpoons \vert - \rangle$  and the dwell time of electrons in the double quantum dot: $c_{\pm,\mp}^{-1}\ll\Delta t\ll a_{js}^{-1}, b_{js}^{-1}$.  Therefore, the stochastic process admits a reduced description in terms of the probability 
\be
p_1(n,t) = p_+(n,t)+p_-(n,t)
\ee
that the double quantum dot is occupied and the probability $p_0(n,t)$ that it is empty.  The probabilities of the states $\vert\pm\rangle$ are obtained as
\be
p_{\pm}(n,t) = p_1(n,t) \, P_{\pm\vert 1}
\ee
in terms of the conditional probabilities of the states $\vert\pm\rangle$ given that the double quantum dot is occupied:
\bea
&& P_{+\vert 1} = \frac{c_{+-}}{c_{+-}+c_{-+}} , \\
&& P_{-\vert 1} = \frac{c_{-+}}{c_{+-}+c_{-+}} .
\eea
These conditional probabilities are normalized according to $P_{+\vert 1}+P_{-\vert 1}=1$.
The master equation (\ref{master-m}) - (\ref{Ldqdqpc}) thus reduces to
\be
\left(
\begin{array}{c}
\partial_t\, p_{0}(n,t) \\
\partial_t\, p_{1}(n,t) 
\end{array}
\right)
=
\left(
\begin{array}{cc}
-a_{1}-a_{2} & b_{1}\,\hat{E}^+ +b_{2} \\
a_{1}\,\hat{E}^- +a_{2} & -b_{1}-b_{2}
\end{array}
\right)
\left(
\begin{array}{c}
p_{0}(n,t) \\
p_{1}(n,t) 
\end{array}
\right)
\label{L-01}
\ee
with the coefficients
\bea
&& a_j \equiv a_{j+}+a_{j-}  ,\\
&& b_j \equiv b_{j+} P_{+\vert 1}+b_{j-} P_{-\vert 1} = \frac{b_{j+}c_{+-}+b_{j-}c_{-+}}{c_{+-}+c_{-+}} ,
\eea
for $j=1,2$.

The cumulant generating function is here given by
\be
\mathcal{G}(\lambda) =  \frac{1}{2} \left[ 
a_1+a_2 +b_1+b_2 -
\sqrt{\left(a_1+a_2 -b_1-b_2 \right)^2
+ 4 \left( a_1\,{\rm e}^{-\lambda}+a_2\right)\left(b_1\,{\rm e}^{+\lambda}+b_2\right)}\right].
\label{Q_red}
\ee
The symmetry (\ref{FT-Q}) is again satisfied with the effective affinity:
\be
\tilde A= \ln \frac{a_1b_2}{a_2 b_1} = \ln\frac{(a_{1+}+a_{1-})\left(b_{2+}c_{+-}+b_{2-}c_{-+}\right)}{(a_{2+}+a_{2-})\left(b_{1+}c_{+-}+b_{1-}c_{-+}\right)},
\label{Aff_eff_red}
\ee
which can be obtained from Eq.~(\ref{Aff_eff}) in the limit (\ref{limit_red}).

\begin{figure}[h]
\centerline{\includegraphics[width=9cm]{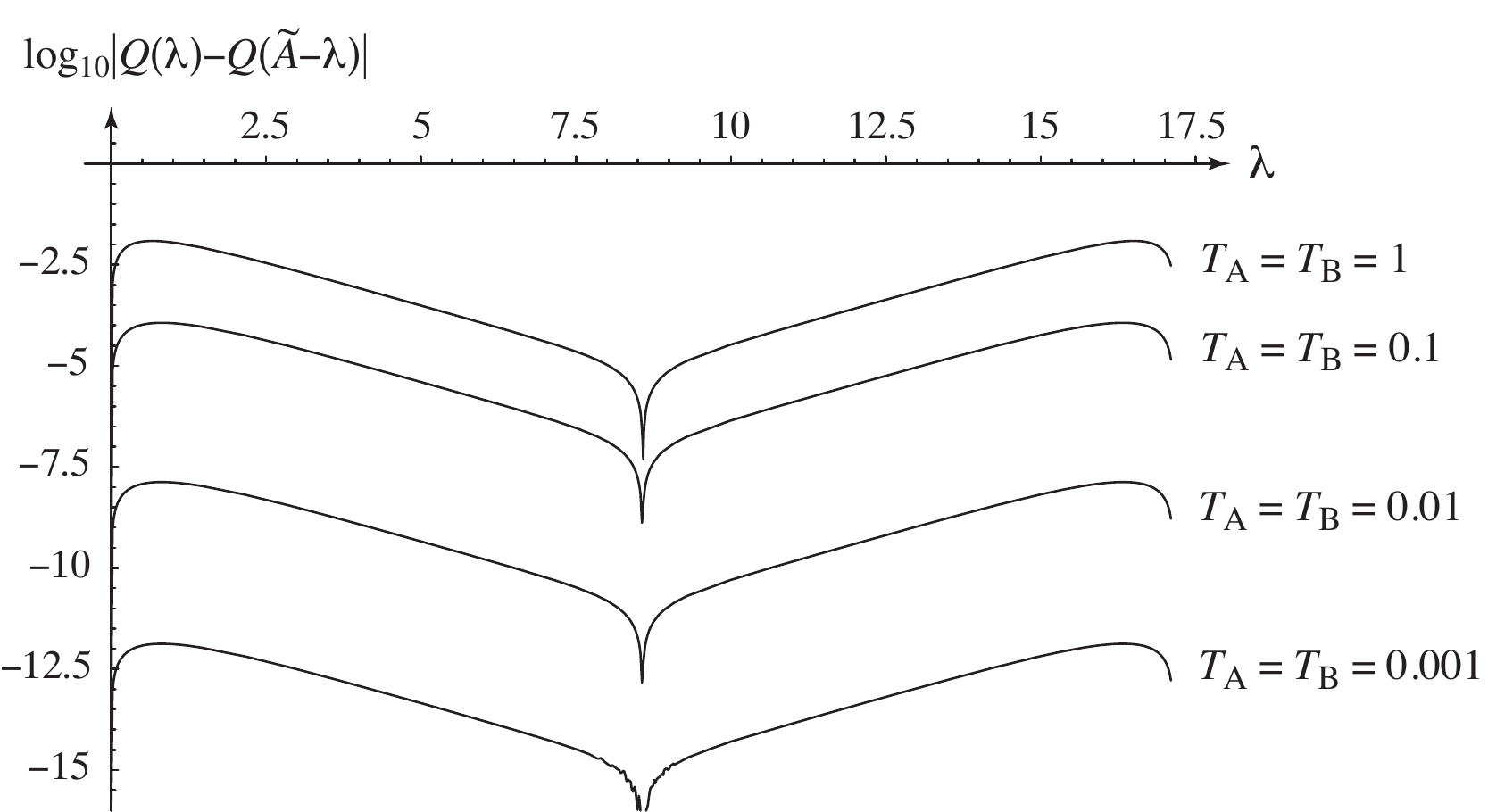}}
\caption{The difference between both sides of Eq.~(\ref{FT-Q}) in absolute value versus the counting parameter $\lambda$ for smaller and smaller values of the tunneling amplitudes $T_{\rm A} = T_{\rm B}$.  The other parameters are $\mu_1=3$, $\mu_2=1$, $\Delta\mu_{\rm C}=5$, $\beta =10$, $\epsilon_{\rm A} = 0.7$, $\epsilon_{\rm B}  = 1.2$, $T=0.5$, $g\,U_{\rm A} =-0.21$, $g\,U_{\rm B} = -0.06$, and $\kappa = 0.2$.  For $\sqrt{g}\, T_{\rm A} = \sqrt{g}\,T_{\rm B}=0.001$, the effective affinity is equal to $\tilde A=17.1397$. The generating function $\mathcal{G}(\lambda)$ has the units of $gT_{\rm A}^2$.  The dips in the middle of the curves have the same origin as in \textsc{Figure}~\ref{fig85}.}
\label{fig95}
\end{figure}

\textsc{Figure}~\ref{fig95} shows that the symmetry (\ref{FT-Q}) is well satisfied in the limit (\ref{limit_red}) as the tunneling amplitudes $T_{\rm A}$ and $T_{\rm B}$ between the double quantum dot and the reservoirs $j=1,2$ are decreased.


\subsection{Thermodynamic implications}


A consequence of the fluctuation theorem is that the product of the effective affinity with the average value of the current is always non negative:
\be
\tilde A \; \langle J_{\rm D}\rangle \geq 0 ,
\label{inequal}
\ee
where $\langle J_{\rm D}\rangle=\lim_{t\to\infty}\langle n\rangle_t/t$ with $\langle n\rangle_t = \sum_{n=-\infty}^{+\infty} n \, p(n,t)$ \cite{E12}. The inequality (\ref{inequal}) constitutes a lower bound on the thermodynamic entropy production
\be
\frac{1}{k_{\rm B}}\frac{d_{\rm i}S}{dt} = A_{\rm C} \langle J_{\rm C}\rangle+A_{\rm D} \langle J_{\rm D}\rangle \geq  \tilde A \; \langle J_{\rm D}\rangle \geq 0
\label{inequal2}
\ee
as demonstrated in Appendix~\ref{AppendixD}.  

This lower bound reduces the thermodynamic efficiency of the energy transduction processes that the system could perform.  In particular, the Coulomb drag of the QPC may drive the double quantum dot current against the applied voltage.  During this process, the QPC provides energy that accumulates between the reservoirs of the double quantum dot circuit.  To characterize the balance of energy per unit time, we introduce the powers $\Pi_{m}=V_{m}I_{m}=k_{\rm B}T A_{m}\langle J_{m}\rangle$ consumed by the circuits $m={\rm C},{\rm D}$.  The power of the QPC  is positive, $\Pi_{\rm C}>0$, although the power of the double quantum dot is negative $\Pi_{\rm D}<0$.  The thermodynamic efficiency of the process can be defined as
\be
\eta \equiv - \frac{\Pi_{\rm D}}{\Pi_{\rm C}} = -\frac{A_{\rm D} \langle J_{\rm D}\rangle}{A_{\rm C} \langle J_{\rm C}\rangle}
\label{eta},
\ee
which is positive in the regime where the Coulomb drag drives the double quantum dot current against the applied voltage, $\eta> 0$. The general non-negativity of the thermodynamic entropy production implies the well-known upper bound $\eta\leq 1$.

Now, in the regimes where the single-current fluctuation theorem (\ref{FT1}) holds, the thermodynamic entropy production has the lower bound (\ref{inequal2}) so that the efficiency is bounded as
\be
\eta \leq \frac{1}{1-\tilde A/A_{\rm D}} < 1 \qquad\mbox{if}\quad \tilde A/A_{\rm D}<0
\label{eta-bound}.
\ee
The thermodynamic efficiency is thus reduced if the back-action of the QPC is strong enough to reverse the effective affinity $\tilde A$ with respect to the value $A_{\rm D}$ fixed by the voltage applied to the double quantum dot.

The different cases where the single-current fluctuation theorem (\ref{FT1}) holds are presented in the following subsections.


\subsection{Dependence of the effective affinity on the quantum dot energies}


The gate voltages applied to the quantum dots control their energy levels.  Therefore, varying the energies $\epsilon_{\rm A}$ and $\epsilon_{\rm B}$ corresponds in the present model to changing the gate voltages of the quantum dots A and B.

\begin{figure}[h]
\centerline{\includegraphics[width=9cm]{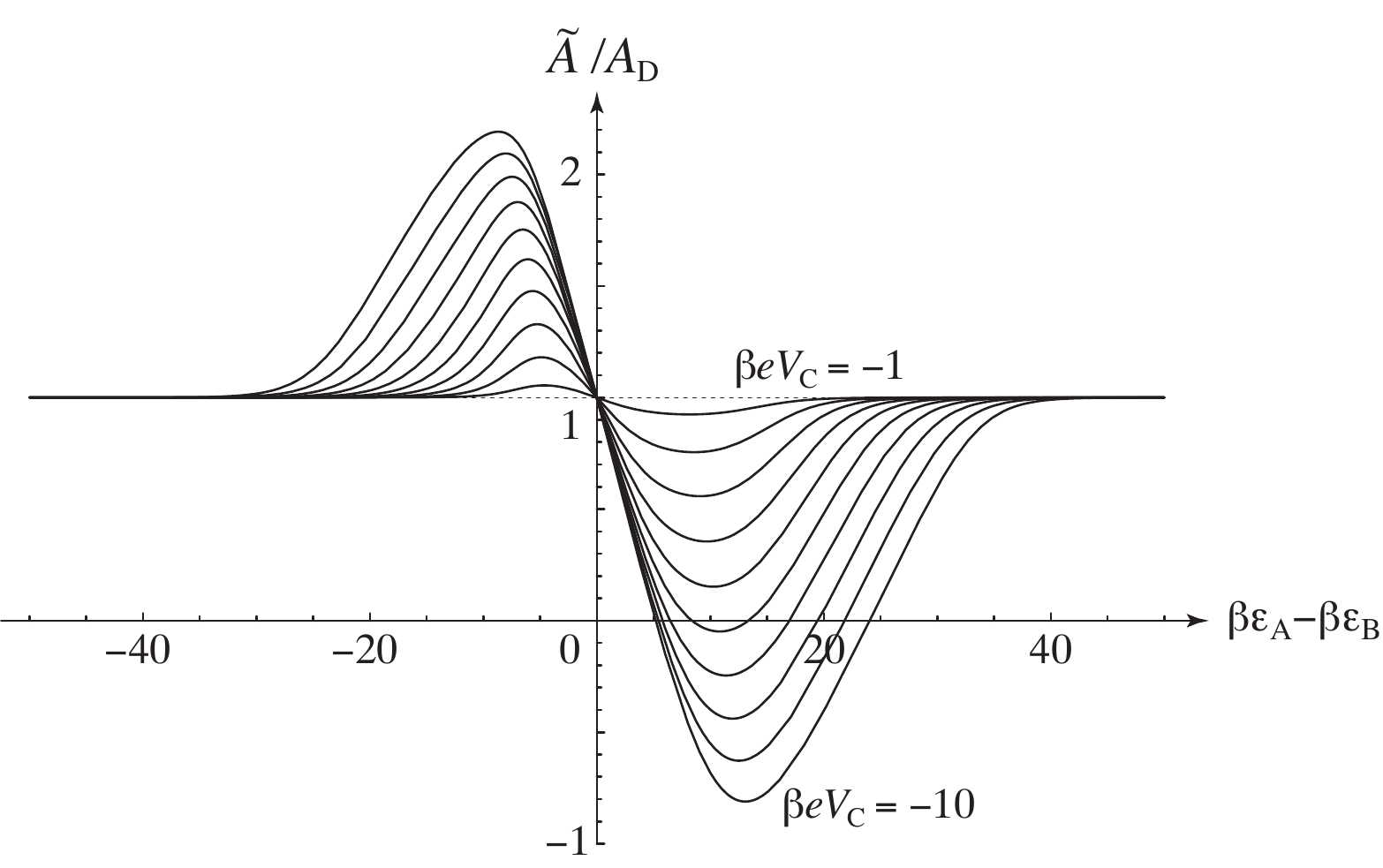}}
\caption{The effective affinity (\ref{Aff_eff}) rescaled by the actual affinity (\ref{A_D}) versus the energy difference $\beta\epsilon_{\rm A}-\beta\epsilon_{\rm B}$ in the DQD for several values of the potential difference applied to the QPC: $\beta\Delta\mu_{\rm C}=\beta eV_{\rm C}=-1,-2,-3,...,-10$.  The other parameters are $\beta\epsilon_{\rm A}+\beta\epsilon_{\rm B}=5$, $\beta\mu_1=2$, $\beta\mu_2=-2$, $\beta T=0.1$, $\sqrt{g}\, T_{\rm A} = \sqrt{g}\, T_{\rm B}=0.1$, $g\,U_{\rm A} =-1.2$, $g\,U_{\rm B} =-1.8$, and $\kappa = 0.2$.  The effective affinity is equal to the value $A_{\rm D}$ at $\beta\epsilon_{\rm A}-\beta\epsilon_{\rm B}\simeq-0.00367$ in the present conditions.}
\label{fig105}
\end{figure}

In \textsc{Figure}~\ref{fig105}, the effective affinity (\ref{Aff_eff}) is depicted as a function of the energy difference $\epsilon_{\rm A}-\epsilon_{\rm B}$ in the double quantum dot for several values of the affinity (\ref{A_C}) in the QPC circuit.  The back-action of the QPC onto the double quantum dot circuit manifests itself by the deviations of the ratio $\tilde A/A_{\rm D}$ from unity.  As expected, the back-action gets larger as the QPC is driven further away from equilibrium by increasing the absolute value of its affinity.  Although the effective affinity of the double quantum dot is nearly equal to its actual value (\ref{A_D}) if the QPC is close to equilibrium for $\Delta\mu_{\rm C}=-1$, they may significantly differ from each other if the QPC is far from equilibrium.  Under some conditions, the sign of the effective affinity may even be reversed with respect to the actual value (\ref{A_D}).  In these circumstances, the fluctuation theorem (\ref{FT1}) predicts that the average current in the double quantum dot should also be reversed.  Indeed, the average number of electrons that are transferred during the time interval $[0,t]$ is given by
\bea
\langle n\rangle_t &=& \sum_n n \, p(n,t) \nonumber\\
&=& \sum_{n>0} n \, \left[ p(n,t)-p(-n,t)\right] \nonumber\\
&=& \sum_{n>0} n \, p(n,t) \, \left(1-{\rm e}^{-\tilde A n}\right)
\eea
as the consequence of the fluctuation theorem (\ref{FT1}).  Therefore, if the effective affinity is negative $\tilde A<0$, the average number of transferred electrons is also negative $\langle n\rangle_t<0$.  This result confirms the inequality (\ref{inequal}).  This change of sign of the double quantum dot current is induced by the Coulomb drag effect due to the QPC.  In this regime, the thermodynamic efficiency (\ref{eta}) is bounded according to Eq.~(\ref{eta-bound}).  

In \textsc{Figure}~\ref{fig105}, we also observe that the effective affinity $\tilde A$ converges to its basic value $A_{\rm D}$ for $\vert\epsilon_{\rm A}-\epsilon_{\rm B}\vert \gg \vert\Delta \mu_{\rm C}\vert$.   The reason is that, in this limit, the rates of the transitions populating the state $\vert + \rangle$ are vanishing: $a_{1+}=a_{2+}=c_{+-}=0$.  Accordingly, the state $\vert + \rangle$ is never populated and it gets out of the dynamics ruled by the master equation (\ref{master-m})-(\ref{Ldqdqpc}): $\lim_{t\to\infty}p_+(n,t)=0$.  In this case, the effective affinity (\ref{Aff_eff}) becomes $\tilde A = \ln(a_{1-}b_{2-})/(b_{1-}a_{2-})=\beta(\mu_1-\mu_2)=A_{\rm D}$.  We point out that, if each quantum dot had more than the sole energy level assumed in the present model, the effective affinity would become more complicated for $\vert\epsilon_{\rm A}-\epsilon_{\rm B}\vert \gg \vert\Delta \mu_{\rm C}\vert$.

\begin{figure}[h]
\centerline{\includegraphics[width=8cm]{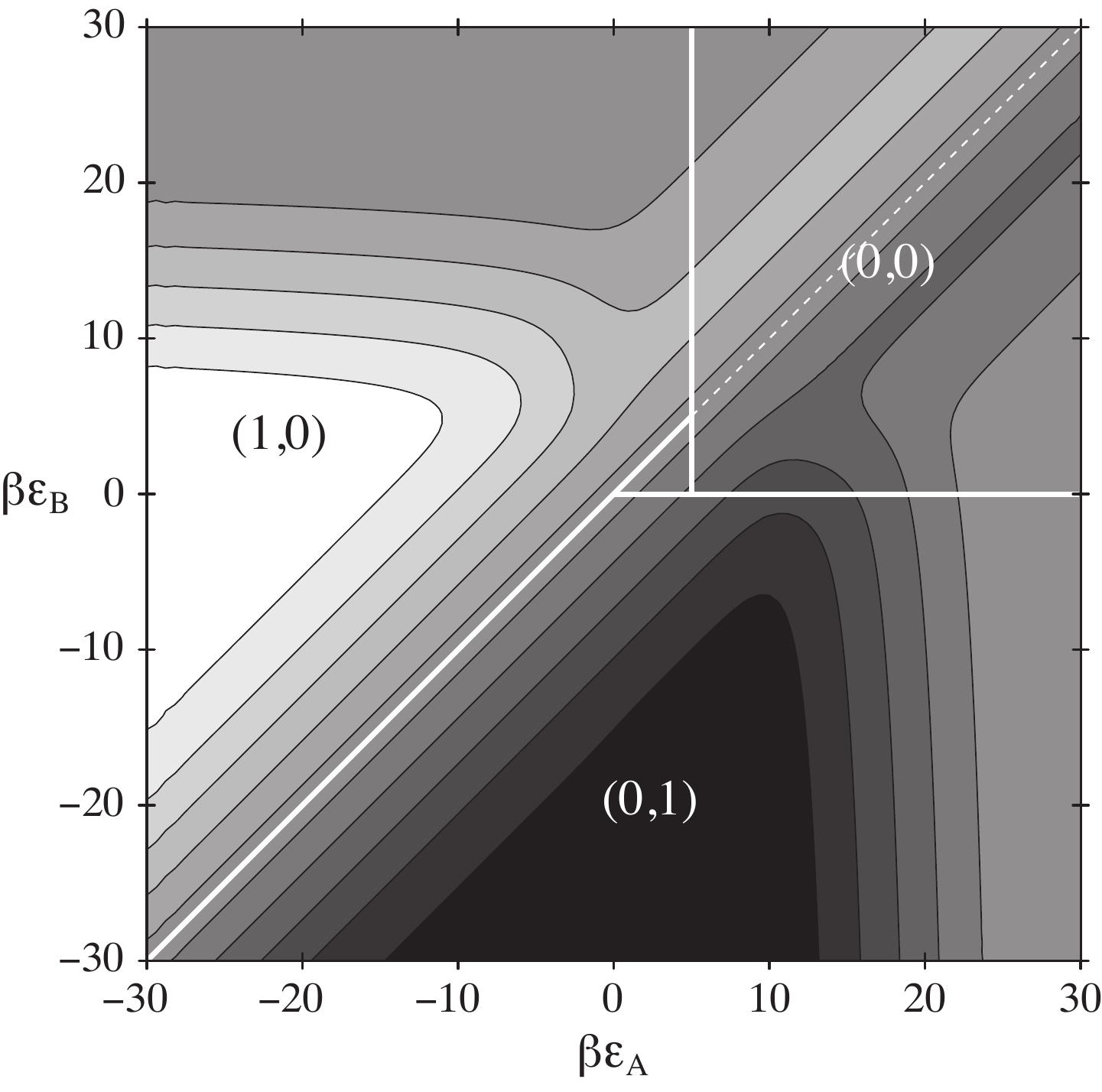}}
\caption{The effective affinity (\ref{Aff_eff}) rescaled by the actual affinity (\ref{A_D}) versus the energies $\beta\epsilon_{\rm A}$ and $\beta\epsilon_{\rm B}$ of the quantum dots.  The parameter values are $\beta\mu_1=5$, $\beta\mu_2=0$, $\beta\Delta\mu_{\rm C}=14$, $T=0.1$, $T_{\rm A} = T_{\rm B}=0.1$, $U_{\rm A} =-0.06$, $U_{\rm B} =-0.15$, $\kappa = 0.05$, and $g = 1$.  The white lines are the borders between the domains where the quantum dots are dominantly occupied according to $(n_{\rm A},n_{\rm B})\simeq(0,0),(0,1),(1,0)$.  We notice that the effective affinity is equal to the actual one, $\tilde A=A_{\rm D}$, along the diagonal $\beta\epsilon_{\rm A}=\beta\epsilon_{\rm B}$.  The effective affinity reaches the value $\tilde A/A_{\rm D}\simeq 3.5$ in the white area and $\tilde A/A_{\rm D}\simeq -1.5$ in the dark area.  The values of the parameters $U_{\rm A}/U_{\rm B}$ and $\kappa$ are estimated from experimental conditions.\cite{gustavsson2009}  In the present conditions, the effective affinity $\tilde A$ is equal to $A_{\rm D}$ near $\beta\epsilon_{\rm B}\simeq \beta\epsilon_{\rm A}+0.00024$ at $\beta\epsilon_{\rm A}=-10$, near $\beta\epsilon_{\rm B}\simeq \beta\epsilon_{\rm A}+0.00099$ at $\beta\epsilon_{\rm A}=0$, and near $\beta\epsilon_{\rm B}\simeq \beta\epsilon_{\rm A}+0.00166$ at $\beta\epsilon_{\rm A}=10$.}
\label{fig115}
\end{figure}

\textsc{Figure}~\ref{fig115} shows the ratio of the effective affinity (\ref{Aff_eff}) to the actual affinity (\ref{A_D}) in the plane of the quantum dot energies $(\epsilon_{\rm A},\epsilon_{\rm B})$ in comparison with the domains where the quantum dots have the dominant occupancies $(0,0)$, $(0,1)$, or $(1,0)$ \cite{RevModPhys.75.1}. These domains are delimited by three straight lines shown in \textsc{Figure}~\ref{fig115}.  The double quantum dot is empty in the domain where $\epsilon_{\rm A}>\mu_1$ and $\epsilon_{\rm B}>\mu_2$.  The dot A is empty while the dot B is occupied by one electron in the domain where $\epsilon_{\rm A}>\epsilon_{\rm B}$ and $\epsilon_{\rm B}<\mu_2$. The dot A is occupied by one electron while the dot B is empty in the domain where $\epsilon_{\rm A}<\mu_1$ and $\epsilon_{\rm B}>\epsilon_{\rm A}$.  In the triangle between these three domains, the double quantum dot is excited in its upper state $\vert +\rangle$ and its current is maximal, as expected \cite{RevModPhys.75.1}.  The effective affinity takes its actual value $\tilde A=A_{\rm D}$ very close to the diagonal $\epsilon_{\rm A}=\epsilon_{\rm B}$, which explains that the frontier between the domains is a favorable region to minimize the deviation of the effective affinity with respect to the value (\ref{A_D}) fixed by the voltage across the double quantum dot \cite{PhysRevX.2.011001}. Besides, the effective affinity (\ref{Aff_eff}) can significantly differs from its actual value (\ref{A_D}).  In \textsc{Figure}~\ref{fig115}, the effective affinity ranges from $\tilde A\simeq 3.5\times A_{\rm D}$ in the white area down to $\tilde A\simeq -1.5\times A_{\rm D}$ in the dark area.  In particular, the effective affinity drops from its actual value in the domain $(0,1)$ away from the frontier between $(0,1)$ and $(1,0)$.


\section{Summary of the results}


In this Chapter, we presented a study of electronic transport properties in a double quantum dot (DQD) circuit capacitively coupled to a QPC.  The QPC plays the role of detector for the single-electron transfers in the DQD and also affects this latter because of the back-action due to its noise.

The system is modeled by a simple Hamiltonian capturing the main features of such circuits.  The reservoirs in contact with the DQD as well as the QPC itself are described by tight-binding quadratic Hamiltonians that are exactly solvable, allowing the non-perturbative analysis of the QPC in arbitrary nonequilibrium states.  The tunneling between the DQD and its reservoirs, as well as the capacitive coupling between the DQD and the QPC are treated at second order of perturbation theory together with the rotating-wave and the wide-band approximations.  The electron transport in the QPC is supposed to behave faster than in the DQD.  The double occupancy of the DQD is assumed to lie at high enough energy to be neglected.

In this way, a Markovian master equation is obtained for the stochastic process of electron transfers across the DQD.  This master equation holds for the QPC in regimes arbitrarily far from equilibrium.
Under the assumption that the DQD is slower than the QPC, the current in the QPC is described by a Landauer-B\"uttiker formula depending on the time-dependent quantum state of the DQD.

The asymmetry of the capacitive coupling required for the bidirectional counting of electron transfers with the QPC is also responsible for its backaction onto the DQD current.  This backaction induces transitions between the internal states of the occupied DQD and, consequently, the Coulomb drag of the DQD current by the QPC if this latter is out of equilibrium.  Remarkably, a current is induced in the DQD if the voltage applied to the QPC exceeds a threshold given by the internal energies of the DQD, which is consistent with experimental observations \cite{PhysRevLett.99.206804}.

On the basis of the master equation, the full counting statistics is established for electron transport in the DQD.  Thanks to  its cumulant generating function, an effective affinity is introduced that characterizes the nonequilibrium driving of the DQD not only by the voltage applied to it, but also by the capacitively coupled QPC.
The value of this effective affinity differs from the value fixed by the voltage bias across the DQD if the QPC is driven out of equilibrium.

In the present Chapter, our main result is the establishment of single-current fluctuation theorems for the DQD current under specific conditions.  On fundamental ground, a bivariate fluctuation theorem is known to hold for both DQD and QPC currents.  However, the sole current across the DQD does not generally obey a fluctuation theorem because of the capacitive coupling to the QPC.  Therefore, it is surprising that there exist conditions under which single-current fluctuation theorems can nevertheless be established for the sole DQD current.  Our analysis shows that a single-current fluctuation theorem is valid under every one of the following conditions:

(1) If the QPC is at equilibrium, in which case the effective affinity remains equal to its value fixed by the voltage applied to the DQD.

(2) If the tunneling amplitude $T$ between the quantum dots composing the DQD is smaller than the difference between their internal energies $\epsilon_{\rm A}$ and $\epsilon_{\rm B}$: $\vert T\vert \ll \vert \epsilon_{\rm A}-\epsilon_{\rm B}\vert$.  In this limit, each dot is essentially at equilibrium with its next-neighboring reservoir and the DQD circuit behaves as another quantum point contact.

(3) If the asymmetry of the capacitive coupling to the QPC is stronger than the tunneling of the DQD to its reservoirs: $\vert U_{\rm A}-U_{\rm B}\vert \gg \vert T_{\rm A}\vert, \vert T_{\rm B}\vert$.  In this case, the QPC induces transitions between the two single-electron internal states of the DQD that are faster than its charging or discharging.  Therefore, the stochastic description reduces to a process for only two internal states: the empty and the singly occupied states.  

Moreover, the effective affinity is analyzed for its dependence on the internal energies of the quantum dots.  In the present model, they constitute the control parameters that are analogue to the gate voltages of the quantum dots.  Interestingly, the effective affinity is shown to remain close to the DQD voltage bias at the frontier between the domains of single occupancy of the DQD $(n_{\rm A},n_{\rm B})\simeq(0,1),(1,0)$ and to deviate from this value away from this frontier.

Besides, in the regimes where an single-current fluctuation theorem holds and the Coulomb drag may reverse the DQD current, the thermodynamic entropy production turns out to have a positive lower bound equal to the product of the effective affinity with the average DQD current.  In these regimes, the thermodynamic efficiency of electron pumping by Coulomb drag is limited by an upper bound lower than unity in terms of the ratio of the effective affinity to the voltage applied across the DQD.

To conclude, the effective affinity can be directly measured if a single-current fluctuation theorem is observed to hold experimentally.  Under such circumstances, the effective affinity can be used to characterize the full counting statistics of electron transport and the mechanisms driving the circuit out of thermodynamic equilibrium.

%% file: Chapter6.tex

\chapter{Nonequilibrium thermodynamics of a double quantum dot coupled to a quantum point contact} 

\label{Chapter6} 

\lhead{Chapter 6. \emph{Nonequilibrium thermodynamics of a DQD coupled to a QPC}} 


In Chapter \ref{Chapter5}, we performed the counting statistics of the particle flux in the double quantum dot channel. In this Chapter we perform the full-counting statistics of the energy and matter transfers in each of the four reservoirs of the setup by using the modified master equation formalism \cite{2004PhRvB70k5327R, 2008PhRvL100o0601F, 2005EL69475F, 2005PhRvB72p5347W, 2006PhRvL96b6805B, 2006PhRvE73d6129E, 2006PhRvB73c3312K, 2006RvMaP18619D, 2007PhRvB76p1404E, 2007PhRvB75o5316E, PhysRevB.76.085408, PhysRevB.77.195315, RevModPhys.81.1665} introduced in Chapter \ref{Chapter3}. In contrast to Chapter \ref{Chapter5}, the quantum point contact (QPC) transparency is here assumed to be low enough to permit the study of electron tunneling to second order in perturbation theory.

We show that the transition rates of the resulting master equation obey a Kubo-Martin-Schwinger (KMS) condition \cite{Kubo1998,RevModPhys.81.1665}. Most remarkably, this result is shown to hold for the transition rates describing the exchange of energy between the double quantum dot and the out of equilibrium QPC. With these relations, we obtain a fundamental symmetry \cite{andrieux2009monnai} for the cumulant generating function of the energy and particle fluctuations.

Subsequently, we use these results to perform the thermodynamic analysis of our model. The aforementioned symmetry relation implies the positivity of the irreversible entropy production in the system. We further consider our setup as a thermal machine where the double quantum dot performs work against a chemical potential bias as a result of the heat exchange with two thermal baths. Our setup is shown to attain highest efficiencies at maximal output power by fine tuning the spectrum of the double quantum dot.


\section{Hamiltonian}


The Hamiltonian we consider in this Chapter is identical the that of Chapter \ref{Chapter5}. However, the tunneling Hamiltonian in the QPC is now going to be treated to second order in consistency with the assumption of weak transparency. In order to do so, we rearrange the Hamiltonian of Chapter \ref{Chapter5} so that the total Hamiltonian is given as the sum
\be
H = H_S + \sum_{j=1}^{4} H_{j} + V_{1} + V_{2} + V_{34},
\ee
where each term is explicited below.

The free Hamiltonian of the system $H_S$ is given in its eigenbasis by (\ref{eigenbas5}) - (\ref{HSeigen}). The reservoir Hamiltonians are given by 
\be \label{res6}
H_j = \sum_{k} \epsilon_{j k} c_{jk}^{\dagger} c_{jk}
\ee
in terms of the creation and anihilation operators in the diagonal basis of reservoir $j=1, \dots, 4$. In the case of one-dimensional tight-binding Hamiltonians of the form (\ref{H_j}), the eigenvalues and corresponding eigenstates are given respectively by (\ref{E_k}) and (\ref{loc5}) with (\ref{phi_k}) in terms of the local basis eigenstates. The sum appearing in (\ref{res6}) becomes an integral over the quantum number $k$ ranging from $0$ to $\pi$.

The tunneling Hamiltonians $V_1$ and $V_2$ of the double quantum dot channel are expressed as
\be
V_j = \sum_{k} T_{js}^{k} \, |s \rangle \langle 0 | \, c_{jk} + T_{js}^{k} \, c^{\dagger}_{jk} \, |0 \rangle \langle s | 
\ee
in terms of the real\footnote{The tunneling amplitudes can be assumed real in absence of external magnetic field.} tunneling amplitudes $T_{js}^{k} $ for $j=1$ and $2$. Again, the link with the tunneling Hamiltonians of Chapter \ref{Chapter5} is made by use of the eigenfunctions of the reservoir Hamiltonians. For a one-dimensional tight-binding Hamiltonian, the tunneling amplitudes can be expressed as
\be
T_{js}^{k} = \phi_{k} (0)   T_{js}
\ee
in terms of the eigenfunction (\ref{phi_k}) at $l=0$ and the tunneling amplitudes $T_{js}$ given in (\ref{T1+}).

The interaction between the QPC and the double quantum dot is now given by 
\be \label{intQPCDQD6}
V_{34} =\sum_{s,s'} \sum_{kk'} V_{ss'}^{kk'}  \, c_{3 k}^{\dagger} c_{4 k'} |s \rangle \langle s' |+   V_{ss'}^{kk'} \,  c_{4k'}^{\dagger} c_{3k}|s' \rangle \langle s |
\ee
in terms of the interaction parameters $V_{ss'}^{kk'} $. The interaction Hamiltonian (\ref{VABC}) of Chapter \ref{Chapter5} can be written in this form with the corresponding tunneling amplitudes given by
\be
V_{ss'}^{kk'} = \left( T_C \, \delta_{ss'} \, \delta_{kk'}+ U_{ss'} \phi_k (0) \phi_{k'} (0) \right)\, \vert s \rangle \langle s' \vert ,
\ee
where the Kronecker delta symbol is denoted by $\delta_{mn}$, and in terms of the state-independent tunneling amplitude $T_{C}$ and the tunneling amplitudes $U_{ss'}$ given in (\ref{U++}) - (\ref{U--}). One should note at this point that the tunneling term of the QPC appearing in (\ref{H_C}) is now included in this interaction and will be treated perturbatively.

We are going to perform the counting statistics of the energy and particle number operators within each reservoir given respectively by (\ref{res6}) and
\be
N_{j} \equiv \sum_{k} c_{jk}^{\dagger} c_{jk}.
\ee
Accordingly, the modified Hamiltonian (\ref{modifiedHam3}) is here given by
\be \label{68}
H ( \xi_j , \lambda_j ) \equiv \mbox{e}^{i \sum_{j=1}^{4} \left( \frac{\xi_j}{2} H_j + \frac{\lambda_j}{2} N_j \right)} \, H \,\mbox{e}^{-i \sum_{j=1}^{4} \left( \frac{\xi_j}{2} H_j + \frac{\lambda_j}{2} N_j \right)},
\ee
where we introduced the energy and matter counting fields respectively as $\xi_j$ and $\lambda_j$ for $j=1, \dots, 4$. The Hamiltonians of the double quantum dot and the reservoirs are left unaffected by this transformation. However, the interaction Hamiltonians change according to
\be  \label{69}
 V_{j} ( \xi_j , \lambda_j ) = \sum_{s=+,-} \sum_{k} \mbox{e}^{-i \left(  \xi_{j} \epsilon_{k}+ \lambda_{j} \right) } \, T_{js}^{k} \, |s \rangle \langle 0 | \, c_{jk} +\mbox{e}^{i \left(  \xi_{j} \epsilon_{k}+ \lambda_{j} \right) } \,  T_{js}^{k} \, c^{\dagger}_{jk} \, |0 \rangle \langle s |  \\
\ee
and
\begin{multline} \label{611}
V_{34}( \xi_j , \lambda_j )  \\
= \sum_{s,s'} \sum_{kk'} V_{ss'}^{kk'} \left( \mbox{e}^{i \left(  \xi_{3} \epsilon_{k} + \lambda_{3} \right)}\mbox{e}^{-i \left(  \xi_{4} \epsilon_{k'} + \lambda_{4} \right)} c_{3 k}^{\dagger} c_{4 k} +   \mbox{e}^{-i \left(  \xi_{3} \epsilon_{k} + \lambda_{3} \right)}\mbox{e}^{i \left(  \xi_{4} \epsilon_{k'} + \lambda_{4} \right)}c_{4k}^{\dagger} c_{3k}\right).
\end{multline}
All those definitions are now ready in order to obtain a modified master equation on the double quantum dot by the procedure described in Chapter \ref{Chapter3} and already applied in Chapters \ref{Chapter4} and \ref{Chapter5}.


\section{Stochastic description}


Herein we apply the results of Chapter \ref{Chapter3} in order to get a modified master equation characterizing the evolution of the double quantum dot and the transport processes driven by the thermodynamic affinities. The transition rates are shown to obey a KMS condition leading to a fluctuation symmetry for the cumulant generating function of the fluxes.


\subsection{Modified master equation}


The starting point of this analysis is the modified reduced density matrix (\ref{reducedens}) which is here given by
\be \label{moddens6}
\rho_S (\xi_j , \lambda_j ,t ) = \mbox{Tr}_R \left\{ \mbox{e}^{-i \sum_{j=1}^{4} \left( \frac{\xi_j}{2} H_j + \frac{\lambda_j}{2} N_j \right)} \, \rho(t) \,\mbox{e}^{-i \sum_{j=1}^{4} \left( \frac{\xi_j}{2} H_j + \frac{\lambda_j}{2} N_j \right)} \right\},
\ee
the trace being taken over the reservoirs $j=1, \dots , 4$ and $\rho(t)$ denoting the density matrix of the whole system at time $t$.

The reservoirs composing the environment are initially in grand-canonical equilibrium states characterized by the inverse temperatures $\beta_{j}$ and chemical potentials $\mu_{j}$ for $j=1, \dots , 4$. Additionally, the double quantum dot is assumed to be  initially in a statistical mixture $\rho_S (0)$. The initial density matrix of the total system is thus given by
\be
\rho(0)= \rho_{\rm S} (0) \prod_{j} \otimes \, \, {\rm 
e}^{-\beta(H_j-\mu_{j} N_j - \phi_{j})},
\label{rho_0}
\ee
where 
\be
\phi_j = -\beta_{j}^{-1} \ln \left[ \mbox{Tr} \left\{ {\rm 
e}^{-\beta_{j}(H_j-\mu_{j} N_j )} \right\} \right]
\ee
denotes the thermodynamic grand-potential of the reservoir $j$. For our purpose, we assume the double quantum dot spectrum to be limited to the empty and single charged states (\ref{eigenstates5}) with (\ref{theta5}).

As we did in previous Chapters, we assume the reservoirs to be good macroscopic baths and the interaction parameters $\{ T_{js}^{k} , V_{ss'}^{kk'} \}$ small enough to be treated perturbatively. As mentioned earlier, the effect of the environment on the dynamics of the reduced density matrix (\ref{moddens6}) is thus characterized by its correlation functions (\ref{corfunc}).

In the present case, the reservoirs $j=1$ and $2$ affect the double quantum dot dynamics to which they are directly connected through the charging and discharging rates
\bea \nonumber
a_{js} & = & \Gamma_{js}   f_{j}(\omega_{s0}) , \\ \label{chargedischarge}
b_{js} & = & \Gamma_{js}  (1 -  f_{j}(\omega_{s0}) ),
\eea
for $s=+$ or $-$, and in terms of the Bohr frequencies of the system $\omega_{ss'} =\frac{ E_{s}-E_{s'}}{\hbar}$. The Fermi distribution of the reservoir $j$ with inverse temperature $\beta_{j}$ and chemical potential $\mu_{j}$ is given by $f_{j}(x) = (1+\exp \beta_{j}(x-\mu_{j}))^{-1}$. The rate constants are given at second order of perturbation theory by
\bea
 \Gamma_{js}&  = & \frac{2 \pi}{\hbar^{2}} \sum_{k} \delta (\epsilon_{k}^{j} -\omega_{s0})  | T^{k}_{j s} |^{2} \\
& = & \frac{2 \pi}{\hbar^{2}} \rho_{j}(\omega_{s0}) | T_{js}(\omega_{s0}) |^{2} ,
\eea
where we made the assumption of a continuum spectrum for the electron states in the reservoirs. The energy-resolved tunneling amplitudes are denoted by $T_{js}(\epsilon)$, while $\rho_{j}(\epsilon)$ is the density of electron states with energy $\epsilon$ in the reservoir $j$. This continuum approximation for the reservoirs spectra is justified when they can be considered macroscopic as compared to the intermediate quantum system.

On the other hand, the QPC also induces transitions between the double quantum dot states. Though there is no exchange of electrons between the QPC and the double quantum dot, electrons tunneling between the electrodes $j=3$ and $4$ may exchange energy with the double quantum dot, thus driving transitions between the single-charged states $|+ \rangle$ and $|- \rangle$. These processes are characterized by the transition rates
\bea
c_{jj'} (\epsilon) =\Gamma_{jj'} (\epsilon)  \, f_{j} (\epsilon) \, (1-f_{j'} (\epsilon - \omega_{+-})) \nonumber , \\
d_{jj'} (\epsilon) = \Gamma_{jj'} (\epsilon)  \, (1- f_{j}(\epsilon)) \, f_{j'} (\epsilon - \omega_{+-}) , \label{QPCrates6}
\eea
for $jj'= 34$ and $43$, and where $f_{j}(x)$ are the Fermi distributions of electron energies in reservoirs $j=3$ and $4$ of the QPC. The energy-dependent rate constants $ \Gamma_{jj'}(\epsilon)$ are calculated to second order of perturbation theory yielding
\bea
\Gamma_{34} (\epsilon)& \equiv & \frac{4 \pi}{\hbar^{2}}  \sum_{kk'} |V^{kk'}_{-+} | ^{2} \delta(\epsilon-\epsilon_{3k}) \delta(\epsilon - \omega_{+-} -\epsilon_{4k'}) \\
 & = &  \frac{4 \pi}{\hbar^{2}} \, |V_{-+} (\epsilon, \epsilon  - \omega_{+-})|^{2} \rho_{3}(\epsilon ) \rho_{4} (\epsilon-\omega_{+-} )  , \\
\Gamma_{43} (\epsilon)& = & \frac{4 \pi}{\hbar^{2}} \sum_{kk'} |V^{kk'}_{+-} | ^{2} \delta(\epsilon - \omega_{+-}-\epsilon_{3k}) \delta(\epsilon-\epsilon_{4k'}) \\
 & = &  \frac{4 \pi}{\hbar^{2}} \, |V_{+-} (\epsilon - \omega_{+-}, \epsilon )|^{2} \rho_{3}(\epsilon - \omega_{+-}) \rho_{4} (\epsilon ) , \\
\eea
where we took the continuum limit in the spectra of the reservoirs on the second and fourth lines, denoting the energy resolved interaction parameters by $V_{ss'}(\epsilon, \epsilon')$ and the density of electron states in the reservoir $j$ by $\rho_{j}(\epsilon)$.

Just as we did in Chapters \ref{Chapter4} and \ref{Chapter5}, we consider the double quantum dot dynamics on a time scale that is intermediate between the correlation times of the interaction operators and the relaxation times induced by the macroscopic reservoirs composing the environment. By further assuming fast oscillations due to the free system dynamics, we can perform the Born-Markov as well as the rotating wave approximations in order to obtain a closed set of dynamical equations for the diagonal elements of the modified reduced density matrix
\be
g_{s} (\xi_{j}, \lambda_{j} ,t) \equiv \langle s| \rho_S (i \xi_j , i \lambda_j , t) | s \rangle
\ee
for $s\in \{ 0, + ,- \}$. As shown in Chapter \ref{Chapter3}, these quantities provide a full characterization of the transport processes within the mentioned approximations. 

By defining the vector
\be 
{\bf g}( \xi_j , \lambda_j ,t)=
\left(
\begin{array}{c}
g_{0}(\xi_j , \lambda_j,t) \\
g_{+}(\xi_j , \lambda_j,t) \\
g_{-}(\xi_j , \lambda_j,t) 
\end{array}
\right),
\ee
our modified master equation reads
\be \label{modeq6}
\dot{{\bf g}}( \xi_j , \lambda_j ,t) = {\bf W} ( \xi_j , \lambda_j ) \cdot {\bf g}( \xi_j , \lambda_j ,t) 
\ee
in terms of the modified rate matrix $ {\bf W} ( \xi_j , \lambda_j )$ whose components are expressed here below in terms of the rates (\ref{chargedischarge}) and (\ref{QPCrates6}).

Tunneling events between the double quantum dot and the reservoirs $j=1$ and $2$ contribute to the rate matrix through the matrix elements
\bea 
\left[ {\bf W}_{j} ( \xi_j , \lambda_j ) \right]_{s0} & \equiv & \sum_{j=1,2} a_{js} \, \mbox{e}^{ -\left( \xi_{j} \omega_{+0} + \lambda_j \right)} \label{DQDrates16} , \\ 
\left[ {\bf W}_{j} ( \xi_j , \lambda_j ) \right]_{0s}& \equiv & \sum_{j=1,2}  b_{js} \, \mbox{e}^{\left( \xi_{j} \omega_{+0} +\lambda_j \right)}   \label{DQDrates26} ,
\eea
while the contribution from the QPC is given by the components
\bea 
 \left[ {\bf W} ( \xi_j , \lambda_j ) \right]_{+-}   = \sum_{jj'} \int d\epsilon \,   c_{jj'} (\epsilon) \, \mbox{e}^{-\xi_j \epsilon +\xi_{j'} (\epsilon - \omega_{+-}) - \lambda_{j} + \lambda_{j'} } ,  \label{QPCrate16}
  \\
 \left[ {\bf W} ( \xi_j , \lambda_j ) \right]_{-+}   = \sum_{jj'} \int d\epsilon  \, d_{jj'} (\epsilon) \, \mbox{e}^{-\xi_{j'} (\epsilon - \omega_{+-}) +\xi_j \epsilon  - \lambda_{j'} + \lambda_j }  \label{QPCrate26},
\eea
where the sum in these last two equalities runs over $jj'=34$ and $43$.

Finally, the diagonal elements of the rate matrix are given by
\be \label{statedepCGF}
\left[ {\bf W} ( \xi_j , \lambda_j ) \right]_{ss}  = - \sum_{s'\neq s }  \left[ {\bf W} ( 0 , 0 ) \right]_{s's} + G_{s} ( \xi_j , \lambda_j ) .
\ee
The first term in the right-hand side of this equation ensures the conservation of the probability for the occupation probabilities in the double quantum dot when the counting fields are set to zero. The second one accounts for the tunneling of electrons in the QPC, which do not exchange energy with the double quantum dot. As a matter of fact,  the quantity $G_{s} ( \xi_j , \lambda_j ) $ is the generating function of the energy and particle transfer in the QPC given that the double quantum dot is in state $|s\rangle$
\begin{multline}
G_{s}  ( \xi_j , \lambda_j ) = \int \, d\epsilon \, \gamma_{s}(\epsilon)  \left[ f_{3}(\epsilon) (1-f_{4} (\epsilon)) \left(1- e^{ -(\lambda_{3} - \lambda_{4})}e^{ -\epsilon (\xi_{3} - \xi_{4})} \right) \right. \\
\left.+  f_{4}(\epsilon) (1-f_{3} (\epsilon)) \left(1- e^{(\lambda_{3} - \lambda_{4})}e^{ \epsilon (\xi_{3} - \xi_{4})} \right)  \right] \label{diagorate} .
\end{multline}
It is straightforward to show that this is the Levitov-Lesovik formula (\ref{levles1}) \cite{levitov1993charge, levitov1996electron, S07, klich2003elementary, GK06} to second order in the tunneling amplitude $T_{s}(\epsilon)$, or equivalently, to first order in
\be
 \gamma_{s} (\epsilon) \equiv 2\pi |T_{s}(\epsilon)|^{2} \rho_{3} (\epsilon) \rho_{4} (\epsilon).
\ee
This results from the fact that we treated the interaction (\ref{intQPCDQD6}) perturbatively. We note that this term vanishes when the counting fields are set to zero, showing that they do not enter the modified master equation of the double quantum dot system if no counting is done on the QPC as was the case in Chapter \ref{Chapter5}.

By applying a Fourier transform to the modified rate equation (\ref{modeq6}), we get a master equation describing the dynamics of the double quantum dot as well as the transport processes between the reservoirs. As a result, we obtain the stochastic master equation
\be \label{rateeq6}
\dot{{{\bf p}}} (\Delta E_j , \Delta N_j , t) = \int \prod_j \left[ d\delta E_j \right]\left[ d\delta N_j \right]  \hat{{\bf W}} (\delta E_j , \delta N_j ) \cdot {\bf p}(\Delta E_j - \delta E_j, \Delta N_j -\delta N_j, t) 
\ee
for the probability distribution $\left[ {\bf p}(\Delta E_j , \Delta N_j, t)  \right]_{s} = p_s(\Delta E_j , \Delta N_j , t)$ of observing the double quantum dot in state $|s\rangle$ and a variation of energy $\Delta E_j$ and particle number $\Delta N_j$ in the reservoirs $j=1,\dots,4$ during time $t$. The rate matrix $\hat{{\bf W}} (\delta E_j , \delta N_j )$ is obtained as the Fourier transform of the modified rate matrix
\be \label{ratematenpart}
\hat{{\bf W}} (\delta E_j , \delta N_j ) = \int \prod_j \left[ \frac{d\delta \xi_j }{2 \pi}\right]\left[ \frac{d\delta \lambda_j }{2 \pi}\right]  \, e^{-i \sum_{j} \left( \xi_j \delta E_j + \lambda_j \delta N_j \right)}  \, {\bf W} ( -i \xi_j , -i \lambda_j ).
\ee
The Fourier transform of the modified rate matrix is easily taken by using the relations
\bea
\int_{-\pi}^{\pi} \frac{dx}{2 \pi} \mbox{e}^{i x \alpha} & = & \delta_{\alpha ,0}  , \\
\int_{-\infty}^{\infty} \frac{dx}{2 \pi} \mbox{e}^{i x \alpha} & = & \delta (\alpha ) ,
\eea
in terms of the Kronecker delta symbol $\delta_{\alpha ,0}$ and the Dirac delta distribution $\delta (\alpha)$. Accordingly, the rate matrix $\hat{{\bf W}} (\delta E_j , \delta N_j ) $ is obtained by making the following substitutions
\bea
\mbox{e}^{\pm i \xi_j \alpha_j} & \rightarrow & \delta (\delta E_j \pm \alpha_j) \nonumber , \\
 \mbox{e}^{\pm i \lambda_j } & \rightarrow & \delta_{\delta N_j , \pm 1},
\eea
in the modified rate matrix elements (\ref{DQDrates16})-(\ref{statedepCGF}).

By integrating equation (\ref{rateeq6}) over the energy and particle fluctuations $\Delta E_j$ and $\Delta N_j$, or equivalently by setting the counting fields to zero in the modified rate equation (\ref{modeq6}), we obtain a stochastic master equation for the occupation probabilities in the double quantum dot. This equation can be written as
\be 
\dot{{{\bf p}}}(t) = {\bf W} \cdot {\bf p}( t) 
\ee
with the rate matrix given by
\be \label{ratemat6}
 {\bf W} = \left( \begin{array}{ccc}
-a_{1+}-a_{2+} - a_{1-} - a_{ 2-} & b_{1+}+b_{2+} & b_{1-}+b_{2-} \\
a_{1+}+a_{2+}& -b_{1+}-b_{2+} -d_{34} - d_{43}& c_{34}+c_{43}\\
 a_{1-} + a_{ 2-} & d_{34} + d_{43}& -c_{34}-c_{43} -b_{1-}-b_{2-} \\
\end{array} \right),
\ee
where we defined
\bea
c_{jj'} = \int d\epsilon \,   c_{jj'} (\epsilon) , \\
d_{jj'} = \int d\epsilon \,   d_{jj'} (\epsilon).
\eea

With these results, we are now in position to study the stochastic thermodynamics of our model. In the next section, we show that the rates introduced above satisfy the KMS condition involving the irreversible entropy production associated with specific microscopic processes.

\subsection{Local detailed balance and fluctuation theorem}


As noticed earlier in Chapters \ref{Chapter4} and \ref{Chapter5}, the charging and discharging rates (\ref{chargedischarge}) do satisfy the KMS condition \cite{Kubo1998,RevModPhys.81.1665}
\be \label{kms6}
\ln \frac{a_{js}}{b_{js}} =-\beta_j (\omega_{s0} - \mu_j)
\ee
in terms of the inverse temperature $\beta_j$ and chemical potential $\mu_j$ of the reservoir involved in the transition. The fundamental element behind these relations is the assumption of initial equilibrium on the reservoirs and the weak influence of the double quantum dot on them. We further note that the two rates appearing on the left-hand side of equation (\ref{kms6}) correspond to a pair of processes related by time reversal. Indeed, as can be seen from (\ref{chargedischarge}), $a_{js }$ accounts for the electrons leaving the reservoir $j$ with energy $\omega_{s0}$ while $b_{js}$ accounts for the electrons entering the same reservoir with the same energy.

In previous sections, we assumed the interaction between the double quantum dot and the macroscopic electrodes of the QPC to be weak in order to obtain the rates (\ref{QPCrates6}). We thus expect similar relations to hold for the pairs of transition rates of processes related by time reversal. It is straightforward to show that 
\be \label{kms62}
\ln \frac{c_{jj'}}{d_{jj'}} = -\beta_j (\epsilon - \mu_j)+\beta_{j'} (\epsilon-\omega_{+-} - \mu_{j'}).
\ee

By further noting that the heat flow out of reservoir $j$ is given by
\be \label{heatflow}
Q_j = \delta E_{j} - \mu_j \delta N_j
\ee
in terms of the variations of energy $\delta E_j$ and particle number $\delta N_j$ in the reservoir $j$, we readily identify the right-hand side of equations (\ref{kms6}) and (\ref{kms62}) as the irreversible entropy production associated to the transition.

\begin{figure*}[htbp]
\centerline{\includegraphics[width=15cm]{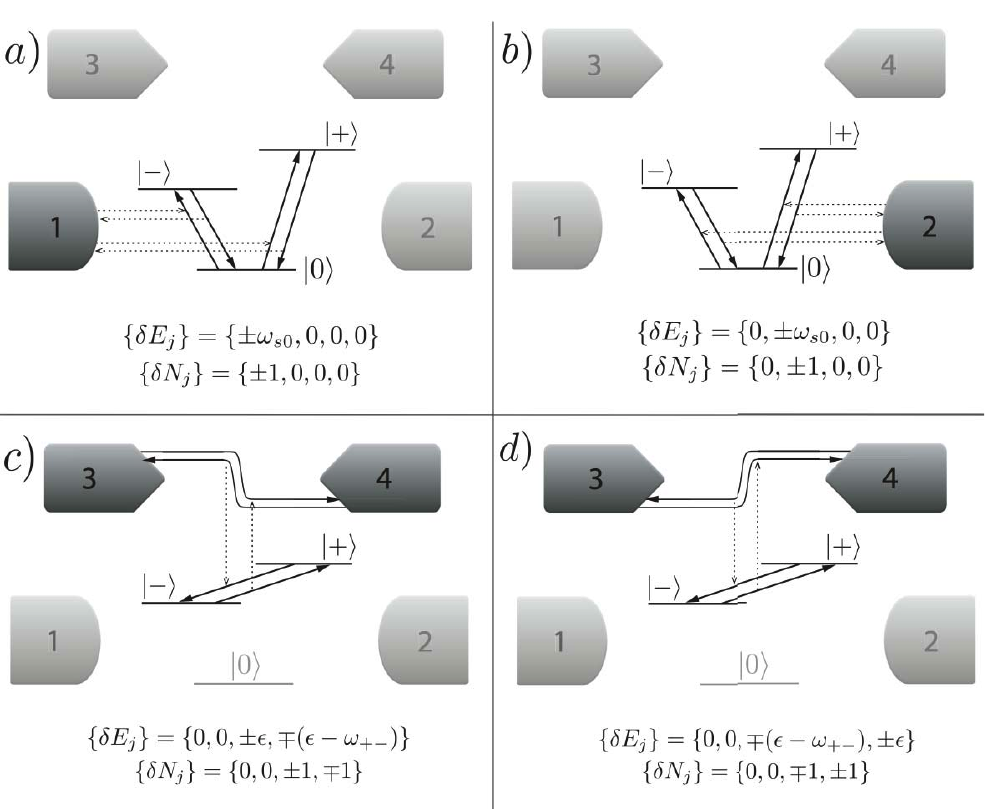}}
\caption{Illustration of the different microscopic processes responsible for transitions in the double quantum dot.}
 \label{DQDtrans}
\end{figure*}

These considerations enable us to summarize these relations in terms of the rate matrix $\hat{{\bf W}} ( \delta E_j , \delta N_j)$ introduced in equation (\ref{ratematenpart}) as
\be \label{genKMS}
\ln \frac{ \left[ \hat {\bf W} (\delta E_j , \delta N_j) \right]_{ss'} }{\left[ \hat {\bf W}  (-\delta E_j ,- \delta N_j) \right]_{s's} } = -\sum_{j} \beta_j Q_{j},
\ee
where $Q_j$ is the heat (\ref{heatflow}) flowing out of the reservoir $j$ because a transition takes place between states $|s \rangle$ and $| s' \rangle$ together with the changes $\delta E_j$ and $\delta N_j$ of the energy and particle number in the reservoir $j$. As noted above, the transition rate $\left[\hat  {\bf W}  (-\delta E_j ,- \delta N_j) \right]_{s's}$ corresponds to the time-reversed process of transition rate $\left[ \hat {\bf W} (\delta E_j , \delta N_j) \right]_{ss'}$. These pairs of microscopic process are depicted in \textsc{Figure} \ref{DQDtrans} for the particular model studied in this Chapter. For each of these microscopic processes the changes of energy and particle number in each reservoir are arranged as
\bea
\left\{  \delta E_j \right\} & = & \left\{ \delta E_1 , \delta E_2 , \delta E_3 , \delta E_4 \right\}  ,\\
\left\{  \delta N_j \right\} & = & \left\{ \delta N_1 , \delta N_2 , \delta N_3 , \delta N_4 \right\} .
\eea
Figures \ref{DQDtrans} a) and \ref{DQDtrans} b) illustrate the transitions induced by charging and discharging of the double quantum dot with reservoirs $j=1$ and $2$. Figures \ref{DQDtrans} c) and \ref{DQDtrans} d) illustrate the two kinds of transitions induced by the QPC on the double quantum dot states $|+ \rangle$ and $| - \rangle$.
 
Conversely, these relations these relations can be translated to the modified transition rates (\ref{DQDrates16})-(\ref{statedepCGF}) as
\be
\left[{\bf W} (\xi_j , \lambda_j)  \right]_{ss'}= \left[{\bf W} (\beta_j - \xi_j , -\beta_j \mu_j - \lambda_j)  \right]_{s's}.
\ee

In turn, these relations ensure the invariance of the characteristic polynomial of the matrix ${\bf W} (\xi_j , \lambda_j)  $ under the transformations
\bea
 \xi_j & \rightarrow & \beta_j - \xi_j  , \\
\lambda_j & \rightarrow  & -\beta_j \mu_j - \lambda_j.
\eea
Besides, we noted in Chapter \ref{Chapter3} that the cumulant generating function
\be \label{CGF6}
\mathcal{G} (\xi_j, \lambda_{j}) = -\lim_{t \rightarrow \infty} \frac{1}{t} \ln \langle \mbox{e}^{- \sum_{j=1}^{4}  \left( \xi_j H_{j} - \lambda_{j} N_{j} \right)}  \rangle_t
\ee
is obtained as the largest eigenvalue of the modified rate matrix ${\bf W} (\xi_j , \lambda_j)  $. As a consequence, the invariance of the characteristic polynomial under the above transformations leads to the symmetry
\be \label{sym6}
\mathcal{G} (\xi_j, \lambda_{j}) = \mathcal{G} ( \beta_j -\xi_j, -\beta_j \mu_j - \lambda_{j})
\ee
for the cumulant generating function.

The use of conservation laws for the energy and particle number is needed in order to get a fluctuation theorem involving the thermodynamic forces. In the long-time limit, the changes of energy and particle number in the double quantum dot are negligible as compared to the changes in the macroscopic reservoirs. By denoting the energy and particle changes between initial time and time $t$ in the reservoir $j$ respectively as $\Delta E_{j}$ and $\Delta N_j$, conservation laws can be written as
\bea
\sum_{j=1}^{4} \Delta E_j & \approx & 0 \label{energycons} , \\
\sum_{j=1}^{4} \Delta N_j & \approx & 0 ,
\eea
with equality holding the the long-time limit. Furthermore, the absence of electron transfers between the QPC and the double quantum dot gives the additional conservation laws
\bea
\Delta N_1 + \Delta N_2 & \approx & 0 \nonumber  ,\\
\Delta N_3 + \Delta N_4 & \approx & 0 \label{particlecons} ,
\eea
again with equality holding in the long-time limit. As shown in Chapter \ref{Chapter2} and references \cite{RevModPhys.81.1665}, these relations can be used together with the symmetry (\ref{sym6}) to establish a fluctuation theorem of the form (\ref{QEFTgen}) in terms of the thermodynamic affinities $A^{\epsilon}_{j} $ and $A^{n}_{j} $ defined in (\ref{affen}) and (\ref{affpart}).

\section{Non-equilibrium thermodynamics}


In this section, we give explicit expressions for the mean currents of energy and particles which will be used to evaluate the efficiency and power extraction of the thermal machine studied in section \ref{maxpowersection}. We also express the irreversible entropy production in the double quantum dot in terms of these processes. The positivity of the entropy production is established as a consequence of the symmetry relation (\ref{sym6}).


\subsection{Mean currents}\label{meancur6}


Recalling equations (\ref{genfuncvect}) and (\ref{longtimecgfvect}), the cumulant generating function of the energy and particle current fluctuations can be written as
\be \label{generating}
\mathcal{G}( \xi_j , \lambda_j  ) = -\lim_{t \rightarrow\infty} \frac{1}{t} \ln {\bf 1 }^{\top} \cdotp \mbox{e}^{ t {\bf W} ( \xi_j , \lambda_j )}  \cdotp {\bf p }_0
\ee
in terms of the modified rate matrix and the initial probability distribution $ {\bf p }_0$ over the double quantum dot. This relation enables us to write the mean currents given by the first derivatives of the cumulant generating function at zero counting fields as
\bea \label{imatter}
\langle J_{E}^{j} \rangle& \equiv &  \lim_{t \rightarrow \infty} \frac{\langle \Delta E_{j} \rangle }{t} \nonumber \\
& = &   \partial_{\xi_{j}} {\bf W} ( 0,0) \cdot {\bf P} , \\ \label{ienergy}
\langle J_{N}^{j}  \rangle & \equiv &  \lim_{t \rightarrow \infty} \frac{ \langle  \Delta N_{j} \rangle }{t} \nonumber \\
& = &  \partial_{\lambda_{j}} {\bf W} ( 0,0) \cdot {\bf P}  ,
\eea
where $\langle J_{E}^{j} \rangle$ and $\langle J_{N}^{j} \rangle$ are respectively the energy and particle currents out of reservoir $j$ while
\be
{\bf P}  = \lim_{t\rightarrow \infty} \mbox{e}^{ t {\bf W}}  \cdotp {\bf p }_{0}
\ee
denotes the vector of steady-state occupation probabilities $ P_s = \left[ {\bf P}  \right]_{s} $ on the double quantum dot. The steady-state probabilities are directly obtained by solving equation
\be
{\bf W}  \cdotp {\bf P } =0
\ee
with the rate matrix ${\bf W}$ given by (\ref{ratemat6}). We note that conservation laws (\ref{energycons}) and (\ref{particlecons}) can be expressed in terms of the mean currents as
\bea
\sum_{j=1}^{4} \langle J_{E}^{j}\rangle & =  &0 \label{conscure} , \\
\langle J_{N}^{1}\rangle & = & - \langle J_{N}^{2}\rangle  \label{conscure2} , \\
\langle J_{N}^{3}\rangle & = & - \langle J_{N}^{4}\rangle \label{conscurn}.
\eea

In the present model, the energy and matter currents are calculated from (\ref{imatter}) and (\ref{ienergy}) to give
\bea \label{materDQD}
\langle J_{N}^{j}  \rangle  & = & \sum_{s=+,-} \left( a_{js} P_{0} - b_{js} P_{s}  \right)  , \\
\langle J_{E}^{j}  \rangle& = & \sum_{s=+,-} \omega_{s0} \left( a_{js} P_{0} - b_{js} P_{s}  \right) ,
\eea
for the outgoing currents of particles and energy from reservoirs $j=1$ and $2$. The corresponding currents for reservoirs $3$ and $4$ are given by
\bea
\langle J_{N}^{3}  \rangle  & = &  \int d \epsilon \, \langle J_{N}(\epsilon) \rangle  =-\langle J_{N}^{4}  \rangle , \\
\langle J_{E}^{3} \rangle& = & \int d \epsilon \, \epsilon \, \langle J_{N}(\epsilon) \rangle+  \omega_{+-} \left( c_{43}  P_{-} - d_{43} P_{+} \right) \label{energyQPC2}  , \\
\langle J_{E}^{4} \rangle & = & - \int d\epsilon \, \epsilon \, \langle J_{N}(\epsilon) \rangle+  \omega_{+-} \left( c_{34}  P_{-} - d_{34} P_{+} \right)  \label{energyQPC} ,
\eea
where we defined the energy resolved matter current in the QPC as
\be
\langle J_{N} (\epsilon ) \rangle= \sum_{s} \gamma_{s}(\epsilon) \left( f_{3} (\epsilon) - f_{4} (\epsilon) \right)  + \left( c_{34} (\epsilon) P_{-} - d_{34} (\epsilon) P_{+} + d_{43} (\epsilon) P_{+} - c_{43} P_{-} \right).
\ee


\subsection{Irreversible entropy production}


The irreversible entropy production $\dot{ S}_{i}$ in the whole setup can be expressed as
\be
\dot{ S}_{i} = \dot{ S} + \dot{ S}_r,
\ee
where $ S$ and $  S_{r}$ denote respectively the system and environment entropies. The double quantum dot being finite, the rate of change of its entropy $\dot{S}$ vanishes at steady state. On the other hand, the entropy rate of the environment $\dot{S}_{r}$ is given by the sum of the entropy rates in each reservoir so that
\be \label{entrchangeenv}
\dot{S}_{r} = -\sum_{j=1}^{4} \beta_{j} (\langle J_{E}^{j} \rangle- \mu_{j} \langle J_{N}^{j} \rangle)
\ee
since we consider grand-canonical reservoirs. As a consequence, the irreversible entropy production at steady state can be linked to the energy and matter currents by
\bea \label{rel16}
\dot{S}_{i} & = & \dot{S}_{r} \\ \label{rel26}
& = & -\sum_{j=1}^{4} \beta_{j} (\langle J_{E}^{j} \rangle- \mu_{j} \langle J_{N}^{j} \rangle) .
\eea

By further using conservation laws (\ref{conscure}) - (\ref{conscurn}), this equation can be rewritten as
\begin{multline}
\dot{S}_{i} = -\left( \beta_{1} - \beta_{2} \right) \langle J_{E}^{1} \rangle + \left( \beta_{1} \mu_{1} - \beta_{2} \mu_{2} \right) \langle J_{N}^{1} \rangle  \\ - \left( \beta_{3} - \beta_{2} \right) \langle J_{E}^{3} \rangle
 - \left( \beta_{4} - \beta_{2} \right) \langle J_{E}^{4} \rangle + \left( \beta_{3} \mu_{3} -\beta_{4} \mu_{4} \right) \langle J_{N}^{3} \rangle \label{entropprod} ,
\end{multline}
which now makes explicitly appear the thermodynamic affinities driving the non equilibrium fluxes of energy and matter across the setup.

In the following, we write the entropy production as the first moment of a cumulant generating function satisfying a fluctuation symmetry. This will enable us to establish the positivity of the irreversible entropy production $\dot{S}_{i}$.

First, we note that the entropy change in the environment (\ref{entrchangeenv}) can be obtained as the first cumulant of the statistics of operator
\be \label{entropyop}
O = - \sum_{j=1}^{4} \beta_{j} \left( H_{j} - \mu_{j} N_{j} \right).
\ee
The fluctuations of this operator in the long-time limit are captured by the cumulant generating function
\bea \label{cumgenentr}
\mathcal{G}_{O} (\alpha) & = & -\lim_{t \rightarrow \infty} \frac{1}{t} \ln \langle \mbox{e}^{-\alpha  \Delta O}  \rangle_t \\
& = & -\lim_{t \rightarrow \infty} \frac{1}{t} \ln \langle \mbox{e}^{-\alpha \sum_{j=1}^{4} \beta_{j} \left( \Delta E_{j} - \mu_{j} \Delta N_{j} \right)}  \rangle_t ,
\eea
the average being over a solution of the stochastic equation (\ref{rateeq6}) at time $t$. From this last equality and definition (\ref{CGF6}), we see that the cumulant generating function $\mathcal{G}_{O} (\alpha)$ is simply obtained from the cumulant generating function of the energy and particle fluctuations by making the substitutions
\be
 \xi_{j} \rightarrow \beta_{j} \alpha  \qquad \mbox{and} \qquad \lambda_{j} \rightarrow -\beta_{j}\mu_{j} \alpha 
\ee
so that
\be
\mathcal{G}_{O}(\alpha) = \mathcal{G}( \beta_{j} \alpha, -\beta_{j}\mu_{j} \alpha  ) ,
\ee
where $\alpha$ is now the counting field associated with the counting of operator $O$. By using (\ref{rel16}) and (\ref{rel26}), the irreversible entropy production can then be expressed as the first moment of this cumulant generating function
\bea
\dot{S}_i & = & \partial_{\alpha} \mathcal{G} (0) \\
& = & \langle \dot{O} \rangle .
\eea

 We futher note that this generating function has the symmetry
\be
\mathcal{G}_{O}(\alpha) = \mathcal{G}_{O}(1 - \alpha)
\ee
as a consequence of (\ref{sym6}). By using Jensen's inequality for the exponential $\langle \mbox{e}^{ \Delta O} \rangle \geq \mbox{e}^{\langle \Delta O \rangle}$ and the fact that $\mathcal{G}_{O}(0)=1$, we then show that
\bea
0 & = &- \lim_{t \rightarrow \infty}  \frac{1}{t} \ln \langle \mbox{e}^{-  \Delta O } \rangle \\
& \leq & -\lim_{t \rightarrow \infty}  \frac{1}{t} \ln  \mbox{e}^{- \langle \Delta O \rangle } \\
& = &   \left( \lim_{t \rightarrow \infty}\frac{\langle \Delta O \rangle}{t}  \right)  \\
& = & \langle \dot{O} \rangle ,
\eea
thus establishing the second law of thermodynamics for our system
\be \label{posit}
0\leq \dot{S}_i.
\ee


\subsection{Thermal machine and efficiency at maximum power}\label{maxpowersection}


In this section we consider a thermal machine in which the roles of hot and cold reservoirs are played respectively by the QPC and the reservoirs $j=1$ and $2$ of the double quantum dot. A fraction of the heat flow from the QPC is then converted by the double quantum dot into work in the form of current against a bias applied between the reservoirs $1$ and $2$.

The inverse temperature $\beta_{h} = \beta_3 = \beta_4$ in the QPC is assumed to be lower than the inverse temperature in the double quantum dot channel $\beta_c = \beta_1 = \beta_2$, that is
\be \label{tempgrad}
\beta_h <\beta_c.
\ee
We assume a vanishing bias in the QPC, $\mu_{3} = \mu_4$, while a finite bias is applied through the double quantum dot channel $\Delta \mu = \mu_2 - \mu_1$ such that
\be \label{biasmach}
0< \Delta \mu .
\ee

Under these conditions, the irreversible entropy production in the system (\ref{entropprod}) reads
\be
\dot{S}_{i} = -\beta_c \Delta \mu \langle J_{N}^{1} \rangle  + \left( \beta_c -\beta_h \right)  \left( \langle J_{E}^{3} \rangle +\langle J_{E}^{4} \rangle \right) \geq 0
\label{entropprodthermalmach}
\ee
in terms of the matter and energy currents introduced in section \ref{meancur6}.
The steady-state rate of work, or output power $\dot{\mathcal{W}}$, performed by the double quantum dot against bias $\Delta \mu$ and the rate of heat flow $\dot{\mathcal{Q}}$ from the QPC to the double quantum dot are respectively given by \cite{esposito2009thermoelectric}
\bea
\dot{\mathcal{W}} & = & \Delta \mu \langle J_{N}^{1} \rangle \label{outputpower} , \\
\dot{\mathcal{Q}} & = & \langle J_{E}^{3} \rangle +\langle J_{E}^{4} \rangle.
\eea

The efficiency $\eta$ of the heat to work conversion process described above is defined by the ratio of the output power $\dot{\mathcal{W}}$ over the input heat flow $\dot{\mathcal{Q}}$ from the hot reservoir so that
\bea
\eta & \equiv & \frac{\dot{\mathcal{W}}}{\dot{\mathcal{Q}}} \\
& = &\frac{\Delta \mu \langle J_{N}^{1} \rangle}{\langle J_{E}^{3} \rangle +\langle J_{E}^{4} \rangle } . \label{effint6}
\eea
The second law of thermodynamics puts a fundamental limit on the efficiency of all thermal machines. In the present case, this can be seen by rewritting the efficiency (\ref{effint6}) in terms of the irreversible entropy production (\ref{entropprodthermalmach}) so that
\be
0 \leq \eta = \eta_C - \frac{\dot{S}_{i}}{\beta_c\left( \langle J_{E}^{3} \rangle +\langle J_{E}^{4} \rangle \right)} \leq \eta_C \label{effboundcarnot} ,
\ee
where $\eta_C$ is the ideal Carnot efficiency
\be \label{carnot}
\eta_C \equiv 1- \frac{\beta_h}{\beta_c} .
\ee
As can be seen from (\ref{effboundcarnot}), the Carnot efficiency is reached for a thermal machine working reversibly, i.e. satisfying $\dot{S}_{i} = 0$. However, such machines work infinitely slowly so that the extracted power vanishes in the limit of a reversible engine.

This issue led to the question of the efficiency  at maximum output power in thermal engines \cite{curzon1975efficiency,tu2008efficiency,schmiedl2008efficiency,esposito2009thermoelectric,esposito2010efficiency,sanchez2011optimal}. In particular, the Curzon-Ahlborn efficiency
\be \label{caeff}
\eta_{CA} = 1- \sqrt{\frac{\beta_h}{\beta_c}}
\ee
has recently been shown to convey a universal upper bound on the efficiency at maximum power in machines working in the linear regime \cite{van2005thermodynamic}. Here below we investigate the efficiency at maximum output power for the thermal engine described above and show that the Curzon-Ahlborn efficiency is reached in the strong coupling limit between the energy and matter currents \cite{esposito2010efficiency}.

\begin{figure*}[htbp]
\centerline{\includegraphics[width=15cm]{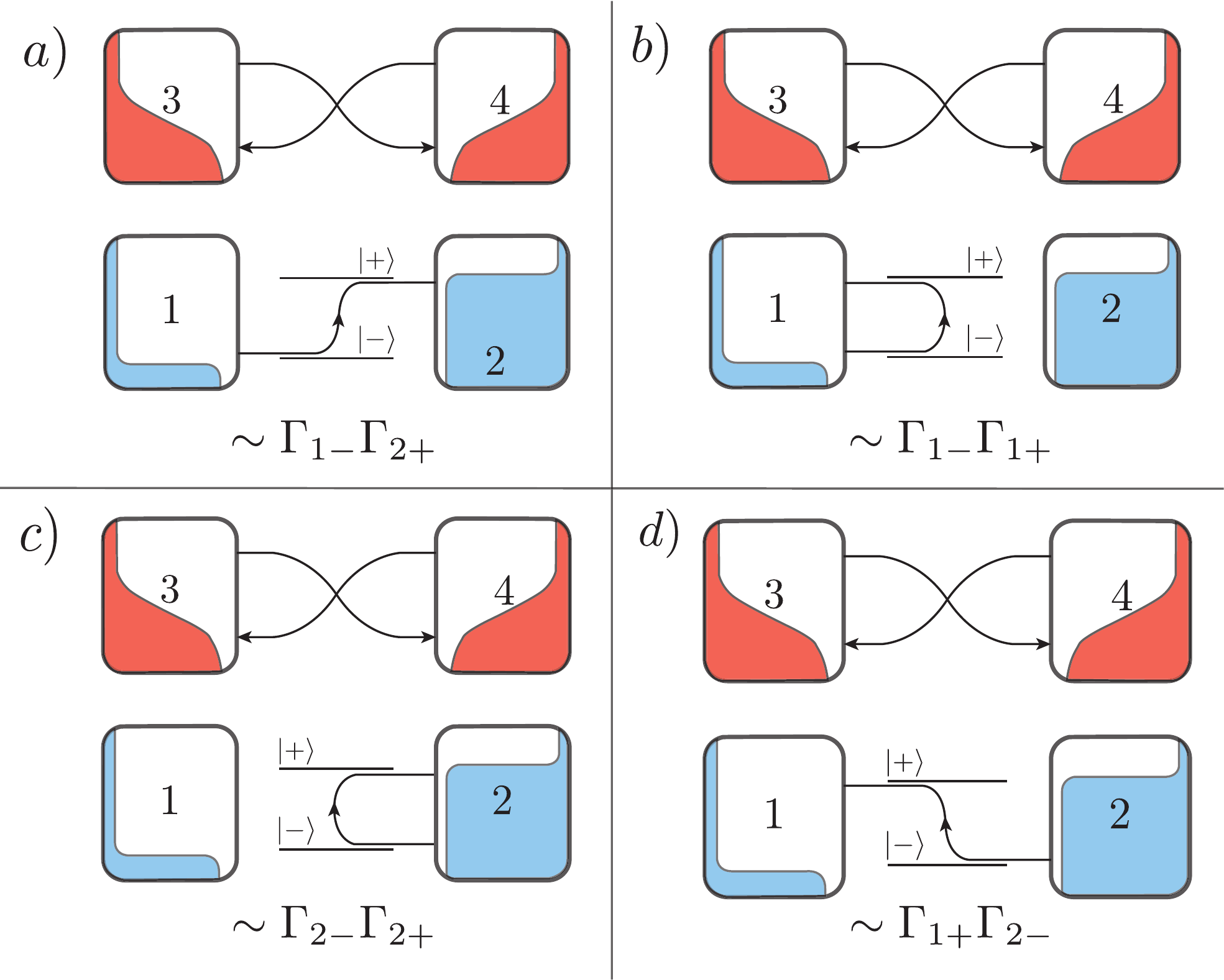}}
\caption{Illustration of the processes involving a transtion from state $|-\rangle$ to state $|+\rangle$.}
 \label{dragtrans}
\end{figure*}

In the presence of the temperature gradient (\ref{tempgrad}), the QPC will preferentially give energy to the double quantum dot in the form of transitions from state $|-\rangle$ to state $|+\rangle$. The microscopic process involving these kinds of transitions and leading to a net flow of charge against bias (\ref{biasmach}) is depicted in \textsc{Figure} \ref{dragtrans} a). However, without specific conditions, this process is not more likely than the other processes depicted in \textsc{Figure} \ref{dragtrans}, which do not involve a charge transfer in the desired direction thus lowering the efficiency and power of the heat to work conversion.

A way out of this issue is to introduce an asymmetry \cite{sanchez2011optimal} in the tunneling amplitude of the double quantum dot, which could be accommodated for example by tuning the electron densities of the reservoirs. The right choice in our situation is given by
\be \label{condmach}
\Gamma_{1+}, \Gamma_{2-} \ll \Gamma_{1-},\Gamma_{2+}
\ee
so that the process depicted in \ref{dragtrans} a) is now the dominant one among the four processes depicted on the same Figure.

As expected, the matter current through the double quantum dot and the energy current from the QPC become strongly correlated in this limit so that
 \be
\langle J_{E}^{3} \rangle +\langle J_{E}^{4} \rangle =  \omega_{+-}  \langle J_{N}^{1} \rangle,
\ee
which can be verified by using the explicit expressions for the currents given in section \ref{meancur6}. Consequently, the efficiency takes the simple form
\be \label{effmach2}
\eta = \frac{\Delta \mu}{\omega_{+-}} ,
\ee
which is comparable to the result of ref \cite{sanchez2011optimal}. This similarity stems from the quantized energy exchange between the reservoirs and the work converter. In the present case, the QPC exchange energy with the double quantum dot in the form of quantas whose energy is given by $\omega_{+-}$.

In order for the double quantum dot to perform usefull work, the outupt power $\dot{\mathcal{W}}\equiv \Delta \mu \langle J_{N}^{1} \rangle$ should be a positive quantity. The sign of the mean current out of reservoir $1$ can be deduced by rewritting (\ref{materDQD}) in the limit (\ref{condmach}) as
\bea
 \langle J_{N}^{1} \rangle & = & \kappa \left( a_{1-} b_{2+} (c_{34} + c_{43}) - a_{2+} b_{1-} (d_{34}+ d_{43} )  \right) \\
& = & \kappa \, a_{1-} b_{2+} (c_{34} + c_{43})  \left(1 - \mbox{e}^{(\beta_h - \beta_c) \omega_{+-} - \beta_c (\mu_1 - \mu_2) } \right) \label{secondmach} ,
\eea
where $\kappa$ is a positive proportionality constant. In obtaining this result, we used the KMS conditions (\ref{kms6}) and (\ref{kms62}), considering homogeneous temperature in each channel and zero bias in the QPC, $\mu_{3} = \mu_{4}$.

We readily identify from equation (\ref{secondmach}) the condition for a positive rate of work $\dot{\mathcal{W}}$ done against bias $ 0 \leq \Delta \mu $ as
\be \label{boundmach}
0\leq \frac{\Delta \mu}{\omega_{+-}} \leq \eta_c .
\ee

\begin{figure*}[htbp]
\centerline{\includegraphics[width=11cm]{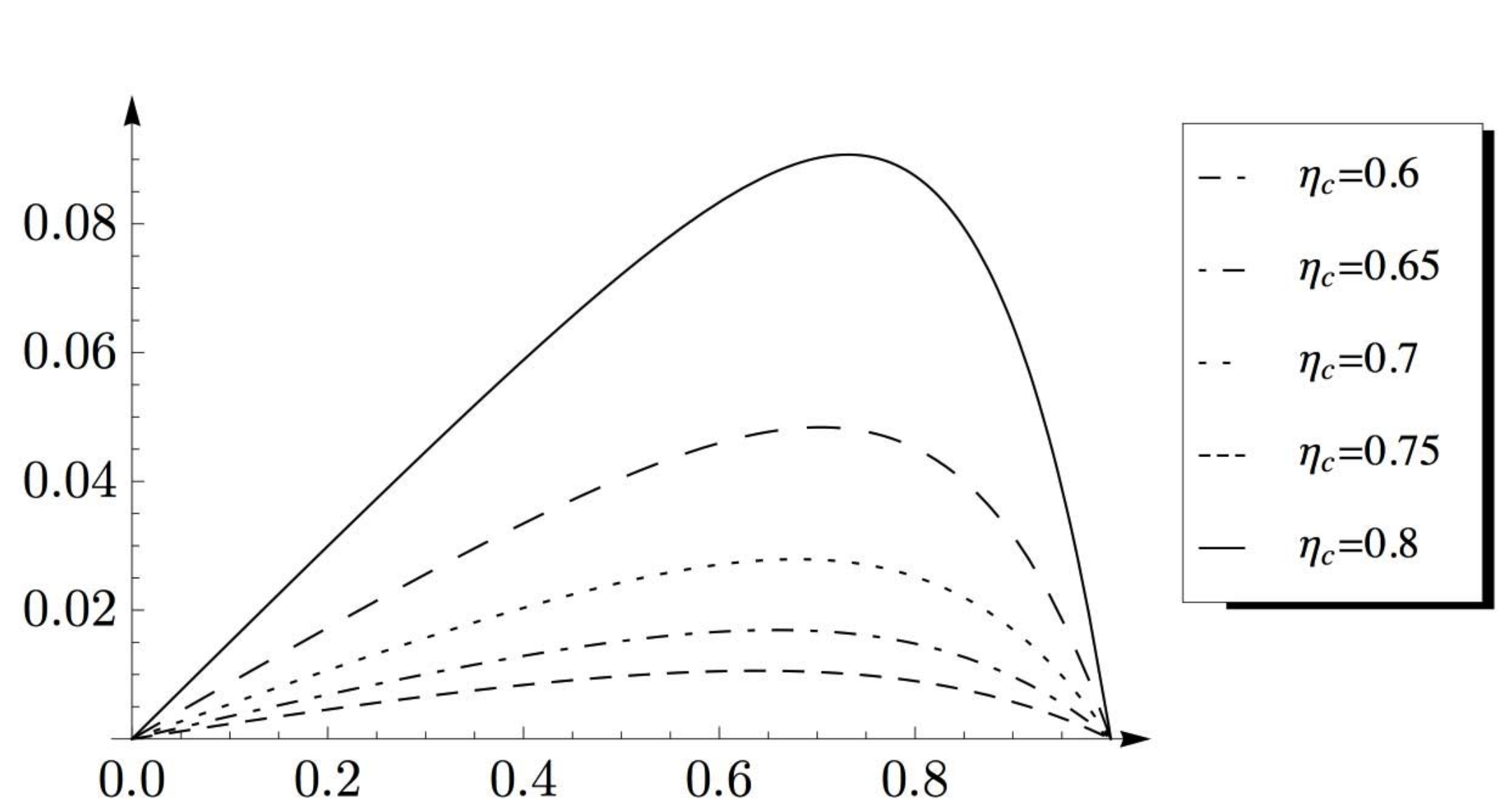}}
\caption{Power extraction as a function of the rescaled efficiency $x=\eta/\eta_C$ bounded by $0 \leq x \leq 1$ as deduced from (\ref{boundmach}). Parameters were chosen as $\beta_c=1$, $E_0 = 0$, $V_3 = V_4 = 0$, $(\mu_1 + \mu_2)/2 = 1.5$, $\Gamma_{34} = \Gamma_{43}= \Gamma_{1-} = \Gamma_{2+} =0.1$ and $\Gamma_{1+} = \Gamma_{2-} = 0$. The energies $E_+$ and $E_-$ of the single-occupied states of the double quantum dot were numerically adjusted in order to get a maximum output power $\dot{\mathcal{W}}$ (see \textsc{Figure} \ref{spectraopt}).}
 \label{powerfig}
\end{figure*}

The parameters of the setup can always be adjusted to reach Carnot efficiency for the heat conversion. However, in the limit of Carnot efficiency, the extracted power goes to zero and the extracted work consistently vanishes in the long-time limit. This point is illustrated in \textsc{Figure} \ref{powerfig} where we see that the output power vanishes systematically as the Carnot efficiency is reached \cite{van2007carnot,humphrey2002reversible,humphrey2005reversible}. We note that the output power is bounded and shows a maximum in the region (\ref{boundmach}).

\begin{figure*}[htbp]
\centerline{\includegraphics[width=11cm]{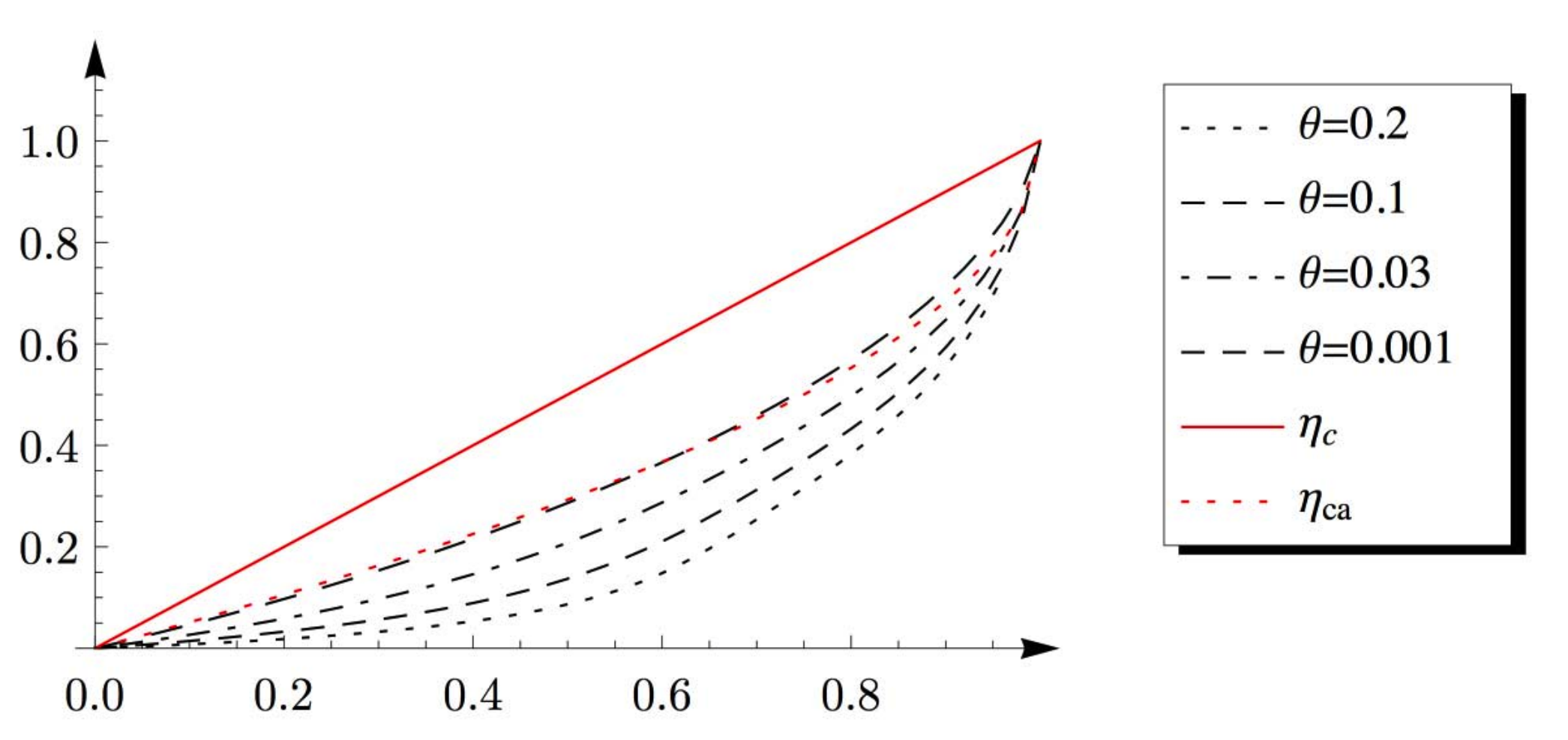}}
\caption{Efficiency at maximum power for different values of the asymmetry parameter $\theta$ defined in (\ref{assympar6}). The continuous red line is for Carnot efficiency (\ref{carnot}) while the dotted one is for the Curzon-Ahlborn efficiency (\ref{caeff}). Parameters were chosen as in \textsc{Figure} \ref{powerfig} except for the tunneling rates $\Gamma_{1+} = \Gamma_{2-} = \theta \Gamma_{1-} =\theta \Gamma_{2+} $ with $\Gamma_{1-} = \Gamma_{2+} =0.1$ and where $\theta$ takes the values given in the legend.}
 \label{effmaxfig}
\end{figure*}

In \textsc{Figure} \ref{effmaxfig}, we show curves of the efficiency at maximum output power for different values of the ratio
\be \label{assympar6}
\theta \equiv \frac{\Gamma_{1+}}{\Gamma_{1-}} = \frac{\Gamma_{2-}}{\Gamma_{2+}} 
\ee
characterizing the degree of asymmetry (\ref{condmach}) in the double quantum dot. We see that as this ratio increases, the efficiency is lowered as resulting from the contributions of the undesired processes of \textsc{Figure} \ref{dragtrans}.
In the numerical optimization, the double quantum dot energies $E_+$ and $E_-$ have been adjusted in order to provide the maximal efficiency. As a result, the thermodynamic efficiency attains values above $\eta_C/2$ and close to $\eta_{CA}$.

\begin{figure*}[htbp]
\centerline{\includegraphics[width=12cm]{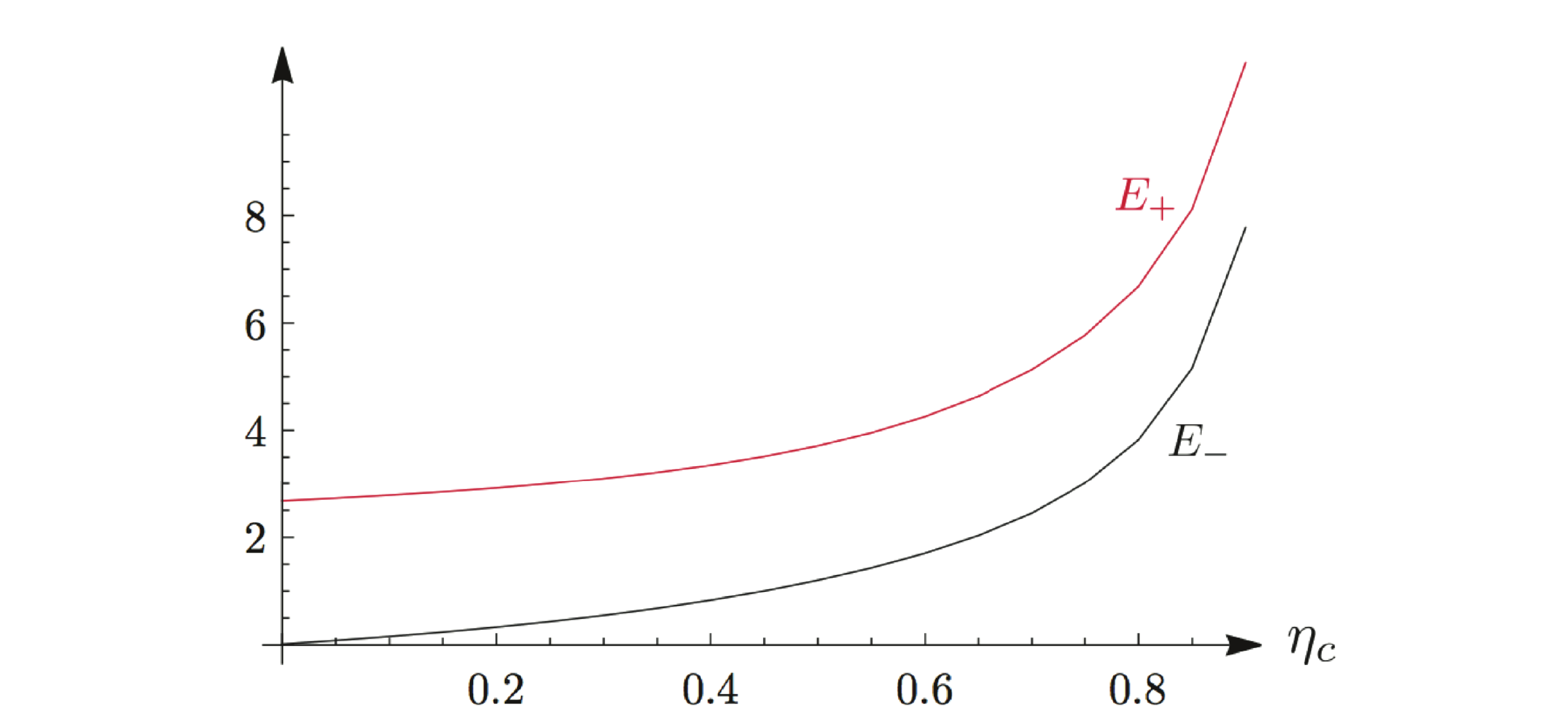}}
\caption{Energies of the single occupied states $E_+$ and $E_-$ of the double quantum dot optimized numerically to yield a maximum output power (\ref{outputpower}). The numerical optimization was done with parameters were chosen as in \textsc{Figure} \ref{effmaxfig}.}
 \label{spectraopt}
\end{figure*}


\section{Summary of the results}


In the present Chapter, we used a stochastic master equation approach to describe the nonequilibrium thermodynamics transport processes in a nanostructured double channel circuit. The model that is considered consists of a double quantum dot exchanging electrons with two reservoirs and energy with a QPC channel.

We showed that the transition rates of processes related by time reversal obey a KMS condition. This condition is shown to hold for the tunneling rates of electron exchange between the double quantum dot and its reservoirs as well as for the transition rates of the energy exchange processes with the out of equilibrium QPC. These results express the asymmetry between two processes related by time reversal in terms of the irreversible entropy production resulting from the transition.

These relations have been shown to imply a fluctuation theorem for the heat and matter currents. Thereafter, the irreversible entropy production was shown to be positive definite by writing it as the first cumulant of a generating function satisfying a fluctuation symmetry.

Finally, the setup is considered as a thermal machine. We numerically show that the energies of the double quantum dot eigenstates can be tuned to reach the highest efficiencies at maximum output power in the non-linear regime.

%% file: Chapter7.tex

\chapter{Conclusions and perspectives} 

\label{Chapter7} 

\lhead{Chapter 7. \emph{Conclusions and perspectives}} 

A main concern of this thesis has been the study of current fluctuations in out of equilibrium nanostructured devices. Particular attention has been given to the characterization of the mutual influence of capacitively coupled channels in composite circuits.

\section{Summary of the thesis}

In the Chapters \ref{Chapter2} and \ref{Chapter3} of this thesis we reviewed important results and methods in the field of nonequilibrium statistical physics. The basic ingredients needed in order to establish a fluctuation theorem in out of equilibrium quantum systems have been presented in Chapter \ref{Chapter2}. We applied these results to circuits composed of two coupled channels and obtained a fluctuation theorem involving their thermodynamic forces in the long-time limit. Such fluctuation theorems put constraints on the probability distribution of the current fluctuations which, equivalently, take the form of symmetry relations for the cumulant generating function.

In subsequent Chapter \ref{Chapter3}, we have given an account of recent advances in the topic of counting statistics. In particular, we introduced the formalism of the modified master equation which has been successfully applied to the study of current fluctuations in nanoscale devices. A general derivation of the stochastic master equation within the Born-Markov and rotating wave approximation is given. Thereafter, we expressed the cumulant generating function of the currents in terms of the resulting modified rate matrix.

These results have been applied in Chapter \ref{Chapter4} to a model of quantum electron transport with capacitively coupled quantum dot channels. Most of the results exposed in this Chapter have been reported in Ref. \cite{cuetara2011fluctuation}. The full counting statistics of the number of electron transfers in each channel has been performed by using the modified density matrix approach to second order in the tunneling amplitudes. This is used in order to establish a fundamental bivariate fluctuation theorem  for the cumulant generating function, depending on the thermodynamic affinities (\ref{A1}) and (\ref{A2}) applied to both channels. Though this relation is valid regardless of the initial conditions in the long-time limit, this is not true in general for the characteristic function of the transport processes at finite times. Nevertheless, a criteria on the initial condition for the observation of a fluctuation theorem at finite times is exposed.

As mentioned in the introduction, one of the main goals in this thesis is to give an account of the back-action of the detector on the probed quantum dots circuits typically used in counting statistics experiments. In these experiments, the detector current is macroscopic as compared to the probed circuit. In this context, a single-current fluctuation theorem is established in the large current ratio limit between the two channels of our model in consistency with experimental observations. Within this limit, the single-current fluctuation theorem is expressed as a symmetry of the cumulant generating function involving an effective affinity depending on the parameters of the model as well as the thermodynamic affinities applied to both channels. A detailed account of the dependence on these parameters has been given. Furthermore, the considerations on the finite-time counting statistics mentioned above are extended to the case of the single-current fluctuation theorem.

In Chapter \ref{Chapter5}, we performed a similar analysis on a system composed of a quantum point contact (QPC) detector and a double quantum dot channel. This model serves as a theoretical description of the experimental setup used in Ref. \cite{fujisawa2006}. The results obtained in this Chapter have been reported in Ref. \cite{cuetara2013effective}. The full counting statistics is again obtained by using the modified density matrix approach. However, the out of equilibrium correlation functions in the QPC are evaluated non-perturbatively in its tunneling parameter. As a result of these calculations, the transition rates induced by the QPC on the double quantum dot have been shown to take the form of Bose-like random transitions.

The mean current in the QPC undergoes discrete jumps resulting from the random transitions in the double quantum dot. On the other hand, the QPC can drag a current in the double quantum dot even though no bias is applied to the double quantum dot channel. A detailed description of this effect has been given in agreement with experimental observations. In particular, the existence of a threshold in the QPC bias for the drag to take place is  accurately reproduced and explained within our approach.

By calculating the cumulant generating function of the electron transfers in the double quantum dot channel, we showed that a single-current fluctuation theorem is present in particular limits which are relevant to experimental situations. In these regimes, the efficiency of the drag effect of the QPC on the double quantum dot turns out to have a upper bound resulting from the single-current fluctuation theorem. This bound is shown to depend on the ratio of the effective affinity over the thermodynamic affinity applied to the double quantum dot channel. Moreover, the dependence on the double quantum dot energies of the effective affinity has been investigated. As a result, the case of close to symmetric double quantum dot, i.e. $\epsilon_A = \epsilon_B $, is shown to be favorable in order to minimize the deviations of the effective affinity with respect to the value fixed by the voltage bias across the double quantum dot.

Finally, we performed in Chapter \ref{Chapter6} the thermodynamic analysis of the model presented in Chapter \ref{Chapter5} by using the modified density matrix formalism introduced in Chapter \ref{Chapter3}.  However, the tunneling amplitudes of the QPC were treated to second order in perturbation theory contrary to what we did in Chapter \ref{Chapter5}. Within this approximation, we showed  that the transition rates do satisfy a Kubo-Martin-Schwinger condition involving the irreversible entropy production associated to the transition. This relation is used in order to establish the positivity of the irreversible entropy production in our setup through the introduction of the cumulant generating function  (\ref{cumgenentr}).

We concluded by considering a thermal machine in which the roles of hot and cold reservoirs is played respectively by the QPC and the double quantum dot reservoirs. The question of the efficiency at maximum power has been investigated in the non-linear regime. The aforementioned constrains on the transition rates are used in order to determine the region of the parameter landscape supporting a positive output power. Thereafter, we showed that a fine tuning of the double quantum dot spectrum leads to highest efficiencies at maximum power close to the Curzon-Ahlborn efficiency.

To summarize, our work gives a better understanding of the statistical fluctuations of the transfer processes taking place in multi-terminal circuits at the mesoscale. In particular, we theoretically established the single-current fluctuation theorems observed experimentally for quantum dot circuits probed by a secondary out of equilibrium circuit. More generally, this work gives insight in the interplay between coupled nonequilibrium circuits subject to both matter and heat currents.

\section{Perspectives}

In Chapter \ref{Chapter6} we calculated the full counting statistics for a model composed of two channels with different structures. This approach is particularly interesting for the investigation of an asymmetry in the mutual influence between two such circuits. As an example, the study of the effective affinities of both channels in the large current ratio limits would give some insight on this issue with consequences on the efficiencies of work extraction by both channels. Furthermore, the full counting statistics of the energy and particle numbers opens the road to the investigation of the interplay between heat and matter current fluctuations \cite{sanchez2013correlations}. On the other hand, thermodynamic  machines similar to the one studied in Chapter \ref{Chapter6} may be studied in the same spirit. The investigation of such devices lies at the forefront of present day research in the field of nonequilibrium thermodynamics \cite{PhysRevLett.111.060802, esposito2009thermoelectric, tu2013bounds, apertet2012efficiency, sanchez2011optimal, esposito2010efficiency}.

Throughout this thesis, the analysis of the out of equilibrium quantum processes is performed by using the stochastic master equation presented in Chapter \ref{Chapter3}. Within this approach, the interaction between the subsystem and the environment is assumed weak and treated to second order in perturbation theory. A key result obtained with this method is the local detailed balance or Kubo-Martin-Schwinger condition (\ref{kms6}) or its more general version we have established in Eq. (\ref{genKMS}). Still, extending this result to higher orders in the interaction between the subsystem and the environment is still and open issue which is of uttermost importance in the description of strongly interacting systems. A step forward may be taken by using a perturbative master equation approach beyond second order in perturbation theory \cite{timm2011time} or alternative methods such as the path-integral formalism \cite{feynman1963theory, jin2010non}.

Regarding the fluctuation theorems, an open question remains as for the validation of the work fluctuation theorems in the quantum regime and for strongly interacting systems \cite{RevModPhys.83.771}. Recent advances in the manipulation of cold atoms with lasers have led to the realization of quantum systems with unprecedented control and tuning possibilities \cite{trotzky2012probing, schneider2012fermionic, endres2013single}. The ability to perform work on these systems and to adjust the interaction potentials between the constituents may prove of great value in the validation of the work fluctuation theorems in the aforementioned regime. The theoretical description of such systems would require the use of numerical tools such as the density matrix renormalization group method in order to accurately treat the strong interactions between the constituents of the system considered.

The investigation of nanoscale systems within statistical mechanics has proven to be fruitful during the last decades. Still, technical advances continuously challenge theoreticians leading us to a better understanding of these kind of systems.

%% file: AppendixA.tex

\chapter{Microreversibility of the quantum dynamics} 

\label{AppendixA} 

\lhead{Appendix A. \emph{Microreversibility of the quantum dynamics}} 

By using Eqs. (\ref{unitaryint2}) and (\ref{timerevU2}), the evolution and time-reversed evolution operators between times $t=T_0$ and $t=T$ can be written respectively as
\bea
U(T,T_0) & = & \mbox{T}_{+} \exp{ \left[ -\frac{i}{\hbar} \int_{T_{0}}^{T} d\tau \, H(\tau)  \right] } ,\\
\tilde{U}(T,T_0) & = & \mbox{T}_{+} \exp{ \left[-\frac{i}{\hbar} \int_{T_{0}}^{T} d\tau \, H(T- (\tau -T_0 )) \right] }.
\eea
By using (\ref{symham}), we have that
\bea
\Theta^{-1} \tilde{U} (T,T_0) \Theta & = & \Theta^{-1}  \mbox{T}_{+} \exp{ \left[ -\frac{i}{\hbar} \int_{T_{0}}^{T} d\tau \,  H(T- (\tau -T_0 )) \right]  }  \Theta \nonumber \\
& = & \mbox{T}_{+} \exp{ \left[ \frac{i}{\hbar} \int_{T_{0}}^{T} d\tau \, \Theta^{-1} H(T- (\tau -T_0 )) \Theta \right]  }   \nonumber \\
& = & \mbox{T}_{+} \exp{ \left[ \frac{i}{\hbar} \int_{T_{0}}^{T} d\tau \, H(T- (\tau -T_0 )) \right]  }   \nonumber \\
& = &  \mbox{T}_{-} \exp{ \left[ \frac{i}{\hbar} \int_{T_{0}}^{T} d\tau \,  H(\tau) \right]  }   \nonumber \\
& = & U^{\dagger} (T,T_0) \nonumber
\eea 
in terms of the anti-time ordering operator $T_{-}$ which orders time-dependent operators on its right in such a way that they appear with decreasing time arguments from right to left. Recalling the unitarity of the time evolution operator, $U^{\dagger}(T,T_0) U (T,T_0) = I$, this establishes the results (\ref{premicro2}) and (\ref{microreversibility}) of Chapter \ref{Chapter2}.

%% file: AppendixB.tex

\chapter{Initial condition for a finite-time fluctuation theorem} 

\label{AppendixB} 

\lhead{Appendix B. \emph{Initial condition for a finite-time fluctuation theorem}} 

The rate matrix (\ref{ratedec4app}) considered in Chapter \ref{Chapter4} can be decomposed as
\be 
{\bf W} (\lambda_{1} , \lambda_{2})\equiv {\bf W}_{{\rm L}} (\lambda_{1} , \lambda_{2}) +  {\bf W}_{\rm R} 
\ee 
with both components given in (\ref{ratedec4app1}) and (\ref{ratedec4app2}). The right component $ {\bf W}_{\rm R} $ obeys the symmetry relation
\be
\mbox{\bf M}^{-1} \cdot \mbox{\bf W}_{\rm R} \cdot \mbox{\bf M}
= \mbox{\bf W}_{\rm R}^{\top}
\ee
with the matrix ${\bf M}$ given in (\ref{Msym}). By using this symmetry and recalling definition (\ref{idealinitial}) we can write
\bea
{\bf W}_{{\rm R}} \cdot {\bf p}^{st}_{0}  & = & \left( \mbox{Tr} \left\{ {\bf M} \right\} \right)^{-1}  {\bf W}_{{\rm R}} \cdot {\bf M} \cdot {\bf 1} \\
& = & \left( \mbox{Tr} \left\{ {\bf M} \right\} \right)^{-1}  {\bf M} \cdot  {\bf W}_{{\rm R}}^{\top} \cdot {\bf 1} \\
& = & 0
\eea
since the rows of the rate matrix sum up to zero $ {\bf W}_{{\rm R}}^{\top} \cdot {\bf 1}  = 0$, thus preserving the normalization of probabiliy (this can be directly verified whether from the general expression of the rate matrix given in Section \ref{generalsolutionof} or directly for the particular expression (\ref{ratedec4app2})). A similar derivation leads to the result (\ref{eqap4}) with (\ref{incondcoarse}) as the initial condition for the single-current fluctuation theorem at finite times.

%% file: AppendixC.tex

\chapter{Two-point correlation functions of non-interacting fermions} 

\label{AppendixC} 

\lhead{Appendix C. \emph{Two-point correlation functions of non-interacting fermions}} 

We consider a reservoir in the grand-canonical equilibrium
\be
\rho = \, \frac{ {\rm 
e}^{-\beta(H-\mu N ) }}{\mbox{Tr} {\rm 
e}^{-\beta(H-\mu N )}}
\ee
with its quadratic Hamiltonian given by
\be \label{hamapp}
H = \sum_{k} h_k = \sum_j \epsilon_k c_k^{\dagger} c_k.
\ee
The creation and annihilation operators $c_k^\dagger$ and $c_k$ do satisfy the anti-commutation relations
\be
\left\{  c_k , c_{k'}^\dagger \right\} = \delta_{kk'},
\ee
where $\delta_{kk'}$ is the Kronecker delta symbol.

As a result, the density matrix can be factorized as 
\be
\rho = \prod_k \frac{\mbox{e}^{-\beta (\epsilon_k - \mu) c^{\dagger}_k c_k}}{z_k}
\ee
with
\bea
z_k & = & \mbox{Tr}\left\{ \mbox{e}^{-\beta (\epsilon_k - \mu) c^{\dagger}_k c_k} \right\} \\
& = & \langle 0 | 0 \rangle + \langle k | k \rangle \mbox{e}^{-\beta (\epsilon_k - \mu)} \\
& = & 1+ \mbox{e}^{-\beta (\epsilon_k - \mu)} ,
\eea
where $| 0 \rangle$ and $| k \rangle \equiv c^{\dagger}_k | 0 \rangle$ denote, respectively, the empty and occupied eigenstates of the single-particle Hamiltonian $h_k$.

We note that the average of the creation-annihilation operators do vanish identically since
\bea
\langle c_k^{(\dagger)} \rangle& = & \langle 0 | c_{k}^{(\dagger)} | 0 \rangle + \langle k | c_k^{(\dagger)} | k \rangle \mbox{e}^{-\beta (\epsilon_k -\mu)} \nonumber \\
& = & 0 . \label{vanishingapp}
\eea

Moreover, the creation-annihilation operators evolve in the Heisenberg picture corresponding to the Hamiltonian (\ref{hamapp}) according to
\bea
c_k (t) & = & \mbox{e}^{i H t} \, c_k \, \mbox{e}^{-i H t} = \mbox{e}^{-i \epsilon_k t} c_k , \\
c^{\dagger}_k (t) & = & \mbox{e}^{i H t} \, c_k^{\dagger} \, \mbox{e}^{-i H t} = \mbox{e}^{i \epsilon_k t} c_k^{\dagger}.
\eea

We define the lesser Green functions as
\be
G^{<}_{kk'} (\tau) \equiv \langle c_k^{\dagger} (\tau) c_{k'}  \rangle .
\ee
For $k\neq k'$ this function vanishes since
\bea
\langle c_k^\dagger (\tau) c_{k'} \rangle & = & \mbox{e}^{i \epsilon_k \tau} \,  \langle c_k^{\dagger} \rangle \langle c_{k'}  \rangle \nonumber \\
& = & 0,
\eea
where we used (\ref{vanishingapp}) in obtaining the last equality.

By using the above results, we thus obtain the lesser Green function as
\bea
G^{<}_{kk'} (\tau) & = & \mbox{e}^{i \epsilon_k\tau} \langle c_k^{\dagger}  c_{k'}  \rangle \delta_{kk'} \nonumber \\
& = & \mbox{e}^{i \epsilon_k \tau} z^{-1}_k \mbox{Tr} \left\{ c_k^{\dagger}  c_k \mbox{e}^{-\beta (\epsilon_k - \mu) c_k^\dagger c_k} \right\}   \delta_{kk'} \nonumber \\
& = & \mbox{e}^{i \epsilon_k\tau} z^{-1}_k \frac{d}{d\lambda} \left. \left( \mbox{Tr} \left\{  \mbox{e}^{\lambda \, c_k^\dagger c_k} \right\}   \right) \right|_{\lambda = -\beta (\epsilon_k - \mu)}  \delta_{kk'} \nonumber \\
& = & \mbox{e}^{i \epsilon_k \tau} z^{-1}_k \frac{d}{d\lambda} \left. \left( 1+\mbox{e}^{\lambda}   \right) \right|_{\lambda = -\beta (\epsilon_k - \mu)}  \delta_{kk'} \nonumber \\
G^{<}_{kk'} (\tau) & = &  \mbox{e}^{i \epsilon_k \tau}  f(\epsilon_k) \delta_{kk'} \label{lessapp},
\eea
where the Fermi-Dirac distribution in the reservoir is given by
\be
f(\epsilon_k) = \frac{1}{1+\mbox{e}^{\beta (\epsilon_k - \mu)}}.
\ee

The greater Green function
\be
G^{>}_{kk'} (\tau) \equiv \langle c_{k'} (\tau) c_k^{\dagger}   \rangle
\ee
is deduced from (\ref{lessapp}) by using the fact that
\be
\langle c_k^{\dagger} c_{k'} \rangle= \delta_{kk'} - \langle c_{k'}  c_k^{\dagger}\rangle,
\ee
so that
\be
G^{>}_{kk'} (\tau) = \mbox{e}^{-i \epsilon_k \tau} (1 - f(\epsilon_k) ) \delta_{kk'} .
\ee

By defining the Fourier transforms
\bea
\hat G^{<}_{kk'} (\omega) & = &  \int \frac{d\tau}{2 \pi } \, \mbox{e}^{-i \omega \tau} \, G^{<}_{kk'} (\tau)  , \\
\hat G^{>}_{kk'} (\omega) & =  & \int \frac{d\tau}{2 \pi }  \, \mbox{e}^{i \omega \tau} \,  G^{>}_{kk'} \tau  ,
\eea
we also get
\bea
\hat G^{<}_{kk'} (\omega) & = & \delta (\omega - \epsilon_k) \, \delta_{kk'} \, f(\epsilon_k) ,\\
\hat G^{>}_{kk'} (\omega)&   = & \delta (\omega - \epsilon_k) \, \delta_{kk'} \, (1 -  f(\epsilon_k) ) ,
\eea
which are directly used to calculate the charging and discharing rates of Chapters \ref{Chapter4} to \ref{Chapter6}.

%% file: AppendixD.tex

\chapter{Inequalities deduced from the fluctuation theorems} 

\label{AppendixD} 

\lhead{Appendix D. \emph{Inequalities deduced from the fluctuation theorems}} 

Here, the inequalities (\ref{inequal}) and (\ref{inequal2}) are proved using Jensen's inequality according to which
\be
\langle f(X)\rangle \geq f(\langle X\rangle)
\ee
for any convex function $f(X)$ and any statistical average $\langle\cdot\rangle$ over the probability distribution of the random variables $X$ \cite{cover2012elements}.  The convex function is here taken as $f(X)=\exp X$.

For $X=-\tilde A \, n$ and the statistical average $\langle\cdot\rangle_t=\sum_{n} p(n,t)(\cdot)$ over the probability distribution of the number $n$ of electrons transferred in the DQD, we find
\be
\langle {\rm e}^{-\tilde A \, n}\rangle_t  \geq  {\rm e}^{-\tilde A \langle n\rangle_t}.
\ee
By the univariate fluctuation theorem (\ref{FT1}), we have that
\bea
\langle {\rm e}^{-\tilde A \, n}\rangle_t  &=& \sum_n p(n,t) \, {\rm e}^{-\tilde A \, n} \nonumber\\
&\simeq& \sum_n p(-n,t) = \sum_n p(n,t)=1
\eea
hence the inequality (\ref{inequal}).

The other inequality (\ref{inequal2}) results from the bivariate fluctuation theorem of fundamental origin
\be
\frac{p(n,n_{\rm C},t)}{p(-n,-n_{\rm C},t)}\simeq {\rm e}^{A_{\rm D} n+A_{\rm C}n_{\rm C}} \qquad\mbox{for}\quad t\to\infty ,
\label{FT2}
\ee
where $n=n_{\rm D}$ is the number of electrons transferred during the time interval $[0,t]$ in the DQD and $n_{\rm C}$ in the QPC, while $A_{\rm D}$ and $A_{\rm C}$ are the basic affinities (\ref{A_D})-(\ref{A_C}) of both circuits.

Since the univariate fluctuation theorem (\ref{FT1}) is here supposed to hold jointly with the bivariate theorem (\ref{FT2}), we get
\be
\sum_{n_{\rm C}} {\rm e}^{-A_{\rm D} n-A_{\rm C}n_{\rm C}} p(n,n_{\rm C},t) \simeq \sum_{n_{\rm C}} p(-n,-n_{\rm C},t)= p(-n,t) \simeq  {\rm e}^{-\tilde A \, n} p(n,t)
\ee
after summing only over $n_{\rm C}$.  Multiplying by $\exp(\tilde A \, n)$ and summing also over $n$, we find
\be
\sum_{n,n_{\rm C}} {\rm e}^{(\tilde A-A_{\rm D}) n-A_{\rm C}n_{\rm C}} p(n,n_{\rm C},t) \simeq \sum_n p(n,t)= 1 .
\label{eq_F1}
\ee  
Jensen's inequality with $X=(\tilde A-A_{\rm D}) n-A_{\rm C}n_{\rm C}$ and the statistical average over the probability distribution $p(n,n_{\rm C},t)$ reads
\be
\langle {\rm e}^{(\tilde A-A_{\rm D}) n-A_{\rm C}n_{\rm C}}\rangle_t  \geq  {\rm e}^{(\tilde A-A_{\rm D})  \langle n\rangle_t-A_{\rm C} \langle n_{\rm C}\rangle_t}.
\ee
Since $\langle {\rm e}^{(\tilde A-A_{\rm D}) n-A_{\rm C}n_{\rm C}}\rangle_t \simeq 1$ by Eq.~(\ref{eq_F1}), we obtain the inequality
\be
A_{\rm D}  \langle n\rangle_t + A_{\rm C} \langle n_{\rm C}\rangle_t \geq  \tilde A \langle n\rangle_t
\ee
from which Eq.~(\ref{inequal2}) is deduced after dividing by the time interval $t$ and taking the limit $t\to\infty$.